\documentclass[a4paper,10pt,twocolumn,useAMS,usenatbib]{mn2e}

\usepackage{graphics}
\usepackage{psfig}
\usepackage{amsmath}
\usepackage{amssymb}
\usepackage{upgreek}
\usepackage{longtable}
\usepackage[mathscr]{eucal}
\usepackage[dvips]{graphicx}
\usepackage{longtable}

\newcommand{\aap}{A\&A}
\newcommand{\aaps}{A\&AS}
\newcommand{\aj}{AJ}
\newcommand{\apj}{ApJ}
\newcommand{\apjl}{ApJL}
\newcommand{\apjs}{ApJS}
\newcommand{\apss}{ApSS}
\newcommand{\araa}{ArA\&A}
\newcommand{\mnras}{MNRAS}
\newcommand{\pasp}{PASP}

\newcommand{\pasj}{PASJ}
\newcommand{\nat}{Nature}

\begin{document}
\pagestyle{plain}
\pagenumbering{arabic}

\twocolumn[
{\centering{\Huge OB associations and their origins \par}\vspace{0.7cm}
{\Large Nicholas J. Wright\\}
\vspace{0.2cm}
{\normalsize Astrophysics Group, Keele University, Keele, ST5 5BG, UK\\}
\vspace{0.7cm}
OB associations are unbound groups of young stars made prominent by their bright OB members, and have long been thought to be the expanded remnants of dense star clusters. They have been important in astrophysics for over a century thanks to their luminous massive stars, though their low-mass members have not been well studied until the last couple of decades. This has changed thanks to data from X-ray observations, spectroscopic surveys and astrometry from {\it Gaia} that allows their full stellar content to be identified and their dynamics to be studied, which in turn is leading to changes in our understanding of these systems and their origins, with the old picture of \citet{blaa64} now being superseded. It is clear now that OB associations have considerably more substructure than once envisioned, both spatially, kinematically and temporally. These changes have implications for the star formation process, the formation and evolution of planetary systems, and the build-up of stellar populations across galaxies. \newline \vspace{0.25cm} \newline {\bf Keywords:} OB associations, star clusters, young stars, star formation, stellar kinematics and dynamics. \vspace{1.5cm} }]

\section{Introduction}

OB associations are gravitationally unbound groups of young stars, typically containing many prominent OB stars as well as numerous low-mass stars. Their low space densities ($< 0.1$~M$_\odot$ pc$^{-3}$) make them dynamically unstable to Galactic tidal forces and therefore over time they should disperse. The fact that they exhibit some spatial and kinematic concentration (despite being unbound) and contain short-lived OB stars implies that they must be young, a fact first realised by \citet{amba47}, which provided the first evidence that star formation was still ongoing in the Galaxy. OB associations continue to be a valuable tracer of the distribution of young stars and the star formation process. Their dimensions can range from a few to a few hundred parsecs, while their total stellar masses cover a thousand to tens of thousands of solar masses (systems less massive than this typically don't contain any OB stars and are thus known as T associations after the T-Tauri stars that are often used to identify them).

OB associations are important objects to study because they represent a transitional phase between the birth environment of stars (star forming regions or star clusters) and the field star population in galaxies. They are therefore useful for studying the star formation process (as they probe the spatial and kinematic configuration of stars at birth) and for understanding how and why young star clusters disperse and how the field star population is built up. There are debates as to whether most stars form in dense clusters that are then disrupted and briefly visible as dispersing OB associations \citep{lada03,krou11} or whether a large fraction of stars are born in unbound, low-density groups like OB associations \citep{mill78,krui12}.

Historically OB associations were vital for identifying groups of OB stars and calibrating their luminosity scale \citep[e.g.,][]{morg53,hump78}. In addition OB associations provide large samples of unobscured young stars that are useful for studies of the initial mass function, the frequency and properties of multiple systems, protoplanetary disks and planetary systems \citep[e.g.,][]{mass95,kouw07,kala15b}. They also allow us to trace star formation over large areas and timescales and thus study the propagation of star formation and the role of feedback in triggering or halting star formation.

\subsection{Historical Summary}

The term ``association'' was coined by \citet{amba47} to describe the low-density groups of O- and B-type stars that had been known about for many years \citep[e.g.,][]{kapt14,eddi14,pann29}, placing them alongside open and globular clusters as a new type of stellar grouping. The term association would later be separated in OB- and T-associations based on their most-prominent members \citep{amba68}. At the time it was thought that only OB stars formed (and existed within) OB associations, with T-Tauri stars restricted to T-associations and open clusters. This view was challenged by \citet{mill78} who argued that unless OB associations also included low-mass stars then there were insufficient numbers of birth sites for low-mass stars to explain the observed field star population.

Concerning the stability of OB associations, \citet{bok34} had already shown that low-density systems (space densities $< 0.1$ M$_\odot$ pc$^{-3}$) were unstable against disruption by galactic tidal forces, which then led \citet{amba47} to realise that this meant such system must be young and in the process of expanding. This implied that OB associations and the stars within them had only recently formed, which provided one of the first pieces of evidence that stars were still forming within the galaxy today, long before the discovery of molecular clouds \citep{amba47}.

Since OB associations are gravitationally unbound, it was reasonable to assume that they were in the process of expanding. \citet{amba49} estimated a typical expansion velocity of $\sim$5~km~s$^{-1}$ for OB associations, based on the balance between their initial expansion and the Galactic tidal forces that would be needed to produce the slightly flattened morphology observed in some associations at the time. A similar view was suggested by \citet{blaa56} who argued that the stars in associations ``may have originated within a much smaller volume than that occupied by the group at the present''. Observations showing that the degree of proximity to the interstellar medium decreased with the increasing size of associations, and that the more dispersed parts of associations had older OB stars within them, solidified the expansion model \citep{blaa64}.

The origin of this expansion remained unclear. \citet{opik53} and \citet{oort54} suggested the idea that an unbound group of stars must form from an unbound cloud of gas and thus be born in a state of expansion. An alternative formation model was first put forward by \citet{tutu78} in which a gravitationally bound cluster of stars, embedded within a molecular cloud, becomes unbound and begins to expand when the gas is dispersed by feedback processes. This idea was consistent with the growing observational evidence that many giant molecular clouds were gravitationally bound \citep[e.g.,][]{duer82} and that O-type stars can efficiently disperse the residual gas in molecular clouds \citep[e.g.,][]{whit79}. Later studies, including numerous N-body simulations of the reaction of the stellar system to gas expulsion, reinforced this idea \citep{lada87,good97,krou01,baum07}.

The next major advance in the study of OB associations was brought about by data from the astrometric satellite {\it Hipparcos} \citep{perr97}, thanks in part to the work of the SPECTER consortium to get many thousands of candidate members of known OB associations included in the {\it Hipparcos} input catalogue. These data lead to a landmark census of the nearby OB associations presented by \citet{deze99}, which included significant improvements in the definition, membership, and characterisation of these associations. This work, and others driven by {\it Hipparcos} data, has since lead to many advances in the study of OB associations, including the first studies of the 3D structure of OB associations \citep{debr99}, the identification of runaway OB stars ejected from associations \citep{hoog00}, the discovery of new OB associations \citep{deze99} and studies of Galactic structure derived from OB association kinematics \citep{meln09}.

One of the major limitations of {\it Hipparcos} for the study of OB associations was its depth, confining samples to only the brightest (and most massive) OBA-type members of even the closest associations. Follow-up surveys to identify the less-massive (and more numerous) members of these associations has been possible, but has, in general, been hindered by the large areas over which OB associations span on the sky and the difficulty of separating young, low-mass association members from field stars \citep{brow99,bric07}. In this regard, astrometry from {\it Gaia}, the successor to {\it Hipparcos} has quickly become invaluable for the identification and study of huge numbers of low-, intermediate- and high-mass stars in OB associations. It is against this backdrop that this review is presented, with the study of OB associations at the cusp of a revolution in our understanding of these important systems.

\subsection{Scope of this review}

The purpose of this review is to summarise our current knowledge of the properties and origin of OB associations, focussing mainly on observational results, but considering recent ideas from theory and simulations where relevant. In particular I focus on work from the past two decades since the last major reviews on OB associations \citep{blaa91,brow99}. This period includes work that has built up on the shoulders of the {\it Hipparcos} studies and leads into early work with {\it Gaia}.

This review is limited to OB associations in the Milky Way, because while similar systems have been identified and studied in other galaxies \citep[e.g.,][]{vand64,efre87}, with NGC 604 worthy of note as the prototype extragalactic OB association \citep{maiz04}, the tools used to study these systems and the results derived from their study are necessarily different. For a recent review of these systems we refer the reader to \citet{goul18}.

Section~2 outlines the various methods used to identify OB association members, which are necessarily different and more complex to those used to identify the members of compact star clusters. Section~3 summarises the properties of some of the most prominent and well-studied OB associations in the Solar vicinity, while Section~4 includes a discussion of these properties in a broader sense, taking into account large-scale studies of multiple associations. In Section~5 these results are discussed and their implications for our understanding of the origin, formation and evolution of OB associations considered. Finally, a summary is provided in Section~6 with some concluding remarks.

\section{Observing OB associations and identifying members}

One of the obstacles in studying OB associations is the difficulty of reliably identifying their members. While high-mass stars can be identified just from photometry, young low-mass stars can be difficult to distinguish from the older (and more numerous) field stars. When studying star clusters the density of members on the sky is significantly higher than the background density that members can be reasonably effectively identified simply from photometry, but the stellar density in OB associations is sufficiently low that this is not possible. The fact that the stars in OB associations are typically older than stars in star forming regions means that the use of H$\alpha$ excess emission or infrared excesses to identify members is inefficient and could certainly lead to an age bias in the resulting sample.

The key methods used for the last few decades have included X-ray observations or multi-object spectroscopy, though recently it has been possible to identify members based on the enhanced luminosity of pre-MS stars using {\it Gaia} parallaxes. Even when young stars are identified and confirmed, assigning them into different OB associations or groups is still a challenging task requiring spatial and kinematic information. This section summarises some of the different techniques used to identify young stars, confirm their youth, and assign them into different groups or associations.

\subsection{Identifying young stars}

The first task in identifying members of OB associations is to efficiently select candidate young stars. For high-mass stars this can usually be done simply with photometry, but for less massive stars that can photometrically appear very similar to their older counterparts it requires some indicator of youth such as elevated luminosity, enhanced X-ray emission, or rapid rotation.

\subsubsection{Photometric identification of O and B-type members}

\begin{figure*}
\begin{center}
\includegraphics[width=18cm]{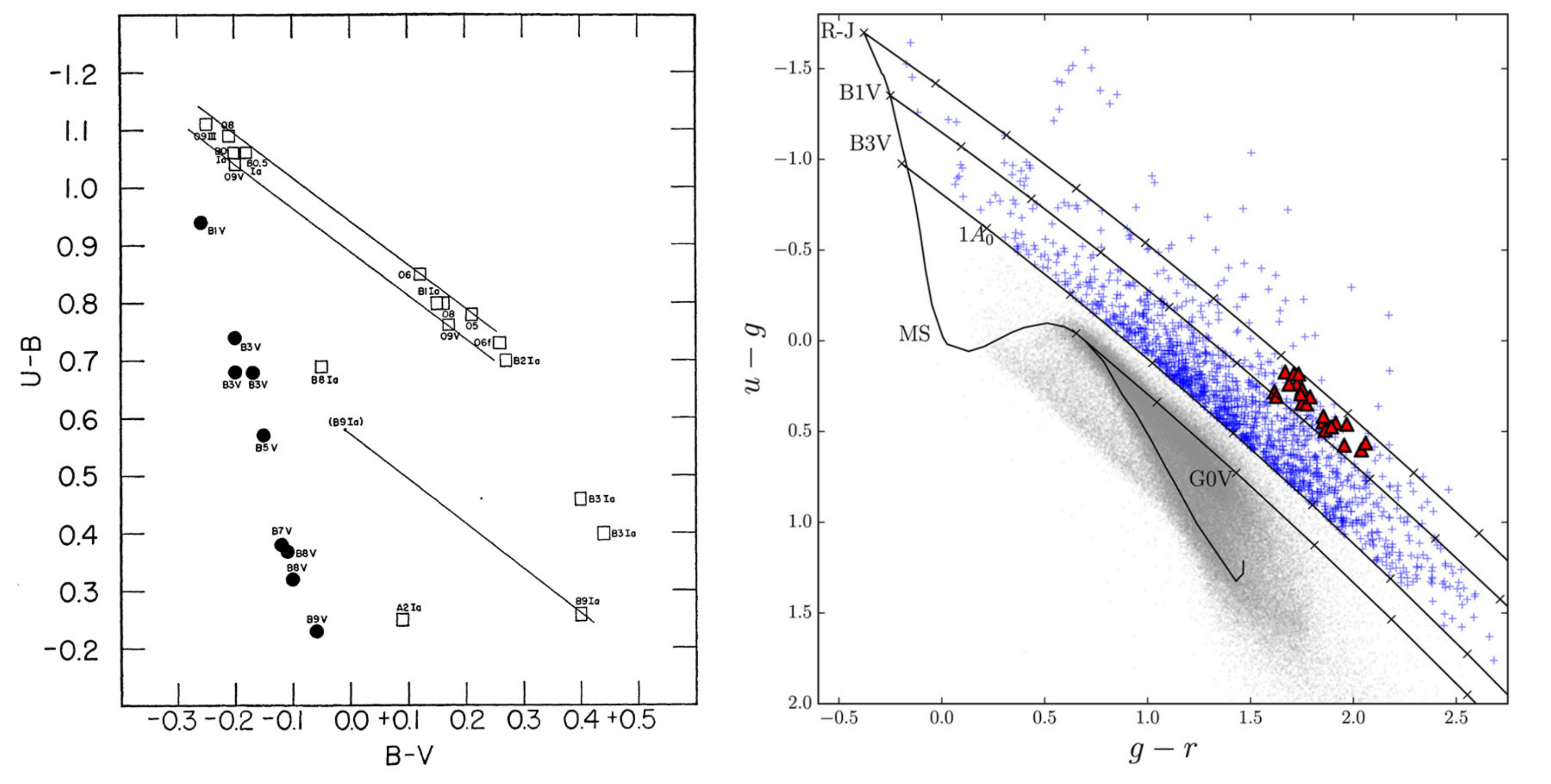}
\caption{Photometric selection of OB stars. The left panel shows the original $(U-B, B-V)$ colour-colour diagram from \citet{john53} with O- and B-type stars shown as open squares and filled circles respectively. An unreddened main sequence can be traced out following the B-type stars while reddening vectors are shown extending to redder colours. The right panel shows a more modern representation of this method in the $(u-g, g-r)$ colour-colour diagram from \citet{mohr15}. The unreddened main sequence can be seen extending to later spectral types, with reddening curves (for $R_V = 3.8$) extending from spectral types B1V, B3V, G0V, and an ideal Rayleigh-Jeans tail spectrum. Grey points show all stars in the field of view, blue crosses show candidate OB stars selected from this method, and red triangles show known OB stars in the field of view.}
\label{OB_selection}
\end{center}
\end{figure*}

The process for photometrically identifying OB stars was developed and pioneered by \citet{john53} in their landmark paper on the $UBV$ photometric system. They showed that field OB stars could be identified (and their reddening determined) using the $Q$ method in the $U-B$ versus $B-V$ colour-colour diagram (see Figure~\ref{OB_selection}). This method is effective because in this diagram early-type stars do not redden onto themselves (or onto late-type stars) and thus trace out a unique region of the colour-colour diagram (in part due to the response of the $U-B$ colour to the strength of the Balmer break in hot stars). The method is quite reliable and contamination rates are low (e.g., \citealt{mohr17} show that when using modern photometry the method is reliable 97\% of the time) and as such validation of photometrically-identified candidates is useful, but not necessary.

This work, combined with the absolute-magnitude calibration of B-type stars \citep{john53}, allowed the identification of OB stars in open clusters and associations and the determination of distances to these systems. This allowed numerous authors to identify groups of OB stars across the sky and thus the first detailed cataloguing of OB association high-mass members \citep[e.g.,][]{john54,harr56,blaa56,blaa64,rupr66}. These membership lists were instrumental in determining the fundamental properties of these stars \citep[e.g.,][]{hump78} and estimating turn-off ages \citep[e.g.,][]{dege89} - for more details see Section~\ref{s-ages}.

While photometric identification of field OB stars is most effective in the blue part of the optical spectrum, within the Galactic plane high extinction can make this approach difficult. In regions of high extinction identifying OB stars in the near-IR is sometimes necessary. For example, \citet{come02} used a combination of near-IR colour-magnitude and colour-colour diagrams to identify 46 new candidate OB stars in Cygnus OB2, significantly increasing the census of such objects in the association.

The availability of wide-field, deep and multi-band optical photometry across the Galactic plane \citep{drew05,drew14} allowed \citet{mohr15} to extend the $Q$ method of \citet{john53} to the large datasets of the modern survey era. They adapted the standard $Q$ method to the Sloan filter system, selecting OB stars in the $u-g$ versus $g-r$ colour-colour diagram (see Figure~\ref{OB_selection}) and used follow-up spectroscopy to confirm their method was successful 97\% of the time \citep{mohr17}. In addition to identifying large numbers of early-type stars across Carina, these authors have also shown how this method can be effectively used to identify new OB associations \citep{mohr17,drew18}.

\subsubsection{X-ray identification of FGKM members}

Young stars are highly luminous X-ray sources, typically hundreds to thousands of times more luminous than older, main-sequence stars \citep{prei05}. For solar and late-type stars these X-rays are emitted from a magnetically-confined plasma at temperatures of millions of Kelvin known as a corona. The corona is heated by the stellar dynamo, which itself is driven by stellar rotation, as evidenced by the strong relationship between rotation and X-ray activity \citep{wrig11b}. Since stars spin down as they age they also gradually become less X-ray luminous, with X-ray to bolometric luminosity ratios dropping from $\sim$10$^{-3}$ for pre-main sequence stars to $10^{-8}$--$10^{-4}$ for older main-sequence stars \citep[e.g.,][]{feig04,wrig10b}. This significant X-ray luminosity difference between young stars and older field stars makes X-rays a powerful tool for identifying young stars.

Observations from the {\it Einstein Observatory} and the ROSAT All-Sky Survey uncovered many thousands of X-ray sources associated with young stars in star forming regions and associations \citep[e.g.,][]{walt88,neuh95,prei98,webb99,nayl99,pozz00}, though the limited sensitivity and low spatial resolution prevented it from being used to identify large samples of low-mass association members. These limitations were overcome following the launch of the {\it Chandra} and XMM-Newton X-ray observatories, but the small fields of view and high demand for observing time with these facilities limited their application to the wide areas needed for fully observing most OB associations \citep[though compact observations of parts of OB associations were still possible, e.g.,][]{argi06,getm06,wrig09a}. For more distant and compact OB associations, such as Cyg~OB2, tiled surveys do facilitate coverage of the entire association \citep[e.g.,][]{wrig14c}. Upcoming all-sky data from eROSITA \citep{merl20} will allow members of nearby associations to be identified, though this will mostly be limited to the brighter solar-type stars in the closest ($d < 500$~pc) associations.

While X-ray observations are very efficient for identifying young stars, there are also many other types of object that are luminous in X-rays and that can act as contaminants in studies of young clusters or associations. These can include foreground main-sequence stars, background giants, X-ray binaries and cataclysmic variables, as well as extragalactic sources such as quasars. The frequency of these different contaminants can depend on the sightline towards the region targeted as well as the background galactic population and any obscuring material that might limit the level of extragalactic contamination. \citet{getm11} show that many of these contaminants can be identified either from their significantly harder X-ray spectra or from the lack of an optical counterpart, particularly for extragalactic contaminants. However spectroscopic follow-up is still useful for identifying stellar contaminants such as foreground or background sources.

\subsubsection{Luminosity identification of association members}
\label{s-astrometric_young_stars}

\begin{figure}
\begin{center}
\includegraphics[width=7.2cm,angle=270]{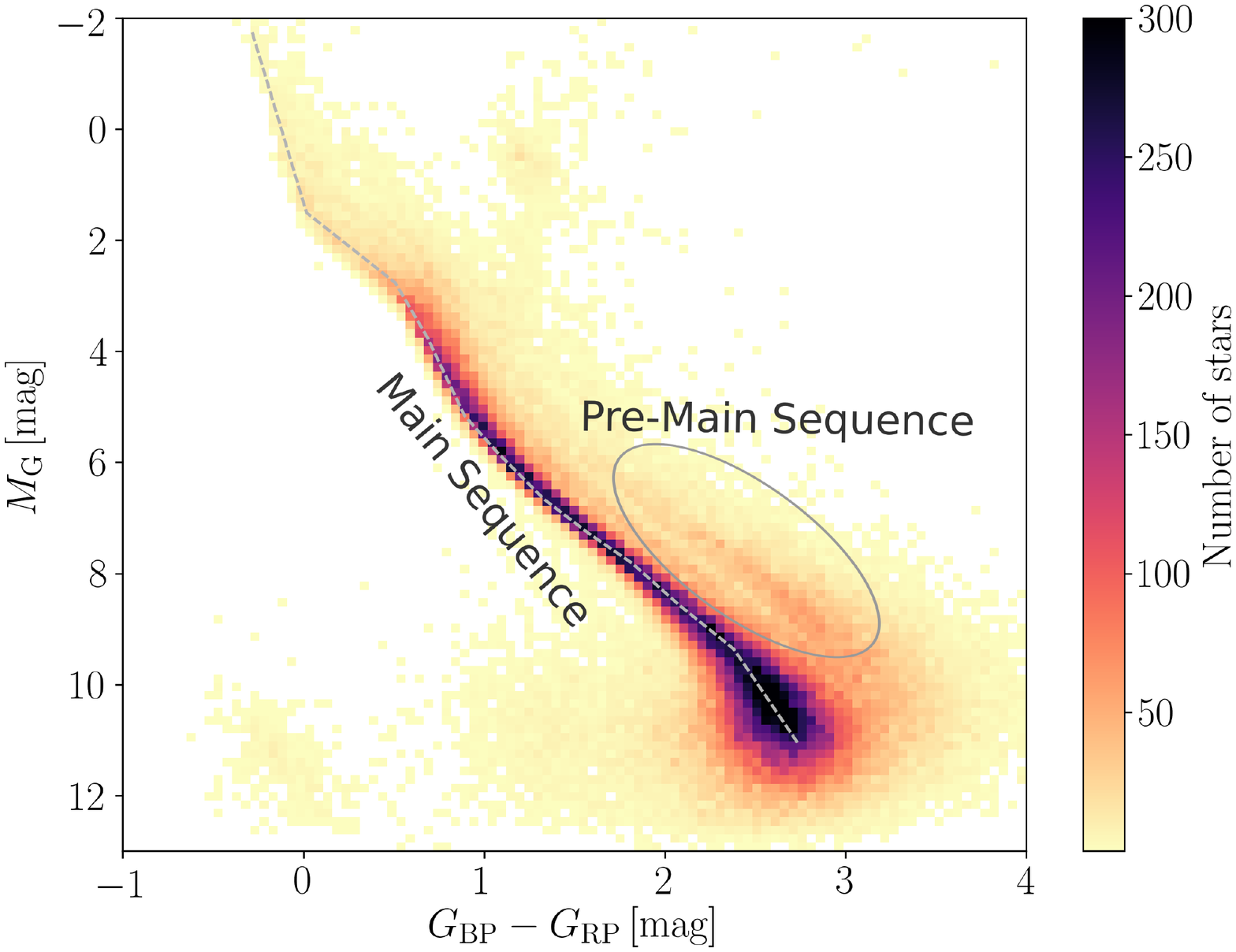}
\caption{Luminosity identification of young stars in the $M_G$ vs. $G_\mathrm{BP} - G_\mathrm{RP}$ absolute colour-magnitude diagram. An unreddened main sequence is visible across the diagram with a pre-main sequence clearly visible above it. Figure adapted from \citet{zari19}.}
\label{astrometric_selection}
\end{center}
\end{figure}

Pre-main sequence stars are more luminous than main-sequence stars and therefore in a colour -- absolute magnitude diagram they stand out above both the single-star and binary star sequences (see Figure~\ref{astrometric_selection}), allowing them to be identified and separated. Until recently this method was not easy to utilise due to the dearth of parallaxes for pre-MS stars \citep[though it could be used for individual star clusters where the distance was known, e.g.,][]{hern14}, but following {\it Gaia}'s second data release in April 2018 this approach can be used to identify most young stars ($<$10--20~Myrs, dependent on colour) with reasonably high-precision parallaxes (roughly within 500~pc for {\it Gaia} DR2 data).

\citet{zari18} used this method with the {\it Gaia} $M_G$ vs. $G_{BP} - G_{RP}$ diagram (Figure~\ref{astrometric_selection}) to identify stars younger than 20~Myrs within 500~pc of the Sun, mapping the 3D distribution of the known YSO populations in the solar neighbourhood. \citet{dami19} combined this technique with a proper motion selection to map out the stellar population of the Scorpius-Centaurus OB association. This approach can reduce the level of contamination, particularly for groups of stars with distinct proper motions relative to the field star population (such as Sco-Cen), though it could also introduce a kinematic bias that should be considered if the sample is to be used for any structural or kinematic study. Recently \citet{zari19} used this method to select members of the Orion OB1 association and map out its 3D structure.

This method does require a careful correction for interstellar extinction to produce an accurate colour -- absolute magnitude diagram \citep[e.g.,][]{zari18}, and can introduce biases to the sample, such as an age--mass bias (since low-mass pre-MS stars are more over-luminous at a given age than more massive stars, and therefore easier to identify) and both distance and magnitude biases (since a cut is often placed on the relative parallax error, $\sigma_\varpi / \varpi$, which typically increases with increasing distance and apparent magnitude). There is also the risk of contamination from binaries or from field stars with uncertain parallaxes, though such contamination is expected to be minor and spectroscopic verification may only be necessary for pre-MS intermediate-mass stars that are less offset from the main sequence than their lower-mass siblings. Despite this, the all-sky availability of {\it Gaia} data and the improved precision expected from future data releases suggests this method is likely to be the most effective for stars within (at least) 1~kpc in the future.

\subsubsection{Other methods}

In addition to the methods discussed above, there are a number of other techniques that can or have been used to identify young stars in OB associations. Historically one of the most common methods has been the use of objective prism surveys or H$\alpha$ photometry to identify young stars from the excess H$\alpha$ emission due to accretion or chromospheric activity \citep[e.g.,][]{liu81,wira89,mika01}. The recent H$\alpha$ Galactic plane surveys IPHAS and VPHAS+ \citep{drew05,drew14} have facilitated deep, large-area surveys for young stars with the H$\alpha$ excess method, with recent applications in Cyg OB2 and Cep OB2 \citep{vink08,bare11}. However, this method has a strong bias towards younger Classical T-Tauri stars, relative to older Weak-lined T-Tauri stars, and therefore carries with it an age bias. There is also the risk of contamination from foreground dMe stars \citep{bric01} and therefore spectroscopic verification can be useful.

Infrared surveys are also effective for identifying young stars. {\it Spitzer Space Telescope} photometry has been used to identify many thousands of young stars in star forming regions and clusters \citep[e.g.,][]{povi13}, though the limited time available hindered its application to wide-field studies of OB associations outside of the Galactic plane or major star-forming complexes \citep[e.g.,][]{guar13}. The availability of all-sky infrared images from the {\it Wide-Field Infrared Survey Explorer} (WISE) has facilitated large-area surveys for young stars in OB associations \citep[e.g.,][]{azim15,fisc16}. Contamination from star-forming galaxies, asymptotic giant branch stars and shock emission regions is possible however, though cuts in magnitude or colour space can be used to reduce the contamination rates to $\sim$10-20\% \citep[e.g.,][]{fisc16}. Spectroscopic verification is therefore useful.

Another method is to identify young stars based on their rapid rotation, since stars spin-up as they form and then gradually spin-down over many Gyrs. Young ($\lesssim$50--100~Myrs) stars can therefore be identified from their rapid \citep[$<$3--5~day,][]{patt96} rotation periods. This has become feasible in recent years due to the availability of time-domain, high-precision photometric surveys designed to detect transiting extrasolar planets, e.g., {\it Kepler}, SuperWASP, and very recently, TESS (e.g., \citealt{bink15} used this method to identify several thousand rapidly-rotating young FGK stars from SuperWASP data). Contamination from tidally-locked binaries can be significant however, and so spectroscopic follow-up to confirm the youth of the stars is necessary.

Multi-epoch photometry can also be used to identify young stars from their inherent variability \citep[e.g.,][]{carp01}, though \citet{bric09} find this method needs spectroscopic verification of the members and has about a 50\% success rate. Certain types of stars can be identified from single-epoch photometry, for example \citet{drew08} used H$\alpha$ photometry to identify young A-type stars and \citet{dami18} used optical and near-IR photometry to identify pre-main sequence M-type stars. Despite this, single epoch photometry is generally used to measure the surface density of young stars above the galactic background level, rather than actually identifying individual young stars \citep[e.g.,][]{alve12,zari17,arms18}.

\subsection{Confirming the youth of stars}

Once candidate young stars have been identified it is often necessary to confirm their youth using other means. None of the methods mentioned above are perfect and can lead to contamination from other objects, either stellar or extra-galactic. Candidate young stars can be verified by obtaining multiple indicators of youth from those listed above, particularly if the contaminants of each method are mutually exclusive (e.g., most contaminants to infrared surveys for young stars are likely to be in the background and so parallax measurements can be effective to confirm such sources). However the most effective approach is to obtain spectroscopy and use one or more spectroscopic indicators of youth.

\subsubsection{Lithium in young stars}

\begin{figure}
\begin{center}
\includegraphics[width=8.5cm]{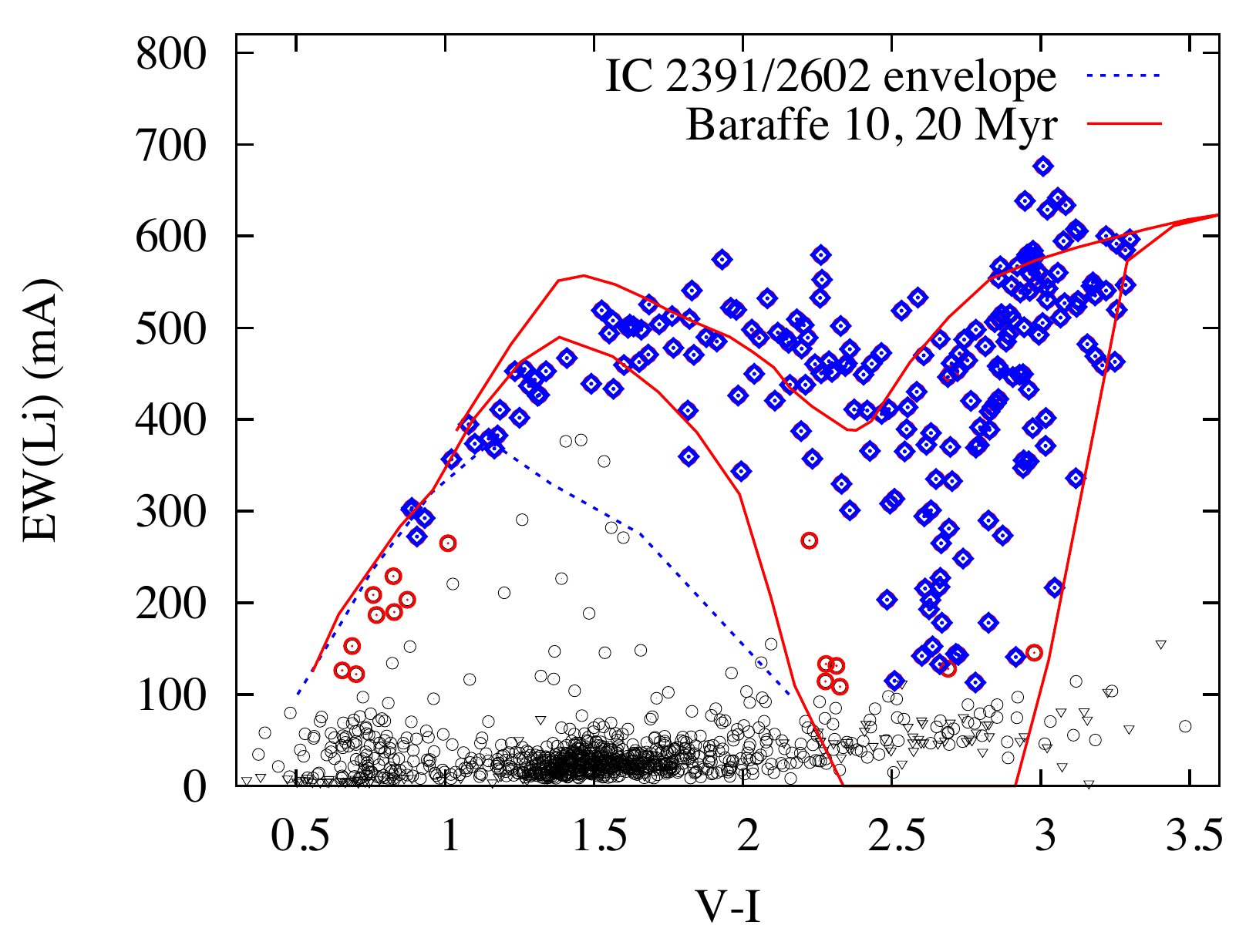}
\caption{Equivalent width of the Li 6708 \AA\ feature versus $V-I$ colour for stars towards the Gamma Velorum cluster in Vela OB2 from \citet{jeff14}. The solid red lines show predictions from the models of \citet{bara98} at 10 (upper) and 20 (lower) Myr, while the dashed blue line shows the upper envelope for EW(Li) for IC~2391 and IC~2602 at $\simeq$50~Myr. Objects whose lithium content is consistent with being a young star are shown in blue and red, with all other stars in the field of view shown as black circles.}
\label{lithium_youth_confirmation}
\end{center}
\end{figure}

Lithium is probably the most effective indicator of youth in K and early M-type stars because it is depleted during their pre-main sequence phase \citep[e.g.,][]{dant94,sode10}, making it a commonly-used spectroscopic indicator of youth \citep[e.g.,][]{prei98,jeff14,peca16}. Lithium is burnt at approximately 3 million K, a temperature that is reached in the core and at the bottom of the convective zone during pre-main sequence evolution. For a short period of time, low-mass stars deplete lithium in their cores whilst mixing Li-depleted material to the surface of the star. This continues until the expansion of the radiative core causes the temperature at the base of the convection zone to drop below 3 million K, bringing lithium depletion to a halt. This point is reached at a later time in lower-mass stars due to their slower pre-MS evolution, and in stars with $M < 0.6$~M$_\odot$ total lithium depletion will eventually occur \citep{bara98}.

The main Li feature in stellar spectra is the (unresolved) doublet at 6708 \AA, which is relatively easy to detect and measure, having an equivalent width of $>$100~m\AA\ for young stars. The strength of this absorption line decreases during the early life of a star, with a timescale for significant depletion of $\sim$10--20~Myrs for mid-M stars, $\sim$100~Myrs for K-type stars, and up to $\sim$1~Gyr for G-type stars. This makes the presence of a strong lithium absorption line in stellar spectra a clear indicator of youth for K- and M-type stars, but less effective for G-type stars. However, since older stars up to the age of the Pleiades can exhibit Li absorption it is necessary to define a threshold equivalent width as a function of colour or mass and use this to select stars \citep[though rotation and other factors may also determine the lithium abundance in stars, e.g.,][]{sode93}. Fortunately, lithium equivalent widths have been measured in many young clusters, providing a valuable calibration for this technique \citep[e.g.,][]{rand97}. Figure~\ref{lithium_youth_confirmation} shows lithium equivalent width as a function of colour for young stars in Vela~OB2 from \citet{jeff14} showing how this can be used to identify young stars.

There are very few types of contaminant for a sample of lithium-rich stars. A small number of field stars may be young enough to have photospheric lithium (1\% for K-type stars and 10\% for G-type stars, based on the depletion timescales), but these objects are rare. Even rarer are lithium-rich field giants, of which $\sim$1\% of G/K giants are expected to have sufficient lithium to reach most detection thresholds for young stars \citep{brow89}.

\subsubsection{Surface gravity indicators}

Another valuable spectroscopic method for confirming the youth of stars is the use of spectral lines that are dependent on surface gravity. Pre-main sequence stars have surface gravities intermediate between those of main-sequence stars and giants. Notable examples of such gravity-sensitive lines include the Na~{\sc i} doublets at 5889, 5896 \AA\ and 8183--8195 \AA\ or the Ca~{\sc i} lines at 6102, 6122 and 6162 \AA. \citet{moha04} present a selection of spectroscopic features in M-type stars that can be used to infer the surface gravity to within 0.25~dex. Some of these features are degenerate with effective temperature and therefore a selection of features are needed to measure surface gravity, possible with sufficiently high-resolution spectroscopy.

\citet{dami14} introduce a gravity-sensitive index $\gamma$ as part of the analysis of spectra from the {\it Gaia}-ESO Survey, based on various sets of gravity-sensitive lines around 6490--6500 and 6760--6775 \AA. With high-resolution spectroscopy the index allows log $g$ to be inferred to a precision of 0.2~dex, sufficient to allow young stars to be separated from giants and main-sequence stars at temperatures $\lesssim$ 5000~K.

\subsubsection{Other spectroscopic indicators of youth}

Other spectroscopic indicators of youth include emission lines due to strong accretion in very young stars, such as He~{\sc i} 5876, 6678 \AA\ or the Ca~{\sc ii} triplet at 8500, 8544, and 8665 \AA\ \citep[e.g.,][]{bric19} or Balmer lines in emission from active late-type stars with strong chromospheric activity \citep[though this is also seen in field stars up to $\sim$1~Gyr old,][]{stau86}.

\subsection{Assigning young stars into groups}

Once a sample of young stars have been identified and confirmed it is often necessary to divide them into groups or identify contaminating young stars. Even within a compact area of the sky there may be young stars that are not part of the group under study, either young field stars or stars from other nearby groups. Stars in clusters or associations have very similar space motions and therefore members of a given group can be identified as a coherent structure in velocity space, for which there are various different methods that have been used over the last few decades that we will discuss here. It is worth noting that the use of proper motions (or parallaxes for nearby systems) to assign stars into groups can introduce a spatial or kinematic bias to the sample, which should be considered when studying it.

\subsubsection{Classical methods not requiring radial velocities}

\begin{figure*}
\begin{center}
\includegraphics[height=10cm]{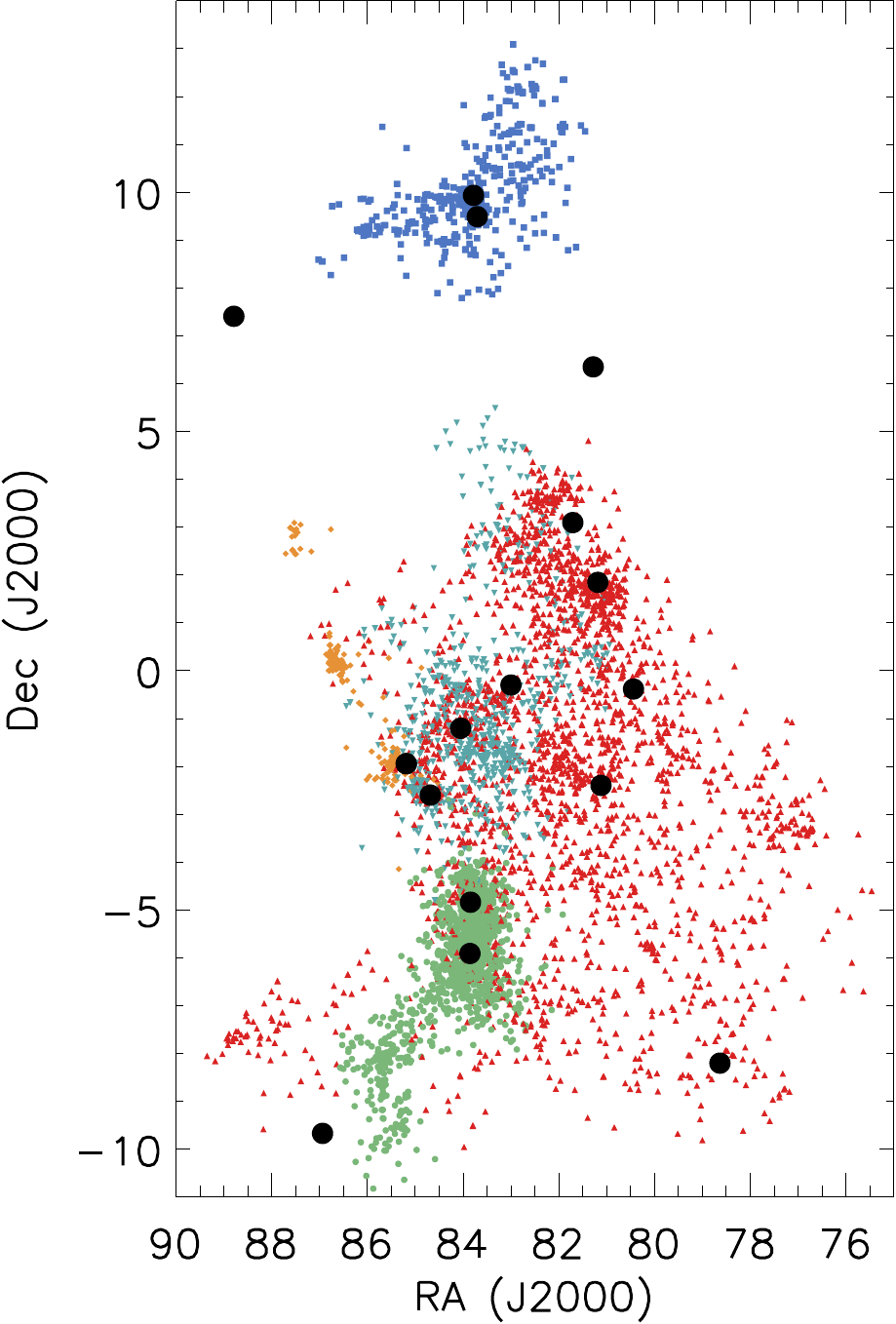}
\includegraphics[height=10cm]{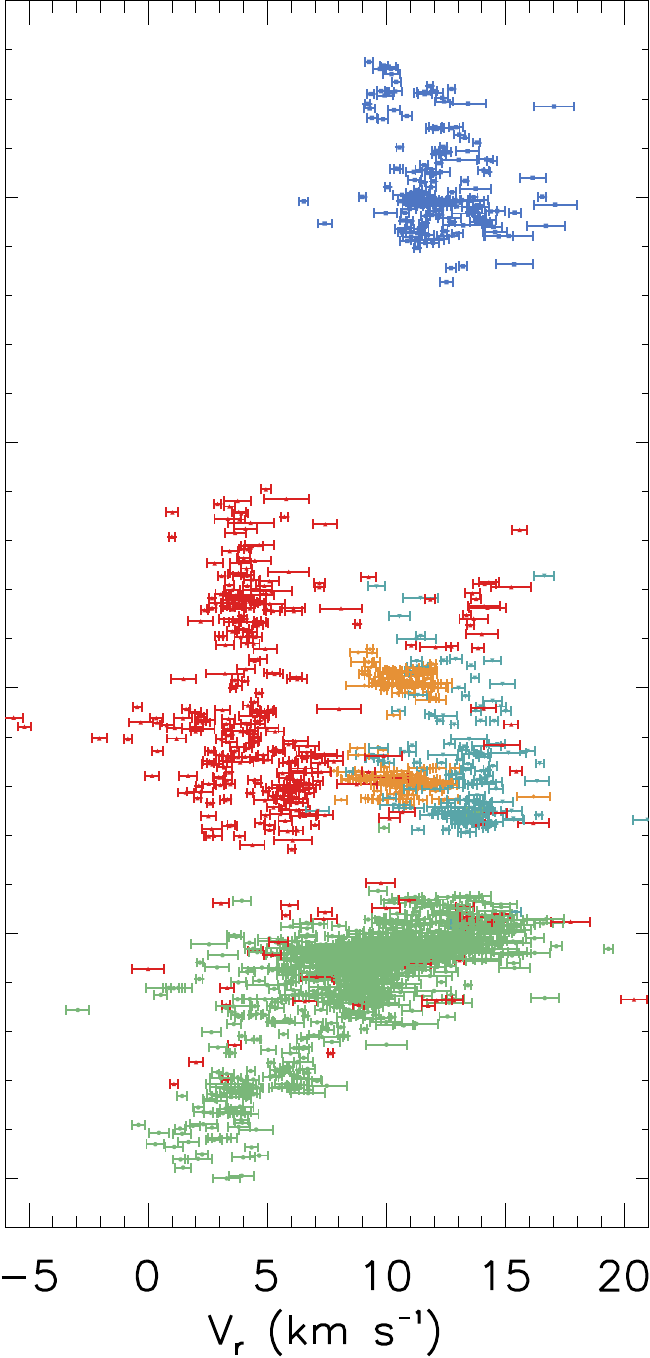}
\includegraphics[height=10cm]{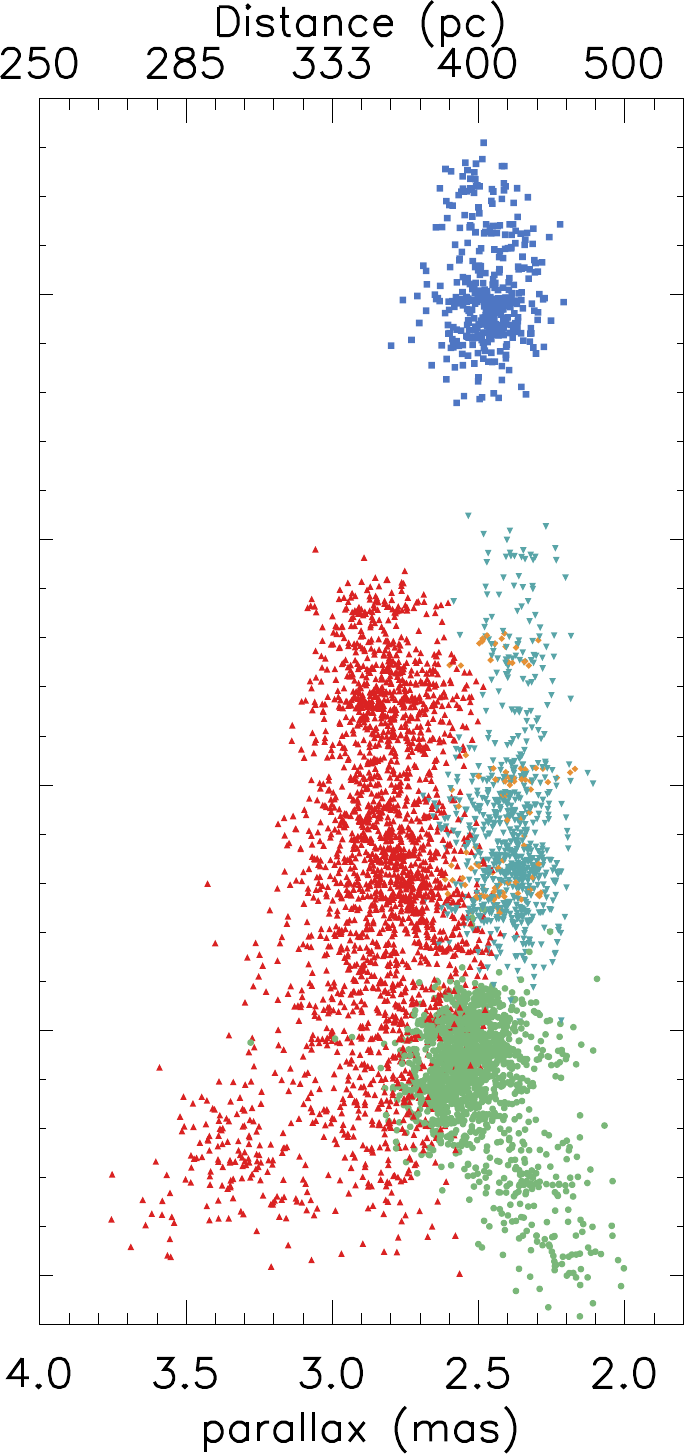}
\caption{Division of young stars in Orion~OB1 into subgroups from \citet{koun18} using hierarchical clustering. The clustering algorithm was applied to 10,248 stars and resulted in the identification of 190 groups across the association, which were then manually combined into five larger structures shown here: Orion A (green; known group), Orion B (orange; known group), Orion C (cyan; new group), Orion D (red; new group) and $\lambda$ Ori (blue; known group). In the left panel, black circles show the position of the major bright stars in Orion with Orion's Belt across the centre of the image.}
\label{orion_kounkel}
\end{center}
\end{figure*}

The classical method for identifying members of a nearby moving group or association is the convergent point method, which is based on the fact that stars with a common space motion will have proper motions that appear to converge from or to a point on the sky \citep[depending on whether the group is moving towards or away from us, e.g.,][]{jone71,debr99b}. The convergent point for a group of stars is found by minimising the proper motion perpendicular to the direction towards the convergent point (accounting for measurement uncertainty and an intrinsic velocity dispersion). New members of the group may be added to the group if their motion perpendicular to the convergent point is below some threshold. This method has been valuable because it only requires proper motions and sky positions, and not parallaxes or radial velocities that are generally harder to obtain. This method was extensively used by \citet{deze99} to perform a census of nearby OB associations using {\it Hipparcos} data. They also used the ``Spaghetti method'' of \citet{hoog99}, which uses parallaxes as well as positions and proper motions, and identifies velocity cylinders or ``spaghetti'' that must overlap in velocity space for all stars that are members of a group.

\subsubsection{Modern methods using clustering algorithms}

The availability of radial velocities for many {\it Hipparcos} stars has allowed the membership of OB associations to be determined in 6-dimensional position and velocity space. This can be achieved using a model for the spatial and kinematic distribution of the group, such as a 6-dimensional Gaussian or a mixture of Gaussians if one or more subgroups are known. \citet{rizz11} applied such a method to re-establish the membership of Sco-Cen using linear models in various dimensions of 6-dimensional space and calculating membership probabilities for individual sources.

The availability of {\it Gaia} data for billions of stars provided 5-dimensional (position, parallax and proper motions) data suitable for the application of more advanced clustering algorithms. These include techniques such as $k$-means clustering (where the user assumes a number of groups, but not their positions, and the data is then iteratively divided into these groups), mixture models (where the data is modelled as a collection of distributions -- usually multi-dimension Gaussians -- and each source is assigned a probability of belonging to a group based on this distribution), and hierarchical clustering (where sources are divided into groups that have source-to-source separations less than some threshold value).  Many of these techniques are {\it unsupervised}, meaning that they do not require pre-existing identification of the clusters and involve a minimum of human intervention. Each of these methods has its own advantages and disadvantages in identifying groups of stars and rejecting non-clustered field stars, including whether or not the number of groups must be assumed a-priori, whether the groups have to be modelled in a certain way (e.g., as a collection of Gaussians), how to determine the dividing line between different groups, and how incomplete data is accounted for (such as the lack of radial velocities for many stars).

A review of the various methods can be found in \citet{feig12}, though we highlight some of them and their recent applications here. \citet{wilk18} apply the Density-Based Spatial Clustering of Applications with Noise algorithm \citep[DBSCAN,][one of the most commonly-used clustering algorithms in the literature]{este96} to {\it Gaia} DR1 data in Upper Sco, increasing the known members by $\sim$50\%. \citet{cant19a} used the classification scheme UPMASK \citep[Unsupervised Photometric Membership Assignment in Clusters,][]{kron14}, which uses principal component analysis and $k$-means clustering to identify groups of stars, finding 7 groups across Vela and Puppis with distinct ages and kinematics.

\onecolumn
\setlength{\LTcapwidth}{7.2in}
\begin{longtable}{lccccccll}
\caption[]{List of known OB associations compiled from \citet{rupr66}, \citet{deze99} and \citet{brow99}. The list is divided into well-studied associations, those that are clearly identified but have not been extensively studied, and those for which very little is known (including some which may not be real associations and are labelled with "?"). Distances and ages are gathered from the literature. Positions are taken from \citet{rupr66} and updated from \citet{deze99} where possible. Distances are taken from the source with the smallest uncertainty, preferring {\it Gaia} measurements or single-association studies where possible. For Sco-Cen and Orion OB1 the subgroups are listed instead of the entire associations due to their prominence (the Sco-Cen subgroups are also known as Sco OB2\_2, OB2\_3 and OB2\_4).}\\
\hline
Name			& l 		& b		& Distance	& Ref.	& Age		& Ref.	& Clusters			& Nebulae	 \\
			& $[$deg.$]$ & $[$deg.$]$ & $[$pc$]$	& 		& $[$Myr$]$	& 		& 				& 		\\
\hline
\endfirsthead
\caption[]{\emph{continued}}\\ 
\hline
Name			& l 		& b		& Distance	& Ref.	& Age		& Ref.	& Clusters			& Nebulae	 \\
			& $[$deg.$]$ & $[$deg.$]$ & $[$pc$]$	& 		& $[$Myr$]$	& 		& 				& 		\\
\hline
\\
\endhead
\\
\hline
\multicolumn{9}{r}{\emph{continued on next page}}\\
\endfoot 
\endlastfoot 
\\
\multicolumn{8}{l}{\emph{Well-studied OB associations:}}\\
\\
Sco-Cen: US		& 351.5	& +20.0	& $143 \pm6$	& W18	& $10 \pm 7$	& P16	& 				& \\
Sco-Cen: UCL 		& 331.0	& +12.5	& $136 \pm5$	& W18	& $16 \pm 7$	& P16	& 				& \\
Sco-Cen: LCC 		& 298.5	& +5.5	& $115 \pm4$	& W18	& $15 \pm 6$	& P16	& 				& \\
Ori OB1a			& 201.0	& -17.3	& $\sim$360	& K18	& 8--12		& B94	& 25~Ori			& \\
Ori OB1b			& 205.0	& -18.0	& 360--420	& K18	& 2--8		& B08	& $\epsilon$~Ori	& \\
Ori OB1c			& 211.3	& -19.5	& $\sim$385	& K18	& 2--6		& B08	& NGC 1980, NGC 1981 &\\
Ori OB1d			& 209.0	& -19.5	& $\sim$380	& Z17	& 1--2		& Z17	& Orion Nebula Cluster & \\
Vela OB2			& 262.8	& -7.7	& $411 \pm12$	& DZ	99	& 10--30		& CG19	& Gamma Vel, P Puppis & \\
Trumpler 10		& 262.8	& +0.7	& $372 \pm23$	& DZ99	& 45--50		& CG19	& 				& \\
Cyg OB2			& 80.2	& +0.8	& 1350--1750	& B19	& 1--7		& W15	&				& \\
\\
\hline
\\
\multicolumn{8}{l}{\emph{High-confidence OB associations that have not been extensively studied:}}\\
\\
Ara OB1			& 338.0	&  0.0	& $\sim$1100	& M73	& $\sim$2		& A87	& NGC 6193		& \\ 
Aur OB1			& 173.1	& -1.6	& $\sim$1060	& M17	& 11-22		& T10	& NGC 1912, NGC 1960 & Sh2-227 \\
Aur OB2			& 173.0	& +0.1	& $\sim$2420	& M17	& $\sim$5.5	& T10	& NGC 1893, Stock 8& IC 410, IC 417\\
Cam OB1			& 142.5	& +2.0	& $\sim$800	& M17	& 7--14		& S85	& NGC 1502		& S202 \\ 
Car OB1			& 286.5	& -0.5	& $2300 \pm50$ & S06	& 1--10		& D01	& NGC 3293, Tr 14--16 & NGC 3372 \\
Car OB2			& 290.4	& +0.1	& $\sim$1830	& M20	& $\sim$4		& G94	& 				& \\
Cas OB6			& 135.9	& +1.3	& $\sim$1750	& M17	& $\sim$4		& T10	& IC 1805, IC 1848	& W3, W4, W5 \\
Cas-Tau 			&169.0	& -16.5	& 125--300	& DZ99	& $\sim$50	& DZ99	& 				& \\ 
Cen OB1			& 304.2	& +1.4	& $\sim$1920	& M17	& 6--12		& K94	& Stock 16		& RCW 75 \\
Cep OB1			& 104.2	& -1.0	& $\sim$2780	& M17	& 1--5		& C11	& NGC 7380		& \\
Cep OB2			& 102.1	& +4.6	& $\sim$730	& M20	& 5			& DZ99	& Tr 37, NGC 7160	& IC 1396 \\ 
Cep OB3			& 110.4	& +3.0	& $\sim$700	& M17	& 5--8		& J96	& Cep OB3b		& S155 \\ 
Cep OB4			& 118.3	& +5.3	& $\sim$660	& M17	& 1--6		& M68	& B59			& S171 \\ 
Cep OB6			& 105.1	& +0.1	& $270 \pm12$	& DZ99	& $\sim$50	& DZ99	&				& \\ 
CMa OB1			& 224.0	& -1.3	& $\sim$1200	& Z20	& 1--10		& S18	& NGC 2353, NGC 2327 & IC 2177, Sh2-296 \\ 
Collinder 121		& 235.7	& -10.0	& $543 \pm23$	& DZ99	& 5			& DZ99	& Collinder 121		& Sh2-306 \\ 
Cyg OB1			& 75.5	& +1.1	& $\sim$1460	& M17	& 6--8		& R08	& NGC 6913, IC 4996 & \\
Cyg OB3			& 72.9	& +1.9	& $\sim$1830	& M17	& 2--12		& R08	& NGC 6871, NGC 6883 & \\
Cyg OB4			& 82.8	& -7.6	& $\sim$800	& M17	& $\sim$8.3	& U01	& 				& \\ 
Cyg OB5			& 67.1	& +2.1	& $\sim$1610	& R66	&			&		& 				& \\
Cyg OB6			& 86.0	& +1.0	& $\sim$1700	& R66	&			&		& 				& \\
Cyg OB7			& 89.0	& 0.0		& $\sim$630	& M17	& 1--13		& U01, W13 & 				& Northern Coalsack \\
Cyg OB8			& 77.9	& +3.4	& $\sim$1830	& M17	& 4--6		& M15	& 				& \\
Cyg OB9			& 77.7	& +1.9	& $\sim$960	& M17	& 2--4		& M15	& 				& Sh2-108 \\
Gem OB1			& 189.1	& +1.0	& $\sim$1210 	& M20	& $\sim$9		& T10	& NGC 2175		& IC 443 \\
Lac OB1			& 96.7	& -17.6	& $368 \pm17$	& DZ99	& 2--25		& C08	& 				& S126 \\
Mon OB1			& 202.1	& +1.0	& $\sim$580	& M17	& 1--10		& F99	& NGC 2264		& NGC 2264, Mon. Ring\\
Mon OB2			& 206.3	& -2.1	& $\sim$1210	& M17	& 2--15		& T76	& NGC 2244		& Rosette Neb.\\
Per OB2			& 159.2	& -17.1	& $296 \pm17$	& DZ99	& 1--10		& A15	& IC 348, NGC 1333	& \\
Per OB3			& 147.0	& -5.5	& $175 \pm3$	& DZ99	& 50			& DZ99	& $\alpha$ Per		& \\
Pup OB1			& 243.5	& +0.3	& $\sim$2010	& M17	& $\sim$4		& H72	& NGC 2467		& S311 \\
Sco OB1			& 343.3	& +1.2	& $1560 \pm 35$ & Y20	& 1--10		& D18	& NGC 6231, Tr 24	& G345.45+1.50, \\
				&		&		&			&		&			&		&				& Gum 55, IC 4628\\
Sct OB2			& 23.1	& -0.4	& $\sim$1600	& M09	& $\sim$6		& S85	& NGC 6705		& \\
Ser OB1			& 16.7	& 0.0		& $\sim$1530	& M17	& 8--13		& T10	& NGC 6611		& M16, M17 \\
Ser OB2			& 18.2	& +1.7	& $\sim$1600	& M17	& $\sim$4.5	& T10	& NGC 6604		& Sh2-54\\
Sgr OB1			& 7.6		& -0.9	& $\sim$1260	& M17	& 5--8		& S85	& NGC 6530, Col. 367 & M20 \\
\\
\multicolumn{8}{l}{\emph{OB associations for which very little is known:}}\\
\\
Aql OB1 (?)		& 37.3	& -0.6	& $\sim$2750	& R66	&			&		& 				& \\
Cam OB3			& 147.0	& +3.0	& $\sim$2650	& M17	& $\sim$11	& T10	& Alicante 1		& \\
Cas OB1			& 124.1	& -1.4	& $\sim$2010	& R66	& $\sim$10	& L86	& 				& \\
Cas OB2			& 112.0	& 0.0		& $\sim$2100	& M17	& $\sim$10	& L86	& 				& NGC 7538, Sh1-57\\
Cas OB4			& 120.1	& -0.3	& $\sim$2300	& M17	& $\sim$8		& L86	& 				& \\
Cas OB5			& 116.2	& -0.5	& $\sim$2010	& M17	& 6--8		& S85	& 				& \\
Cas OB7			& 122.8	& +1.2	& $\sim$2010	& M17	& $\sim$8		& L86	& 				& Sh2-180 \\
Cas OB8			& 129.2	& -1.1	& $\sim$2300	& M17	& $\sim$20	& T10	& NGC 581, NGC 663 & \\
Cas OB9 (?)		& 113.5	& -2.5	& $\sim$800	& R66	&			&		& 				& \\
Cas OB10	 (?)		& 130.8	& -6.3	& $\sim$3800	& R66	&			&		& 				& \\
Cas OB14			& 120.4	& +0.7	& $\sim$880	& M17	& $<$10		& T10	& 				& \\
Cen OB2			& 294.3	& -1.0	& $\sim$2100	& R66	& 3--10		& B14	& IC 2944			& IC 2948 \\
Cep OB5			& 108.5	& -2.8	& $\sim$2090	& H78	& $\sim$10	& L86	& 				& \\
Cir OB1			& 315.5	& -2.8	& $\sim$2010	& M17	&			&		& Pismis 20		& \\
Cor. Aust. (?)		& 0.0		& -18.0	& $\sim$130	& DZ99	&			&		& Coronet cluster	& NGC 6726, NGC 6729\\
Cru OB1 (?)		& 294.9	& -1.1	& $\sim$2010	& M17	& 5--7		& K94	& 				& IC 2944 \\
Mon OB3			& 217.6	& -0.3	& $\sim$2420	& M17	& $\sim$7		& T10	&				& S287 \\
Nor OB1			& 328.0	& -0.9	& $\sim$2780	& M17	&			&		& 				& \\
Ori OB2 (?)		& 192.6	& -11.6	& $\sim$3240	& R66	&			&		& 				& \\
Per OB1			& 134.7	& -3.2	& $\sim$1830	& M17	& 8--11		& M87	& $h$ and $\chi$ Persei & \\
				&		&		&			&		&			&		& (NGC 869 \& 884) & \\
Pup OB2 (?)		& 244.6	& 0.7		& $\sim$3180	& M17	& $\sim$2		& H72	& Ruprecht 20		& \\
Pup OB3			& 254.0	& 0.0		& $\sim$1460	& M17	& $\sim$4		& W63	& 				& RCW 19 \\
Sco OB4 (?)		& 352.4	& +3.4	& $\sim$960	& M17	&			&		& 				& NGC 6334, Sh2-10 \\
Sct OB3			& 17.3	& -0.8	& $\sim$1330	& M17	& $\sim$4.5	& T10	& 				& Sh2-50 \\
Sgr OB4			& 12.2	& -1.0	& $\sim$1920	& M17	& $<$10		& T10	& 				& \\
Sgr OB5			& 359.9	& -1.2	& $\sim$2420	& M20	& 6--12		& S85	& 				& Sh2-15 \\
Sgr OB6			& 14.2	& +1.2	& $\sim$1600	& M17	& $<$10		& T10	& 				& \\
Sgr OB7			& 10.7	& -1.5	& $\sim$1390	& M17	& 4--5		& T10	& 				& \\
Vela OB1			& 265.0	& -0.7	& $\sim$1460	& M17	& $\sim$20	& T79	& NGC 2659		& \\
Vul OB1			& 60.3	& +0.1	& $\sim$1600	& M17 	& 10--16		& T10	& NGC 6823		& NGC 6820 \\
Vul OB2 (?)		& 64.7	& +1.8	& $\sim$4130	& R66	&			&		& 				& \\
Vul OB4			& 60.5	& +0.5	& $\sim$800	& M17	& $\sim$10	& T80	& 				& \\
\\
\hline
\label{list_of_associations} 
\end{longtable}
\vspace{-0.8cm}
\noindent References: A15 \citep{azim15}, A87 \citep{arna87}, B08 \citep{ball08}, B14 \citep{baum14}, B19 \citep{berl19}, B59 \citep{blaa59}, B94 \citep{brow94}, C08 \citep{chen08}, C11 \citep{chen11b}, CG19 \citep{cant19b}, D01 \citep{degi01}, D18 \citep{dami18}, DZ99 \citep{deze99}, F99 \citep{flac99}, G94 \citep{garc94}, H72 \citep{havl72}, H78 \citep{hump78}, J96 \citep{jord96}, K08 \citep{kalt00}, K18 \citep{koun18}, K94 \citep{kalt94}, L86 \citep{lozi86}, M09 \citep{meln09}, M15 \citep{mahy15}, M17 \citep{meln17}, M20 \citep{meln20}, M68 \citep{macc68}, M73 \citep{moff73}, M87 \citep{maed87}, P16 \citep{peca16}, R08 \citep{reip08}, R66 \citep{rupr66}, S06 \citep{smit06}, S18 \citep{sant18}, S85 \citep{schi85}, S97 \citep{sung97}, T10 \citep{tetz10}, T76 \citep{turn76}, T79 \citep{turn79}, T80 \citep{turn80}, U01 \citep{uyan01}, vG84 \citep{vang84}, W13 \citep{wolk13}, W15 \citep{wrig15a}, W18 \citep{wrig18}, W63 \citep{west63}, Y20 \citep{yaly20}, Z17 \citep{zari17}, Z20 \citep{zuck20}.\twocolumn

The rich young stellar population and significant substructure across the Orion OB1 association has lead many authors to apply clustering algorithms to trace its star formation history. \citet{koun18} use a form of hierarchical clustering to identifying clusters in the Orion OB1 association from 6-dimensional spatial and kinematic data (using radial velocities), applying the algorithm to $\sim$10,000 stars and identifying 190 groups that they then manually combine into 5 larger structures (see Figure~\ref{orion_kounkel}). \citet{chen19} used machine learning algorithms to identify stellar groups in Orion using {\it Gaia} DR2 astrometry, finding 21 spatially- and kinematically-coherent groups, only 12 of which were previously known or associated with known clusters. \citet{zari19} used DBSCAN to identify 17 groups in Orion, including foreground populations.

The wealth of new, multi-dimensional data and advanced statistical techniques for identifying clusters will be critical in the future for identifying new OB associations and star clusters, or ascertaining the membership of known groups. However, detailed testing of these techniques in their application to highly complex datasets with correlated uncertainties will be necessary to fully understand their limitations as well as any biases they introduce to the spatial or kinematic study of the resulting groups.

\section{Properties of individual OB associations}
\label{s-individual}

\begin{figure*}
\begin{center}
\vspace{-2cm}
\includegraphics[height=25cm]{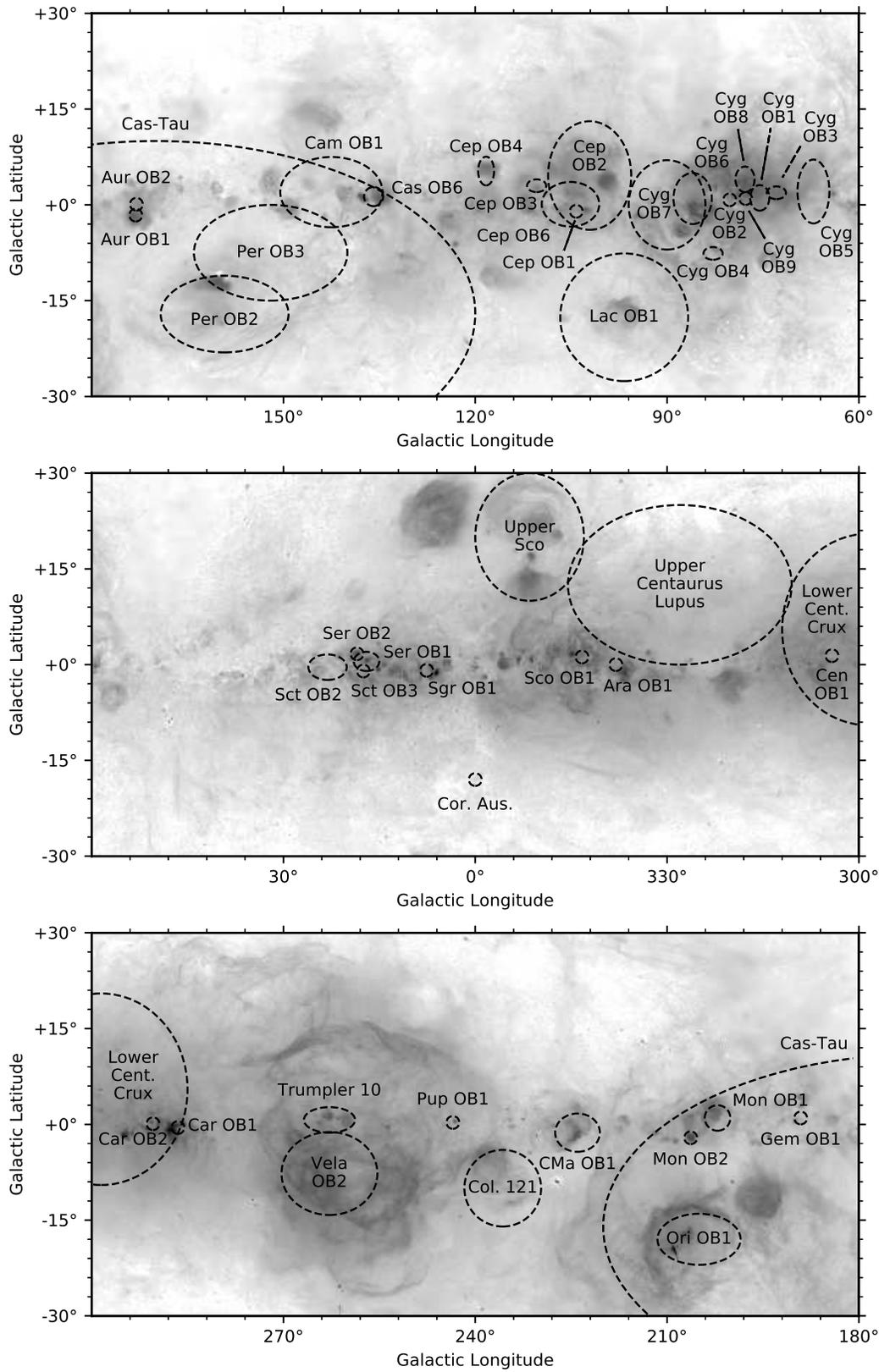}
\vspace{-1.5cm}
\caption{Map of the most well-studied OB associations from Table~\ref{list_of_associations} projected on an inverted Galactic plane H$\alpha$ image \citep{fink03}.}
\label{associations_map}
\end{center}
\end{figure*}

The number of known OB associations runs into the dozens, though the majority of these have been poorly studied. The most well-studied associations are probably Scorpius OB2 (Scorpius-Centaurus, the nearest OB association to the Sun), Orion OB1 (the nearest OB association with active high-mass star formation within it), Vela OB2 (one of the associations with numerous well-studied open clusters), and Cygnus OB2 (which while distant is probably the most massive of the known OB associations). In this section I summarise the properties of these associations, as well as some of the less well studied associations, highlighting their features that can help us understand the properties and formation of OB associations.

The cataloguing of OB associations began in the 1950s by various authors motivated to group OB stars into clusters or associations to better constrain their fundamental properties. By the 1960s various different lists of OB associations existed \citep[e.g.,][]{morg53,schm58} with some studies even beginning to subdivide these into subgroups \citep[e.g.,][]{blaa64}. \citet{rupr66} was the first to gather these different lists into a single list, codifying the nomenclature and defining their boundaries. The previous numbering system using Roman numerals was changed to the one we are familiar with today that uses Arabic numbers affixed to the letters OB and their respective constellation (for example I~Ori became Ori~OB1). Table~\ref{list_of_associations} provides an updated list of known OB associations with coordinates, distances, ages and associated clusters and nebulae gathered from the literature. Figure~\ref{associations_map} provides an observer's view of the distribution of OB associations on the sky in Galactic coordinates.

The high-mass membership of these associations was expanded and refined with improved photometry over the following decades, leading to membership lists that have been used for many decades \citep[e.g.,][]{hump78,garm92}. The availability of {\it Hipparcos} astrometry further refined these membership lists for nearby ($d < 500$~pc) associations \citep{deze99}. At the same time attempts have been made to refine the existing lists of OB associations using either new clustering algorithms \citep{meln95} or improved astrometry for members \citep{deze99}.

\subsection{Scorpius-Centaurus}

\begin{figure*}
\begin{center}
\vspace{-8cm}
\hspace{-1cm} \includegraphics[width=18.5cm]{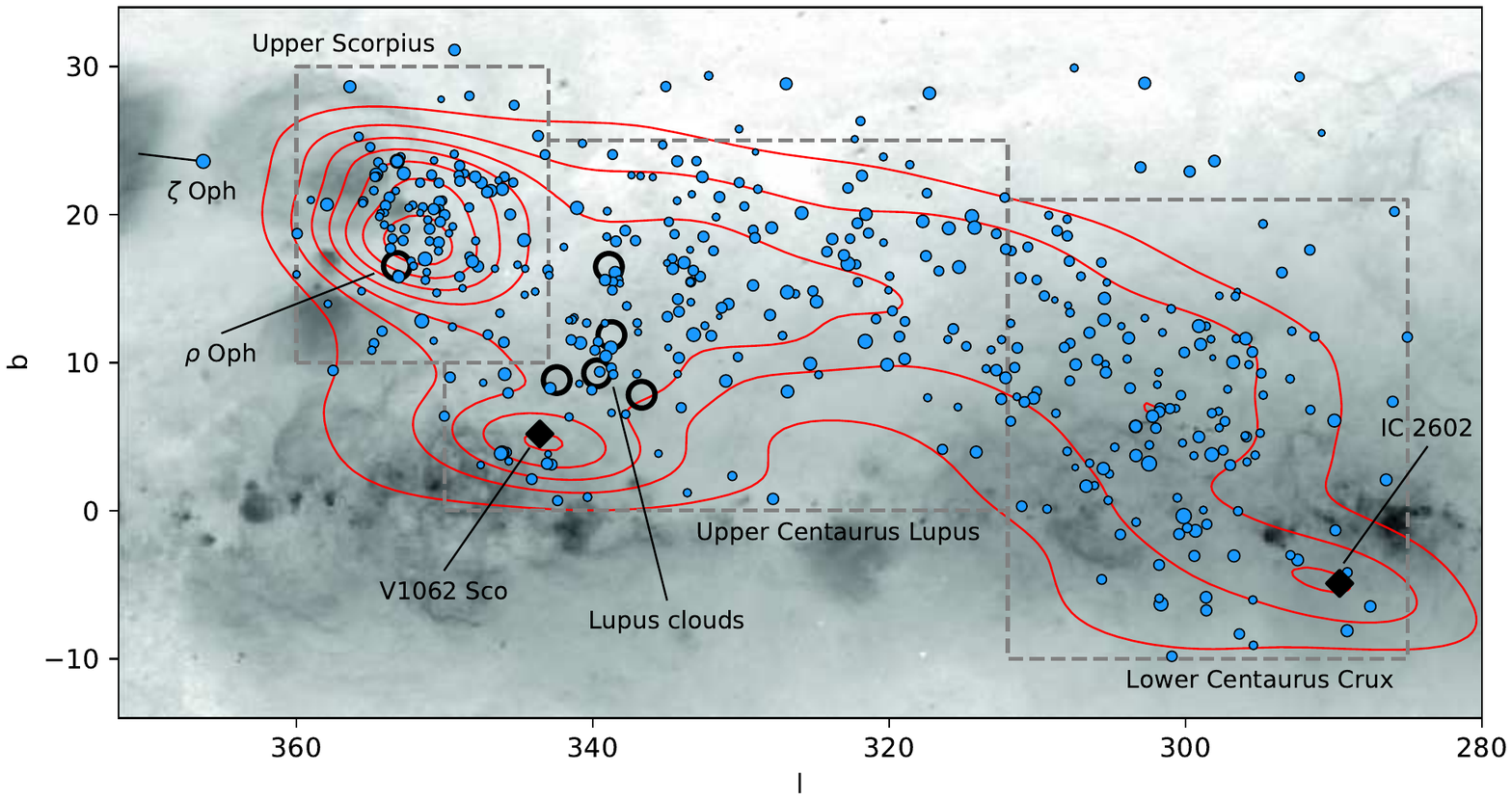}
\vspace{-7.2cm}
\caption{A wide-field view of the Sco-Cen association projected onto an inverted H$\alpha$ image from \citet{fink03}. Blue circles show the classical {\it Hipparcos} members of the association identified by \citet{rizz11}, with symbol size inversely proportional to their $V$ magnitude (brighter stars have larger symbols). The runaway star $\zeta$~Oph is also shown as a blue circle, with a {\it Gaia} DR2 proper motion vector covering 0.5~Myr of motion extending from it. The red contours show the spatial distribution of {\it Gaia}-selected pre-main sequence and upper main sequence stars from \citet{dami19}. The grey dashed rectangles show the historical division \citep{blaa46} of the association into three subgroups from \citet{deze99}.  Black diamonds show the IC~2602 cluster and the V1062~Sco moving group \citep{rose18}, while empty black circles show the locations of the $\rho$~Ophiuchus star forming region and the Lupus I--V clouds.}
\label{scocen_map}
\end{center}
\end{figure*}

The Scorpius-Centaurus (Sco-Cen or Scorpius OB2) association is the nearest OB association to the Sun at a distance of 100--150~pc \citep{deze99}. Due to its proximity and $>$100~pc size the association spans over 90$^\circ$ on the sky, extending over the constellations Scorpius, Lupus, Norma, Centaurus, Circinus, Musca, and Crux. The association contains $\sim$150 B-type stars and (currently) a single O-type star, the runaway O9V star $\zeta$ Ophiuchi \citep{blaa64,dege89,brow98,deze99,hoog01}, as well as a rich population of lower-mass stars \citep{peca16,dami19}. The total stellar content of the association has been estimated as anywhere from 6,000 \citep{prei08} to 13,000 stars \citep{dami19}, with a total stellar mass of $\sim$4,000~M$_\odot$ \citep{wrig18}. Figure~\ref{scocen_map} shows the distribution of the known high- and low-mass population of Sco-Cen, as well as notable star forming regions and young clusters in their vicinity.

The association has historically been divided into three subgroups (Figure~\ref{scocen_map}) known as Upper Scorpius (US), Upper Centaurus-Lupus (UCL) and Lower Centaurus-Crux (LCC). Of these US is the youngest, with a median age of 11~Myr, while UCL and LCC are older with ages of 16 and 17~Myr respectively \citep{peca16}. Historically US has been the most well-studied due to it being the youngest, the most compact, and the furthest from the contaminating effects of the Galactic Plane, though this is changing thanks to data from {\it Gaia}.

The division of Sco-Cen into these subgroups was first defined by \citet{blaa46}, based on the positions and velocities of the brightest members, and has since been used by many authors, including the study by \citet{deze99} that defined the {\it Hipparcos} membership of the association. The revised {\it Hipparcos} data allowed \citet{rizz11} to revisit the membership and borders of the subgroups, favouring a more continuous distribution rather than a division into subgroups. Recent temporal \citep{peca16} and kinematic \citep{wrig18} studies have supported this view that the historical divisions do not necessarily reflect the real substructure of the association \citep[see also][]{vill18,dami19}.

\subsubsection{The high-mass stars}

The association and its three subgroups have historically been delineated by their bright B-type members, the first identification of which as a moving group dates back to \citet{kapt14}, though was not fully confirmed until \citet{blaa46}. The modern census of high-mass members of Sco-Cen can be attributed to \citet{dege89}, \citet{deze99}, and since refined by \citet[][see Figure~\ref{scocen_map}]{rizz11}. There are approximately 150 main-sequence B-type stars across the association, the late O-type runaway star $\zeta$ Ophiuchi \citep[first suggested to have originated in Sco-Cen by][]{blaa52}, and a number of evolved high-mass stars (such as the M1.5 supergiant Antares, $\alpha$ Sco, and the B1.5III giant, $\alpha$ Lup) that define the main-sequence turn-offs in each subgroup \citep{dege89}. Many of the LCC high-mass members have been debated \citep[e.g.,][]{hoog00}, with the recent study by \citet{rizz11} adding the bright B-type stars $\alpha$ Cru and $\beta$ Cru as members.

\citet{dege92} was the first to estimate the number of supernovae to have exploded across parts of the association, estimating $6 \pm 3$ past supernovae in UCL, which \citet{prei08} support with an estimate of $\sim$7 supernovae. Since most estimates put the total stellar content of LCC at just over half that of UCL \citep[e.g.,][]{mama02}, this suggests $\sim$4 past supernovae in this subgroup. For US there is one strong candidate for a past supernova, that of the progenitor of the pulsar PSR J1932+1059, which \citet{hoog01} suggest was an O5-O6 star that exploded 1.5~Myr ago to leave this remnant and the runaway star $\zeta$ Ophiuchi. This explanation for the origin of $\zeta$~Oph was questioned by \citet{chatt04} with improved astrometric data. \citet{neuh19} used {\it Gaia} DR2 astrometry to kinematically connect $\zeta$~Oph and the pulsar PSR~B1706-16, suggesting they were ejected from a supernova in a binary system $1.78 \pm 0.21$~Myr ago. \citet{tetz10} also identified the isolated neutron star RX~J1856-3754 as probably originating from Upper-Sco $\sim$0.3~Myr ago, with an O-type progenitor. Together these studies suggest the entire association might once have contained $\sim$13 O-type (primary) stars.

\subsubsection{Ages and age spreads}

\begin{figure*}
\begin{center}
\vspace{-2cm}
\includegraphics[width=18cm]{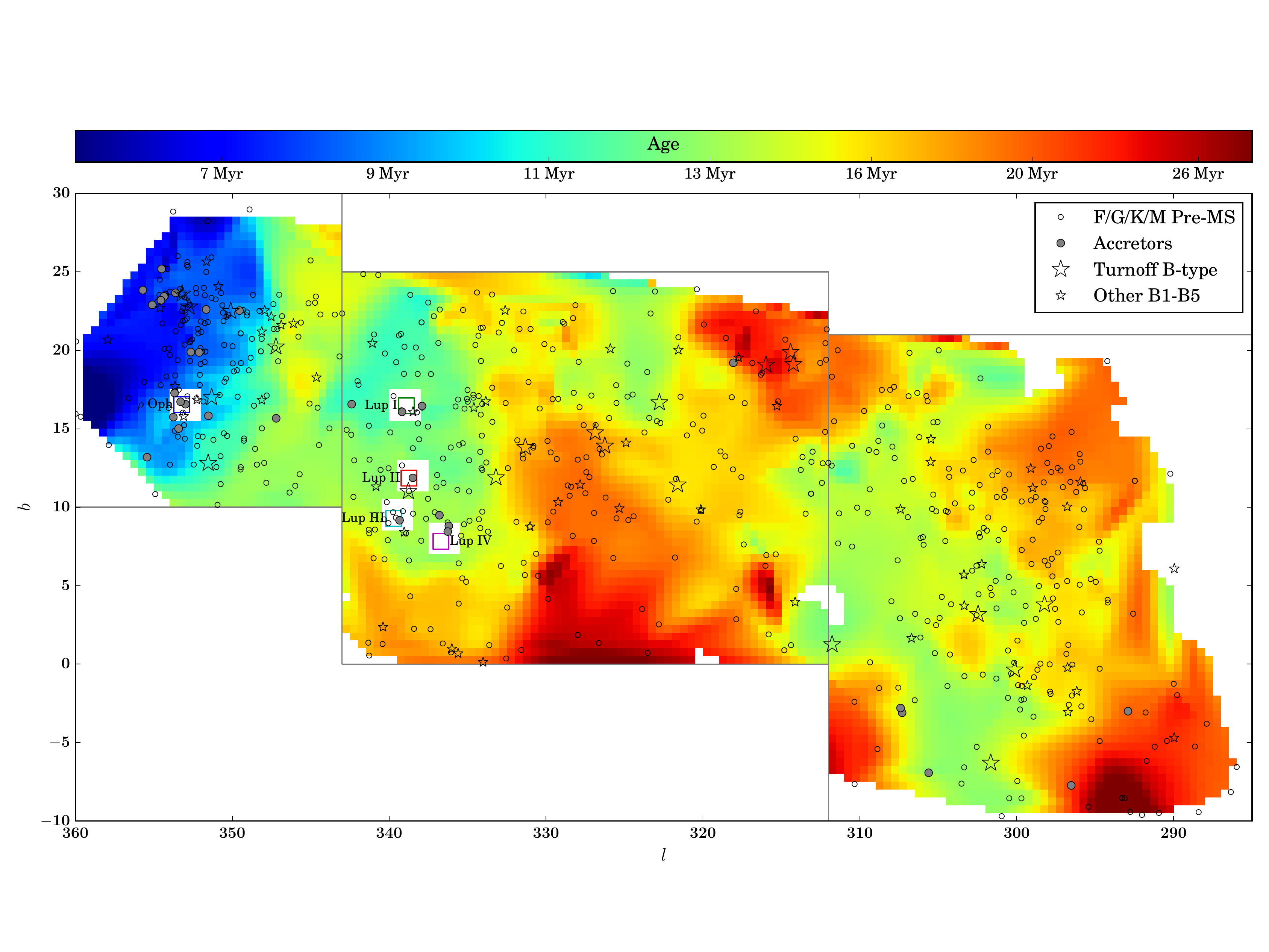}
\vspace{-1cm}
\caption{A spatially averaged `age map' of Sco-Cen constructed from the ages of 657 F/G/K/M-type pre-MS members of the association, with the luminous early-type stars also shown. The three boxes delineate the classical boundaries of the three subgroups US (left), UCL (centre) and LCC (right) that have median ages of 10, 16 and 15~Myr, respectively. The very young regions of $\rho$~Ophiuchus and the Lupus clouds, which are projected against Sco-Cen, are masked over in this map. Figure from \citet{peca16}.}
\label{pecaut_agemap}
\end{center}
\end{figure*}

The Sco-Cen association exhibits a considerable range of stellar ages. The youngest stars are typically associated with US, with stars as young as $\sim$5~Myr at the northern end of the subgroup, though the lack of dense molecular material or embedded stars in the association indicates that star formation has certainly finished. The median age of US is approximately 10~Myr \citep{peca16}, while UCL and LCC are older, with ages of 10--20~Myr \citep{dege89,mama02,peca16}.

The ages estimated for the three subgroups of Sco-Cen have generally increased over time as evolutionary models have improved. For example, early estimates put the age of US at approximately 5--6~Myr, as estimated from both the high- \citep{dege89} and low-mass stars \citep{prei99,prei02}, and from kinematic traceback \citep{blaa78}. Such studies also favoured a very short age spread of, at most, 1--2 Myr, which lead to suggestions that star formation across US had been triggered by some external event \citep[e.g.,][]{prei99}. More recent studies suggest a median age of $\sim$10~Myr for US for both the early- and late-type members of the association \citep{sart03,peca12,peca16,feid16}. This change has followed a general increase in pre-MS stellar ages derived from late-type stars \citep[e.g.,][]{bell13} that has been attributed to inaccuracies in the radii of pre-main-sequence M-type stars due to either magnetic inhibition of convection \citep{feid16} or starspots \citep{jack14}. Sco-Cen has proved to be a valuable calibrator for evolutionary models due to its proximity, rich low-mass population, and observations of eclipsing multiple systems from {\it Kepler}/K2 \citep[e.g.,][]{krau15}.

Figure~\ref{pecaut_agemap} shows an age map of Sco-Cen from \citet{peca16} that was produced by spatially-averaging the ages of 657 pre-MS members of the association calculated relative to an empirical isochrone. There are clear age spreads of $\pm$6--7~Myr across each of the subgroups \citep{peca16}, which are larger than previous measurements of the age spread \citep[e.g.,][]{prei02,sles06} due to a combination of the increased median age (which increases all ages and age spreads) and the wider areas studied by recent surveys compared to the small regions studied previously. These age spreads appear to be real, as evidenced by the clear age substructure seen in Figure~\ref{pecaut_agemap} that suggests groups of stars exist within the association with distinct ages relative to the age spreads across each subgroup. This was confirmed by \citet{dami19} who found that the more clustered stars within the association were younger than the more distributed population.

\subsubsection{Structure and kinematics}

The 3D shape of Sco-Cen is that of a highly elongated ellipsoid with approximate dimensions of $100 \times 100 \times 50$~pc with the three subgroups arranged in a row from north to south, with US the furthest away with a median distance of 143~pc followed by UCL at 136~pc and LCC the closest at 115~pc \citep{wrig18}. The line-of-sight dispersion of the subgroups UCL and LCC was first partially studied by \citet{debr99} using {\it Hipparcos} data, but it wasn't until {\it Gaia} DR1 that the 3D internal structure of the subgroups could be properly resolved \citep[e.g.,][]{wrig18,gall18}. The availability of {\it Gaia} DR2 facilitated more detailed structural studies of the entire association, including the low-mass population for the first time, and revealed a highly complex spatial structure \citep{dami19}.

The first estimation of the velocity dispersion of Sco-Cen came from \citet{debr99} who used kinematic modelling of {\it Hipparcos} data to estimate 1D velocity dispersions of 1.0--1.5~km~s$^{-1}$ for each subgroup. These estimates were supported by \citet{mads02} who calculated 1D velocity dispersions of 1.1--1.3~km~s$^{-1}$ for each subgroup using astrometric radial velocities calculated from {\it Hipparcos} data. The first direct measurement of the velocity dispersion came from \citet{wrig18} who combined {\it Gaia} DR1 astrometry with literature radial velocities to calculate 3D velocity dispersions in the Galactic cartesian system. They found that the subgroups were highly anisotropic, with 1D velocity dispersions varying from 0.5--2.5~km~s$^{-1}$ in each dimension, thought notably lower in all three subgroups in the $V$ dimension (the direction of Galactic rotation) than in the $U$ or $W$ dimensions. They calculated average 1D velocity dispersions of $1.86 \pm 0.21$ (US), $1.38 \pm 0.21$ (UCL), and $1.21 \pm 0.28$~km~s$^{-1}$ (LCC), which are approximately consistent with previous studies, with the exception of US. They attribute this disagreement to their inclusion of radial velocities and the differences in the assumed membership of the association. Their 3D velocity dispersions are $3.2 \pm 0.2$ (US), $2.5 \pm 0.2$ (UCL), and $2.2 \pm 0.3$~km~s$^{-1}$ (LCC), for which they calculate virial masses\footnote{The virial mass is defined as the mass enclosed within a given radius for a system to be gravitationally bound. It is given by a variant of the virial equation, $M_{vir} = \frac{2 r \sigma^2}{G}$, where $r$ is the radius of the system, $\sigma$ is its velocity dispersion, and $G$ is the gravitational constant \citep[e.g.,][]{port10}.} that are an order of magnitude larger than the estimated stellar mass and conclude that the three subgroups are gravitationally unbound.

\citet{wrig18} also find evidence for kinematic substructure within each of the subgroups in the form of small groups of spatially-associated stars that have more similar kinematics than the subgroup as a whole, particularly in UCL and LCC. The V1062 Scorpii moving group (see Figure~\ref{scocen_map}) identified independently by \citet{rose18} and \citet{zari18} within UCL is an example of this kinematic substructure, being a closely co-moving group of $\sim$60 stars with a 1D velocity dispersion $<$1~km~s$^{-1}$ \citep[see also][]{gold18}. These kinematic substructures appear to be related to the spatial and temporal substructure that has already been identified in Sco-Cen.

\subsubsection{Expansion}

Due to its proximity Sco-Cen is one of the most ideal associations to search for evidence of expansion \citep[studies of its expansion date back to][]{blaa52}, but it also makes it difficult to reliably measure its expansion, for two reasons. The first is that its large area on the sky (almost 90$^\circ$) means that proper motions or radial velocities do not probe the same dimension across the association, while the second and more important issue is that any radial motion of the association towards (or away from) the observer will cause a {\it virtual expansion} (or {\it contraction}), even when none is actually present. Both of these issues mean that a reliable assessment of the expansion of a nearby association requires radial velocities for some or all association members.

\citet{blaa64b} studied proper motions and radial velocities for $\sim$20--30 members of each subgroup and concluded that their kinematics were more consistent with linear expansion of the association than with no expansion. Blaauw estimated an expansion age of 20~Myr based on an assumed distance of 200~pc (10 or 15~Myr if scaled to a distance of 100 or 150~pc). \citet{blaa78} then followed this up by estimating an expansion age of $\sim$5~Myr for US. Using radial velocities and revised {\it Hipparcos} proper motions \citet{peca12} found a lower limit on the kinematic expansion age of 10.5~Myrs (99\% confidence), though their results are consistent with no expansion.

\begin{figure*}
\begin{center}
\hspace{-2cm} \includegraphics[height=14cm]{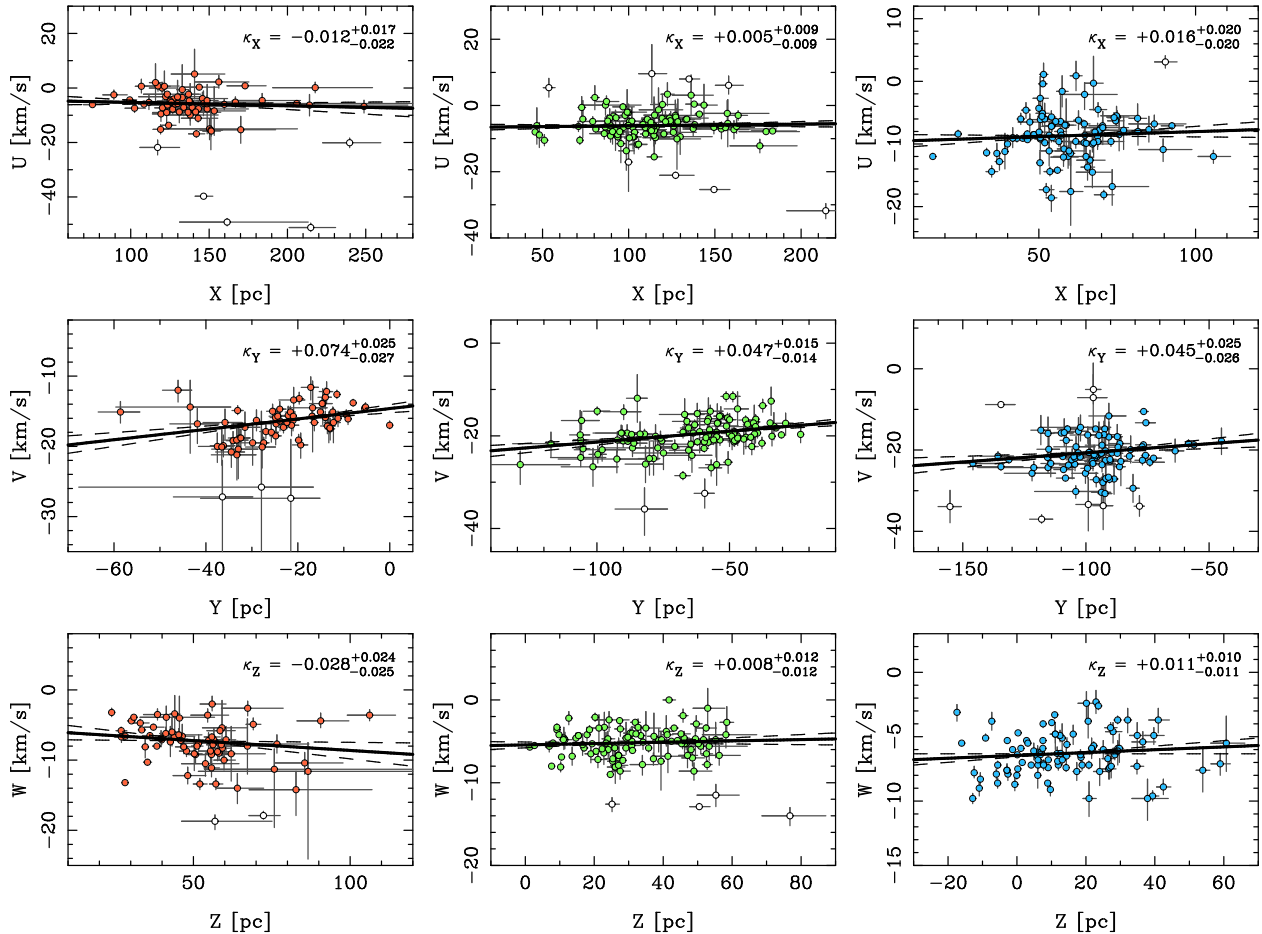}
\caption{Positions versus velocities along the three Galactic Cartesian axes $XYZ$ for each of the three subgroups of Sco-Cen (US on the left, UCL in the centre, and LCC on the right). 1$\sigma$ error bars are shown for all sources. The best-fit linear relationship between the plotted quantities shown as a solid line with the best-fitting slope, $\kappa$, noted in each panel (positive slopes imply expansion).}
\label{wright_3Dexpansion}
\end{center}
\end{figure*}

Using {\it Gaia} DR1 proper motions \citet{wrig18} investigated the expansion of the three subgroups of Sco-Cen using numerous techniques to explore evidence for expansion but could find no evidence for coherent expansion patterns in any of the subgroups. When analysing the expansion of the association in the 3D Galactic cartesian system (Figure~\ref{wright_3Dexpansion}) they found no evidence for expansion in the $X$ and $Z$ dimensions, but notably found significant ($\sim$3$\sigma$) evidence for expansion in the $Y$ dimension (the direction of Galactic rotation) in all three subgroups. They suggested that this expansion could be due to the effects of Galactic shear on the parental molecular cloud from which the subgroups formed and that has now been inherited by the stars of the association. \citet{ward18} also used {\it Gaia} DR1 proper motions for stars across the entire Sco-Cen association to assess evidence for expansion using multiple kinematic diagnostics, but could find no evidence for expansion from either a single or multiple expanding clusters. 

\citet{gold18} used {\it Gaia} DR2 astrometry to identify a large moving group within LCC with approximately 1800 members, comparable in size to the entire LCC subgroup and therefore potentially one of the larger kinematic substructures in the region. They find significant evidence that this group is expanding in all three dimensions with an expansion age of 8--10~Myr. This age is consistent with the isochronal age \citet{gold18} estimate, but different from the age of $\sim$15~Myr calculated by \citet{peca16} for LCC as a whole. This would imply that this group is one of the younger substructures within LCC, supporting the findings of \citet{dami19} that the clustered populations in Sco-Cen are younger than the diffuse population.

\citet{boby20} used a sample of 700 low-mass stars selected by \citet{zari18} to study the expansion of Sco-Cen. After correcting for the influence of the Galactic spiral density wave (which previous studies of the region have not), they measure a linear expansion rate of $39 \pm 2$ km~s$^{-1}$~kpc$^{-1}$ across the entire association. However, it is worth noting that the magnitude of the correction applied is $\sim$50 km~s$^{-1}$~kpc$^{-1}$, larger than the magnitude of the expansion signal measured. If true, this implies that the expansion of the association has been entirely hidden by the spiral density wave perturbation, since the observed, uncorrected kinematics do not show any evidence for expansion on the scale of the entire association. This raises questions of whether such corrections are necessary to expose the initial expansion of the association, and if so whether the calculations for such corrections should be based on the initial positions of the stars, rather than their current positions (as this study has done).

\subsubsection{The Greater Sco-Cen Complex}

The Sco-Cen association is not an isolated group of young stars as there are many young groups of stars and small star-forming regions in the local vicinity that appear to be related, such as $\rho$ Ophiuchus, the Lupus clouds, and R Corona Australis (see Figure~\ref{scocen_map}). \citet{mama99} discovered a cluster of young ($\sim$8~Myr) stars in the vicinity of the B8 star $\eta$ Cha whose Galactic motion suggests that the cluster, as well probably as members of TW Hya and $\epsilon$ Cha, probably originated in or near the same giant molecular cloud (GMC) that formed UCL and LCC 15~Myr ago \citep{mama00}. Going even further, \citet{mama01} consider that $\beta$~Pic and other nearby $\sim$10~Myr old stars may also have formed in this large star-forming event.

This extended region of young stars has been referred to as either the Oph-Sco-Cen association \citep{blaa91} or Greater Sco-Cen \citep{mama01} and is usually considered to include the three OB subgroups, surrounding molecular clouds, nearby star-forming regions such as $\rho$ Ophiuchus and the Lupus clouds, and the outlying associations and moving groups that have recently been uncovered. Other nearby groups or associations have been reported as possible extensions of Sco-Cen \citep[e.g.,][]{egge98,maka00}, though these have not generally stood up to further investigation \citep{prei08}.

\subsubsection{Feedback and effects on the local ISM}

The interstellar medium surrounding Sco-Cen shows numerous large, loop-like structures and shells that suggest significant interaction between the association and the surrounding medium. They are most clearly seen in H{\sc i} observations \citep[e.g.,][]{weav77,dege92,robi18}, with small bubbles centred around US and LCC and a larger bubble encasing the entire association and approximately centred on UCL \citep{dege92}. The total H{\sc i} mass of these bubbles has been estimated to be $\sim 5 \times 10^5$~M$_\odot$ and suggested to be composed of gas swept up by stellar winds and supernovae from the progenitor giant molecular cloud that Sco-Cen formed from, as well as the ambient ISM \citep{dege92}.

\citet{dege92} calculated the total kinetic energy of the bubbles to be $10^{51}$~erg and argued that the energy input from the existing massive stars coupled with past supernova activity across Sco-Cen would be sufficient to produce the observed network of bubbles. \citet{krau18} verify this by using a 3D hydrodynamic simulation to show that the observed massive star population could reproduce the size and morphology of the observed superbubbles. The presence of soft X-ray emission from the US bubble coupled with the detection of $^{26}$Al confirms that there has been at least one relatively recent ($\lesssim$1~Myr) supernova from within this subgroup \citep{robi18,krau18}. 

These supernovae may have played some role in the propagation of star formation through Sco-Cen, an idea advocated by \citet{prei99} to explain the small age spread measured in US at the time. This idea was challenged by the observation of larger age spreads by \citet{peca16}, with suggestions that the subgroups of Sco-Cen probably do not represent distinct star formation events. \citet{krau18} put forward an alternative model wherein the expanding superbubbles originating from one of the subgroups triggers star formation in the next subgroup by surrounding and squashing the primordial dense gas. Whatever the mechanism, this propagation of star formation appears to have continued beyond Sco-Cen itself and may have been responsible for triggering star formation in Lupus, Ophiuchus, Corona Australis and Chameleon \citep{mama01}.

\subsection{Orion OB1}

\begin{figure}
\begin{center}
\includegraphics[width=8.7cm]{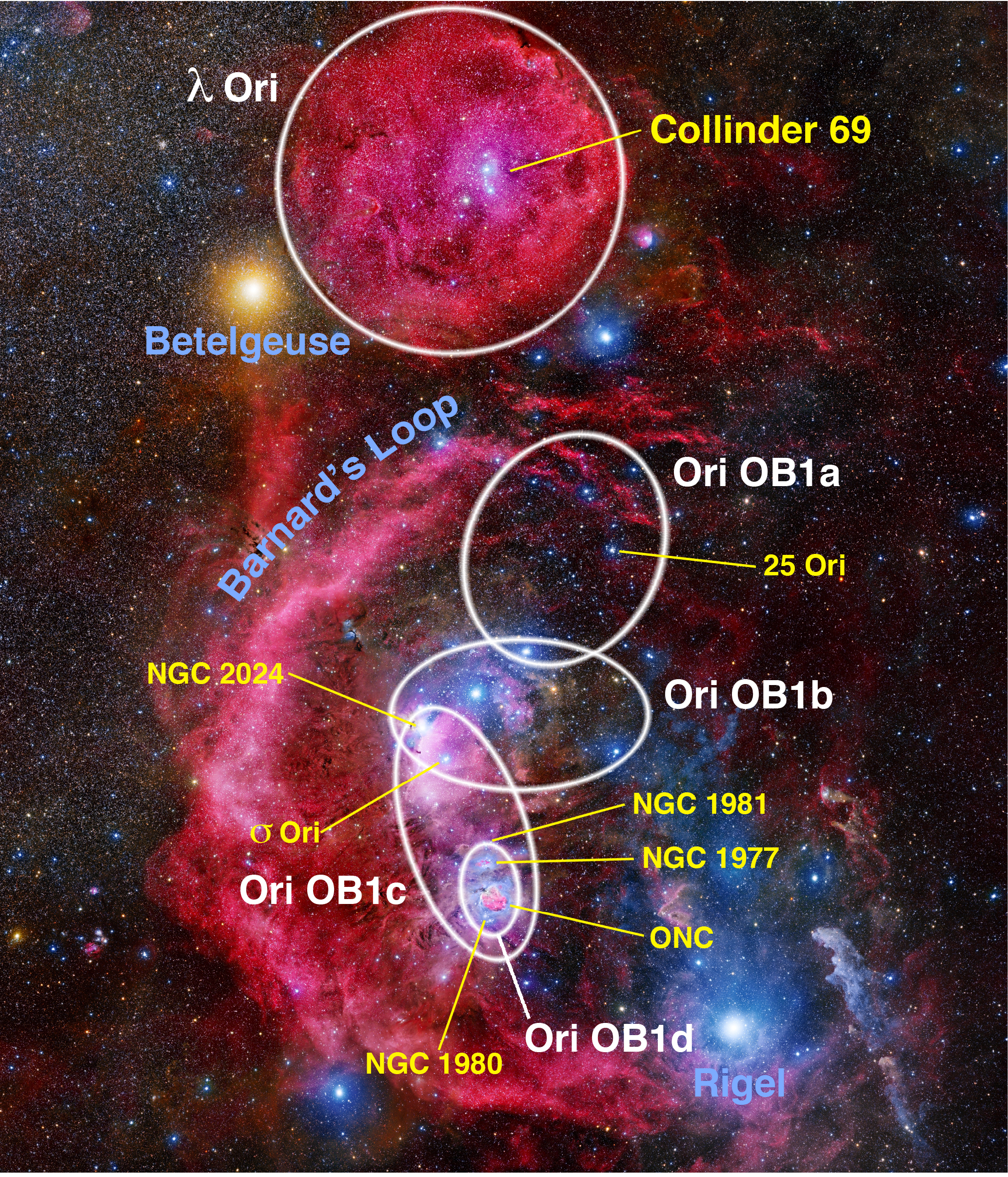}
\caption{Wide-field map of the Orion constellation showing the locations of the four subgroups of the Orion OB1 association as well as the $\lambda$ Ori region (white ellipses and labels). Also shown are prominent star clusters across the region (in yellow), Barnard's Loop and the bright stars Betelgeuse and Rigel (light blue). Background astrophotograph courtesy of Stanislav Volskiy.}
\label{orion_structure}
\end{center}
\end{figure}

The Orion OB association (Orion OB1) is the richest association in the Solar neighbourhood and one of the few still closely associated with a giant molecular cloud. It spans over 200~deg$^2$ on the sky across the constellation of Orion at distances between 300--450~pc \citep{brow94,zari18}. It contains at least 56 massive stars of spectral type B2 and earlier \citep{blaa64} including the bright stars $\zeta$ and $\delta$ Ori and $\theta^1$ Ori~C, the most massive stars of the trapezium in the Orion Nebula Cluster (ONC), part of Orion OB1d. Figure~\ref{orion_structure} shows the structure of the Orion OB1 association, its division into subgroups, and some of the notable young clusters in its vicinity.

\citet{blaa64} divided the association into four subgroups labelled $a$ to $d$ (see Figure~\ref{orion_structure}). Orion OB1a is generally considered to be the oldest subgroup, aged approximately 8--12~Myr \citep{brow94}, and lying north-west of Orion's Belt. It includes the cluster 25~Ori \citep{bric07}, aged 7--10~Myr, which may be the youngest component of this subgroup. Subgroups 1b and 1c have approximately similar ages of 2--8 and 2--6~Myr, respectively \citep{ball08}, with different authors placing them in different chronological order \citep[e.g.,][]{blaa64,brow94}. Orion OB1b includes the stars of Orion's Belt as well as the cluster $\epsilon$~Ori \citep{kubi17}, while Orion OB1c includes the stars around Orion's Sword as well as the clusters NGC~1980 \citep[considered by some to lie in the foreground,][]{alve12} and NGC~1981. $\sigma$~Ori \citep{walt08} is projected close to Orion OB1b but its age is closer to that of Orion OB1c, so is often considered part of the latter. Finally, Orion OB1d is the youngest of the subgroups and is associated with the molecular gas in the region (the Orion A and B filaments). It includes the ONC, M43, NGC~1977, NGC~2024 clusters and the Orion molecular clouds \citep[note that while the NGC~2024 cluster is projected against Orion OB1b and c, its very young age makes it more consistent with OB1d, e.g.,][]{levi06}. These subgroups have been defined almost entirely based on position on the sky, and while important from a historical perspective, recent work suggests that sub-dividing the association using position, parallax and kinematics, providing a more accurate view of the star formation history of the region \citep[e.g.,][]{zari17,koun18,chen19,zari19}.

In addition to the four subgroups defined by \citet{blaa64} there are numerous outlying groups associated with the same star formation events that formed the main OB association. These include $\lambda$~Ori, which while spatially offset from the main association has an age similar to the 1b and 1c subgroups, so appears to be related to the association (see Figure~\ref{orion_structure}). \citet{zari17} identified further young stars in the vicinity of $\lambda$~Ori that appear to be associated with the H$\alpha$ bubble surrounding the main cluster, suggesting further star formation across this area.

There are very few estimates of the total stellar mass of the Orion OB1 association in the literature. \citet{ball08} estimate that 30--100 stars more massive than 8~M$_\odot$ have formed across the association. Assuming a standard \citet{krou01a} initial mass function (IMF) as formulated by \citet{masc13} such stars constitute approximately 0.4\% of the full IMF, implying a total stellar population of 7,500--25,000 stars. Assuming a mean stellar mass of 0.36~M$_\odot$ \citep{masc13} and an average binary fraction of 50\% this equates to a total stellar mass of 4000--13,500~M$_\odot$.

\subsubsection{The high-mass stars}

There are a considerable number of high-mass OB stars within the Orion OB1 association. \citet{blaa64} identify 53 OB stars of spectral type B3 and earlier, specifically within the subgroups a, b and c. The most massive of these are found in subgroup 1b and include $\zeta$ Ori A (O9.7Ib, $\sim$49~M$_\odot$), $\delta$ Ori A (B0III, $\sim$45~M$_\odot$) and $\epsilon$~Ori A (B0Ia, $\sim$42~M$_\odot$), with masses from \citet{lame93}. Other notable massive stars include $\eta$~Ori \citep[B1V, subgroup 1a,][]{warr78}, $\iota$ Ori A \citep[O9III, subgroup 1c,][]{lame93}, and the stars of the Trapezium system in the ONC ($\theta^1$ Ori C, O6V, being the most massive of them). The latter in particular have been the focus of numerous searches for binary companions. $\theta^1$ Ori C was resolved by \citet{weig99} as a close binary, the companion having a mass of $\sim$5~M$_\odot$ compared to the primary's mass of $\sim$45~M$_\odot$. The other O-type star in the ONC, $\theta^2$ Ori A (O9.5V) is known to be a triple system composed of a spectroscopic binary with a mass ratio of 0.35 \citep{abt91} and a slightly less massive, resolved companion \citep[approximately 3--7~M$_\odot$,][]{prei99b}. See also the recent interferometric study by \citet{karl18} that uncovered additional companions to many of the OB stars in Orion.

There are many runaway massive stars known in Orion, most notable of which are AE~Auriga and $\mu$~Columbae, which are moving in opposite directions at speeds of 150 and 117~km$^{-1}$, respectively \citep{blaa91}. \citet{hoog01} showed that these two stars and the massive binary system $\iota$~Ori, were at the same location in the sky $2.6 \pm 0.05$~Myrs ago, and may be the product of a dissolved double-binary system. Another runaway, 53~Ari, may have been ejected from subgroup 1c 7.3~Myrs ago \citep{brow94} during a supernova explosion.

In addition to runaway massive stars, the Orion association is probably the birthplace of a number of ejected neutron stars. \citet{pell05} studied the origin of the Geminga neutron star and estimate an origin in either the Orion OB1 or the Cas-Tau association, while \citet{koun20} used the proper motion of the pulsar at the centre of the Monogem supernova remnant \citep{gold05} to suggest that it may have been ejected from the $\lambda$ Ori cluster approximately 1.7 Myr ago.

\subsubsection{Structure and age}

The spatial, kinematic and temporal structure of Orion OB1 is highly complex. The classical subdivision of the association into predominantly four subgroups was based entirely on sky position, but the availability of radial velocity data \citep{jeff06} and later from {\it Gaia} parallaxes and proper motions have suggested a revision of this structure that takes into account the full 6-dimensional spatial and kinematic information \citep{kubi17,koun18,zari19}.

\begin{figure}
\begin{center}
\includegraphics[width=8.5cm]{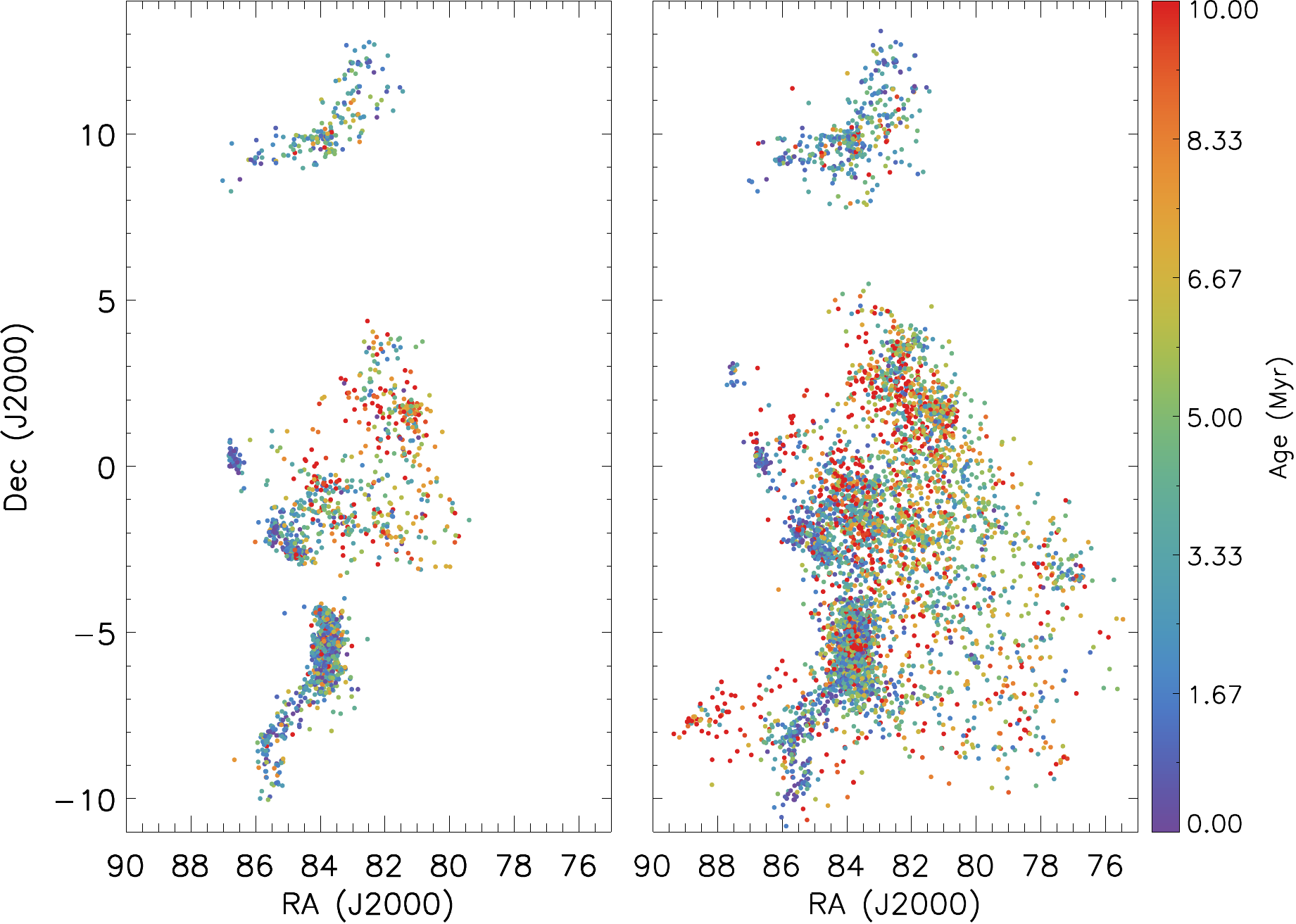}
\caption{Estimated stellar ages across the Orion OB1 association from \citet{koun18}. The left-hand panel shows ages derived using spectroscopic effective temperature and bolometric luminosities in the HR diagram, while on the right are shown ages derived using just photometry in the colour magnitude diagram (assigning distances according to the average distance of stars in each group).}
\label{orion_ages}
\end{center}
\end{figure}

\citet{koun18} performed a clustering analysis of APOGEE and {\it Gaia} DR2 data of 10,248 young stars across Orion, identifying 190 groups of stars. They associated many of these groups with either the young populations in Orion A or B, the cluster $\lambda$~Ori, or two new progenitor groups dubbed Orion C and D (to extend the existing naming scheme). These two new groups have similar distributions on the sky, both broadly tracing the areas of the OB1a and OB1b subgroups, though they have distinctly different radial velocities and parallaxes (see Figure~\ref{orion_kounkel}). Orion~D makes up the majority of the OB1a and OB1b subgroups, the 25~Ori cluster, and includes the bright stars $\eta$ and $\psi^2$ Ori. It spans the largest area on the sky of the new groupings, extending as far South as Rigel, Orion X \citep{bouy15} and some of the Orion outlying clouds (though this southern component could be associated with Orion~A). It is aged from 4--7~Myr and for the most part has a radial velocity of 3--5~km~s$^{-1}$ and parallaxes of 2.6--3.0~mas ($d \simeq 360$~pc). Orion~C is mostly composed of the $\sigma$~Ori cluster, but also extends northwards to the Orion Belt population and into the area covered by Orion OB1a. Its age varies from $\sim$7.5~Myrs in the north down to $\sim$2~Myrs around $\sigma$~Ori and has distinctly different radial velocity \citep[as first observed by][]{jeff06} and parallax from the Orion B and D groups it is projected near. Figure~\ref{orion_ages} shows the age distribution across the association, showing its complex physical and temporal structure.

\citet{zari19} noted a particularly high degree of substructure in the region towards the Belt stars, finding multiple groups of stars with different kinematic properties, including groups towards $\sigma$~Ori, $\zeta$~Ori, $\epsilon$~Ori, and $\delta$~Ori. These groups range in age from $\sim$4~Myr (for the groups around $\sigma$~Ori and $\zeta$~Ori) to $>$10~Myr for the more extended groups. The age estimates and rankings found by \citet{zari19} are generally in agreement with those found by \citet{koun18}, with both studies finding a very complex and fragmentary star formation history.

Historically an age gradient was suggested to extend from subgroups OB1a and 1b to the younger OB1d group \citep{brow94}, however the results from {\it Gaia} DR2 do not show evidence for a clear progression of star formation across the association and suggest a much more complex age distribution \citep{koun18,zari19}. {\it Gaia} data has also lead to the discovery of multiple older populations in the Orion OB1 association, including a $15--21$~Myr population connected to the ASCC~20 cluster \citep{kos18,zari19}, another similarly-aged population towards Orion's Belt \citep[group B8 in][]{zari19}, and a $\sim$17~Myr population behind the main association ($d \sim 430$~pc) that is spatially quite extended ($\sim$90~pc in length) but kinematically compact \citep{koun18,zari19,jera19}.

This temporal complexity isn't limited to the widest parts of the association. \citet{becc17} found evidence for multiple pre-main sequences in the colour-magnitude diagram of stars around the Orion Nebula Cluster, which they argued represented multiple distinct episodes of star formation within the cluster, with the youngest populations more concentrated toward the ONC than the older populations. \citet{jera19} verified the existence of these sequences, but found indications that the older population may be slightly in the foreground of the ONC.

\subsubsection{Foreground populations}

There is growing evidence for recent star formation in front of the Orion OB1 association. Some of the brightest stars of the Orion constellation are young and massive stars, such as Betelgeuse ($\alpha$~Ori, M2Iab), Rigel ($\beta$~Ori, B8Iab) and Saiph ($\kappa$~Ori, B0Iab), and are found at distances of 150--300~pc \citep{perr97}. These stars are too far in the foreground to be consistent with being ejected from Orion OB1, and instead suggest recent star formation has occurred in the foreground \citep{ball08}. This was verified by \citet{bouy15} who identified a large population of early-type stars in the foreground of Orion and named it Orion~X, suggesting they constituted a previous generation of star formation to the main Orion OB1 association. \citet{zari17} also found evidence for an older population of stars in this direction from {\it Gaia} DR1 data, but using {\it Gaia} DR2 data \citet{zari18} and \citet{zari19} were not able to find clear evidence for the existence of Orion~X, thus leaving the origin of the brightest stars in the Orion constellation unresolved.

\citet{alve12} identify a rich population of young (4--5~Myr) stars just in front of the Orion Nebula Cluster (ONC) that includes many stars associated with $\iota$~Ori and the NGC~1980 cluster (previously considered part of Orion OB1c, see Figure~\ref{orion_structure}). They argue that NGC~1980 is a relatively massive ($\sim$2000 members) and hitherto unrecognised cluster and, combined with a general foreground populations, these stars include many previously thought to be part of the ONC. \citet{bouy14} and \citet{zari19} both confirmed the existence of a foreground population, with the former estimating it to consist of $\simeq$2600 stars with an age of 5--10~Myr. However, \citet{dari16} and \citet{fang17} found that these stars have similar radial velocities to stars in the ONC and suggested they are likely to be part of the same star formation event. \citet{fang17} calculate a median age of 1--2~Myr, while \citet{koun17} estimate an age of 3~Myr, suggesting the group is similarly-aged to the ONC. It is possible that these foreground populations are related to the older pre-main sequence identified by \citet{becc17} towards the ONC and that \citet{jera19} suggest may be slightly in the foreground of the cluster.

\subsubsection{Kinematics and expansion}

The youngest parts of the Orion OB1 association, particularly subgroups OB1c and OB1d, are still spatially and kinematically associated with the molecular clouds that they formed from. \citet{fure08} and \citet{tobi09} performed wide-field spectroscopic surveys of these young stars and found a correlation between the stellar radial velocities and that of the dense gas they are projected against. They argue that this kinematic coherence implies these stars are still very young and that they haven't yet decoupled from or dispersed the gas that they formed from. \citet{haca16} and \citet{dari17} extended this work with spectroscopic studies across the Orion~A molecular cloud and observed a correlation between stellar and gas velocities over the 5$^\circ$ extent of the cloud. In addition to this correlation, \citet{fure08} found evidence for kinematic substructure within the association, in the form of spatially-coherent groups of stars with similar motions. \citet{dari17} found that these groups are preferentially found in the low-density regions of the cloud, while the denser ONC is less kinematically substructured.

Most measurements of the velocity dispersion within the Orion OB1 association have been limited to the more-easily identified young stars in and around the ONC and the OB1c and OB1d subgroups. \citet{jone88} and \citet{vana88} calculated proper motion velocity dispersions for stars in the ONC of $2.34 \pm 0.09$ and $1.49 \pm 0.2$~km~s$^{-1}$, respectively. The smaller velocity dispersion calculated by \citet{vana88} for a sample of generally brighter stars can be at least partly explained by an anti-correlation between velocity and mass. This was also as found by \citet{hill98} who measured a velocity dispersion of 2.81~km~s$^{-1}$ for stars with $M = 0.1$--0.3~M$_\odot$ and a velocity dispersion of 2.24~km~s$^{-1}$ for stars with $M = 1$--3~M$_\odot$. This anti-correlation suggests partial energy equipartition within the system (i.e., that the massive stars are moving slower than the less massive stars). \citet{dari17} measure a radial velocity dispersion of 2.2~km~s$^{-1}$, larger than their estimated virial velocity dispersion of $\sim$1.7~km~s$^{-1}$, suggesting that the system is slightly supervirial. Recently, \citet{kuhn19} and \citet{kim19} measured the PM velocity dispersion of stars in the ONC using {\it Gaia} DR2 and Hubble Space Telescope + Keck data, respectively. They both found the velocity distributions to be highly anisotropic, \citet{kuhn19} measuring velocity dispersions of $1.51 \pm 0.11$ (RA) and $0.50 \pm 0.12$ (Dec) mas~yr$^{-1}$ from 48 stars, while \citet{kim19} measure dispersions of $0.83 \pm 0.02$ (RA) and $1.12 \pm 0.03$ (Dec) mas~yr$^{-1}$. \citet{kim19} attribute the difference in measured dispersions to differences in the sample size and under-estimated PM uncertainties in {\it Gaia} DR2. On a larger scale, \citet{fure08} measure a radial velocity dispersion of 3.1~km~s$^{-1}$ across the Orion OB1c association.

Numerous attempts have been made to determine whether the Orion OB association subgroups are expanding and to estimate their kinematic ages. \citet{lesh68} studied the expansion of the OB1a subgroup and measured a kinematic age of 4.5~Myr ($\pm$30\%), while \citet{blaa61} estimate a kinematic age of 2.2--4.9~Myrs for subgroup OB1b based on the motions of the three runaway stars AE~Aur, $\mu$~Col, and 53~Ari. \citet{koun18} find that their Orion~D group, which contains subgroups OB1a and OB1b as well as Orion~X exhibits PMs consistent with expansion, while the stars associated with the Orion~B cloud have PMs more consistent with contraction. \citet{koun18} also found clear evidence for expansion for stars in the $\lambda$~Ori cluster, with a correlation between distance from the cluster centre and radial outward velocity that suggests a single trigger of the expansion approximately 4.8~Myr ago.

\subsubsection{Feedback and the Orion-Eridanus Superbubble}

The Orion OB1 association is very prominently surrounded by a bright crescent of H$\alpha$ emission known as Barnard's Loop, which is believed to be the brighter part of a larger bubble that extends 40$^\circ$ west and incorporates the Eridanus Loop \citep{siva74,reyn79,ochs15}. This is the Orion -- Eridanus Superbubble, an approximately $140 \times 300$~pc bubble \citep{ball08} with a 1--$5 \times 10^5$~K interior that has been detected in the UV, X-ray and in the 1.8~MeV $\gamma$-ray line that indicates the presence of the short-lived radioactive species $^{26}$Al \citep[][implying recent supernova activity]{dieh04}. The bubble is surrounded by an expanding \citep[$\sim$15--20 km~s$^{-1}$,][]{reyn79,joub19} shell of ionized gas, which is itself surrounded by an expanding H~{\sc i} shell with a mass of $\sim$300,000~M$_\odot$ \citep{brow95}. \citet{ochs15} studied the Orion-Eridanus superbubble with multi-wavelength data and argue it is both larger and more complex than previously thought, potentially viewed as a series of nested shells superimposed along the line of sight, a picture that was supported by \citet{joub19}.

\citet{ball08} estimate that there have been at least 10--20 SNe in the last 12~Myr that would have released $> 10^{52}$~ergs of kinetic energy, sufficient to form the bubble. Additional energy may be being injected by smaller bubbles within the Orion-Eridanus superbubble, such as the Veil shell that surrounds the ONC and is thought to be powered by stellar winds from $\theta^1$ Ori-C \citep{pabs20}.

The SNe that have driven the superbubble may have contributed to the enrichment of the association subgroups. \citet{cunh94} found that the younger subgroups OB1c and OB1d appear to be more abundant in oxygen and silicon with respect to stars in the older subgroups OB1a and OB1b, while other elements such as carbon, nitrogen or iron do not show such abundance variations. Such a pattern is consistent with being due to supernova ejecta enriching the interstellar gas that went on to form subgroups OB1c and OB1d. However, both \citet{simo10} and \citet{biaz11} find similar abundances of the $\alpha$-elements silicon and titanium across the OB association, arguing against self-enrichment.

\subsection{Vela OB2}

\begin{figure*}
\begin{center}
\includegraphics[width=18cm]{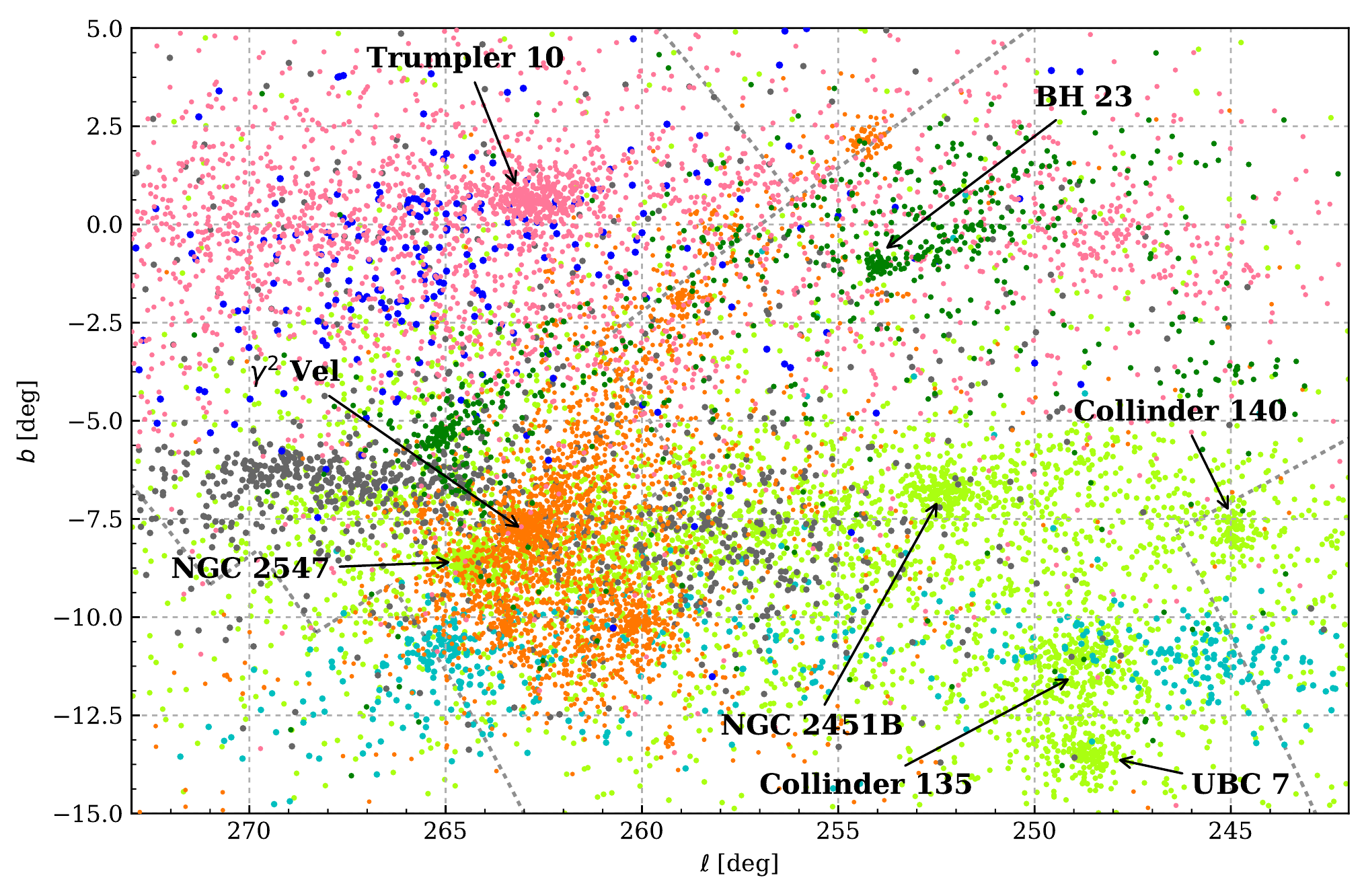}
\caption{Spatial distribution of young stars across Vela~OB2 and surrounding regions from \citet{cant19b}. The stars are coloured according to the groups identified by those authors, from I to VII (blue, pink, grey, light green, turquoise, dark green and orange). Notable clusters in the region are marked.}
\label{vela_structure}
\end{center}
\end{figure*}

The Vela OB2 association was first identified by \citet{kapt14} and included in \citet{blaa64}'s list of associations. At a distance of between 350--400~pc \citep{deze99,zari18}, it spans at least $15 \times 15$ degrees across the constellations of Vela, Puppis and Carina (see Figure~\ref{vela_structure}). Its total mass has been estimated to be between $\sim$1300 M$_\odot$ \citep{arms18} and over 2330~M$_\odot$ \citep{cant19a}, depending on where the borders of the association are drawn. Due to its projection against the Galactic Plane, its relative dearth of massive stars (compared to Sco-Cen and Orion for example), and its older age, it has historically not received as much attention as other nearby associations. However, in recent years and thanks to {\it Gaia} data considerable work has helped uncover its low-mass population \citep[e.g.,][]{zari18,cant19b}.

The brightest star in the association is the Wolf-Rayet WC8 + O9I binary system $\gamma^2$ Velorum, the nearest known Wolf-Rayet star, with masses of $9 \pm 0.6$ and $28.5 \pm 1.1$ M$_\odot$, respectively \citep{nort07b}. Questions have been raised over the membership of the star in Vela OB2, since evolutionary models for the star suggest an age of $5.5 \pm 1$~Myr \citep{eldr09}, while the lower-mass pre-main sequence stars in its vicinity have ages of $\sim$10~Myr \citep{jeff09}. This has led to suggestions of mass transfer within the binary system that might make it appear younger \citep{jeff14}, since {\it Hipparcos} astrometry suggests the star is a member of the association \citep{rate20}. \citet{deze99} identified a further 92 members of the association, predominantly B-type stars, which \citet{arms18} reduced to 81 after excluding photometric and astrometric contaminants.

The association borders the Trumpler~10 region, previously considered an open cluster, but redefined by \citet{deze99} as an association of 22 B-type stars (earliest spectral type B3V) that are slightly older than Vela~OB2 but at a similar distance. Its connection to Vela~OB2 and the wider Vela-Puppis region has recently been confirmed by \citet{cant19b}, as seen in Figure~\ref{vela_structure}.

\subsubsection{Structure and ages}

The most prominent overdensity within the association is the Gamma Velorum cluster that surrounds $\gamma^2$~Vel. It was discovered by \citet{pozz00} from ROSAT X-ray observations that coincided with low-mass, pre-main sequence stars and verified with follow-up photometric and spectroscopic observations \citep{jeff09}. The cluster and the massive binary share a common proper motion and so are strongly believed to be associated with each other and with the wider Vela OB2 association \citep{jeff09}. \citet{jeff14} and \citet{pris16} estimate the mass of the cluster to be $\simeq$100~M$_\odot$. Another cluster within the association, the P~Puppis cluster, was identified by \citet{caba08b} and later by \citet{becc18}.

\begin{figure}
\begin{center}
\includegraphics[width=8.5cm]{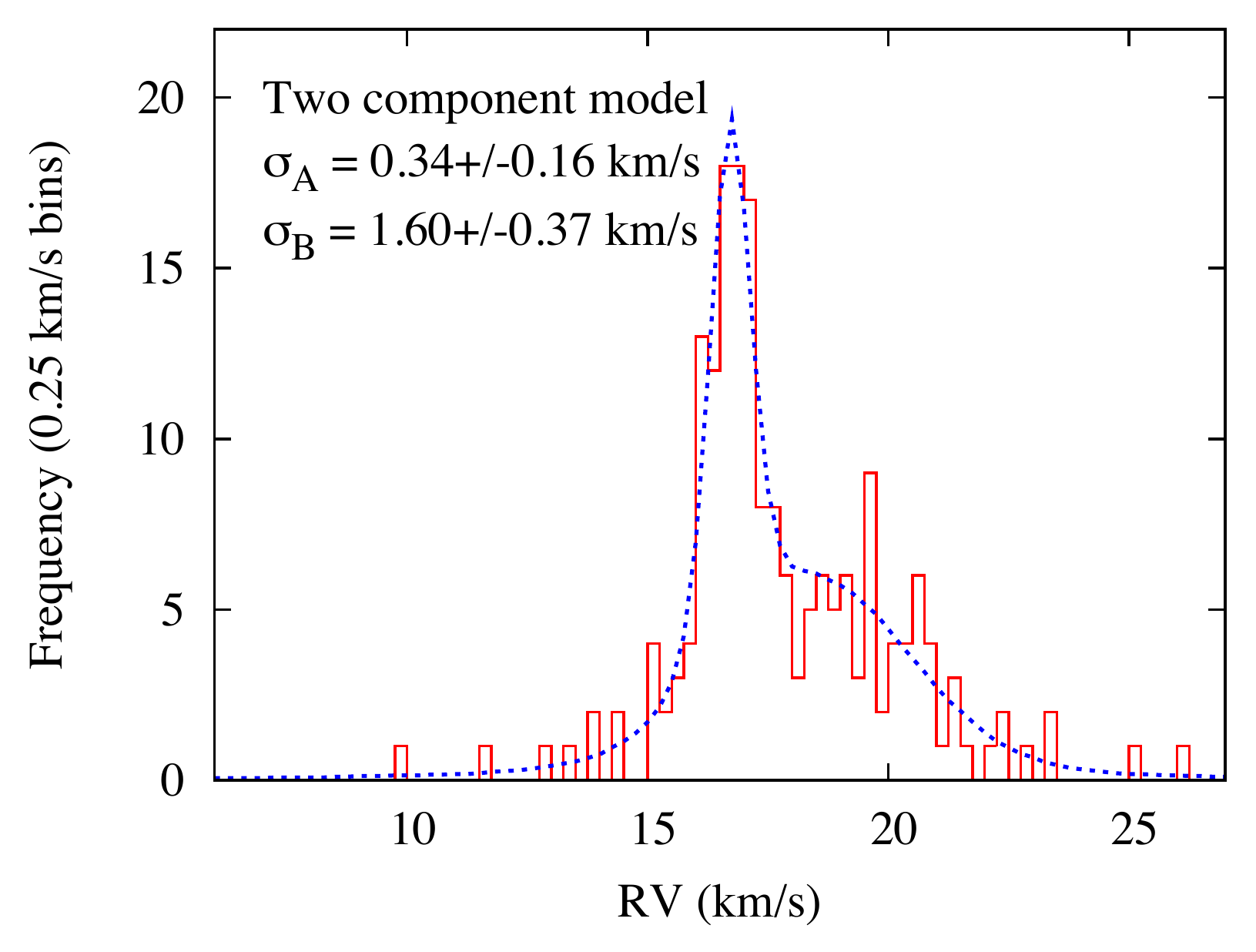}
\caption{Radial velocity distribution of young stars towards Gamma Vel from \citet{jeff14}. The red histogram shows the young stars towards the cluster while the blue dotted line shows the best-fitting two Gaussian component model (velocity dispersions noted on the figure), which takes into account the effect of binaries on the velocity dispersions.}
\label{vela_substructure}
\end{center}
\end{figure}

Using {\it Gaia}-ESO Survey spectroscopy, \citet{jeff14} performed a detailed spectroscopic study of the Gamma Velorum cluster and found evidence for two distinct radial velocity components, one broad and one narrow, as shown in Figure~\ref{vela_substructure}. The narrow velocity component has been associated with the cluster itself, and suggests the system is in virial equilibrium, while the broad component was attributed to the wider Vela OB2 association and seemed to be slightly younger than Gamma Vel. \citet{sacc15} also identified a broad kinematic component projected against the older NGC~2547 cluster, suggesting the young, low-mass population of Vela OB2 was spread over a large area in this region. \citet{bouy15} found hints of this population in the foreground of Vela OB2 and suggested an age gradient existed covering this young foreground population, Gamma Velorum, Trumpler 10, and the older Vela~OB2 association behind it.

The full extent and complex spatial and kinematic structure of Vela~OB2 and the surrounding region has only become fully apparent after the release of {\it Gaia} data. \citet{arms18} used {\it Gaia} DR1 photometry to identify low-mass stars across Vela OB2, uncovering an extended population with concentrations around Gamma Velorum and a distributed population over a wider area that has considerable spatial substructure. \citet{becc18} applied a clustering analysis to {\it Gaia} DR2 astrometry across this area to verify some of these over-densities and identify new possible clusters or substructures within the association \citep[see also][]{dami17,cant19a}.

\citet{cant19b} performed a wide-area study of young stars across the Vela and Puppis regions, selecting young stars using a combination of photometry and astrometry and then using an unsupervised classification scheme to divide their sample into separate populations, finding 7 distinct groups as shown in Figure~\ref{vela_structure}. Three of these populations (labelled III, V and VII by the authors) fall within the classical boundaries of Vela OB2 \citep{deze99}, while two (I and II) fall within the Trumpler~10 region. The other two populations (IV and VI) are found to the west of Vela~OB2 towards the constellation Puppis. Population IV is the most populous of all the groups and contains the open clusters NGC 2547, NGC 2451B, Collinder 135, Collinder 140 and UBC~7, while population VI contains the open cluster BH~23. \citet{cant19b} estimate a range of ages for these populations, from $\sim$10~Myrs for population VII (containing Vela OB2 itself) through to 45--50~Myrs for Populations I and II in the Trumpler~10 region. The authors find no evidence for the spatial gradient with age found by \citet{bouy15} and instead argue for multiple episodes of star formation across the region with a complex star formation history.

\subsubsection{Kinematics and expansion}

\citet{jeff14} calculated radial velocity dispersions of $0.34 \pm 0.16$ and $1.60 \pm 0.37$~km~s$^{-1}$ for the two kinematic components they identified towards Gamma Vel. The first component corresponds to the Gamma Velorum cluster itself, and is consistent with the cluster being in virial equilibirum (given a cluster mass of $\sim$100~M$_\odot$). The second component constitutes some part of the Vela~OB2 association and indicates that it is unbound. \citet{fran18} studied the correlation between radial velocity and parallax of this component and found a gradient that they interpret as a sign of expansion.

\citet{cant19b} measure the 3D expansion in all 7 groups they identify across Vela--Puppis and find evidence for expansion in all of them, predominantly along the Galactic $X$ axis (towards the Galactic centre), though the majority do also show expansion in $Y$ and $Z$. Notably, the expansion rates they measure appear non-isotropic, in that the expansion is occurring at different rates along the different axes, with differences up to a factor of 10 between the expansion rates (one of the groups also shows significant evidence for contraction along the $Z$ axis). This may be due to physical and kinematic substructure within the groups that has yet to be resolved or it may be that the expansion of these groups has been non-isotropic since they started expanding. Due to the anisotropic expansion it is not immediately clear of the legitimacy of calculating kinematic ages for these groups but they vary from $\sim$20 to $\sim$50~Myrs, approximately in line with the evolutionary ages for each population.

\citet{arms20} performed a spectroscopic survey of young stars in Gamma Vel and the region of Vela OB2 in its vicinity, confirming the youth of their targets from the presence of lithium, and separating cluster and association members based on both their positions and kinematics. They combined spectroscopic radial velocities with {\it Gaia} DR2 proper motions to study the structure and 3D dynamics of their targets, finding strong ($5-7 \sigma$) evidence that the association is expanding in all three dimensions. The expansion rates measured are non-isotropic, varying by more than a factor of two between axes.

\subsubsection{Feedback and formation}

\citet{sahu92} discovered a giant ($\sim$10$^\circ$) ring-like structure in IRAS images of Vela and Puppis, part of the Gum nebula. At a distance of $\sim$450~pc it is believed to be connected to the Vela OB2 association and is also discernible in maps of interstellar extinction and neutral hydrogen \citep{test06}. The massive stars $\eta$ Puppis (O4I) and $\gamma^2$ Velorum are thought to be responsible for powering the shell \citep{reyn76}. \citet{sahu92} estimate the energy output of the Vela OB2 stars from both stellar winds and supernovae, and find it to be consistent with that needed to produce the shell, while \citet{higd13} find that only 10\% of the ionizing radiation from the Vela O stars is necessary to power the shell. \citet{cant19a} argue that the distribution of stars in Vela OB2 forms a ring-like structure that follows the Vela shell and may have formed by triggering as part of the shell's expansion.

\subsection{Cygnus OB2}

\begin{figure*}
\begin{center}
\includegraphics[width=12cm]{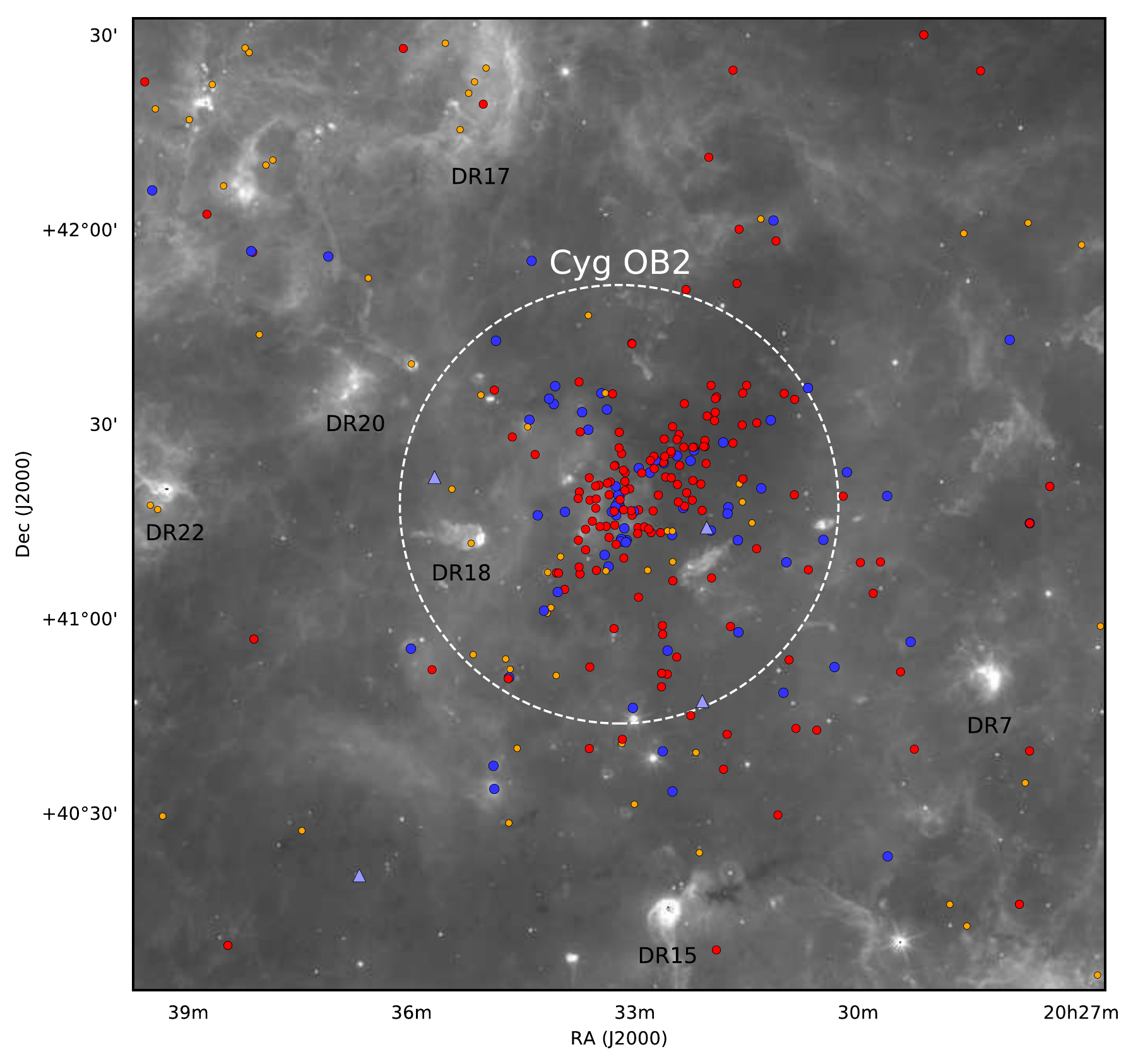}
\caption{Spatial distribution of known high-mass stars in the vicinity of Cygnus OB2 from \citet{wrig15a} and \citet{berl18a}. Blue dots show known O-type stars, red dots show B-type stars, white triangles show Wolf-Rayet stars and orange dots show suspected OB stars \citep[photometrically identified, primarily from][]{come02,come08}. The white circle shows the 1~deg$^2$ area studied by \citet{wrig15a}, which is centred on the `trapezium' of O stars known as Cyg~OB2 \#8. It is clear however that the density of both O and B-type stars is non-zero across the entire Cygnus region. The background image is a mid-IR 12~$\mu$m image \citep{wrig10c} showing emission from the surrounding Cygnus~X giant molecular cloud, with prominent radio sources from \citet{down66} labelled.}
\label{cygnus_map}
\end{center}
\end{figure*}

The Cygnus~OB2 association is probably the most massive association known in our galaxy, with at least 50 O-type stars and a total mass of $\sim$16,500~M$_\odot$ \citep{wrig15a}. While it is more distant than the majority of the other associations discussed here, it has received a comparable level of attention thanks to its impressive contingent of massive stars and its position within the massive Cygnus~X star forming region that spans almost $\sim$200~pc \citep{reip08}. Figure~\ref{cygnus_map} shows the distribution of known OB stars in and around Cygnus OB2, showing a more concentrated distribution than many other OB associations due to its mass and relative youth. It was first identified as a group of early-type stars by \citet{munc53}, labelled as an O-association by \citet[][who obtained some of the first spectral types for the O stars in the region]{john54} and given the name VI~Cygni by \citet{schu56}. As the second of nine OB associations listed by \citet{rupr66} it became Cygnus~OB2 and recognised as the massive association that we know now by \citet{redd67}. 

The distance to the association was originally estimated to be about 1.8~kpc \citep{torr91,mass91}, but was brought down to $\sim$1.4~kpc by \citet{hans03} based on a revised effective temperature scale for OB stars. This was found to be in good agreement with radio parallax measurements towards multiple masers within Cygnus~X that suggested a typical distance of 1.4~kpc for the star forming region \citep{rygl12}. A recent study by \citet{berl19} used {\it Gaia} DR2 parallaxes to constrain the distance to the most massive members of the association, suggesting that the main Cyg~OB2 association was actually at 1.76~kpc, with a smaller, foreground population of massive stars at 1.35~kpc.

\subsubsection{The high-mass stars}

A considerable amount of attention has been focussed on the high-mass stars within Cyg~OB2 since the association provides one of the largest groups of O-type stars in our galaxy. The first detailed study of the high-mass stellar content of Cyg~OB2 was performed by \citet{mass91}, who spectroscopically identified 42 O-type and 26 B-type stars. The association however is highly obscured by the foreground Cygnus Rift, with the majority of its members having an extinction of $A_V = 4$--7~mag \citep{wrig15a}, and therefore further work to expand the census of the association was carried out in the infrared \citep[e.g.,][]{knod00,come02}. \citet{knod00} performed a photometric study of the association that suggested it might include $>$100 O-type stars, and while \citet{come02} and \citet{hans03} were able to verify many of these spectroscopically, their estimates of the high-mass stellar content in the association could not reach these levels.

The current estimate of 55--60 O-type stars \citep[e.g.,][]{wrig15a,berl18a} still makes it comparable to many of the most massive young clusters in our galaxy and yet Cyg~OB2 has the advantage that it is significantly closer and not as crowded, making source confusion less of a problem \citep[e.g.,][]{caba14}. Studies of the high-mass initial mass function in Cyg~OB2 have found evidence for the slope to be both flatter \citep{mass91} and steeper \citep{kimi07} than the canonical Salpeter value, though more recent studies that take a full census and account for the ageing of the population suggest it is consistent \citep[see Section~\ref{s-IMF} and][]{wrig15a}.

The recent census of Cyg~OB2 by \citet[][see Figure~\ref{cygnus_map}]{wrig15a} puts the total number of O-type stars currently in the association at $\sim$52, though new members are still being discovered \citep[e.g.,][]{berl18a,berl20}. This includes two stars in the rare O3 spectral class \citep{walb73,walb02}, the massive main sequence O5.5V star MT516, three Wolf-Rayet stars, and the B-type supergiant Cyg OB2 \#12, one of the most luminous and heavily reddened stars known. \citet{come07} have also identified a massive ($70 \pm 15$M~$_\odot$) runaway O4If star whose bow shock suggests it was ejected from Cyg~OB2 $1.7 \pm 0.4$~Myr ago. If this star was ejected when its binary companion exploded as a supernova \citep{blaa61} it also provides evidence that Cyg~OB2 has already seen its first supernova \citep[for a discussion of evidence for other past supernovae in Cyg~OB2 see][]{wrig15a}. \citet{berl20} performed a spectroscopic survey of the O-type stars in and around Cyg OB2, expanding the census of \citet{wrig15a} and extending it to the south-west of the association where an older population of massive stars had previously been identified \citep{come12}.

\begin{figure*}
\begin{center}
\includegraphics[width=15cm]{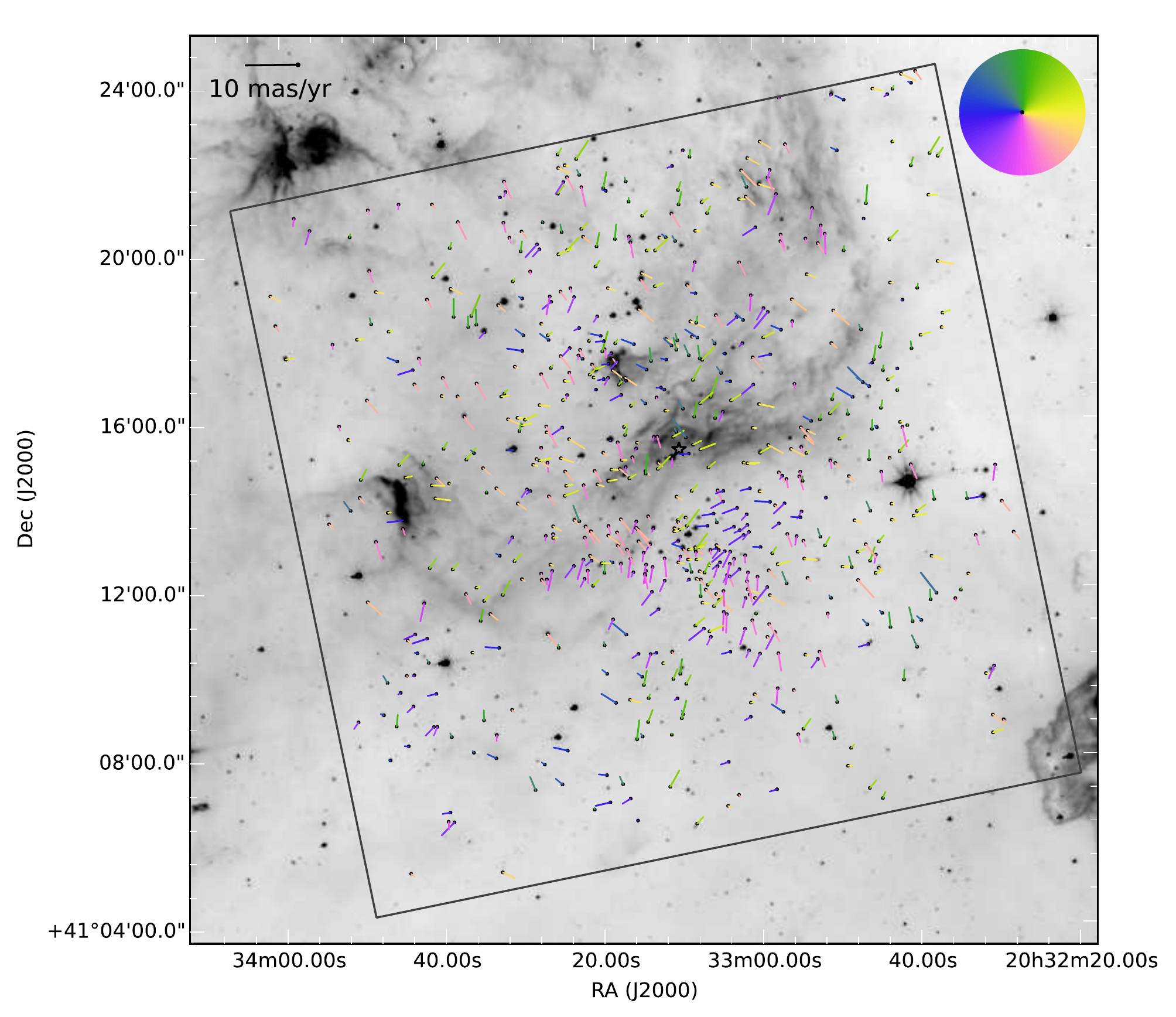}
\caption{Proper motion vector map for X-ray and spectroscopically selected young stars towards the centre of Cyg~OB2 (including 16 O-type stars and 34 B-type stars) from \citet{wrig16}. Dots show the current positions of the stars with the vectors showing their proper motions, colour-coded based on their direction of motion to highlight the kinematic substructure (colour wheel in the top-right corner illustrates this). Kinematic outliers were removed due to their large motions relative to most stars in the sample. The grey box shows the border of the region studied.}
\label{cygnus_substructure}
\end{center}
\end{figure*}

Thanks to the proximity of the association many of these stars are known to be in either close spectroscopic (SB1/2) or visual binary systems and have had their properties studied in considerable detail \citep[e.g.,][]{kimi09,kobu14,caba14}. This includes the massive O3If+O6V binary Cyg OB2 \#22 \citep{walb02}, the colliding-wind binary Cyg~OB2 \#8A \citep[O6I+O5.5III,][]{debe04}, and the quadruple system and contact-binary Cyg OB2 \#5 \citep[O7I + O6I + O9V + B,][]{dzib13}. Based on their accumulated sample, \citet{kimi12b} estimate an upper limit for the intrinsic binary fraction in Cyg~OB2 to be $90 \pm 10$\%, with $\sim$45\% having companions close enough to interact with during their lifetimes. \citet{caba20} performed a high angular resolution survey of 74 O and early B-type stars in Cyg OB2 and found that 47\% of targets had a resolved companion with a $<$1\% chance of not being a companion to the primary star. They noted a particularly large number of targets with wide (100--100,000 AU) binary companions. Including the known spectroscopic binaries from \citet{kobu14}, \citet{caba20} calculated lower limits to the multiplicity fraction and companion frequency of $0.65 \pm 0.05$ and $1.11 \pm 0.13$, respectively.

\citet{hans03} noted that many of the OB stars discovered at that time appeared to be older than the previously-accepted age of $\sim$2~Myrs for the association, and that these stars might represent a `halo' population distributed over a wider area than the stars studied by \citet{mass91}. Recent studies have widened the area surveyed around Cyg~OB2, uncovering many more massive stars over a wider area that are typically older than those in the centre of the association \citep[e.g.,][]{come08,come12,berl18a}. While some of these may be related to the nearby Cygnus~OB9 association, it does suggest that massive star formation has been occurring in Cygnus long before the formation of Cyg~OB2 and that an older population of massive stars now exists across the area \citep{come08,come16}.

\subsubsection{Age and age spread}

Early estimates of the age of the high-mass stars in Cyg~OB2 suggested an age of $2 \pm 1$~Myr, based on the presence of a well-defined upper main sequence and O5V and O5.5V stars \citep{mass91,hans03}. The presence of numerous evolved giant and supergiant stars in the area was seen as contamination from a non-coeval population within the region \citep{hans03}. Later studies of the A-type star population across the region \citep{drew08} and X-ray studies of the low- and solar-mass population in relatively compact areas of the sky \citep{wrig10a} suggested ages of 5--7 and 3--5~Myr, respectively, implying an older population.

Later studies of the high-mass stars in Cyg~OB2 have not distinguished between any young `central' group and an older more dispersed population \citep{come12}. The large census of OB stars by \citet{wrig15a} showed that when stellar ages for the high-mass stars are calculated for more-realistic rotating stellar models (rather than the classical non-rotating models) the ages of these stars shift from 1--4~Myr to 2--6~Myr. Furthermore, when the sample is limited to stars in the mass range of 20--35~M$_\odot$ \citep[at which masses no stars should have exploded as supernovae within 6~Myr,][]{ekst12}, then the star formation history appears broadly constant over the last $\sim$6~Myr, suggesting more or less continuous star formation over this time period across the association. Combining new spectroscopic, {\it Gaia} parallaxes, and by comparison with evolutionary models, \citet{berl20} studied the age distribution of the O-type stars in Cyg OB2 and found evidence for two possible star-forming bursts at $\sim$3 and $\sim$5--6 Myr.

\subsubsection{Structure and kinematics}

In their photometric study of Cyg~OB2, \citet{knod00} identify the association as spherically symmetric and suggest it to be considerably more massive than previously suspected, with a total mass of (4--10) $\times 10^4$ M$_\odot$. They argued that Cyg OB2 should not be classified as an OB association, but as a young globular cluster or super star cluster, but the size of the association ($\sim$30~pc) weakens the argument it is an analogue of the super-star clusters found in other galaxies. A number of studies have found the low- \citep{drew08,vink08} and high-mass \citep{come08} stellar population extends further from the association than previously recognised, particularly towards the south.

\citet{wrig14b} performed the first quantitative study of the 2D structure of Cyg~OB2 by using statistical diagnostics to measure substructure and mass segregation in the association, finding considerable physical substructure but no evidence for mass segregation. Both of these are indications that the association is not dynamically evolved (as mixing and two-body interactions destroy substructure and can lead to mass segregation through energy equipartition). {\it Gaia} DR2 parallaxes allowed \citet{berl19} to extend this work to the third dimension, revealing that the known OB stars constitute at least two separate groups along the line-of-sight. \citet{lim19} studied the spatial distribution of high-mass members of Cyg~OB2 and suggest the association is comprised of two dense clusters and a lower-density halo, though the clusters do not appear to be kinematically distinct and there is no evidence for such clusters in the spatial distribution of low-mass stars \citep[e.g.,][]{wrig14c}.

\citet{kimi07} performed the first kinematic study of Cyg~OB2, measuring radial velocities for 146 OB stars over a 6~year period, allowing them to identify binaries and measure systemic velocities. They derived a radial velocity dispersion of $8.03 \pm 0.26$~km~s$^{-1}$ \citep[see][]{kimi08b} from the systemic velocities of the systems studied (and therefore no further correction due to binarity was applied, noting that this was larger than found for most (typically less-massive) OB associations and open clusters.

\citet{wrig16} conducted a proper motion study of $\sim$900 low- and solar-mass stars in Cyg~OB2, finding dispersions of $13.0^{+0.8}_{-0.7}$ and $9.1 \pm 0.5$~km~s$^{-1}$ in RA and declination, respectively. In addition to producing a 3D velocity dispersion of $17.8 \pm 0.6$~km~s$^{-1}$ (implying that the association is gravitationally unbound) this also revealed the velocity dispersion to be significantly non-isotropic. \citet{wrig16} searched for evidence of expansion in Cyg~OB2, but could find no indication of a global expansion pattern. They also found that their proper motions exhibited significant kinematic substructure, as shown in Figure~\ref{cygnus_substructure}, which they argued echoed the physical structure already observed \citep{wrig14b} and also implied that the association was dynamically unevolved. They estimated that the kinematic subgroups they observed were close to virial equilibrium and had typical masses of $\sim$40--400~M$_\odot$.

\subsubsection{Feedback and the Cygnus Super-Bubble}

The Cygnus Super-Bubble, first detected by \citet{cash80}, is a large ($18^\circ \times 13^\circ$) soft X-ray emitting region that is roughly centred on Cyg~OB2 \citep[see also][]{boch85}. The bubble is believed to be a single massive shell, powered by both Cyg~OB2 and possibly the other OB associations in the region. \citet{murr10} associate one of their radio free-free emission sources with Cyg~OB2 and suggest the source is powered by the O stars in the association, similar to the bubble.

Whether the Cygnus super-bubble is a real, single structure has been a matter of debate for many decades. \citet{uyan01} studied the Cygnus super-bubble and argued that the X-ray emission observed was actually due to a superposition of separate regions at different distances. \citet{alba18} traced the diffuse X-ray emission across Cyg~OB2 that arises from stellar winds from the O-type stars in the association, which showed evidence for the interaction of these winds with the surrounding interstellar medium, and suggest that at least this emitting region is constrained to the area immediately around the association. \citet{kimu13} have however shown that the large-scale X-ray emitting regions in the super-bubble have similar absorbing columns and temperatures that suggests they originate from a single entity and therefore that the bubble is one large structure. Taken together these results imply that if the Cygnus super-bubble is a single structure with a single origin then it might not be related to Cyg~OB2 but to a previous generation of massive stars in the Cygnus complex.

Both \citet{cash80} and \citet{kimu13} have estimated the energy within the bubble to be $\sim$10$^{52}$ erg. \citet{cash80} estimate, based on the radiative criterion of \citet{cox72}, that if the $\sim$450~pc diameter shell were produced by a single, very-powerful supernova then this would require an injection of $10^{54}$ erg to produce such a structure, larger than currently observed in the bubble. Instead they suggest that if the energy were injected steadily over a longer period by a series of 30--100 SNe, then the energy required would be lower. Such a number of recent SNe would not be impossible over the entire Cygnus complex, though it is considerably more than could have originated from Cyg~OB2 itself \citep{wrig15a}. \citet{kimu13} suggested that this level of energy could instead be injected by 2--3 Myr of mechanical wind energy from the massive stars in Cyg~OB2, or alternatively from a hypernova, the explosion of a massive star composed mainly of carbon and oxygen, that \citet{iwam98} have hypothesised could release a sufficient amount of energy. \citet{higd13} estimate that the O stars in Cyg~OB2 could produce an ionizing flux of $10^{51}$ s$^{-1}$ (40 times larger than from the Orion OB1 association), irradiating the bubble and the surrounding Cygnus~X molecular cloud.

\subsection{Other prominent or well-studied associations}

\subsubsection{Perseus OB2}

\begin{figure}
\begin{center}
\includegraphics[width=8.5cm]{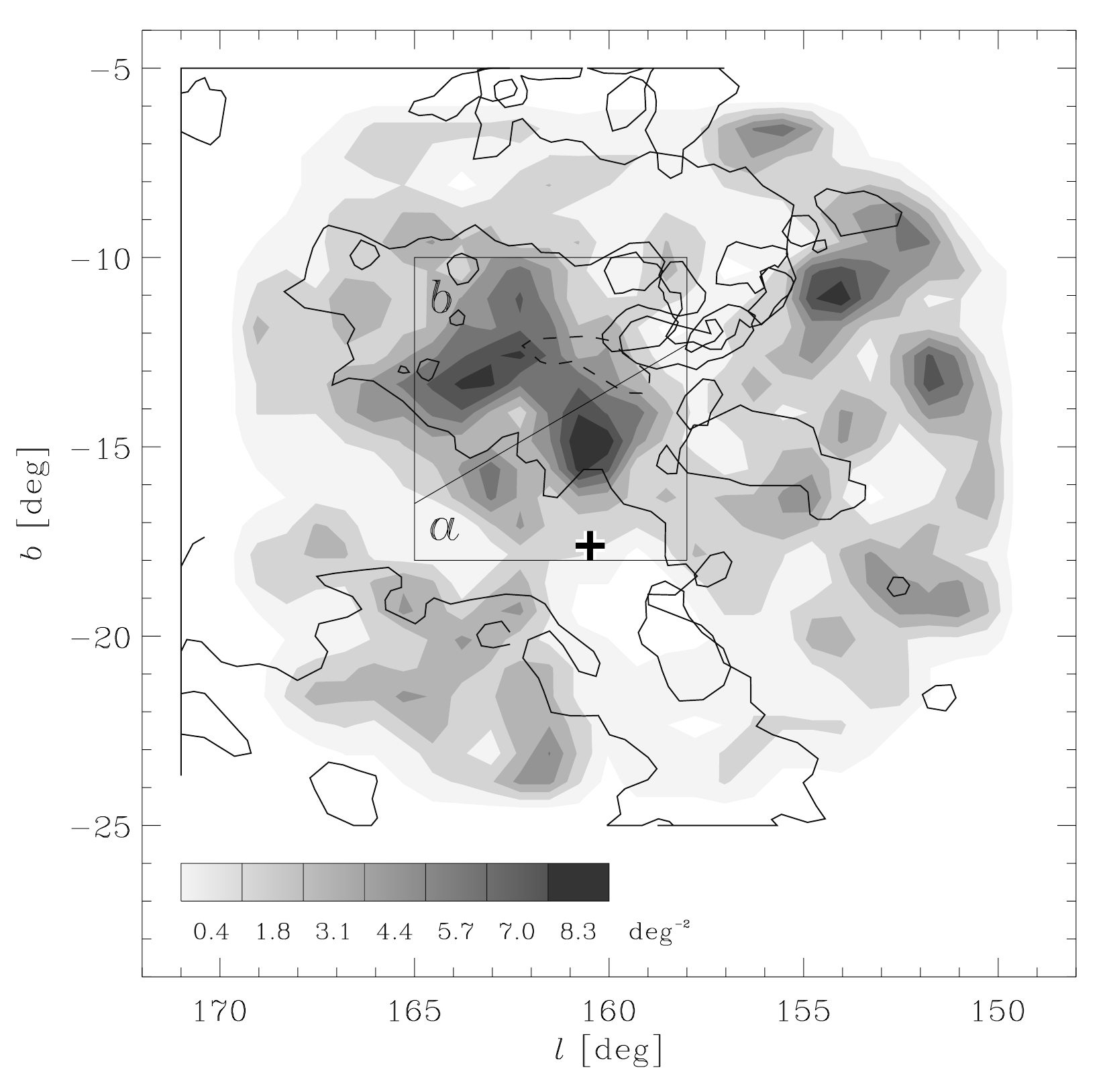}
\caption{Density map of probable members of the Perseus OB2 association from \citet{beli02b} with the main area of the association shown with a rectangle. The greyscale shows the surface density of members while the black solid lines show contours of CO emission. Also shown are the outline of the California Nebula (dashed line), the position of the young embedded cluster IC~348 (plus sign) and the approximate division within the association of the subgroups $a$ and $b$.}
\label{perseus_substructure}
\end{center}
\end{figure}

The Perseus OB2 association was first identified by \citet{blaa44} and is part of a series of star formation events in Perseus that includes multiple young clusters in the Perseus molecular cloud and the quiescent California Nebula \citep[for a review see][]{ball08}. While it is amongst the closest OB association to the Sun \citep[$d = 296 \pm 27$~pc,][]{deze99} it has significantly fewer massive stars compared to Sco-Cen, Orion or Vela, and therefore has not received as much attention. \citet{deze99} identified 17 B-type members of the association, including $\zeta$~Per (B1Ib) and 40~Per (B0.5V), the two brightest members of the association. The runaway O7IIIe star $\xi$~Per is thought to have been ejected from the association following a supernova explosion \citep{blaa61}, which may have played a role in powering a shell observed to be surrounding the association \citep{sanc74}. The total mass of the association has been estimated as $\sim$6000~M$_\odot$ \citep{beli02b}.

\citet{beli02a} and \citet{azim15} have identified many hundreds of candidate low- and intermediate-mass members of the association, tracing out its distribution from the classical region where the high-mass stars can be found out as far as the nearby California Nebula and the Auriga dark cloud (see Figure~\ref{perseus_substructure}. The association is also visible in the pre-MS density map produced by \citet{zari18}, but not in their upper MS density map due to the low number of massive stars. While almost spherical with a diameter of $\sim$40~pc there is evidence for multiple subgroups, both structurally \citep{herb98,beli02b} and from the age distribution of stars \citep{azim15}. \citet{beli02b} also observed an anti-correlation with the distribution of dust clouds in the almost co-spatial Perseus molecular cloud, which may suggest an extinction bias or that the association is excavating a cavity within the molecular cloud.

The association is approximately co-spatial with the Perseus molecular cloud and its embedded clusters IC~348 and NGC~1333 \citep{orti18}, and these are often considered part of Per OB2 given that they each contain two young B-type stars. \citet{azim15} trace a population of young stars aged $\sim$1~Myr within these star forming regions, as well as older populations aged 1--5 and $>$5~Myr, with the latter concentrated towards the north-west of the association.

There have been very few kinematic studies of Per OB2. \citet{stee03} calculated a radial velocity dispersion of $\sim$3.9~km~s$^{-1}$ from the known high-mass stars, but this was not corrected for the influence of binary motions, which will inflate the dispersion. \citet{blaa53} and \citet{lesh69} used proper motions from photometric plates to assess evidence for expansion within the association, calculating an expansion age of $\sim$1.3~Myr, but \citet{beli02b} could find no evidence for expansion from {\it Hipparcos} data.

\subsubsection{Carina OB1}

Despite being distant, the Carina OB1 association is prominent thanks to its very large number of OB stars \citep[e.g.,][]{mass93,walb95} and bright H{\sc ii} regions. The brightest of these is NGC 3572, the Carina Nebula, which sits broadly at the centre of the $4 \times 4$ degree area covered by the association's members \citep{hump78}. The total mass of the association is thought to exceed $2 \times 10^4$~M$_\odot$ \citep{prei11}. The association includes several massive star clusters (Trumplers 14, 15, \& 16, NGC 3293 \& 3324, Bochum 10 \& 11, and Collinder 228\footnote{Collinder 228 may appear as a separate cluster, but it is believed to be part of Trumpler 16, divided from it by an obscuring foreground dust lane.} \& 232), as well as a distributed population of young stars \citep{feig11}. The association, and the Carina Nebula itself, sit at a distance of $2350 \pm 50$ pc \citep{smit06}. The edge of the association is also projected against the massive star cluster Westerlund~2, though this is significantly more distant.

There are at least 70 O-type stars across the association, the majority of which are found in the dense clusters of the Carina Nebula. Of these, Trumpler~16 is by the wealthiest, home to multiple very massive stars such as $\eta$ Carina (a luminous blue variable), the very early-type stars HD 93205 (O3V+O8V) and Tr16-244 (O3/4 If), as well as many late O-type stars \citep{smit08b}. Trumplers 14 and 15 are also relatively massive and contain $\sim$16 O-type stars between them, the latter containing the O2 supergiant HD 93129A \citep{walb02b}. The association contains three late-type WN Wolf-Rayet star members \citep[WR22, WR24 \& WR25,][]{crow95}, and \citet{rate20} suggest that WR18 and WR23 may be possible members of the association as well. In addition to these there are a number of candidate OB stars awaiting spectroscopic confirmation \citep[e.g.,][]{povi11,mohr17}.

There are stars and clusters in the association with ages ranging from 1--10 Myrs. The youngest parts of the association are centred around the Carina Nebula, including the clusters Tr 14, 15 \& 16, aged approximately 1, 5, and 3 Myrs, respectively \citep{smit08b,hur12}. Trumpler 14, the youngest of the clusters, is kinematically associated with the molecular gas in its vicinity, consistent with its young age \citep{kimi18}. Beyond the Carina Nebula, there are some young clusters such as NGC~3324 \citep[1--2~Myr,][]{prei14}, as well as older clusters like NGC~3293 \citep[8--10~Myr,][]{prei17}.

\citet{kimi18} performed a radial velocity survey of the Carina Nebula's O-type stars and measured a velocity dispersion of $< 9.1$~km~s$^{-1}$ (not corrected for binaries), consistent with those measured for other similarly-sized and substructured OB associations such as Cyg OB2. The association, and particularly the well-studied young stellar population in the vicinity of the Carina Nebula, is highly substructured, both spatially \citep{feig11,buck19,reit19}, kinematically \citep{kimi18}, and temporally (see previous paragraph). \citet{meln17} measured the expansion of the association for the first time, deriving a kinematic age of $\sim$8~Myrs along both proper motion axes, consistent with the maximum age of stars in the association.

\subsubsection{Canis Major OB1}

\citet{amba49} was the first to propose the existence of an association of early-type stars in Canis Major, which is thought to contain over 200 B-type stars and a few O-type stars \citep{greg08} at a distance of $\sim$1.2~kpc \citep{zuck20}. The region is dominated by the large H~{\sc ii} region Sh2-296 (The Seagull Nebula), as well as numerous bright-rimmed clouds and an expanding shell of emission nebulae that \citet{reyn78} suggest was produced by stellar winds or supernovae (\citealt{fern19} suggest that Sh2-296 is part of a 60~pc diameter shell that surrounds CMa~OB1 that itself may be nested within a larger, 140~pc superbubble that is visible in H$\alpha$).

Within the boundaries of the association can be found the CMa R1 association and at least 19 open clusters \citep[e.g.,][]{dias02,khar13}, with the 7.6~Myr open cluster NGC~2353 \citep{fitz90} thought to be the nucleus of the association \citep{clar74b}. The brightest and earliest stars in the association are HD~54662 (O7III) and HD~54879 (O9.5V), as well as many early B-type stars \citep{clar74b}. There have been multiple runaway stars connected to the association, including HD~54662 \citep[][O7V]{herb77}, HD~57682 \citep[][O9V]{come98} and HD~53974 \citep{fern19}.

Recent surveys have begun to uncover the low-mass population of the association. \citet{fisc16}, \citet{sant18} and \citet{pett19} used infrared, X-ray and H$\alpha$ observations, respectively, to uncover hundreds of YSOs across the association and in nearby star forming regions. \citet{fisc16} find the spatial distribution of young stars in the association to be highly clumpy, with multiple groups and clusters within it. \citet{sant18} find that the age distribution of the young stars is bimodal, with an older population aged $\sim$10~Myr and a younger population localised to the regions of molecular gas, in agreement with the discovery by \citet{sewi19} of a new star forming region within the association with $\sim$300 very young stars within it.

\subsubsection{Monoceros OB1 and OB2}

The Mon OB1 and OB2 associations are projected close to each other in the constellation Monoceros at distances of $\sim$600 and $\sim$1200~pc, respectively. They each include a bright H~{\sc ii} region, which is the youngest part of each association, the Cone Nebula, the young cluster NGC~2264 and the R association Mon~R1 in Mon~OB1, and the Rosette Nebula and its central cluster NGC~2244 in Mon~OB2. Of the two associations Mon OB2 is richer and more massive, with at least 13 O-type stars \citep[e.g.,][]{john62,hump78}, including 5 in the cluster NGC~2244 that are responsible for illuminating and shaping the Rosette Nebula \citep{ware18}. These stars represent the most recent epoch of star formation in Mon~OB2, with evidence for other subgroups with ages of $\sim$5 and $\sim$15~Myr \citep{turn76,blit80}. Mon~OB1 is less massive, its most massive member being the O7V star S~Monoceros in the young young NGC~2264 cluster, though the region also shows evidence for multiple subgroups of different ages \citep{flac99}. There is also evidence for an older population of stars across Mon OB1 and a possible indication of post supernova activity from the presence of six arc-like molecular structures around the association \citep{oliv96b}.

\subsubsection{Scorpius OB1}

The Sco OB1 association is located on the near side of the Sagittarius spiral arm at a distance of $\sim$1.5~kpc \citep{vang84,sung98,yaly20}, and includes the massive young cluster NGC~6231 at its core, the open cluster Trumpler 24, and the H{\sc ii} regions G345.45+1.50 and IC 4628. The association contains a large number of OB stars, with 28 OB giants and supergiants listed in the census of \citet{hump78}, including two of the most luminous OB stars known, HD 151804 (O8Iaf) and HD 152236 (B1.5Ia), as well as multiple Wolf-Rayet stars.

\citet{dami18} performed a detailed photometric and astrometric study of the low-mass pre-MS population of Sco~OB1, uncovering $\sim$4000 candidate low-mass members that show a highly-substructured spatial distribution with several subgroups and clusters, including NGC~6231 and the young cluster Trumpler~24. They estimate the association has a total mass of $\sim$8500~M$_\odot$. \citet{yaly20} applied a clustering algorithm to stars in the direction of Sco OB1 and identified multiple groups with ages of 3--10~Myr at similar distances with an average of $1560 \pm 35$~pc. They identify these groups as part of Sco OB1, and note that they have similar ages and kinematics to NGC~6231.

NGC~6231 is projected against the centre of the association and is believed to be at the same distance \citep{vang84} and therefore related to it. It contains at least 15 O-type stars \citep{shob83}, including the 6th magnitude star V1007 Sco (HD~152248), an O7.5III+O7.5III binary. Notably, 11 of these are short period spectroscopic binaries, all but one of which have periods less than 10 days \citep{garc01}. The massive runaway star HD~153919 (O6.5Iaf), though 4 degrees away, is suspected to have been ejected from the cluster \citep{fein74}. \citet{dami16} performed an X-ray and photometric study of the cluster and find evidence for a significant age spread amongst the low-mass stellar population with ages of 1--8~Myr.

\subsubsection{Lacerta OB1}

Despite being relatively nearby \citep[$d = 368 \pm 17$~pc,][]{deze99}, Lacerta~OB1 is relatively sparse and therefore has not received significant attention or been highlighted on recent surveys of the solar neighbourhood \citep[e.g.,][]{bouy15}. Thought to be part of the Gould Belt, the association contains a number of B-type stars identified by \citet{deze99}, a single O-type star (10~Lac, O9V), and may have been the birth-place of the runaway B5V+F8V binary system $\nu$~Andromedae \citep{hoog01}. The association was originally divided into two subgroups, OB1a and OB1b, by \citet{blaa58}, though subsequent authors have questioned this \citep[e.g.,][]{deze99}. The age of the association has been estimated to be anything from a few to 25~Myrs \citep{chen08}, though 10~Lac's main-sequence lifetime of $\sim$3.6~Myr \citep{scha97} and the presence of a few small star forming regions (e.g., LBN~37 and GAL~110-13) suggests a large part of the association is at the low end of this distribution. \citet{blaa53} estimate an expansion age of 4.2~Myr, but the quality of data used for this was questioned by \citet{wool58}, and the most recent estimate of the expansion age by \citet{lesh69} is $2.5 \pm 0.5$~Myr.

\subsubsection{Ara OB1}

The Ara OB1 association, at a distance of $\sim$1.1~kpc \citep{moff73}, is a compact ($\sim$1~deg$^2$), but moderately-massive OB association containing at least 7 O-type stars \citep{hump78} and numerous B-type stars \citep[the B0 supergiant HD~156359 may have been ejected from the association,][]{maiz18}. The association contains multiple clusters, including the central cluster NGC~6193 \citep[which contains the O3.5III+O6IV binary HD150136 and the O6.5V star HD150135,][]{sota14} and the star-forming region RCW~108. The association is near a massive H~{\sc i} shell with a diameter of $\sim$100~pc, which is thought to have been formed by stellar winds or supernovae from the massive stars in the nearby, but older, cluster NGC~6167 \citep{arna87}, and may have played some role in the formation of Ara~OB1.

\subsubsection{The Serpens and Scutum OB associations}

The Serpens and Scutum regions contain a number of OB associations at distances of 1--2~kpc \citep[e.g.,][]{amba49,rupr66}, including Sct OB2 (1.6 kpc), Sct OB3 (1.3 kpc), Ser OB1 (1.5 kpc) and Ser OB2 \citep[1.6 kpc,][]{hump78,meln17}. Ser OB2 is probably the most massive of these, with over 100 OB stars identified by \citet{forb00}, including the massive O stars HD 168112 (O5.5IIIf) and HD 167971 \citep[O8Ibf,][]{hump78}, which illuminate the nearby H~{\sc ii} region S54 and likely fuel a thermal `chimney' that extends 200~pc perpendicular to the plane \citep{forb00}. The association is thought to be associated with the young cluster NGC 6604, aged $\sim$4~Myr \citep{khar05}. \citet{deze99} used {\it Hipparcos} data to study the bright stars in this region of the sky, but could not find any kinematic signature of any of these associations, albeit from only 62 stars spread across the area.

\subsubsection{The Cygnus OB associations}

The Cygnus region is rich with young and massive stars, which \citet{hump78} divided up into 9 OB associations at distances of $\sim$1--2~kpc \citep{uyan01}. While Cyg~OB2 is probably the most massive and certainly the most well-studied of these, the others are still quite rich. Cyg~OB1 is likely the second most massive, with 12 O-type stars and at least 50 OB stars \citep{hump78,hump84} spread over an area of $4 \times 3.5$ degrees, and containing multiple open clusters such as NGC 6913 and IC 4996. \citet{cost17} analysed the spatial and kinematic structure of stars in Cyg~OB1 and were able to identify two distinct kinematic groups at different distances, indicating substructure within the association.

\citet{bran75} found a ring of H$\alpha$ emission approximately 4$^\circ$ in diameter centred on the Cyg OB1 association that probably formed by feedback from the OB stars and \citet{sitn19} have identified ongoing star formation on the edge of this shell. Of the other associations, Cyg~OB3 is notable for containing a number of early O-type stars, including the O4 star HD 190429, and the cluster NGC 6871 at its nucleus. \citet{deze99} studied Cyg OB4 and Cyg OB7, the only two associations in Cygnus closer than 1~kpc, but couldn't find any evidence of common motion and suggested they may not be real groups. Further work is certainly needed to disentangle the large number of OB stars in Cygnus.

\subsection{Notes on other associations}

Many other associations have been suggested or identified, though the data on these groups is often limited or out-dated. Some notes are provided here for reference.

\begin{itemize}
\item {\bf Cas-Tau}: First identified by \citet{blaa56} as a loose group of OB stars related to the cluster $\alpha$~Per, at distances of 125--300~pc and aged $\sim$50~Myr \citep{deze99}. The association spans an area of approximately $100^\circ \times 60^\circ$ on the sky. The association is kinematically similar to the (albeit younger) Tau-Aur T association discovered by \citet{walt88}, which is slightly in the foreground of Cas-Tau and related to the nearby Taurus molecular cloud. \citet{deze99} identified 83 members of the Cas-Tau association and found them to have a similar motion to the $\alpha$~Per cluster within the association.
\item {\bf Perseus OB3 (and $\alpha$ Per)}: Perseus OB3 was first discovered as the $\alpha$ Per moving group but listed as Per OB3 by \citet{rupr66}. The association extends beyond the compact centralised $\alpha$~Per cluster as shown by \citet{deze99} who identified 79 members that formed by a central compact cluster and an extended halo that makes up the association. At a distance of $174.9 \pm 3.0$~pc \citep{babu18} the cluster and association are embedded within the Cas-Tau association and form the nucleus of it.
\item {\bf Collinder 121}: An approximately 5~Myr old association with evidence for at least 2 subgroups and a size of $100 \times 30$~pc at a distance of $\sim$540~pc \citep{deze99}. \citet{heil98} identified a superbubble in the direction of Collinder~121 that may be connected to the association, though with a kinematic expansion age of $\sim$20~Myr it would more likely originate from an older generation of stars than those prominently observed within the association. 
\item {\bf Camelopardalis OB1}: Appearing in Ruprecht's (1966) list of associations at a distance of $\sim$800~pc \citep{meln17}, Cam OB1 contains two O8.5-O9 stars as well as at least 35 known early B-type stars. The two A-type supergiants HD 21291 and HD 21389 are sometimes considered part of Cam OB1 or the nearby Cam R1 association \citep{raci68}, the latter being sometimes consumed within Cam OB1. The {\it Hipparcos} PMs did not allow \citet{deze99} to verify the existence of a moving group, but the existence of both O and early B-type stars, variable stars and H$\alpha$ emission-line stars \citep{stra07} and the open cluster NGC~1502 suggests considerable recent star formation in this region. 
\item {\bf Cepheus OB2}: The most prominent of three OB associations in Cepheus at distances of 600-900~pc, Cepheus~OB2 includes 76 members identified by \citet{deze99} and is associated with the open clusters NGC~7160 and Trumpler~37, as well as the H~{\sc ii} region IC~1396 \citep{schu97,bare11}. The association falls within a large, $\sim$120~pc ring-like feature first identified from H$\alpha$ photographs \citep{siva74} and since seen at other wavelengths.
\end{itemize}

\section{Properties of OB associations}

In this section we review the global properties of OB associations, introducing observations of systems not discussed in Section~\ref{s-individual} and, where possible, focussing on global studies that compared the properties of different OB associations to understand their key features.

\subsection{Size and internal structure}
\label{s-structure}

\begin{figure}
\begin{center}
\includegraphics[width=8.5cm]{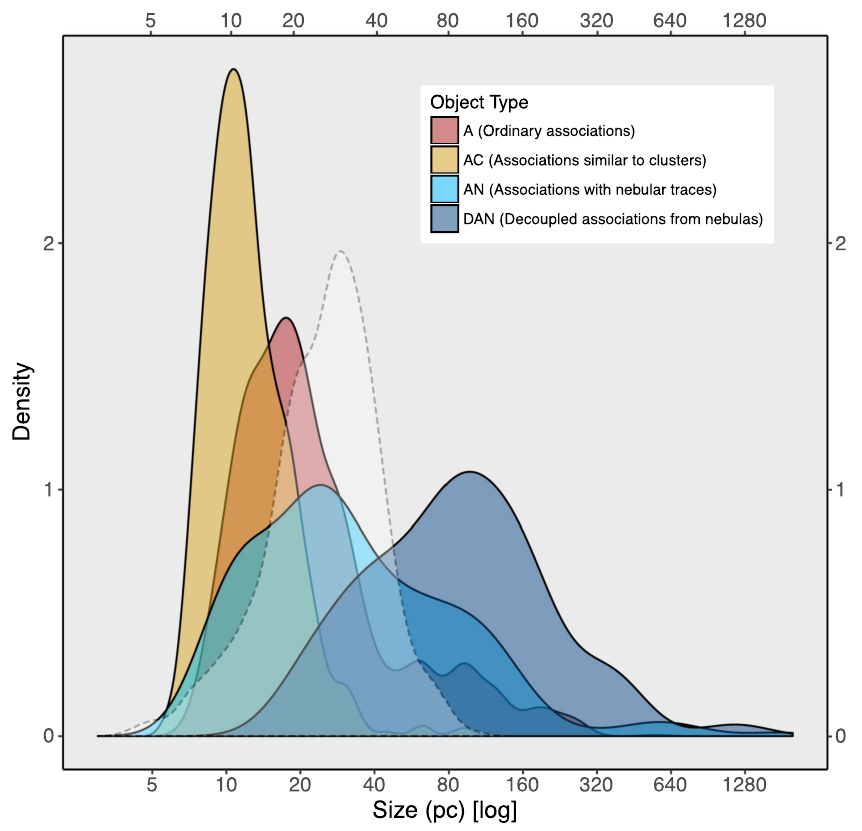}
\caption{Size distribution for different types of OB association in the Large Magellanic Cloud from \citet{goul18} based on the sample of \citet{bica08}. The typical size for associations in each category are 13~pc (associations similar to clusters), 33~pc (ordinary associations), 63~pc (associations with nebular traces) and 160~pc (decoupled associations from nebulae).}
\label{gouliermis_sizes}
\end{center}
\end{figure}

Measuring the sizes of OB associations has always been a difficult task, due to a combination of the difficulty of defining the boundaries of a given association (that can be particularly vague at the largest scales) and the complexity of defining what an association is. As such, catalogued associations range in size from $\sim$10 to several hundred parsecs \citep[e.g.,][]{blaa64,blah89}. Even the typical sizes of OB associations can vary from study to study due to different approaches to defining their borders and membership, e.g., \citet{garm92} measure a mean size of $137 \pm 83$~pc, while \citet{meln95} measure an average diameter of $\sim$40~pc. Extra-galactic studies have found similar size distributions, albeit with some resolution issues. For example \citet{luck70} catalogued 122 OB associations in the Large Magellanic Cloud (LMC) with typical sizes of 15--150~pc and an average of 80~pc, while \citet{goul18} used the better-resolved catalogue of associations in the LMC from \citet{bica08} to measure an average size of 30~pc.

There is strong evidence that the sizes of associations increase with age \citep{blaa64}, which has been interpreted as evidence that the systems are expanding. There is also evidence that associations not connected to nebular material are larger, as shown in Figure~\ref{gouliermis_sizes} \citep{goul18}. Given that the association with nebular material is considered an age proxy (because stars either drift away from the molecular cloud they formed from or disperse it over time), this also suggests that associations increase in size as they age.

The majority of OB associations show considerable internal substructure, which can take various forms such as dense clusters within them \citep{amba49}, or subgroups with different ages and kinematics \citep{blaa64,garm92}. The most well-studied associations have been sub-divided by various authors into ``OB subgroups'', most famous of which are the division of Sco-Cen into its three subgroups (Upper Sco, Upper Centaurus-Lupus and Lower Centaurus-Crux) and Orion OB1 into its four subgroups \citep{blaa64}. These definitions have historically been based on the plane-of-the-sky distribution of stars within the association, but more recent studies have incorporated parallaxes and proper motions to subdivide these structures \citep[e.g.,][]{koun18,cant19b}. Many OB associations also contain open or embedded clusters within them, such as $\gamma$~Vel in Vela~OB2 \citep{jeff14} or NGC~2353 in CMa OB1 \citep{fitz90}, with ages consistent with being part of them.

Large-scale studies of OB associations, both in the Milky Way and in nearby galaxies, have also shown that OB associations typically have some bright central concentration within them \citep{ivan87,meln95}. In some cases this substructure is not obviously apparent but can be revealed through structural studies that quantify physical substructure \citep[e.g.,][]{cart04,schm08,wrig14b}. This substructure has contributed to the difficulty in explicitly defining the boundaries and therefore the scales of OB associations.

\subsection{Velocity dispersion and virial state}

The velocity dispersions of OB associations are typically a few kilometres per second, though there is quite a large variation. For example, \citet{meln17} study 18 OB associations with {\it Gaia} DR1 and measure an average 1D velocity dispersion of 3.9~km~s$^{-1}$ (3.7~km~s$^{-1}$ after correcting for the influence of binaries on the measured proper motions),while \citet{ward18} also study 18 OB associations with {\it Gaia} DR1 and measure velocity dispersions of 3--13~km~s$^{-1}$ with a median of 7~km~s$^{-1}$. \citet{meln20} studied 28 OB associations with {\it Gaia} DR2 data and measure an average velocity dispersion of 4.5~km~s$^{-1}$. These differences do not originate from the data or the method, but more likely from the OB association membership lists, many of which date from the 1980s \citep[e.g.,][]{hump78,blah89} and may need revisiting.

OB association velocity dispersions typically show evidence for significant levels of anisotropy, with velocity dispersion ratios between the two proper motion axes up to $\sim$6 \citep{meln17}, though the median ratios are typically around $\sim$1.5 \citep{wrig16,meln17,ward18}. In 3D kinematic studies a similar trend is found, for example \citet{wrig18} found that for the three subgroups of Sco-Cen the velocity dispersions along the three axes had ratios of $2.2 : 1.4 : 1$, $2.7 : 1.7 : 1$ and $3.7 : 2.1 : 1$. There is some evidence that when association subgroups are identified using kinematic methods (such as the $k$-means clustering tool UPMASK) the velocity dispersion ratios are smaller \citep[e.g.,][find smaller ratios of 1.0--1.6 for the subgroups in Vela OB2]{cant19a}. This may indicate a kinematic bias that has been introduced by this kinematic subgroup identification method, or potentially that association subgroups are truly isotropic and that tools like this are a better way to identify them than traditional methods based on plane-of-the-sky positions.

From the measured velocity dispersions, and with some estimation of the total stellar mass of the association (often extrapolated from the high-mass population), the virial state of the association can be estimated. Since OB associations have a low stellar density compared to gravitationally bound open clusters it is not surprising that all OB associations have been found to be super-virial. \citet{meln17} find that estimates of the ratio of the virial mass to the stellar mass typically range from 10 to 1000 with a median of $\sim$50, similar to estimates from studies of individual associations \citep[e.g.,][]{wrig18}. Note that, as long as no additional forces are acting on the members of the association then the virial mass increases proportional to the radius of the association as it expands. If the expansion is assumed to be linear then the virial mass will increase linearly with time and as the stellar mass remains constant the virial to stellar mass ratio will increase with time as well. There are suggestions of this in Sco-Cen where the young Upper Sco subgroup has a virial to stellar mass ratio of 40 and the older Upper Centaurus Lupus and Lower Centaurus Crux subgroups have ratios of $\sim$70 \citep{wrig18}.

\subsection{Expansion}
\label{s-expansion}

\begin{figure*}
\begin{center}
\includegraphics[width=16.8cm]{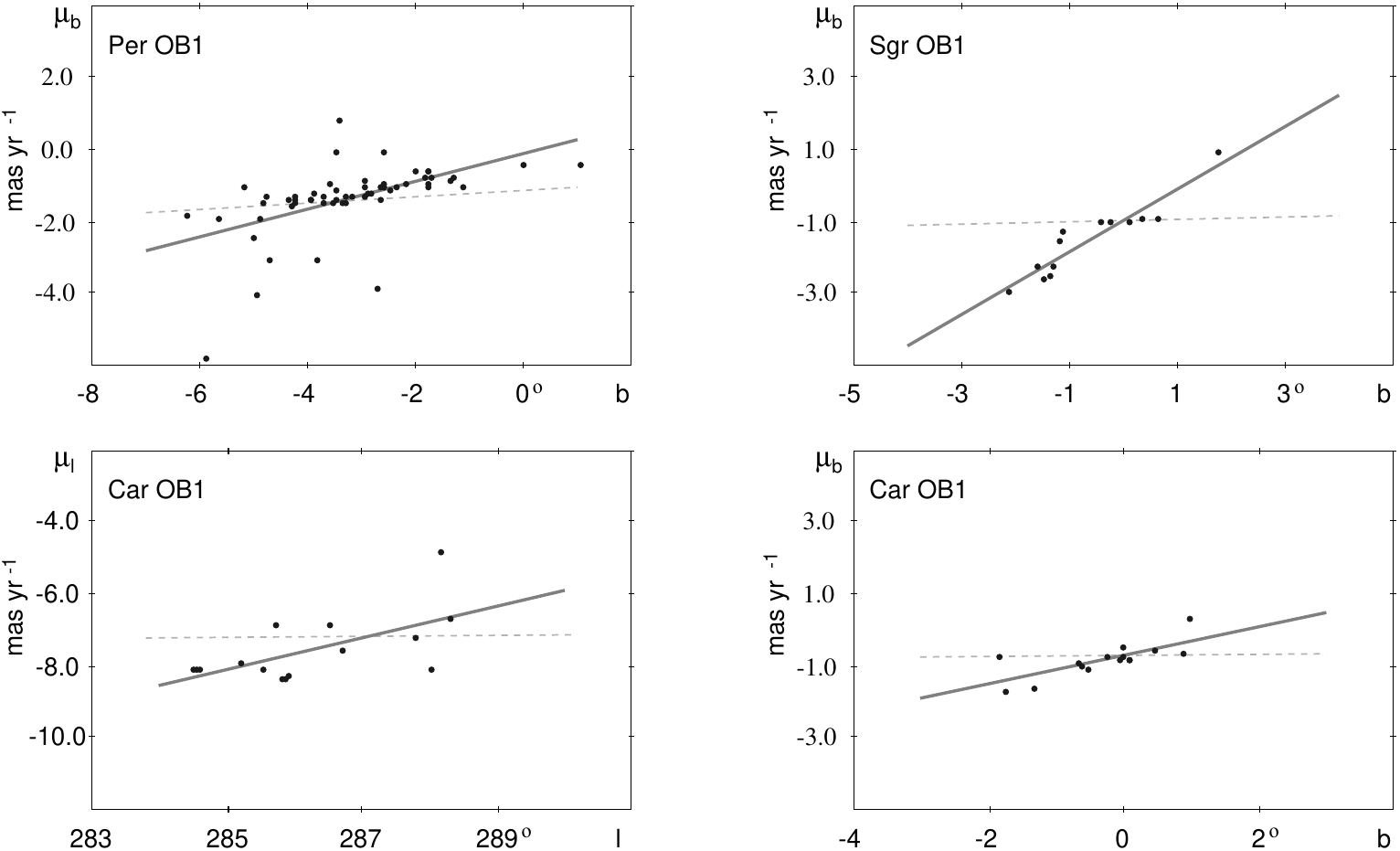}
\caption{Evidence for expansion of the OB associations Per OB1, Sgr OB1 and Car OB1 from \citet{meln17}. Proper motions in Galactic coordinates ($\mu_l$ or $\mu_b$) from {\it Gaia} DR1 are plotted against the corresponding coordinates ($l$ or $b$) for members of these associations from \citet{blah89}. The solid lines show the best fit linear relation between the two quantities, with a positive slope indicating expansion. The dashed line shows the contribution to this apparent expansion that would be expected due to virtual expansion of the association from its radial motion.}
\label{melnik_expansion}
\end{center}
\end{figure*}

Ever since the first studies of OB associations it has been known that they cannot be gravitationally bound and therefore will expand and disperse in the future \citep[e.g.,][]{amba47,amba49,blaa52}, which led many to suggest that they may already have expanded to some extent and would therefore have been more compact in the past \citep[e.g.,][]{amba47,blaa64,lada03}.

This work on the expansion of OB associations led to the development of the linear expansion model, which assumes that OB associations have a small initial size (compared to their present sizes) and have expanded linearly (in time) from a compact configuration \citep{blaa46,blaa64}. Later models took into account the Galactic potential, for example using the Galactic epicycle theory \citep{maka04} or considered the spiral arms and bar components \citep[e.g.,][]{fern08}. Testing this model has been a goal of many studies of OB associations over the last half century, though the difficulties of measuring sufficiently precise proper motions or the lack of radial velocities (needed to account for the radial streaming motions of nearby associations) has limited such work and many studies have not found evidence for expansion in their targets \citep[e.g.,][]{fern08}.

Indirect evidence for the expansion of OB associations comes from observations of a correlation between the radius and density of star clusters and associations, which has been interpreted as evidence that many OB associations were originally more compact, similar to embedded clusters such as the Orion Nebula Cluster \citep{pfal09}. However, many early kinematic studies of OB associations and moving groups found them to have internal velocity dispersions that are too small to explain their present-day size by expansion from a significantly more compact state \citep[e.g.,][]{prei08,torr08}.

One of the difficulties of measuring the expansion of OB associations is that nearby associations can exhibit {\it virtual expansion}, whereby the radial motion of the association towards (away from) the observer can cause an apparent expansion (contraction) on the sky, even when no physical expansion exists \citep{blaa64}. To overcome this one must have some knowledge of the radial velocity of the association, either to perform a simple correction for the virtual expansion \citep{brow97}, or if individual radial velocities are present to test the \citet{blaa64} linear expansion model or (if parallaxes are also available) convert to a Cartesian coordinate system and test the expansion in 3D \citep[e.g.,][]{wrig16}. This was a significant issue for early studies of OB associations, but has generally been overcome thanks to the availability of RVs for stars in most associations.

Searches for evidence of expansion in OB associations using {\it Gaia} data have been mixed. Early studies using {\it Gaia} DR1 found that most OB associations do not show evidence for a coherent radial expansion pattern. \citet{meln17} studied 18 OB associations and found that only three showed evidence for significant ($>$2.5$\sigma$) levels of expansion along one or more axes (Sgr OB1, Per OB1 and Car OB1, see Figure~\ref{melnik_expansion}), though for the first two of these the expansion was only along one axis. Furthermore, three other OB associations actually showed significant evidence of contraction rather than expansion. \citet{meln20} extended this study using {\it Gaia} DR2, confirming their previous results and identifying expansion in Ori OB1, Gem OB1 and Sco OB1, all of which are significantly anisotropic. In contrast, \citet{ward18} searched for expansion in 18 OB associations \citep[5 overlapping with the sample of][]{meln17}, but could not find evidence for expansion in any of their targets. In their study of Sco-Cen, \citet{wrig18} could not find evidence for a coherent 3D expansion pattern in any of its subgroups, though all three subgroups did show significant expansion along the Galactic $Y$ axis (the direction of Galactic rotation, see Figure~\ref{wright_3Dexpansion}). \citet{ward20} studied the kinematics of 110 OB associations selected from {\it Gaia} DR2 data and found that their properties were not consistent with either a monolithic, radial expansion pattern or completely random velocities, but were instead best reproduced by a highly substructured velocity field with some degree of localised expansion from subclusters within the association.

Better evidence for expansion has been obtained from studies that divide OB associations into subgroups based on their spatial and kinematic substructure, rather than just the positions of stars on the plane of the sky. \citet{koun18} find evidence for expansion of their Orion D subgroup, though the significant spatial and kinematic substructure within it suggests it is not expanding from a single compact cluster but from multiple subclusters. \citet{cant19b} find that all 7 of the groups they identify within the Vela-Puppis region are expanding, \citet{arms20} find that the Vela OB2 association is clearly expanding once members of the Gamma Vel cluster are removed from the sample.

In younger, more compact systems (such as star forming regions and star clusters), which may represent the precursors of expanded OB associations, the evidence for expansion is equally mixed. \citet{dzib17} could find no evidence for expansion of the Orion Nebula Cluster (ONC) from their proper motion study, while \citet{dari17} found a correlation between radial velocity and extinction in the ONC that could be caused by expansion. \citet{koun18} found a clear radial expansion pattern in the proper motions of stars in the $\lambda$ Ori cluster and argued it may be due to a supernova. \citet{kuhn19} found that at least 75\% of the young clusters in their sample showed evidence for expansion in the form of positive median outward velocity.

For systems where expansion has been observed there is strong evidence that the expansion is not isotropic. For the three associations with significant levels of expansion identified by \citet{meln17}, two are anisotropic, with only Car OB1 showing isotropic expansion. \citet{wrig18} found that all three subgroups of Sco-Cen exhibited strong expansion along the Galactic $Y$ axis but not along the other two axes (Figure~\ref{wright_3Dexpansion}), while \citet{cant19b} and \citet{arms20} found significant differences in the expansion rates along different axes for the majority of groups they identified in Vela-Puppis. The young cluster NGC~6530 was also found by \citet{wrig19} to have a very strong level of asymmetry in its expansion pattern.

\citet{zamo19} perform magnetohydrodynamic simulations of the radiative feedback-induced expansion of H~{\sc ii} regions around star clusters. As the expanding shells of gas carry the majority of mass in the system, they also drive the gravitational potential around the cluster, which can accelerate the cluster expansion following residual gas expulsion. In the simulations presented by \citet{zamo19}, neither the parental clouds, nor the expanding shells of gas, are symmetrically distributed around the cluster, and therefore the gravitational potential and resulting expansion need not be symmetric. Whether this asymmetry is high enough to explain the strongly asymmetric expansion patterns observed in OB associations remains to be seen.

In summary, the evidence for expansion of OB associations is mixed, with some studies finding evidence or hints of expansion and others not, though there are indications that a better-informed division of associations into subgroups is producing groups that do show evidence for expansion. For all studies however there is clear evidence that the majority of expanding systems are doing so asymmetrically and there is currently very little evidence for the simple picture of isotropic radial expansion.

\subsection{Kinematic ages of OB associations}
\label{s-kinematic_ages}

\begin{figure*}
\begin{center}
\vspace{-6cm}
\hspace{-1cm} \includegraphics[height=14cm]{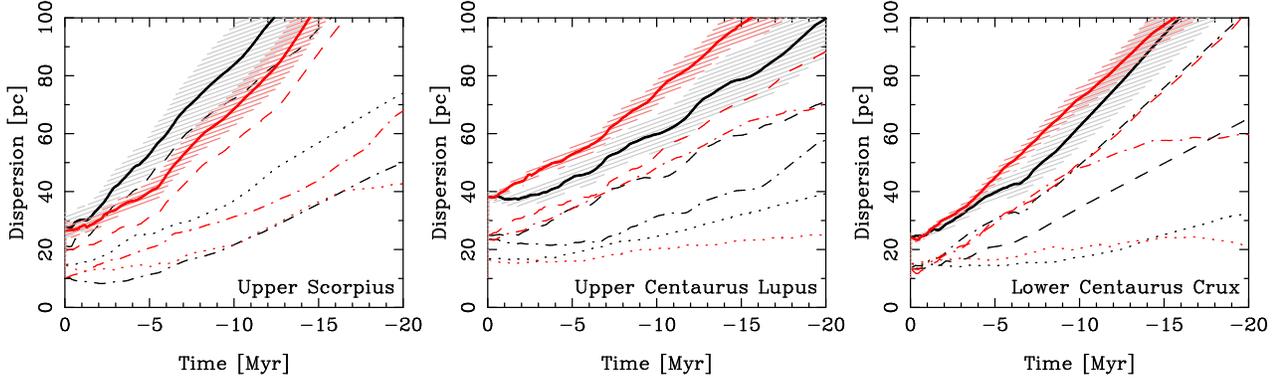}
\vspace{-2cm}
\caption{Application of the 'traceback' approach to the three subgroups of Sco-Cen to determine the smallest volume they occupied in the past and therefore their kinematic ages. The lines show the size (estimated from the 1$\sigma$ dispersion) of each subgroup along each axis of the $XYZ$ coordinate system ($\sigma_X$ shown with a dashed line, $\sigma_Y$ with a dotted line, and $\sigma_Z$ with a dash-dotted line) as well as the quadrature sum of all three dimensions (solid line). Black lines show the results of traceback analysis performed with linear trajectories and the red lines show the analysis performed using the epicycle orbit approximation. Shaded areas around the 3D sum lines show the 90\% confidence interval in the quadrature sum dispersion of each subgroup determined from Monte Carlo simulations exploring the impact of uncertainties in the velocities. That none of the subgroups appear to have been significantly more compact in the past suggests they are not expanding from a more compact configuration. Figure from \citet{wrig18}.}
\label{wright_traceback}
\end{center}
\end{figure*}

Kinematic ages for young, unbound stellar systems such as OB associations are derived by calculating the time the system needs to have expanded, at its current rate, to reach its present size from an initially compact configuration. They can be derived simply by tracing back the motions of a group of stars into the past to estimate when they occupied the smallest volume \citep[e.g.,][see Figure~\ref{wright_traceback}]{blaa78,maka07,wrig18} or alternatively by measuring the linear expansion rate (see e.g., Figure~\ref{melnik_expansion}) and size of the system and thus calculating its age \citep[e.g.,][]{lesh68,peca12}. More recently studies have incorporated orbital integration \citep{mire18} or forward-modelling techniques \citep{crun19}. Since kinematic ages don't rely on any stellar physics (only on astrometry) they are sometimes argued to be more reliable than other age estimates, though they do require a number of assumptions that have yet to be fully tested.

Kinematic ages have been calculated for many OB associations, as well as for nearby moving groups. For many decades these have disagreed with nuclear ages. \citet{blaa84} noted that the majority of kinematic ages were smaller than the nuclear ages known at the time and suggested it might be because OB associations were originally not much smaller than we see them now, making the assumption of expansion from a smaller volume incorrect. In contrast, for moving groups most kinematic ages calculated during the 2000s were greater than the pre-MS evolutionary ages \citep{sode10}, though the latter were calculated before the recent revision of pre-MS evolutionary ages \citep[e.g.,][]{bell13}, which could explain the disagreement. The availability of {\it Gaia} DR2 astrometry coupled with improved kinematic age calculation methods is showing some improvement; the $\beta$ Pic moving group was historically measured to have a kinematic age of 11--12~Myr \citep[][but see also \citealt{mama14}]{orte02}, in disagreement with the lithium depletion boundary age of $21 \pm 4$~Myrs \citep{bink14}, but the kinematic age has since been revised to $18.3^{+1.3}_{-1.2}$~Myr \citep{crun19}, which is now consistent with the latter age.

Despite being a model-free estimate of stellar age there are many reasons to be wary of kinematic ages. Firstly it relies on very good membership information for the systems being studied. For example many different kinematic ages have been estimated for the TW~Hya moving group, ranging from $4.7 \pm 0.6$~Myr \citep{maka05} to $\sim$8.3~Myr \citep{maka01,dela06} and even a 95\% confidence lower limit of 10~Myr \citep{mama05}, all depending on the choice of members used to calculate the kinematic age. There are also many assumptions necessary to apply the technique, including that the time the stars were in closest proximity to each other was the time of their formation \citep[for example some of the stars in the expanding $\lambda$ Ori system appear to be much younger than the kinematic age of the system, suggesting they may have formed from material that was already expanding or was triggered into forming stars,][]{koun18}, that large OB associations have a single age (questionable given the evidence for age spreads within many associations, see Section~\ref{s-ages}), or that OB associations originate from a compact volume of space (if not, that would prevent a kinematic age being easily measurable, see Figure~\ref{wright_traceback}), an issue first noted by \citet{blaa64} -- for more details see the evidence for physical substructure in Section \ref{s-structure}, which argues against a single origin of expansion. The observation of anisotropic expansion is also difficult for the concept of kinematic age dating as it often leads to different ages derived in different dimensions, or potentially a system that is expanding along one axis but contracting along another. 

\citet{brow97} performed N-body simulations to explore how astrometric uncertainties and the different methods used can affect the derived kinematic age. They find that the traceback method of estimating the smallest configuration of an expanding association both underestimates the age and overestimates the initial size of the association, with ages typically converging to $\sim$4~Myr. The alternative approach of comparing velocity with position in a given dimension also leads to considerable uncertainty. Forward-modelling techniques, such as those used by \citet{crun19}, have the potential to overcome some of these biases, but can be computationally time-consuming for large systems.

\subsection{Ages and age spreads}
\label{s-ages}

Most OB associations have ages between 1 and $\sim$20~Myrs \citep{blaa64}, with some containing very young, embedded stars (e.g., Cepheus OB2) and others that have long since dissipated their progenitor clouds (e.g., Sco-Cen and Vela OB2). This age range is effectively defined by our ability to observe and define an association. Younger systems are likely to still be embedded in their parental molecular cloud (and therefore difficult to observe), while older systems will lack the defining O and early B-type stars and will have expanded to the point that (historically at least) their densities are too low to clearly identify. The latter limitation for identifying an OB association is changing thanks to {\it Gaia}, allowing more dispersed systems lacking early-type members to be identified \citep[e.g., see the older parts of the Vela--Puppis groups identified by][]{cant19b}.

The ages of OB associations are usually determined either from the position of low-mass stars on the pre-MS tracks in the Hertzsprung-Russell or colour-magnitude diagrams, the positions of high-mass stars turning off the main sequence, or from kinematic expansion ages (see Section~\ref{s-kinematic_ages}). Evolutionary ages are model-dependent \citep{hill08} and many models have problems predicting the radii (and therefore luminosities) of low-mass stars \citep[e.g.,][]{krau15,feid16}. For high-mass stars the effects of rotation or binarity can also complicate age estimates \citep[e.g.,][]{ekst12}. Absolute ages can therefore be biased and are less accurate than relative ages.

Various effects in a single-age population \citep[such as variability, binarity or non-uniform accretion, e.g.,][]{burn05} are known to lead to luminosity spreads in the Hertzsprung-Russell diagram (at a given temperature) that can be interpreted as an age spread, even when none exists \citep[see the discussion in Section~6 of][]{sode14}. Many OB associations exhibit age gradients or have substructures with different ages, and these can appear as age spreads if the structure of the association is not sufficiently resolved. For example, \citet{peca12} measure an age spread of $\pm$4--7~Myr for Upper Centaurus Lupus, which \citet{peca16} resolved as age substructure using higher spatial resolution measurements. \citet{peca16} found clear age substructure in all three subgroups of Sco-Cen, noting that when integrated this appears as age spreads of $\pm$6--7~Myr in each subgroup (see Figure~\ref{pecaut_agemap}). Across Sco-Cen this age substructure broadly manifests as an age gradient from the north (youngest) to the south (oldest).

In Orion OB1, the individual subgroups have been known to have different ages for many years \citep[e.g.,][]{bric05}, with ages varying from 1--10~Myr in the different subgroups. Recently \citet{koun18} traced an age gradient along their Orion~C subgroup, with ages extending from $\sim$2~Myr in the south to $\sim$7.5~Myr in the north. In Vela--Puppis \citet{cant19b} identified seven populations based on their kinematics with relatively distinct ages that varied from $\sim$10~Myrs (for the main component of Vela~OB2) to 40--50~Myrs for the oldest groups (which includes the Trumpler 10 cluster and Collinder 135). Finally, in Cygnus~OB2 \citet{drew08} and \citet{wrig15a} found evidence for a large spread of ages of 1--7~Myrs from the ages of the O, B and A-type stars across the association, with evidence for older populations away from the main association centre \citep{berl18a,come16}.

\subsection{Distribution of OB associations in the galaxy}

\begin{figure*}
\begin{center}
\includegraphics[width=18cm]{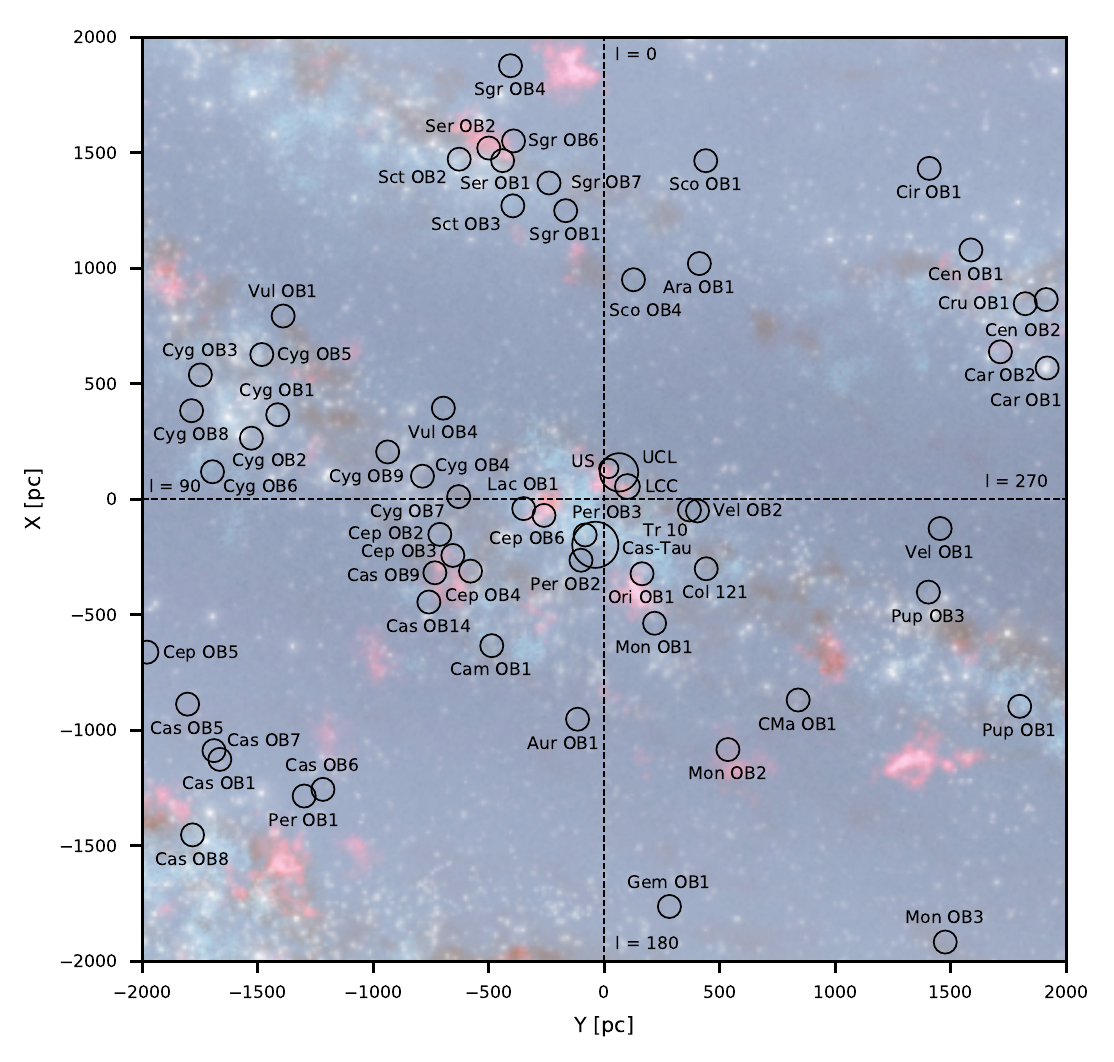}
\caption{Distribution of known OB associations within $\sim$2~kpc from Table~\ref{list_of_associations}, projected onto the Galactic plane in the Galactic $X$-$Y$ coordinate system. The figure is an updated and expanded version of figures by \citet{blaa91} and \citet{deze99} and based on the list of associations in \citet{rupr66}. Distances to OB associations are primarily from {\it Gaia} DR2 studies. The Sun is at the centre of the dashed lines, which give the principal directions in Galactic longitude, $l$. The sizes of each association are kept uniform for clarity, with the exception of the nearby associations Sco-Cen and Cas-Tau. The distribution is projected onto an artist's depiction of the shape and spiral arm structure of the Milky Way galaxy (Credit: NASA/JPL-Caltech/R. Hurt), which notably precedes {\it Gaia} DR2. The spiral arms shown are (from top to bottom) the Sagittarius Arm, the (local) Orion Spur, and the Perseus Arm.}
\label{XY_associations}
\end{center}
\end{figure*}

\begin{figure*}
\begin{center}
\includegraphics[width=18cm]{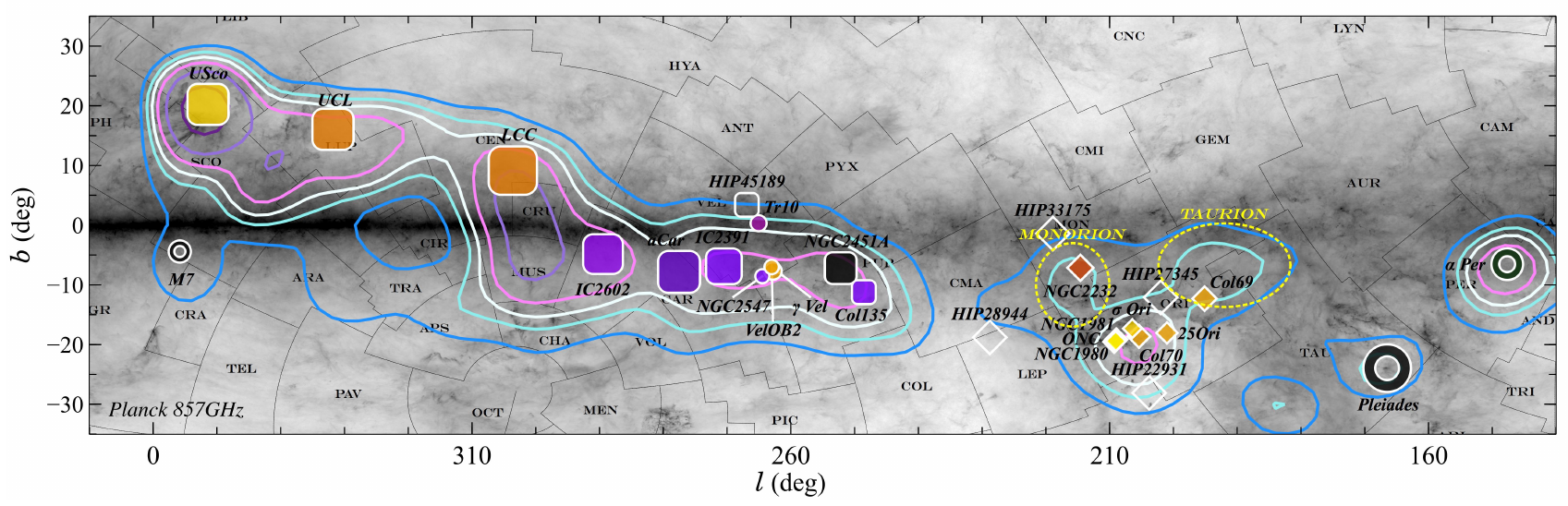}
\caption{Maximum density of OB stars across the blue streams identified by \citet{bouy15}, projected onto a {\it Planck} 857~GHz image of the sky. Clusters and associations belonging to the Sco-CMa, Vela, and Orion streams are represented by squares, dots and diamonds, respectively. Symbol size is inversely proportional to the group distance, while the colour scale is proportional to the age (ranging from 10~Myr in yellow, to 30~Myr in red 50~Myr in purple and $\sim$70~Myr in black). An age gradient is seen across each of the streams and is particularly visible across the Sco-CMa and Orion streams in this figure. Figure from \citet{bouy15}.}
\label{Sco-CMA_stream}
\end{center}
\end{figure*}

Thanks to their high luminosity and not being embedded in molecular clouds, OB associations have been used for many decades to trace the distribution of young stars and hence the spiral structure of our galaxy \citep[e.g.,][]{morg52,vazq08}. Furthermore, since OB associations are often found to be part of larger-scale structures (sometimes referred to as stellar aggregates or stellar complexes) they have been useful for tracing the hierarchical distribution of star formation and young stars \citep[e.g.,][]{blaa64,efre87}.

Using the updated list of OB associations presented in Table~\ref{list_of_associations}, for which distances were gathered from the literature (primarily from {\it Gaia} DR2 studies), Figure~\ref{XY_associations} shows the distribution of OB associations projected onto the Galactic plane. The distribution clearly shows the presence of the three local spiral arms; the Sagittarius (or sometimes Sagittarius-Carina) Arm towards the inner galaxy, the Orion (or sometimes Cygnus-Orion) Spur that the Sun is approximately within, and the Perseus Arm towards the outer galaxy. There are virtually no OB associations known in the inter-arm regions between these three arms, which is particularly striking. The distribution is projected against an artist's impression of the spiral arm structure of the Milky Way, which notably precedes {\it Gaia} DR2 (and therefore the agreement between the spatial distribution of OB associations from {\it Gaia} and the prior knowledge of the spiral structure of our Milky Way is also noteworthy).

A number of studies have linked the local ($d < 500$~pc) distribution of young stars and OB associations into larger structures \citep[e.g.,][]{elia06a,bouy15,zari18,koun19}. \citet{bouy15} mapped out the distribution of OB stars within 500~pc of the Sun and revealed the presence of multiple large and elongated structures with monotonic age sequences that they called `blue streams' (see Figure~\ref{Sco-CMA_stream}). The authors suggest that these streams represent the progression of star formation on large scales, possibly due to triggering. They are associated with the three largest OB associations in the solar neighbourhood, Sco-Cen, Vela OB2 and Orion OB1, but extend over larger areas. For example the authors find that Sco-Cen is the youngest part of a larger structure that they call the Scorpius -- Canis Majoris blue stream, extending 350~pc and as far as the latter constellation. However, the extension of this system beyond the classical extent of the Sco-Cen association is not as well-defined or populated in the {\it Gaia} DR2 distribution of young stars presented by \citet{zari18}.

The two other structures reported by \citet{bouy15} are centred around the Vela OB2 and Orion OB1 associations, and the authors identify possible foreground associations in front of both (with proposed names Vela OB5 and Orion X). The large extent of each of these regions has since been verified by \citet{arms18} and \citet{cant19b} for Vela OB2, and by \citet{koun18} and \citet{zari18,zari19} for Orion OB1, however the existence of the foreground associations for both regions is still debated. Furthermore, the correlation between age and position reported by \citet{bouy15} could not be verified by either \citet{zari18} or \citet{cant19b}.

On larger scales the spatial distribution and properties of OB associations have been traced by using radio emission from H~{\sc ii} regions \citep[e.g.,][]{mcke97,murr10}, though at large distances such techniques do not easily distinguish between dense star clusters and loose OB associations. From their sample of the 13 brightest free-free Wilkinson Microwave Anisotropy Probe sources, \citet{rahm11} identify the most luminous as originating from an OB association at a distance of $\sim$9.7~kpc in the constellation Crux. From near-IR photometry they estimate the association contains at least 400 stars of spectral type O and early B. The authors suggest it to be the most luminous OB association in the galaxy.

\subsection{Feedback, interaction with the ISM, and super-bubbles}

\begin{figure*}
\begin{center}
\includegraphics[height=5cm,trim=0 0 0 0]{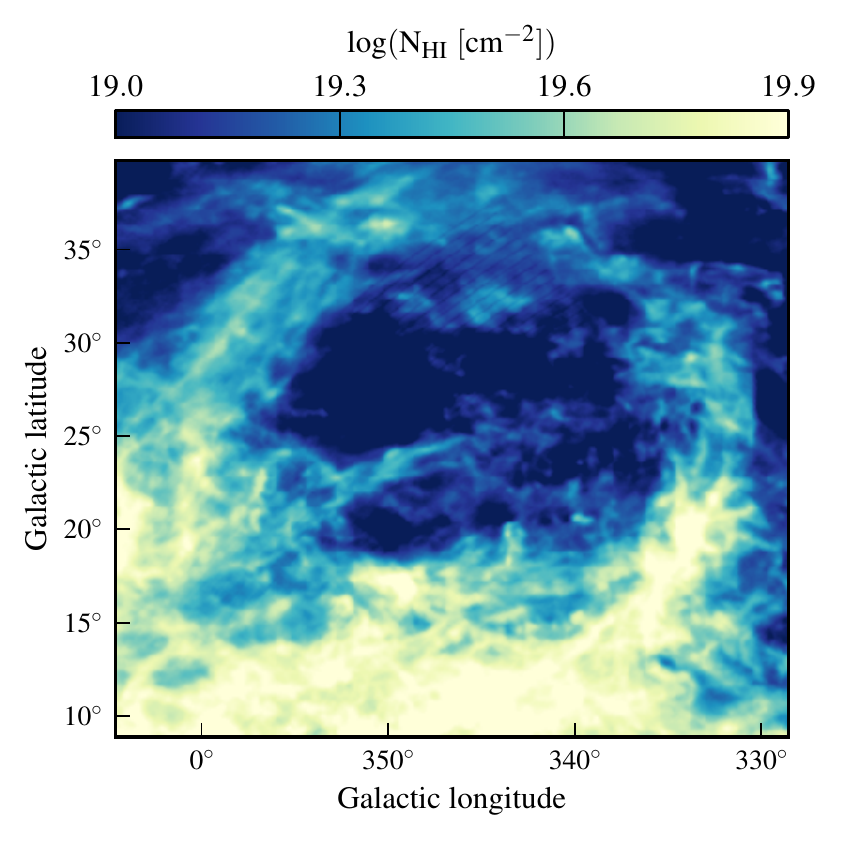}
\includegraphics[height=4.5cm,trim=50 -50 0 100]{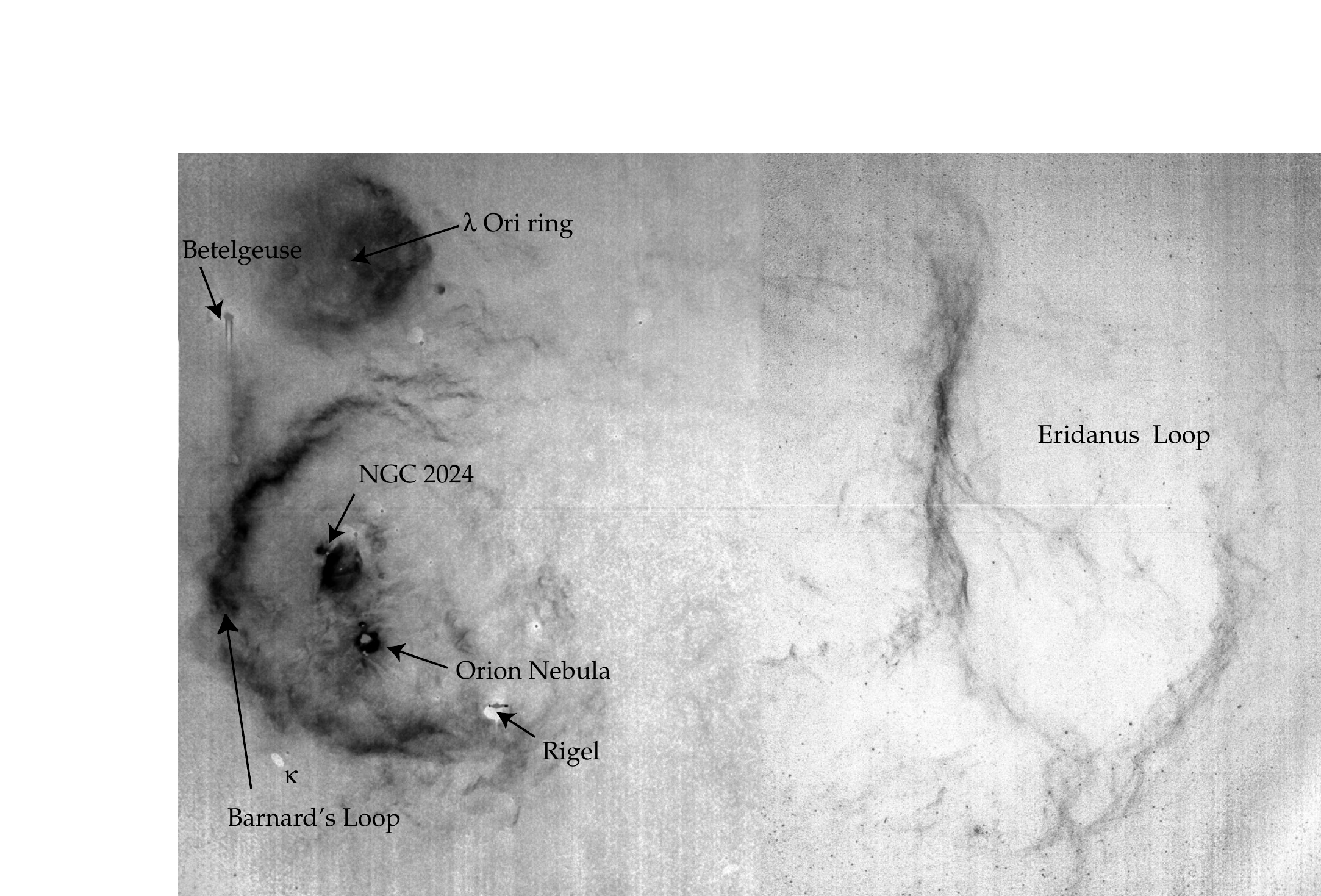}
\includegraphics[height=5cm]{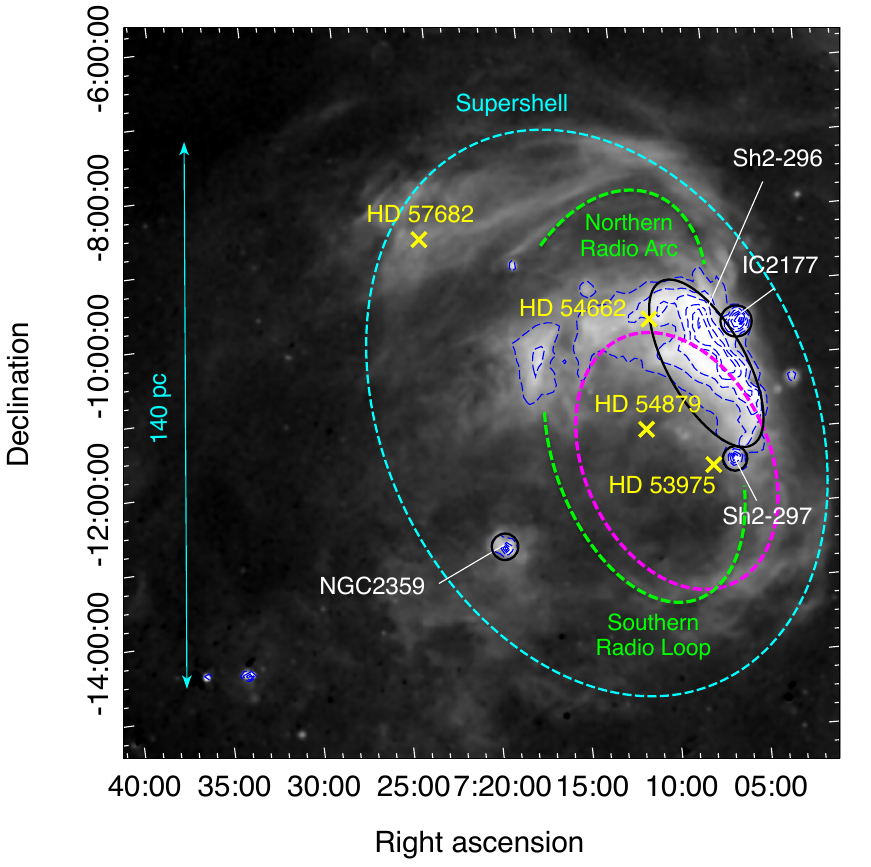}
\caption{Images of three prominent super-bubbles or shells that surround well-known OB associations. {\it Left:} A bubble surrounding the Upper Sco subgroup of the Sco-Cen association, as shown in the light of the 21~cm line of neutral hydrogen over an area of $35 \times 30$ degrees \citep{krau18}. {\it Centre:} The Orion-Eridanus super-bubble that surrounds the Orion OB1 association, as shown in H$\alpha$ over an area of $30 \times 47$ degrees \citep{ball08}. {\it Right:} The super-shell surrounding the Canis Major OB1 association shown in the light of H$\alpha$ over an area of $10 \times 10$ degrees \citep{fern19}. The three bubbles have diameters of approximately 65, 300, and 140~pc, respectively.}
\label{superbubbles}
\end{center}
\end{figure*}

O and early B-type stars produce immense amounts of ionising and dissociating radiation that can heat up surrounding molecular clouds and produce large H~{\sc ii} regions. Strong winds and supernovae can also play a significant role in the dynamics of the local interstellar medium \citep[for a recent review, see][]{dale15b}, with each stellar death releasing as much as 10$^{51}$~erg \citep[e.g.,][]{chev77}. OB associations were amongst the first objects used to show that stars disperse their parental molecular material, as evidenced by the decreasing proximity between associations and interstellar material as a function of association age \citep{blaa64}.

Type II supernovae are produced by the core collapse of massive stars with minimum stellar mass of approximately 8~M$_\odot$ \citep{smar09}, which is equivalent to a main sequence spectral type of approximately B3V. Such stars are abundant within OB associations and live up to $\sim$35~Myrs \citep{higd13} and therefore up to this time an OB association will see multiple supernovae.  This will release a considerable amount of energy, sweeping up the surrounding interstellar material and creating giant ($>$100~pc) cavities of hot ($>$10$^6$~K), low-density ($<$0.01~cm$^{-3}$) gas that are known as ``super-bubbles'' \citep{mccr79,higd13}. These super-bubbles can break out of the Galactic plane and therefore affect the chemistry and dynamics of the gas in the Galactic halo as well.

Super-bubbles surrounding the nearby OB associations are well known and have been studied in some detail. These include the Orion-Eridanus super-bubble surrounding Orion OB1 \citep{reyn79}, the IRAS Vela shell surrounding Vela OB2 \citep{sahu92}, the Cygnus super-bubble surrounding Cyg OB2 \citep{cash80}, and the multiple bubbles surrounding Sco-Cen \citep{weav77,dege92,robi18}. See Figure~\ref{superbubbles} for examples of these bubbles and the different morphologies. Many of these bubbles are distinctly asymmetric, which may be due to asymmetries in the surrounding interstellar medium, or might be related to the asymmetric expansion of their central OB associations (see Section~\ref{s-expansion}). That these well-studied OB associations all have prominent super-bubbles surrounding them suggests that they are probably a ubiquitous feature of the interstellar medium surrounding OB associations. It will therefore be interesting to see if new super-bubbles can be identified in the future around more distant OB associations.

\subsection{The initial mass function}
\label{s-IMF}

The initial mass function (IMF) is an important product of the star formation process \citep[see the recent review by][]{lee20}. For high-mass stars, accretion plays a dominant role, be it in a dense cluster environment \citep[competitive accretion, see][]{bonn07} or in a more isolated configuration \citep[monolithic collapse of a turbulent core, see][]{beut07}. Both scenarios of massive star formation might occur in OB associations, and therefore the resulting IMF may be a mixture of the two distinct modes. For many years there has been a theoretical suggestion that the characteristic mass or possibly the slope of the IMF should vary systematically with environment, especially due to temperature variations and feedback heating in the star-forming clouds (e.g., \citealt{lars85} and recently \citealt{esse20}), but observationally the vast majority of studies have found that the upper IMF in OB associations is consistent with that found in young star clusters and in the field. A possible explanation for the universality of the upper IMF is given in the review on massive star formation by \citet{zinn07}.

\citet{mass95} could find no statistical difference between the high-mass IMF slope in OB associations from that in young star clusters, with an average value of $\Gamma = -1.1 \pm 0.1$ for stars with masses $> 7$~M$_\odot$. Where differences in the high-mass slope have been measured, this may be due to incorrect consideration of the full age distribution of stars in the association and the evolution of the most massive stars to their end states.

For example, \citet{wrig15a} studied the IMF of the OB star population in Cyg~OB2 and found that the derived slope was highly dependent on the assumed age distribution of the population. The observed mass function in Cyg~OB2 was found to have a slope of $\Gamma = 2.06 \pm 0.06$, though limiting the sample to stars younger than 3.5~Myrs old (for which stellar evolution should not bias the results) or modelling the mass distribution of the population and accounting for stellar evolution, leads to a mass function slope consistent with a `universal' IMF. This study highlights how closely intertwined the IMF and the star formation history are when studying a population of high-mass stars more than a few million years old. Other factors such as the predictions of different stellar evolution models \citep[e.g.,][]{berl20} and the role that rotation and mass-loss play on stellar evolution (and therefore on a star's position in the HR diagram) can make the derivation of the IMF slope highly uncertain.

At lower masses many studies have also found the IMF in OB associations to be consistent with that of star clusters and the Galactic field \citep[e.g.,][]{prei02,bric07}. It is therefore well-established at this point, at least down to the hydrogen-burning limit, that there is no evidence for IMF variations between star clusters and OB associations \citep[see][for a more detailed discussion on this topic]{bast10}.

\subsection{Stellar multiplicity}
\label{s-binaries}

Some of the most important clues towards the formation of massive stars come from their high frequency of binary and multiple systems \citep[for an overview of the various possible formation processes, see][particularly the 3-dimensional SPH simulation of the collapse and fragmentation of a $10^6$~M$_\odot$, 100~pc size cloud into an OB association in Figure~3]{zinn07}. OB associations are valuable targets to study stellar multiplicity and the properties of close and wide binary systems in a plentiful, young population. It is well known that both the multiplicity fraction (the fraction of stars that are not single) and the companion frequency (the average number of companion stars per primary star) increase with stellar mass \citep{duch13}, and this has been observed in the field as well as in clusters and associations. For example, the multiplicity fraction of stars in Sco-Cen has been observed to vary from 60--80\% for A/F stars \citep{jans13} to 70--100\% for A and B stars \citep{kouw07,rizz13}, both of which are higher than observed for low-mass and solar-type stars \citep[e.g.,][]{toko20}. The companion frequency has also been observed to increase with stellar mass, going from $\sim$0.3 for low-mass and solar-type stars \citep{toko20} to 1.35 for B-type stars \citep{rizz13} and 1.5--2.0 for O and B-type stars \citep{prei99b,karl18}.

The most thorough investigation of the multiplicity of massive stars is a series of papers by \citet{sana12}. This vast and comprehensive statistical study established that, after correcting for observational selection effects, at least 75\% of massive stars are in multiple systems, often consisting of a close spectroscopic binary with a lower mass distant companion. These systems with periods less than a few years are close enough to experience mass exchange and interactions during the late stages of stellar evolution, and 1/3 of these will be so tight as to undergo a merger between the components. For bright O-type stars (many of which are members of OB associations), we refer to the early spectroscopic survey of \citet{garm80} and also the speckle interferometry survey of \citet{maso98}. The multiplicity of intermediate-mass stars (types B, A, F) in Sco~OB2 was observed and analysed by \citet[][for B-type stars]{shat02} and by \citet[][for A and F stars]{kouw05,kouw07}. Both of these high-resolution imaging surveys concluded that the mass distribution of secondary components is inconsistent with random sampling from an IMF.

There is growing evidence that the frequency and properties of binary systems can be different in low-density associations compared to denser star clusters. Clearly, tight (“hard”) binary systems with orbital velocities larger than the cluster/association velocity dispersion will not be disrupted by stellar encounters, while wide (“soft”) binaries can be. Hard binaries harden, while soft binaries get softer (Heggie’s rule). The dividing line between hard and soft binaries depends on the stellar density, i.e. it is different in clusters and associations. Thus, wide binaries are more likely to be disrupted in dense clusters than in less dense associations \citep[e.g.,][]{krou01b}. 

For example, \citet{duch99} and \citet{krou03} find that the multiplicity frequency of low-mass stars in the Taurus association is almost twice that in dense clusters or for field stars \citep[see also][]{duch13}, a phenomenon that could be explained by the disruption of binary systems in dense clusters and the infrequency of such disruptive close-counters in low-density associations \citep{krou03}. \citet{king12} observed a tentative decline in the multiplicity fraction with increasing stellar density in nearby star forming regions that would support this theory. Recently, \citet{toko20} studied the low-mass (0.4--3 M$_\odot$) binary population in the Upper Sco subgroup using speckle interferometry \citep[see also the earlier work of][on a smaller, X-ray selected sample]{kohl00}. They found that the low-mass binary properties of Upper Sco differ from the binary field star population and that the statistics of young low-mass binaries is not universal and may depend on the dynamical and/or radiative environment. \citet{bran96} also found that low-mass stars in the vicinity of Upper Sco B-type stars had a deficit of visual companions compared to field stars, while further away from the B stars they were more comparable.

For low-mass pre-main sequence stars in clusters and associations, we respectively refer to the binary companion studies of \citet{duch18} in the Orion Nebula Cluster (binaries with separations $10-60$ AU) and to \citet{toko20b} in the Ori OB1 association (binaries with separations $220-7400$ AU). Both investigations conclude that, on the basis of companion statistics, low-mass field stars must originate from a mixture of both star clusters and associations.

The jury is still out as to whether the frequency of OB binaries (as well as their orbital separation and mass ratio distributions) differ between dense clusters and loose OB associations. For example, \citet{chin12} found similar binary fractions in clusters ($72 \pm 13$\%) and associations ($73 \pm 8$\%), though both are larger than in the field ($43 \pm 13$\%) -- in agreement with past findings \citep{gies87,maso09}. 

\citet{garc01} studied the fraction of spectroscopic binaries among O-stars in several clusters by searching for radial velocity variations and found it to be mostly high, but variable, and possibly related to the cluster density. For example, in the loose open cluster NGC 6231 in Sco OB1, 11 out of 14 O-stars are close spectroscopic binaries (with periods less than 10 days). On the other hand, in the very dense cluster Tr 14 in the Carina Nebula, only 1 of the 7 O-stars is a close spectroscopic binary. To explain the dearth of short-period binaries in M17, \citet{sana17} hypothesise that OB star binaries are born with large separations ($\geq$100 R$_\odot$) and are then hardened on a timescale of 1 Myr or less.

The assumption that wide binaries are more likely to be disrupted in dense clusters than in less dense associations is true only if OB associations did not result from denser initial configurations. A case in point may be Cyg OB2, where $44 \pm 8$\% of the observed “complete” sample of 114 OB stars ($8-80$ M$_\odot$) are spectroscopic binaries, most of which have short-periods ($P < 1$ month). This is consistent with an extrapolated frequency of $90 \pm 10$\% if periods up to $10^4$ yr are considered \citep{kimi12,kimi12b}, while observationally $\sim$50\% of the Cyg OB2 O-stars are detected with an abundance of wide ($>100$ AU) companions \citep{caba20}.

\subsection{Protoplanetary discs and their evolution}

As with binary systems, OB associations are useful targets for studying protoplanetary discs, particularly since they typically offer older populations than most young clusters and thus provide clues to the evolution of discs and the formation of planets \citep[e.g.,][]{carp06}. Many studies have searched for evidence of the impact of environment on the properties and evolution of protoplanetary disks, with some studies finding evidence for disk destruction \citep[e.g.,][]{balo07,guar07} and other studies finding no evidence for disk destruction \citep[e.g.,][]{rich15}, though the different spatial scales and ultraviolet radiation fields present in different environments can make comparisons between different studies difficult.

One notable study is that by \citet{guar16}, who find that the disc frequency within the Cygnus OB2 association is inversely correlated with the local ultraviolet radiation field, suggesting that discs are more rapidly evaporated in regions with strong ultraviolet radiation fields. \citet{wint19} went on to use this observation and the known kinematics of the association, to constrain its dynamical history. They found that the observed protoplanetary disc fraction, driven almost entirely by ultraviolet photoevaporation, implies that the gas expulsion process must have finished 0.5~Myr ago, giving the discs 0.5 Myr of exposure to ultraviolet radiation.

Multiple studies have observed a correlation between disc radius and stellar density in young star forming regions. \citet{deju12} observe that, at stellar surface densities greater than $10^{3.5}$ pc$^{-2}$ there is an abrupt change in the protoplanetary disc radius of young stars, with the maximum radius dropping from $\sim$$10^3$ AU to $\sim$200~AU. This effect is primarily driven by stars in the ONC, the only local environment where such stellar densities are reached. \citet{eisn18} performed an Atacama Large Millimeter Array study of the ONC and found most discs had radii $<$50~AU, significantly smaller than found in studies of other nearby star forming regions. However it is unclear if this truncation of protoplanetary disks is caused by close encounters or by photoevaporation from ultraviolet radiation, since the ultraviolet radiation field scales strongly with stellar density \citep[e.g.,][]{wint18b}.

\begin{figure*}
\begin{center}
\includegraphics[height=8cm]{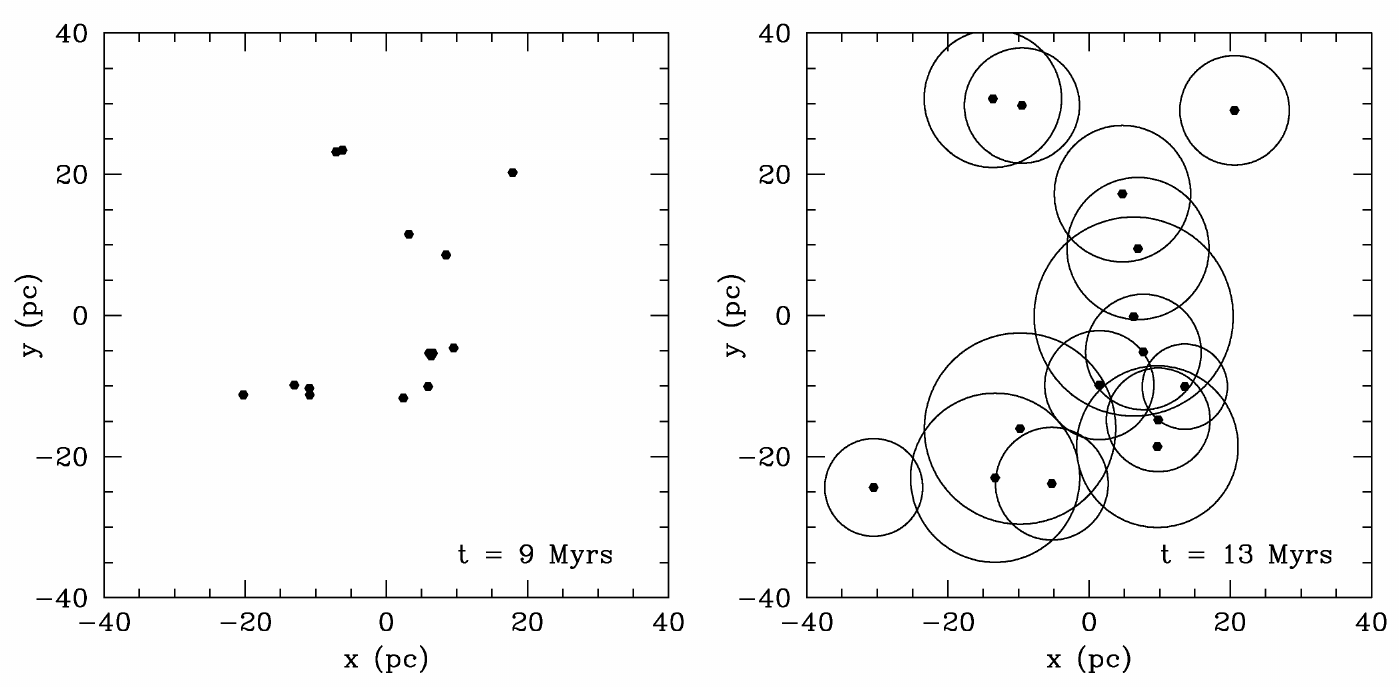}
\caption{Positions of young clusters at ages of 9~Myrs (left, before gas expulsion) and 13~Myrs (right, after gas expulsion) that formed during a smoothed particle hydrodynamics simulation of star formation within a giant molecular cloud. The circles in the right-hand panel show the size of the cluster after 4~Myr of expansion, assuming the clusters expand with their internal velocity dispersion. The simulations result in an extended distribution of young stars with multiple kinematic substructures (the expanding star clusters) within a larger dispersing stellar system (the OB association). Figure from \citet{clar05c}.}
\label{clark2005}
\end{center}
\end{figure*}

\section{Discussion}

Work over the last decade has provided us with a far more detailed picture of OB associations than since the last reviews of the {\it Hipparcos} era \citep[e.g.,][]{brow99,bric07,prei08}, but that we are also on the brink of an even greater revolution in the study of young stars brought about by {\it Gaia}. The availability of parallaxes for low-mass members of OB associations has allowed recent studies to break away from the classical `projected' view of nearby, young stars and provided the first ever 3D views of their true structure. It is clear now that OB associations have considerably more substructure than once envisioned, both spatially, kinematically and temporally. Furthermore, our improved ability to identify low-mass pre-MS stars over large areas of the sky has allowed the true spatial extent of associations to be traced away from their bright, young OB members, and revealed for the first time the large spatial extent and star formation history of these systems. Here we discuss these results and what they mean for our understanding of the formation and evolution of OB associations.

\subsection{Origin of OB associations}

\subsubsection{Substructure in OB associations and its implications}

\begin{figure*}
\begin{center}
\hspace{-1cm} \includegraphics[height=5.5cm]{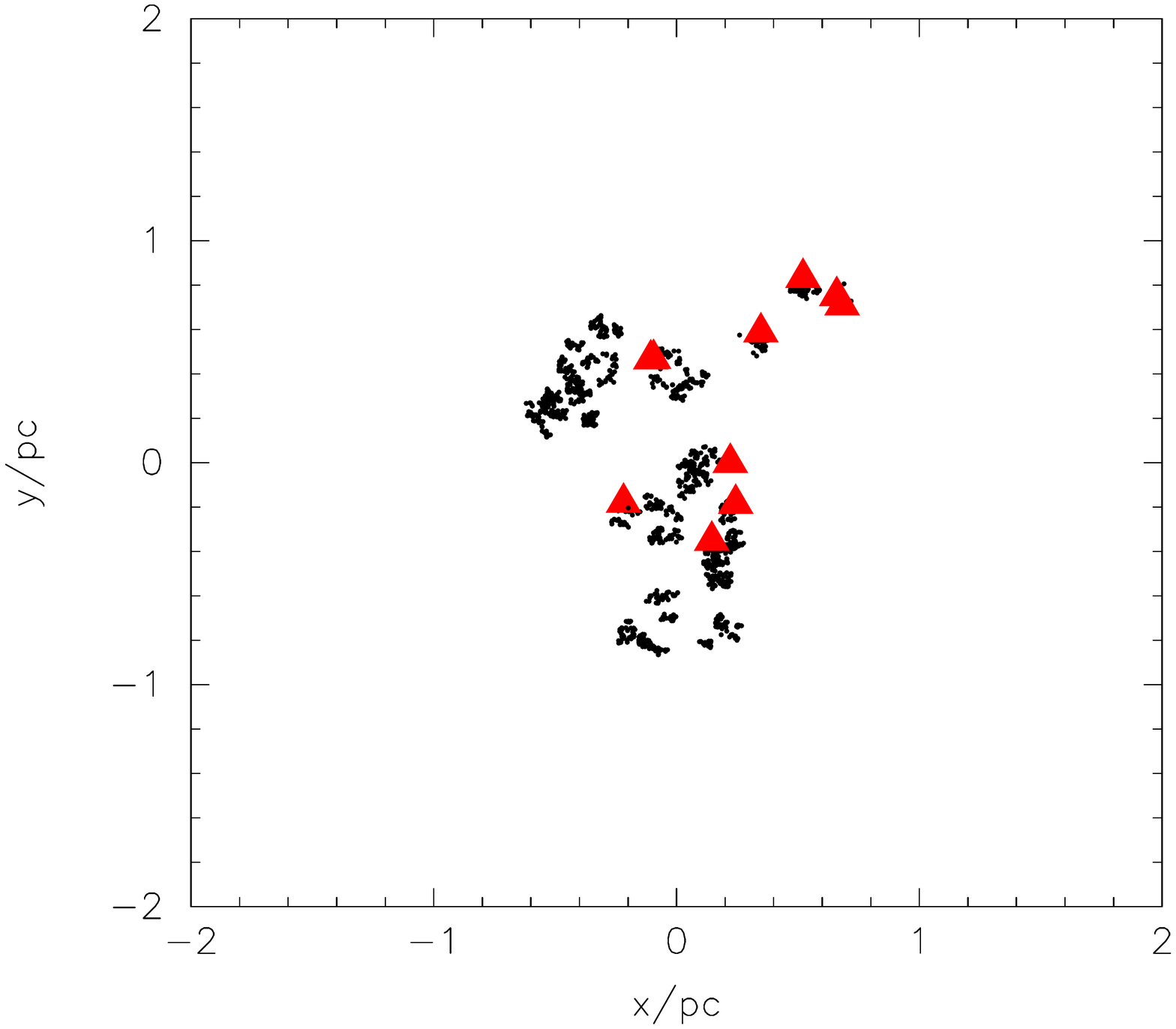} \hspace{-1.5cm} \includegraphics[height=5.5cm]{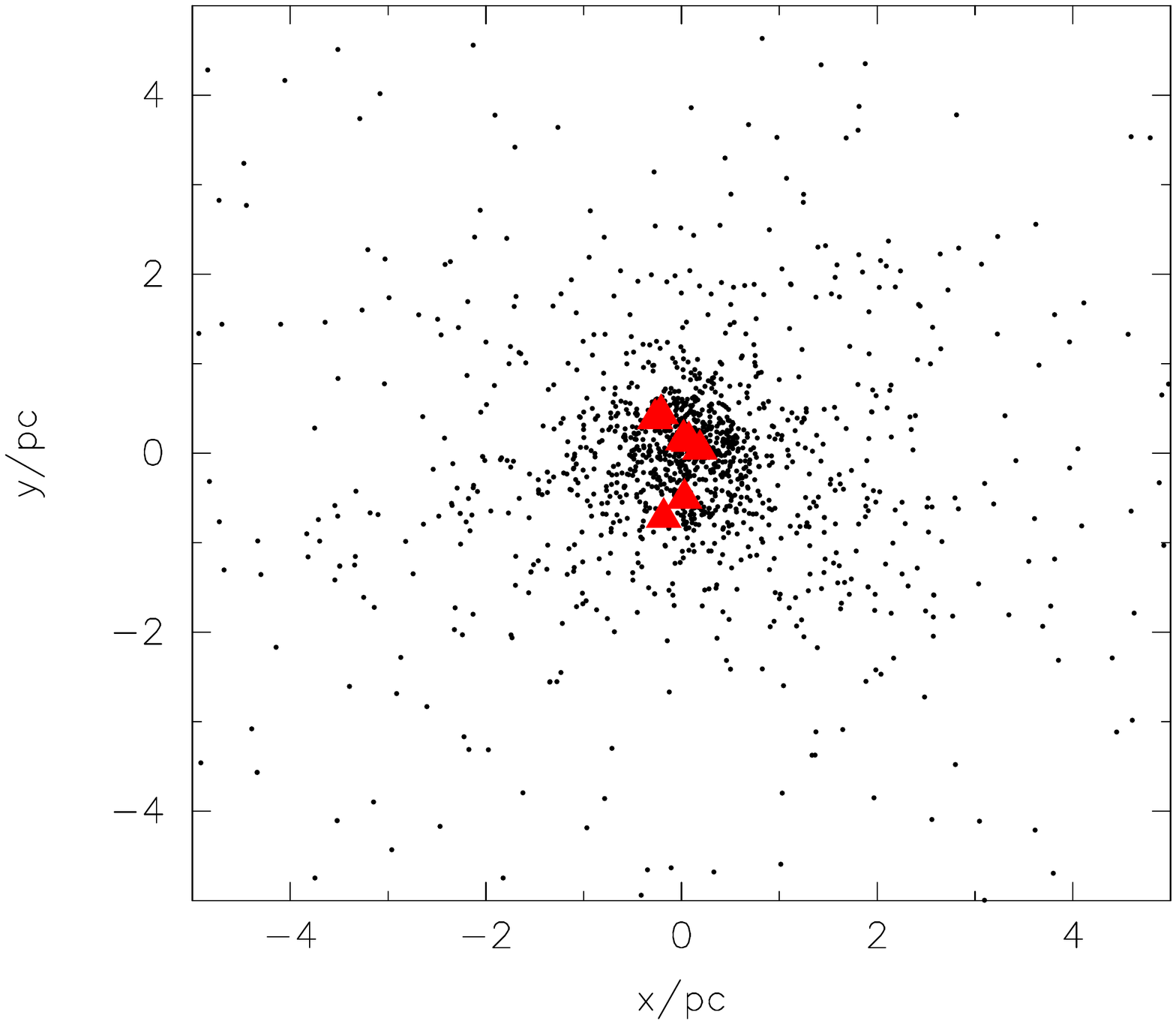} \hspace{-1.5cm} \includegraphics[height=5.5cm]{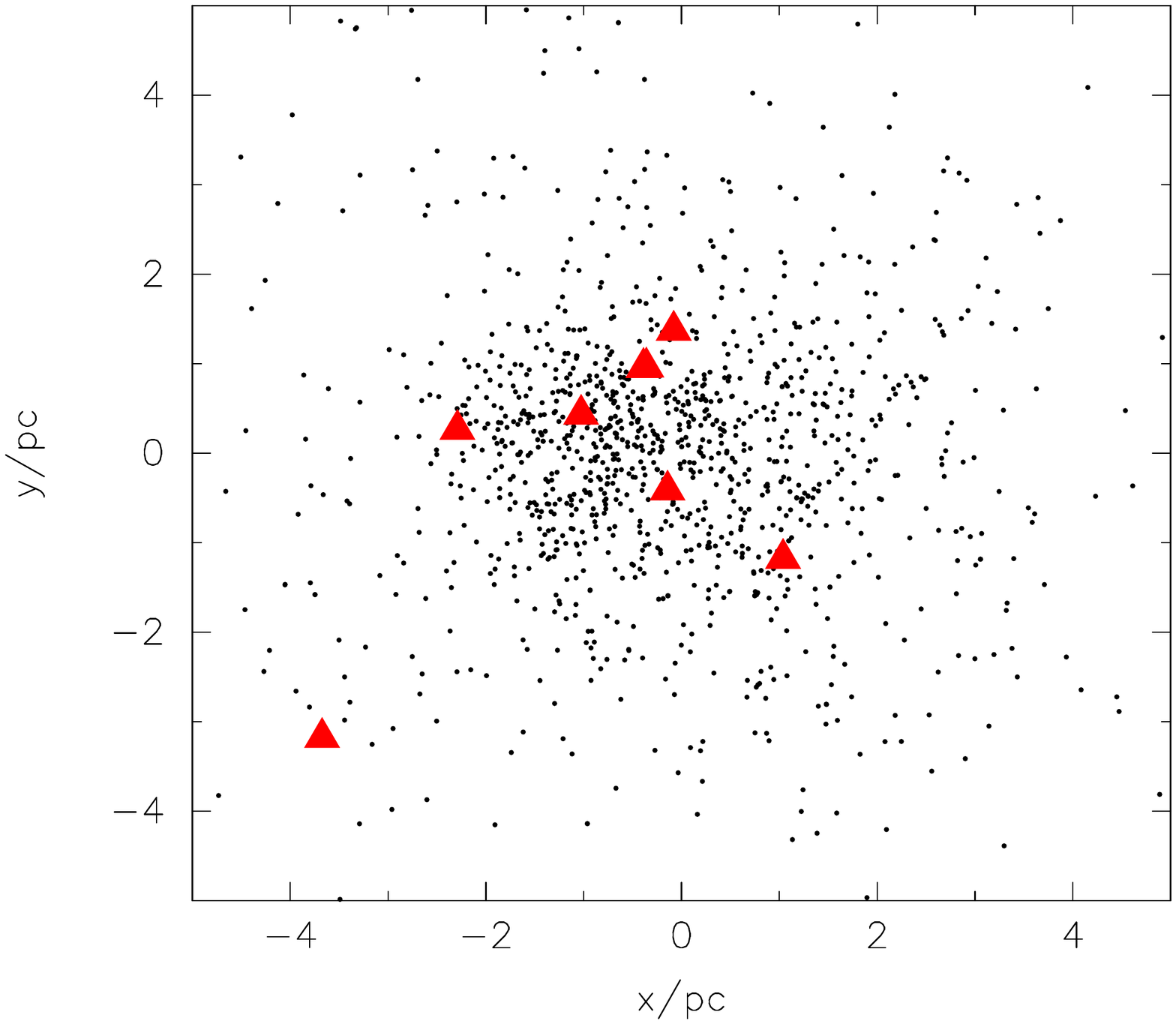} \\
\vspace{0.3cm} \hspace{-1cm} \includegraphics[height=5.5cm]{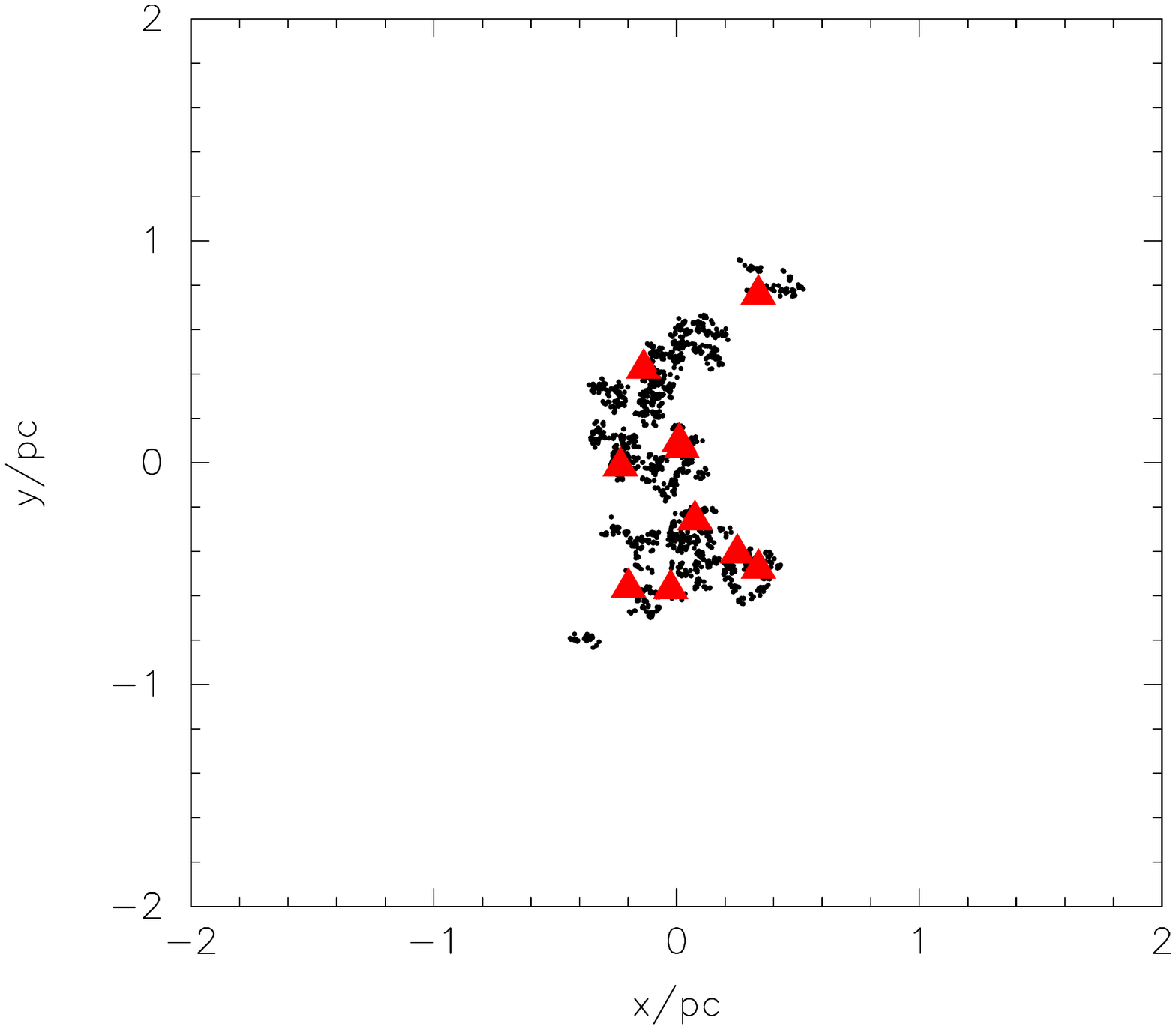} \hspace{-1.5cm} \includegraphics[height=5.5cm]{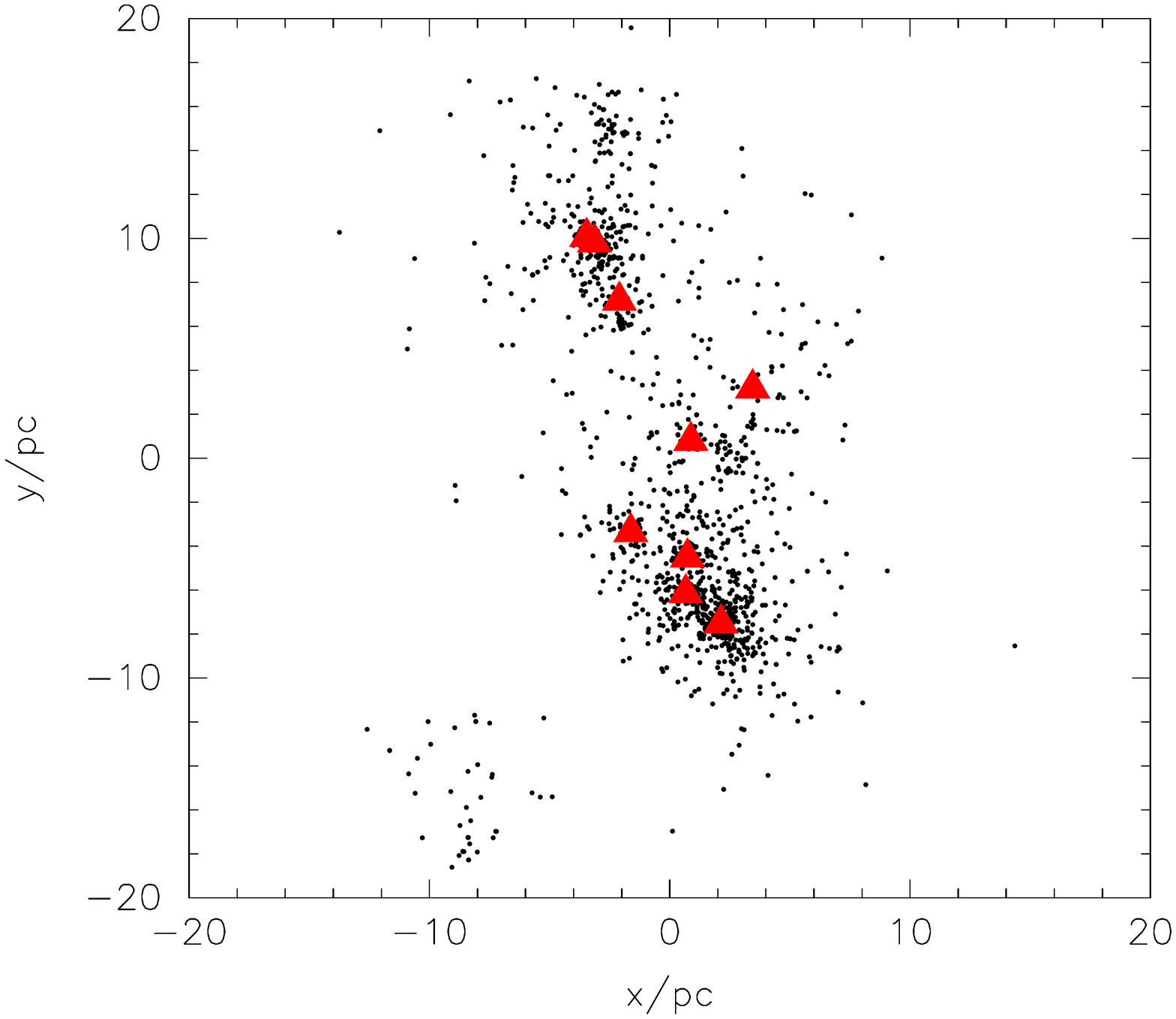} \hspace{-1.5cm} \includegraphics[height=5.5cm]{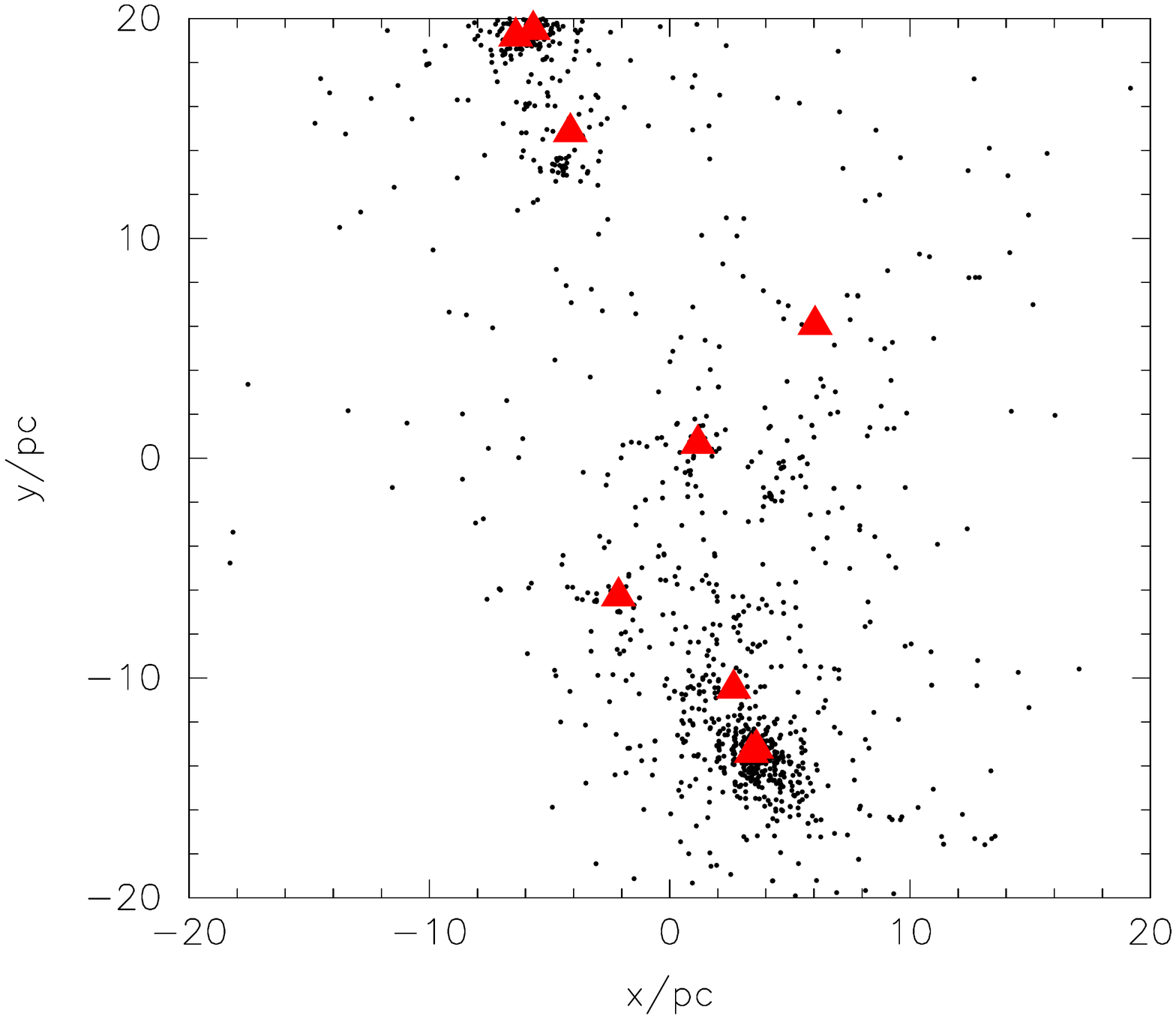}
\vspace{0.6cm} \caption{Evolution of subvirial (top row) and supervirial (bottom) substructured star-forming regions with 1500~stars from an N-body simulation. The morphologies of the regions are shown at 0 (left), 5 (centre) and 10 Myrs (right), with stars shown as black points and the 10 most massive stars as red triangles. In the subvirial simulation the distribution collapses to form a compact cluster, erasing most substructure, while in the supervirial simulation the region expands, retaining the majority of spatial and kinematic substructure. Figure from \citet{park14}.}
\label{parker2014}
\end{center}
\end{figure*}

One of the most significant developments in the study of OB associations over the last few years has been the identification of substructure in these systems. That OB associations have substructure has been known for over half a century \citep{amba49,blaa64,garm94}, but recent studies have revealed how significant and prevalent this substructure is, both physically \citep{prei08,wrig14b}, kinematically \citep{wrig16,wrig18}, and temporally \citep{peca16}. Furthermore, this substructure is connected, with spatially- or kinematically-identified substructures showing distinct age differences \citep[e.g.,][]{koun18,cant19b,zari19}.

Many authors have argued that the existence of this substructure is inconsistent with OB associations being the expanded remnants of single (or a small number of) star clusters \citep[e.g.,][]{prei08,wrig14b}. This is because two-body interactions between stars can disrupt their organised motions and erase substructure \citep[][see Figure~\ref{parker2014}]{park14}. Therefore any system of stars that has been through a densely-clustered phase in its past should not retain significant levels of substructure, even if it was born with substructure. For example, \citet{wrig14b} argued that the substructure in Cyg~OB2 implies it must have been born highly-substructured and with a relatively low volume density, much as we see it today.

An OB association with substructure may have formed as a collection of star clusters that have since dissolved and may be briefly visible as kinematic substructures within the association \citep[e.g.,][]{prei08}. \citet{wrig16} argued that the kinematic substructures or internal `moving groups' seen within Cyg~OB2 could be the remnants of denser structures that existed within the association. They concluded that the association was most likely born highly substructured and globally unbound, with the individual subgroups born in (or close to) virial equilibrium, and that the OB association has not experienced significant dynamical evolution since then.

If OB associations have not passed through a densely clustered phase then some (or most) of the substructure observed within these systems is likely to be primordial, i.e., it could directly correspond to the physical and kinematic groupings and separations predicted to result from the star formation process \citep{elme08}. This could place valuable constraints on the spatial distribution of stars during formation and therefore the star formation process itself \citep[e.g.,][]{bres10}. \citet{wint19} argued that the velocity anisotropy observed within Cyg~OB2 required the presence of primordial substructure (such as large-scale filaments of fractal clumps) from which the substructure within the association has evolved. The presence of velocity anisotropy within a group of stars could therefore be evidence for internal kinematic substructure and insufficient dynamical mixing to erase this anisotropy.

The question this leads to then is whether the observed substructure is primordial and if so, whether we are seeing it in the state it was born in or whether all substructures were initially more compact and dense in the past.

\subsubsection{Evidence (or not) for expansion of OB associations}
\label{discuss-expansion}

N-body simulations of the residual gas expulsion process predict that once the gravitational potential holding the cluster together has been removed then the cluster expands rapidly and that the motions of the stars should exhibit a strong and symmetric radial expansion pattern \citep{baum07}. The first kinematic studies of OB associations capable of resolving such a signal could find no evidence for such radial expansion, indicating that the associations studied were not currently undergoing radial expansion\footnote{Note however that the systems studied are gravitationally unbound and therefore should expand in the future.} \citep{wrig16,ward18,wrig18}. Pre-{\it Gaia} studies of (less massive) moving groups have found similar results \citep{mama05,wein13,mama14}. This suggests that these systems are not expanded remnants of dense star clusters. \citet{ward18} were able to go one step further and show that, with the use of various kinematic diagnostics quantifying the degree of radial expansion, that none of the 18 OB associations they studied showed evidence of expansion from either a single or multiple star clusters.

Despite this, more recent studies that divide OB associations into subgroups based on their kinematics or 3D structure have found evidence of expansion \citep{koun18,cant19b}. This could indicate that our previous division of OB associations into subgroups based primarily on the positions of stars on the plane of the sky was inaccurate and that a spatio-kinematic division is necessary to properly identify the remnants of expanded star clusters. If this is the case it would support a picture whereby a large fraction of stars are born in star clusters (or at least pass through a clustered phase) that then expand and disperse for some reason.

The cause of this expansion is not currently known. Models of residual gas expulsion predict a symmetric radial expansion of stars as the cluster disperses \citep{baum07}, and yet recent 2- or 3-D kinematic studies of OB associations find strong evidence that most expanding associations are doing so asymmetrically \citep{meln17,wrig18,cant19b,arms20}. Even some recent studies of more clustered regions find that where expansion exists it is asymmetric \citep{wrig19}. If residual gas expulsion is not responsible for the observed expansion of these systems, then what process is responsible? One possibility is that star clusters may become unbound by tidal disruption due to a nearby giant molecular cloud \citep[e.g.,][]{elme10,krui11}. This should produce expansion preferentially along a single axis (aligned towards the disrupting body), with expansion along the other axes developing as the cluster becomes unbound. This process could be combined with residual gas expulsion if the molecular cloud that the cluster is embedded in is not symmetrically distributed around the cluster \citep[e.g., the filamentary structure of molecular clouds such as Orion~A that the ONC is in the middle of,][]{hart07} or if its density is sufficiently asymmetric that the dispersal of the gas does not become isotropic . As the cluster becomes unbound, the asymmetric distribution of the molecular cloud mass surrounding the cluster could then lead to asymmetric expansion of the cluster.

A final possibility explored by \citet{wrig19} is that the observed asymmetric expansion of these systems is a relic of their formation process and that the clusters did not survive long enough to be sufficiently mixed to erase it. This might occur following the collision between two molecular clouds or if the region formed from the merger of multiple sub-clusters along a certain axis \citep[e.g.,][]{vazq17,wrig19b}. In both of these circumstances the cluster would form with an excess of kinetic energy along one axis and, unless sufficient mixing occurred to erase this, and once the cluster became unbound, this asymmetry would be carried over to its expansion. The violent relaxation and expansion of a cluster that formed by cool collapse is a possible mechanism for this \citep{park16}.

The key question therefore is whether the asymmetric expansion patterns observed are produced by the cluster disruption process (either residual gas expulsion or tidal disruption) or if they are a relic of an asymmetry that existed before cluster formation (possibly introduced during the cluster formation process) itself. Further simulations that explore these different models would be useful to ascertain if they can reproduce the observed levels of asymmetry.

\subsubsection{Clustered versus hierarchical star formation}

There are two models for the origin of OB associations that are commonly discussed in the literature. The first is that OB associations are the expanded remnants of dense star clusters disrupted by processes such as residual gas expulsion \citep[e.g.,][]{krou01,lada03}, while the second is that OB associations are born out of unbound giant molecular clouds with an initially hierarchical distribution of stars \citep[e.g.,][]{clar05c,elme08,krui12}. This difference is sometimes postulated as the difference between clustered and hierarchical star formation, or alternatively as the difference between star formation taking place in bound or unbound groups. The truth is likely to be somewhere between these extremes, but they represent useful models to explore and compare to observations.

In the clustered star formation model the view is that most (if not all) stars form in dense and gravitationally bound clusters that are born embedded within molecular clouds \citep[e.g.,][]{lada03}. Since the crossing time in such clusters is smaller than the star formation timescale this implies that the cluster is thoroughly mixed before it finishes forming \citep{boil03a}. Feedback from the young stars heats up and disperses the residual gas left over from star formation, removing the gravitational potential holding the cluster in virial equilibirum and unbinding the system \citep{tutu78,hill80,krou01,good06}. Becoming gravitationally unbound the cluster then expands and would briefly be visible as a low-density OB association, before its stellar contents disperse and populate the Galactic field \citep{brow97,baum07}. Despite the popularity of this model, particularly for explaining the significant difference between the number of young, embedded clusters and older exposed clusters \citep{lada03}, recent studies have questioned the effectiveness of residual gas expulsion for disrupting clusters in regions of high star formation efficiency \citep[e.g.,][]{giri12,krui12b,dale15}.

In the hierarchical model of star formation stars form in a substructured and scale-free distribution over a range of densities \citep[e.g.,][]{elme08,krui12}, following the same fractal structure that has been observed in interstellar gas \citep{scal85,elme96b,beut07} and is thought to be shaped by supersonic turbulence \citep{lars81,stut98}. In the high-density regions, dense and bound star clusters form due to a combination of high star formation efficiencies \citep{krui12} and hierarchical collapse \citep{vazq17}, while low-density regions remain substructured and gravitationally unbound. The model of hierarchical star formation does not require GMCs to be gravitationally bound or in virial equilibrium for them to be sites of star formation, supporting observations that many GMCs are unbound \citep[e.g.,][]{heye01}. Once these regions form stars they are already gravitationally unbound and dispersing and therefore processes such as residual gas expulsion are not necessary to produce unbound systems \citep[they may still play a role, though studies suggest the process is less efficient for clumpy initial conditions than for centrally concentrated conditions,][]{smit11}. This model predicts that young stars should be found at a range of densities, from low-density groups to dense star clusters, consistent with observations \citep[e.g.,][]{gome93,gute08,bres10}.

\citet{clar05c} presented simulations of star formation in an unbound and turbulent GMC that are often cited as a possible method for the formation of OB associations. They find that stars form in a series of star clusters that are gravitationally unbound from each other and therefore naturally move apart. The individual clusters do undergo residual gas expulsion and potentially disperse, just as in the clustered star formation model, with these clusters dispersing within the larger system of clusters, potentially representing the subgroups observed within OB associations (see Figure~\ref{clark2005}). These simulations include aspects of both models, with stars forming at a range of stellar densities indicative of the hierarchical star formation model and the formation of dense clusters that get unbound by residual gas expulsion as in the clustered star formation model. \citet{grud20} presented magnetohydrodynamic simulations of star formation in a turbulent GMC, including the effects of radiative feedback. They find that once feedback disperses the cloud, the stars are distributed in both gravitationally-bound star clusters and unbound OB associations, the former often embedded within the latter, as is commonly observed.

To distinguish between these models we need to know whether all the substructures observed within OB associations were originally more compact (and are now expanding) or whether some have not been through a more compact phase in the past and therefore constitute a truly `distributed' form of star formation that occurs at a lower density to that occurring in dense clusters. If all substructures within associations were originally more compact then an OB association may just be considered the expanded remnants of {\it multiple} star clusters. This leads to the question of how OB associations should be defined.

\subsection{Large-scale distribution and propagation of star formation across OB associations}

Two of the most significant results concerning OB associations that have come from {\it Gaia} data over the last few years have been the discovery that OB associations appear to be more spatially extended and with longer star formation histories than previously thought, and that most OB associations do not appear to exhibit a clear age gradient, but instead exhibit a much more complex age - position structure that suggests a more complex and fragmentary star formation history.

\subsubsection{Uncovering the full spatial extent of OB associations}

One of the major results to come from {\it Gaia} so far has been the discovery of extended distributions of young stars around many known star forming regions, clusters and OB associations. This highlights the power that {\it Gaia} parallaxes provide in identifying young stars based on their position above the zero-aged main sequence (see Section~\ref{s-astrometric_young_stars}). Coupling these new samples of young stars with proper motions has allowed many authors to trace out the large-scale distributions of candidate young stars connected to known clusters and associations. For example \citet{zari18} traced out the distribution of young low- and high-mass stars in the Solar neighbourhood and found extended distributions of older stars surrounding clusters of younger stars. This included foreground extensions to some of the known Cygnus and Cepheus associations, as well as a confirmation of the foreground population in front of Orion \citep[previously identified by][]{bouy15}. \citet{cant19b} found that the young stars of the Vela OB2 association were extended over a significantly larger area as well, covering an area at least $200 \times 300$~pc across the Galactic disk and extending into the neighbouring constellation of Puppis. They found diffuse young populations of stars surrounding many of the known open clusters in the area and traced a star formation history over the last 50~Myrs.

\begin{figure}
\begin{center}
\includegraphics[height=8cm]{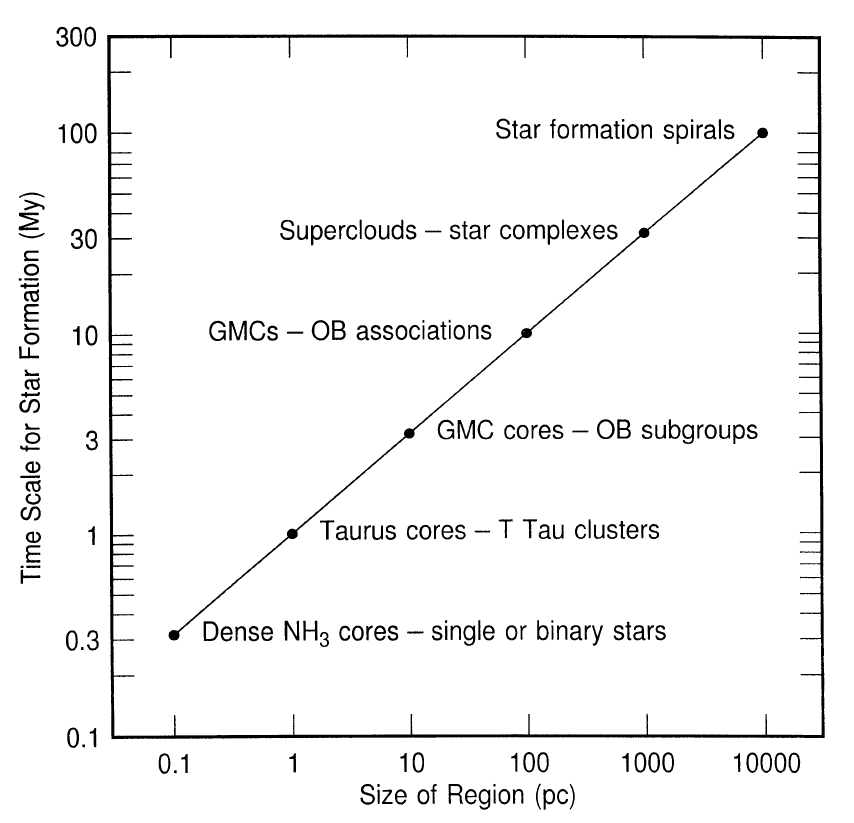}
\caption{Schematic diagram showing the size dependence of the duration of star formation in various regions. Figure from \citet{efre98}.}
\label{efremov1998}
\end{center}
\end{figure}

Extragalactic studies of OB associations have also found that some systems can be very large. \citet{vand64} studied groups of blue stars in the Andromeda Galaxy that he called OB associations, finding $\sim$200 such groups (some up to $\sim$500~pc across) while \citet{hodg86} found 42 associations in Andromeda with a typical size of 300~pc. Such spatial scales should not necessarily be a surprise. Giant molecular clouds can span up to a hundred parsecs and have velocity dispersions of 5--10~km~s$^{-1}$ \citep[e.g.,][]{lars81,myer83}. If these scales and velocities are imprinted on the stellar populations that form then after 10--20~Myr the size of these systems can span hundreds of parsecs. Whether such systems are identified or labelled as OB associations becomes a matter of nomenclature. \citet{mcke97} argue for a physical upper limit to the size of OB associations that is set by the size of molecular clouds. \citet{efre98} suggest that all stellar structures exist on a hierarchy of scales (see Figuire~\ref{efremov1998}) and that OB associations ($\sim$100~pc) are just one step on this hierarchy, larger than OB subgroups ($\sim$10~pc) and smaller than star complexes ($\sim$1000~pc). As our census of young stars improves over the next decade and the spatial extent of the known OB associations is improved it is likely that their borders and subdivisions will need to be redrawn.

This raises the question of how we define an OB association, i.e., where does one association end and another begin? As with molecular clouds, this is likely to be an observational, rather than a physical, definition, as there may be no meaningful physical way to divide up young stars distributed over an extended volume of space.

\subsubsection{Models of triggered star formation in OB associations}

The classical model for the formation of OB associations put forward by \citet{elme77} is that star formation propagates through a molecular cloud by triggering, with photoionization triggering the formation of each new generation of massive stars. This model explains the existence of OB subgroups with different ages and kinematics as the product of each generation of triggered star formation, the triggering process causing the stars to move away from each other and thus creating an association of stars that are gravitationally unbound \citep[extending up to Galactic scales,][suggested that this model of triggering-induced self-propagation of star formation could even explain the spiral structure of galaxies]{gero78}. This model was popular because it explained the age sequences seen in OB subgroups at the time, such as in Cep~OB3 \citep{blaa64} and Cep OB4 \citep{chur70}, though the authors did note that an irregularly-shaped molecular cloud could prevent a clear age sequence from being visible. A downside to this model is that it predicted that OB stars would form by triggering, with low-mass stars forming spontaneously throughout the cloud, in disagreement with more recent observations \citep{dege92,prei99}. Despite this, triggered star formation could still explain the propagation of star formation within OB associations.

There are many different models for triggered star formation, including radiation-drive implosion \citep{sand82,bert89} and the collect-and-collapse process \citep{elme78}. Each of these models makes different predictions for the effectiveness of the triggering process, the types of stars that it forms (low- versus high-mass stars) and therefore the age differences or gradients expected between stars of different mass \citep[see discussion in][]{dale15b}. Many of these predictions have been tested, but so far overwhelming evidence for any one mechanism has not been found. However, the idea of triggered star formation has still proven popular with observers. It is worth noting that star formation may be hindered or halted by feedback as well as triggered by it, and in fact \citet{dale13} estimate that the disruptive effects of feedback typically outweigh the constructive effects. 

\subsubsection{Evidence for (or against) triggered star formation}

The Sco-Cen association has provided an interesting testing ground for theories of triggered star formation and the propagation of star formation in associations. Early observations of the low-mass population of Upper Sco provided no evidence for an age spread \citep{prei98,prei02}, suggesting some external mechanism must have triggered star formation across the subgroup. The most likely candidate for this was a supernova from the nearby Upper Centaurus Lupus subgroup \citep{prei07}. More recent studies of the age distribution and spreads across Sco-Cen show multiple substructures within each subgroup with clear age substructure \citep{peca16}, suggesting a single external triggering agent is not required. Many authors have commented that the classical picture of sequential star formation is not valid for Sco-Cen because the oldest subgroup (Upper Centaurus Lupus) is actually in between the other two \citep[e.g.,][]{dege89}, though other models of triggered star formation do not require linear propagation of star formation \citep[e.g.,][]{krau18}.

In Orion OB1 the positions and relative ages of the various subgroups in the association do not suggest star formation has propagated through the cloud in a linear fashion. The association has been forming stars for 15--20~Myrs (and still continues to do so), but does not show any age gradient or evidence for a linear and sequential pattern of star formation \citep{ball08,zari19}. However, this conclusion is usually based on the current positions of stars and yet these subgroups can have very different kinematics and thus will have been in very different positions (relative to each other) when they formed. This may not reveal an age gradient indicative of triggering, but it would be valuable to see where each subgroup is relative to the other ones at the time they formed. A similar lack of age gradient is seen in Vela OB2 and Puppis \citep{cant19b}, despite forming stars for $\sim$50~Myrs, though again this is based on the current positions of stars and not their positions at the time of formation.

To conclusively address the question of what role triggering has played in star formation within OB associations it will be necessary not only to map out the star formation history of an association, but to do so at the time the stars were born, tracing the motions of the substructures back to their configuration at birth.

\subsection{Implications from our understanding of OB associations}

\subsubsection{Where does star formation take place?}

For many decades OB associations were considered to be the dominant mode of star formation, due to both their ubiquity in the solar neighbourhood and the fact that embedded clusters had yet to be discovered in huge numbers \citep{robe57,mill78}. With the discovery of vast numbers of embedded clusters in the infrared \citep[e.g.,][]{gras73,wilk83,herb86,mcca94} the consensus view was that star formation primarily took place in clusters \citep{lada03,pfal09} and that OB associations were remnants of dissolved embedded clusters. It is here that the term `cluster' becomes vague, as for some this is simply {\it a group of stars whose surface density significantly exceeds that of the field} \citep{lada91}, but for others it implies a dense, gravitationally bound and well-mixed group of stars \citep{baum07}, which can have very different implications \citep[e.g.,][]{zinn07}. For a discussion of the different density thresholds used for defining star clusters, see \citet{bres10}.

Recent studies have suggested that star formation takes place over a wide range of stellar densities, including both dense clusters, low-density OB associations, and potentially in relative isolation \citep[e.g.,][]{bres10,lamb10,wrig14b}. A recent example of this is the study by \citet{rate20} of the membership of known WR stars in OB associations and clusters. They find that 59--75\% of Galactic WR stars are currently isolated (i.e., are not found in any known cluster or association), a distinctly larger fraction than for Galactic O-type stars \citep[$\sim$30\%,][]{maiz13}. Comparing their results to predictions from simulations of dispersing star clusters and OB associations, \citet{rate20} conclude that $\gtrsim$50\% of WR stars formed in low-density, moderately substructured associations that expand during the WR star lifetime and make the WR star appear relatively isolated. The degree to which OB associations are made up of dense clusters is potentially still open for debate, as highlighted by the ongoing search for evidence of expansion in OB associations and their substructures (see Section~\ref{discuss-expansion}), and these systems may include stars that formed at a wide range of stellar densities in groups that are now dispersing and mixing.

One of the implications of such a view is that star formation takes place at different densities and therefore that understanding and quantifying the distribution of star formation densities is important for fully understanding the star formation process. \citet{krui12} present a model whereby stars form over a range of stellar densities, with bound clusters arising at the high-density (and high star formation efficiency) end and the remaining stars forming in unbound, substructured groups that would be observed as OB associations. Models such as this that attempt to quantify the frequency at which star formation takes place at different densities are useful for understanding the relative important of different birth environments, particularly if the products of the star formation process (such as binary or planetary system properties) vary with environment. In their model \citet{krui12} generate a probability density function at which star formation takes place and then integrate this to calculate the fraction of stars that form in bound clusters or in unbound associations. They predict that, over cosmic time, 65--70\% of stars in the Universe formed in associations, which would therefore make it the most common form of star formation. The simulations of star formation presented by \citet{grud20} suggest that the fraction of stars born within bound clusters is related to the overall star formation efficiency in the GMC, that itself is driven by stellar feedback (though with significant scatter due to small-scale variations in the star formation process). This would imply that the fraction of star formation occurring in bound clusters should increase as metallicity decreases (due to weaker OB stellar winds) and therefore bound clusters should be more common in the early Universe.

\subsubsection{Evolution of binary systems}

There is some evidence that the multiplicity fraction of stars in associations is higher than in star clusters or the field and that the properties of binary stars also differ (see Section~\ref{s-binaries}). Given that binary systems can be affected by close encounters with other stars and that the close encounter timescale scales as $\rho^{-1}$ (such that close encounters between stars are less common in associations than in star clusters) then it is not surprising that the multiplicity rate is higher in associations.

If this is the case then binary stars are more likely to originate from associations than from clusters and the multiplicity properties of those systems could vary in the two different environment. This could have implications for the origins of high-mass X-ray binaries, the ejection of runaway OB stars \citep{gies86}, and the evolutionary descendants of high-mass stars in binary systems \citep{sana11}. It could also affect how observers deal with the impact of binary systems, such as the effect on the positions of stars in the colour-magnitude diagram \citep{hurl98} and on measured radial velocity dispersions \citep{cott12b}. It will therefore be necessary to understand how the multiplicity properties of stars varies as a function of their birth environment.

A notable example of the possible impact of environment on the properties of binary systems is the abundance of wide O-type binaries in Cygnus~OB2, where \citet{caba20} find that $\sim$51\% of stars have companions with projected separations $>$100~AU. \citet{grif18} argue that this observation places strong constraints on the number of clusters or bound structures that could have existed within this association, as their simulations show that only one massive wide binary can exist within a cluster, implying that the association must have been made up of at least 30 distinct massive star formation sites \citep[supporting the significant kinematics structure observed within the association,][]{wrig16}.

\subsubsection{Evolution of protoplanetary disks and formation of planetary systems}

\begin{figure}
\begin{center}
\includegraphics[height=8cm]{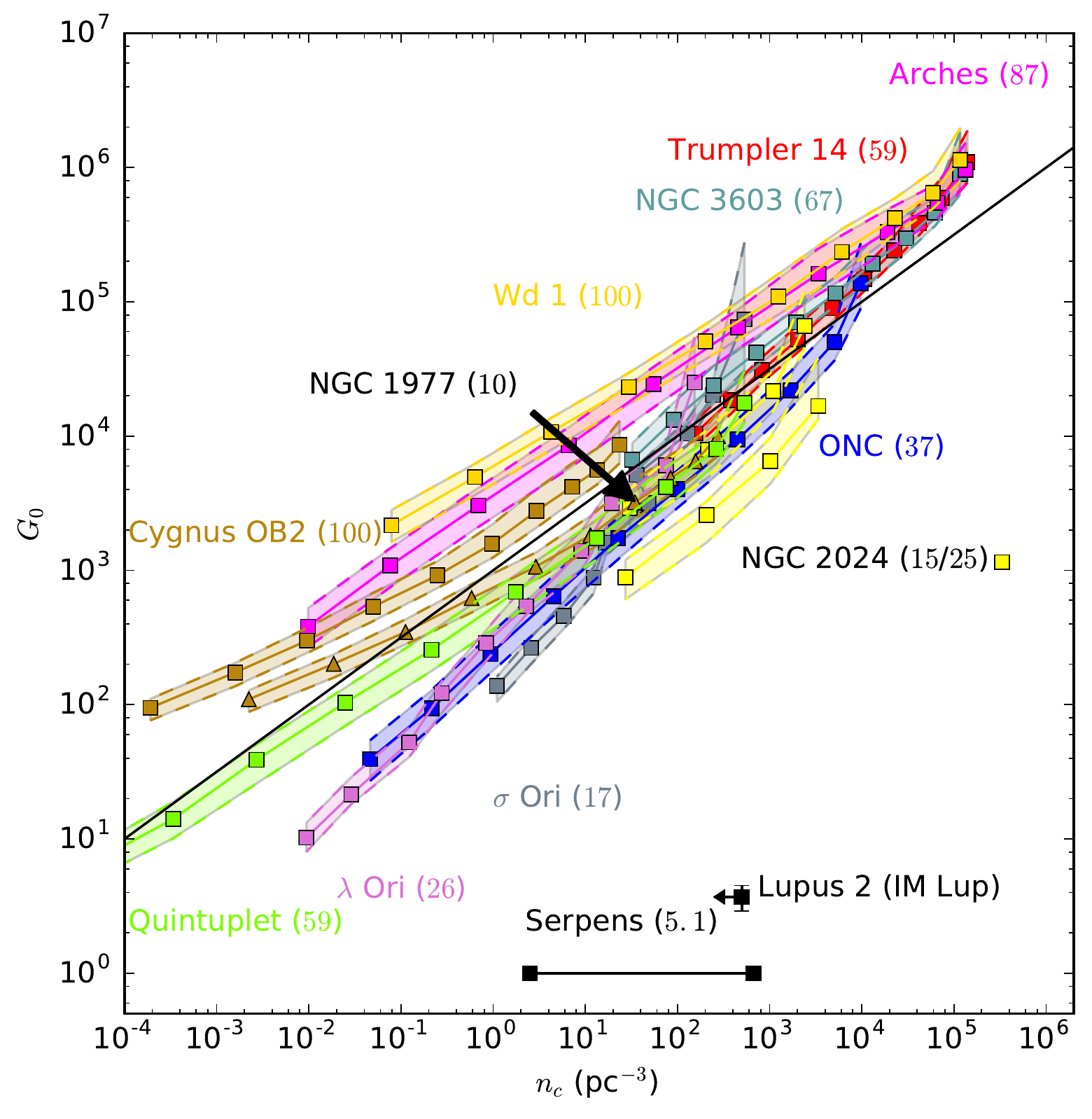}
\caption{Local stellar density, $n_c$, versus far-UV flux, $G_0$, within various star clusters and OB associations. Each region was divided into radial bins and the mean flux and density in that bin is shown by the square markers, except for the Cyg~OB2 association, for which triangles also show the mean flux and density when the substructure within the association is considered. Shaded regions show the standard deviation ($\pm$1$\sigma$) of the flux in each radial bin. The numbers in brackets represent the assumed maximum stellar mass in solar masses for each region. The solid black line follows $G_0 = 10^3 \, (n_c / \mathrm{pc}^{-3})^{1/2}$. Figure from \citet{wint18b}.}
\label{winter2018}
\end{center}
\end{figure}

As with binary systems, protoplanetary disks and planetary systems are strongly affected by close encounters with other stars and by UV photoevaporation due to nearby OB stars \citep[e.g.,][]{john98,scal01,adam04}. For example, \citet{roso14} find that the radii of protoplanetary disks are set by the closest dynamical encounter they experience. \citet{vinc16} also find that in dense clusters ($> 10^3$ M$_\odot$ pc$^{-3}$), disks with radii larger than $\sim$500~AU can be easily affected by fly-bys that truncate their disks down to 100--200~AU in radius. \citet{deju12} show that dynamical interactions between stars and planetary systems can also eject planets from their host star. \citet{vane19} find that in young dense clusters like the Orion Nebula Cluster, 16.5\% of planets will be ejected within 10~Myr due to a combination of strong close encounters with other stars and internal planetary scattering \citep[see also][]{laug98}. Such ejections have implications for the chances of life existing within these systems if planets are ejected from the habitable zones of these systems. Photoevaporation due to far-ultraviolet radiation is also thought to be important in the vicinity of luminous OB stars, inducing mass-loss that can evaporate gas from protoplanetary discs within as little as 1 Myr \citep[e.g.,][]{adam06,nich19}.

All of these effects are strongly dependent on the stellar density. The lower densities in OB associations (typically $\sim$10 M$_\odot$ pc$^{-3}$) means that young stars will (typically) have lower close encounter rates and be exposed to weaker UV fluxes than stars in dense clusters \citep[e.g.,][]{adam06,nich19,wint19}. For example, \citet{wint18b} compare the far-UV fluxes and stellar densities in various clusters and associations and show that stars in Cyg~OB2 experience significantly lower ionising fluxes and stellar densities than stars in comparably-massive star clusters such as Westerlund~1, the Arches cluster or NGC~3603 (see Figure~\ref{winter2018}). One would therefore expect protoplanetary disks in high-density environments to be less frequent, smaller, and less massive, and studies confirm this \citep[e.g.,][]{guar16,deju12,eisn18}.

Smaller disks and more frequent close encounters will lead to shorter disk lifetimes, which may have implications for planet formation. In very high density clusters, such as Westerlund~1, most wide-orbit planets ($a > 20$~AU) will be ejected within 10~Myr, favouring the production of systems with fewer planets, that are preferentially close-in and high mean eccentricities or inclinations \citep{cai19}. This means that most wide-orbit planets, planets with low eccentricities or inclinations, or planetary systems with a high degree of multiplicity are likely to originate in either small star forming regions or low-density OB associations. At first glance this might hinder the formation of planetary systems, though some studies have argued that photoevaporation may help trigger the formation of planetesimals \citep{thro05}. It may also be the case that less massive protoplanetary disks may lead to the formation of less-massive, and therefore more Earth-like, exoplanets. We refer the reader to \citet{park20} for a more detailed discussion on these topics.

\subsection{Future prospects}

Early {\it Gaia} data have already provided a tantalising glimpse of the potential of data from this satellite over the next few years. The existing astrometric data will improve in precision, the current systematic uncertainties in the astrometric solution will hopefully be resolved \citep{lind18}, and additional stellar parameters (such as radial velocities and effective temperatures) will become available for an increasing number of stars. In addition, complementary spectroscopy from wide-field multi-object spectroscopic surveys for millions of stars will become available from facilities such as WEAVE \citep{dalt18} and 4MOST \citep{dejo19}. Rotation periods from time-domain surveys (such as those used to discover exoplanets) and X-ray data from satellites such as eROSITA \citep{merl20} will also help identify young stars for studies of OB associations.

Here we discuss how these developments and new facilities could lead to advances in our understanding and investigation of OB associations.

\subsubsection{Identification of members of nearby associations}

{\it Gaia} data has already proven revolutionary in allowing us to use the luminosity method to identify nearby young stars \citep[e.g.,][]{zari18,dami18,cant19a}. This has so far been effective for associations within $\sim$500~pc where {\it Gaia}'s current parallax uncertainty \citep[0.3~mas random error at $G = 19$~mag, combined with a 0.1~mas systematic uncertainty][]{lind18}, allows identification of unreddened stars above the binary main sequence (0.32~mas at 500~pc equates to a photometric uncertainty of $\sim$0.35~mag). Stellar reddening complicates matters, but this can be offset by using spectroscopic extinctions or applying broad proper motion cuts to isolate the OB association population \citep[e.g.,][]{zari18}. {\it Gaia}'s end-of-mission astrometric performance will improve \citep{prus16} and systematic errors should be resolved, which will extend the viability of this method, potentially out to $\sim$1~kpc (though at these distances the smaller parallaxes and fainter sources limits the range of this approach).

The use of rotation periods to identify young stars based on their rapid rotation will become more effective as more wide-field time-domain surveys come about \citep[e.g., TESS,][]{rick15}. Though these missions are designed to detect exoplanets, the high cadence and precision photometry is ideal for measuring rotation periods, and being wide-field with high-enough spatial resolution makes them well-suited for studying OB associations (the pixel scale of 21$^{\prime\prime}$ for TESS should be sufficient for the brighter members of most nearby associations). New all-sky X-ray satellites such as eROSITA \citep{merl20} will also be valuable for identifying nearby young stars in OB associations. While both of these approaches will not extend as deep as {\it Gaia} astrometry, they will be more effective for older ($>$20~Myr) OB associations members whose luminosity is not sufficiently above the main sequence to facilitate easy identification.

All these techniques will greatly increase our knowledge of the membership of nearby associations. This will facilitate detailed studies of the structure of OB associations to help understand their initial conditions and evolution. Larger and more complete samples of young stars within OB associations will be useful for studying their mass functions, particularly at low-masses, and to determine whether there is any systematic variation of the mass function between stars born in low-density associations and high-density clusters \citep[e.g.,][]{bast10}. Combined with data from infrared or sub-mm surveys these data will be important for measuring the fraction of stars with protoplanetary disks and their properties and to study how they vary with environment \citep{will11}. As new OB associations at different ages are identified they will provide useful samples for studies of stellar evolution, for example to understand how radius inflation varies with stellar mass and age \citep[e.g.,][]{macd13} and to study the evolution of exoplanet systems and atmospheres \citep{riba05}.

\subsubsection{Identification of distant OB associations and members}

For more distant OB associations ($>$1~kpc), the identification of their low-mass members becomes more difficult, due to both their faintness and the reduced precision of {\it Gaia}-derived distances for such stars. Our knowledge of OB associations at such distances is very limited and typically dates back to the works of \citet{hump78} and \citet{garm92}, both of which based their membership only on photometry and sky position, as kinematic information was not available. As such, some of these associations may not prove to be physical entities once analysed with parallaxes and kinematic information.

Identifying the members of more distant associations (and thus mapping out the distribution and properties of OB associations across the galaxy) is best achieved photometrically and focussing on the more massive members of these associations. Recent Galactic Plane photometric surveys such as IPHAS \citep{drew05}, VPHAS$+$ \citep{drew14}, and GALANTE \citep{lore19} allow OB stars to be identified to great distances \citep[e.g.,][]{mohr17}, with their kinematics and parallaxes constrained by {\it Gaia}. \citet{drew18} use these data to study the distribution of massive stars around the star cluster Westerlund~2, uncovering a new OB association projected in its vicinity.

The combination of photometric surveys and {\it Gaia} astrometry offers the potential over the next decade to map out the distribution of OB stars within the majority of the near-side of our galaxy. This will allow our list of known OB associations to be verified, updated and expanded, providing samples for further studies of these systems, recalibration of the upper main sequence in the Hertzsprung-Russell diagram, and for studying the luminosity function and spatial distribution of OB associations in the galaxy. This can be used to address important questions such as whether OB associations differ in and between spiral arms, how much substructure OB associations exhibit at different ages, and how OB associations within large (kpc) volumes, such as in star-forming complexes and spiral arms, are related to each other.

\subsubsection{Youth indicators and age estimates from spectroscopy}

Despite advances in our methods to identify candidate young stars from photometric and astrometric surveys, spectroscopy is still crucial to verify the youth of candidate members of OB associations. The next-generation of multi-object spectroscopic instruments (e.g., WEAVE, 4MOST, and MSE, the Maunakea Spectroscopic Explorer) are ideal for studying OB associations as they have wide fields of view (several square degrees) and can observe a large number of sources in each observation (several thousand in each configuration). They are therefore well suited to the large spatial extent and low stellar densities within OB associations. Spectroscopy from these surveys can provide lithium equivalent widths and surface gravity indicators to verify the youth of low-mass stars, as well as providing effective temperatures, radial velocities, and abundances.

The combination of precise effective temperatures and absolute magnitudes can provide improved age estimates for OB associations and their subgroups. This will help to address the question of whether age spreads exist within OB associations and to what extent such spreads can be explained by the observed physical substructure in associations. On larger scales, stellar ages allow us to trace the star formation history across OB associations and OB complexes, which can be used to address the question of what role triggering may play in the propagation of star formation.

\subsubsection{3D kinematics of OB associations}

The combination of precise astrometry from {\it Gaia} and radial velocities from spectroscopic surveys opens the doors to 3D kinematic studies of OB associations, which will be necessary to constrain their past evolution and understand their dynamics. Combined with parallaxes for stars in nearby associations this will facilitate 6D spatial and kinematic studies. Early examples of such studies have shown their potential for dissecting the complex structure of OB associations and searching for evidence of expansion \citep[e.g.,][]{wrig18,koun18,cant19b}. Extending these studies to more associations and larger samples of stars will allow us to study how the kinematics of different subgroups are related and to determine how OB associations are dispersing and what processes were responsible for disrupting them.

Combining 3D kinematics with 3D positions and age estimates for groups of stars provides 7D spatial, kinematic and temporal structure in the association to be studied \citep[e.g.,][]{cant19b,zari19}. With age and kinematic information we can also trace back the motions of stars in 3D to ascertain their initial arrangement and kinematics at formation, and also look forward in time to determine how long OB associations will retain a distinct phase-space structure and be identifiable to observers.

\subsubsection{Chemistry within OB associations}

The next frontier of OB association studies is to use high-resolution spectroscopy to study their chemistry. Chemical abundances for large numbers of stars will allow us to determine whether OB association subgroups exhibit distinct chemical variations and if so how this compares to their primordial structure and potentially the chemistry within the molecular clouds from which they formed. Such data can also be used to search for abundance gradients across the large scales of OB association complexes, testing evidence for supernova-driven enrichment from stars in one association subgroup to another. A variant of `chemical tagging' may also be possible, allowing the membership of dispersed OB associations to be reconstructed from isolated young stars in the vicinity of OB associations or in the Galactic field.

\section{Summary}

OB associations have long been recognised as important objects in both star formation and the early evolution of stars and stellar systems. There is now growing evidence that star formation takes place in OB associations at lower densities than in dense star clusters, and that the low density in such environments is favourable to the formation and evolution of binary systems and for the survival of protoplanetary disks and young planetary systems. It is therefore important to understand how OB associations form and evolve.

It is clear that OB associations are highly substructured, both spatially, kinematically and temporally, and that this substructure is connected. It appears that these systems were born with at least as much substructures as we seen them now and therefore that this substructure can provide clues to their primordial state. Though OB associations are globally unbound, the evidence for their expansion has so far been mixed, with some studies finding clear expansion patterns and other studies finding no evidence for expansion. These disagreements may originate with the different samples or subdivisions of associations into subgroups used by different authors, with the complex substructure within associations hiding their true expansion patterns. Despite these disagreements, even when expansion is observed and measured there is very little evidence for the simple picture of symmetric radial expansion predicted by models of residual gas expulsion. It is likely that either more complex models of residual gas expulsion or even other models for the dispersal of young stars will need to be considered.

Large-scale studies are also increasing the size of known OB associations beyond the classical OB star markers of these systems. As with the recent discovery that open clusters have a larger physical extent than once thought \citep{vanl16}, OB associations also seem to extend further. These extended regions appear older and more diffuse than the main parts of the association, lacking the bright OB members that would explain why they have only recently been uncovered. These extended regions will be important for studying the propagation of star formation on larger scales, within and between associations.

It is an exciting time to be studying OB associations. {\it Gaia} data is revolutionising our view of these systems, while future spectroscopic and multi-wavelength facilities will only improve the observations. There are still many open questions regarding the initial spatial and kinematic structure of OB associations, their properties, stellar content (including binarity), evolution and dispersal. Whatever the outcome of the studies that address these questions, it is clear the results will only enrich and improve our understanding of the star and planet formation process in the Universe.

\section{Acknowledgements}

I would like to thank Janet Drew, Simon Goodwin, Tim Naylor, Eleonora Zari, and Hans Zinnecker for reading the manuscript and providing comments for its improvement. I am grateful to John Bally, Tristan Cantat-Gaudin, Rob Jeffries, Marina Kounkel, Martin Krause, Mike Masheder, Anna Mel'nik, Richard Parker, Mark Pecaut, Andrew Winter and Eleonora Zari for providing the figures used in this review, with special thanks to Eleonora Zari for adapting and providing Figure 2 for use in this article and to Anthony Brown, Jos de Bruijne and Tim de Zeeuw for helping track down old articles. I would also like to extend special thanks to the anonymous referee, who provided extensive and detailed comments on the paper, including clarifications on key points and suggestions for additional text.

NJW acknowledges an STFC Ernest Rutherford Fellowship (grant number ST/M005569/1). This work has made use of results from the European Space Agency space mission {\it Gaia}. {\it Gaia} data are being processed by the {\it Gaia} Data Processing and Analysis Consortium (DPAC). Funding for DPAC is provided by national institutions, in particular the institutions participating in the {\it Gaia} Multi-Lateral Agreement (MLA). This research has made use of NASA's Astrophysics Data System and the Simbad and VizieR databases, operated at CDS, Strasbourg. The authors would like to thank Rob Jeffries for discussions on this work and the anonymous referee for careful reading of this paper and helpful suggestions to improve the work.

\bibliographystyle{mn2e}

\begin{thebibliography}{500}
\expandafter\ifx\csname natexlab\endcsname\relax\def\natexlab#1{#1}\fi

\bibitem[{{Abt} {et~al}\mbox{.}(1991){Abt}, {Wang}, \& {Cardona}}]{abt91}
{Abt} H.~A., {Wang} R., {Cardona} O., 1991, \apj, 367, 155

\bibitem[{{Adams} {et~al}\mbox{.}(2004){Adams}, {Hollenbach}, {Laughlin}, \&
  {Gorti}}]{adam04}
{Adams} F.~C., {Hollenbach} D., {Laughlin} G., {Gorti} U., 2004, \apj, 611, 360

\bibitem[{{Adams} {et~al}\mbox{.}(2006){Adams}, {Proszkow}, {Fatuzzo}, \&
  {Myers}}]{adam06}
{Adams} F.~C., {Proszkow} E.~M., {Fatuzzo} M., {Myers} P.~C., 2006, \apj, 641,
  504

\bibitem[{{Albacete Colombo} {et~al}\mbox{.}(2018){Albacete Colombo}, {Drake},
  {Flaccomio}, {Wright}, {Kashyap}, {Guarcello}, {Briggs}, {Drew}, {Fenech}, \&
  {Micela}}]{alba18}
{Albacete Colombo} J.~F. {et~al.}, 2018, arXiv e-prints, arXiv:1806.01231

\bibitem[{{Alves} \& {Bouy}(2012)}]{alve12}
{Alves} J., {Bouy} H., 2012, \aap, 547, A97

\bibitem[{{Ambartsumian}(1947)}]{amba47}
{Ambartsumian} V.~A., 1947, {Stellar Evolution and Astrophysics}. Armenian
  Acad. of Sci.

\bibitem[{{Ambartsumian}(1949)}]{amba49}
{Ambartsumian} V.~A., 1949, {Astronomischeskii Zhurnal}, 26, 3

\bibitem[{{Ambartsumian}(1968)}]{amba68}
{Ambartsumian} V.~A., 1968, {Problemy evoliutsii Vselennoi.} {}

\bibitem[{{Argiroffi} {et~al}\mbox{.}(2006){Argiroffi}, {Favata}, {Flaccomio},
  {Maggio}, {Micela}, {Peres}, \& {Sciortino}}]{argi06}
{Argiroffi} C., {Favata} F., {Flaccomio} E., {Maggio} A., {Micela} G., {Peres}
  G., {Sciortino} S., 2006, \aap, 459, 199

\bibitem[{{Armstrong} {et~al}\mbox{.}(2018){Armstrong}, {Wright}, \&
  {Jeffries}}]{arms18}
{Armstrong} J.~J., {Wright} N.~J., {Jeffries} R.~D., 2018, \mnras, 480, L121

\bibitem[{{Armstrong} {et~al}\mbox{.}(2020){Armstrong}, {Wright}, {Jeffries},
  \& {Jackson}}]{arms20}
{Armstrong} J.~J., {Wright} N.~J., {Jeffries} R.~D., {Jackson} R.~J., 2020,
  \mnras, 494, 4794

\bibitem[{{Arnal} {et~al}\mbox{.}(1987){Arnal}, {Cersosimo}, {May}, \&
  {Bronfman}}]{arna87}
{Arnal} E.~M., {Cersosimo} J.~C., {May} J., {Bronfman} L., 1987, \aap, 174, 78

\bibitem[{{Azimlu} {et~al}\mbox{.}(2015){Azimlu}, {Mart{\'\i}nez-Galarza}, \&
  {Muench}}]{azim15}
{Azimlu} M., {Mart{\'\i}nez-Galarza} J.~R., {Muench} A.~A., 2015, \aj, 150, 95

\bibitem[{{Bally}(2008)}]{ball08}
{Bally} J., 2008, {Overview of the Orion Complex}, Vol.~4, {Handbook of Star
  Forming Regions}, p. 459

\bibitem[{{Balog} {et~al}\mbox{.}(2007){Balog}, {Muzerolle}, {Rieke}, {Su},
  {Young}, \& {Megeath}}]{balo07}
{Balog} Z., {Muzerolle} J., {Rieke} G.~H., {Su} K.~Y.~L., {Young} E.~T.,
  {Megeath} S.~T., 2007, \apj, 660, 1532

\bibitem[{{Baraffe} {et~al}\mbox{.}(1998){Baraffe}, {Chabrier}, {Allard}, \&
  {Hauschildt}}]{bara98}
{Baraffe} I., {Chabrier} G., {Allard} F., {Hauschildt} P.~H., 1998, \aap, 337,
  403

\bibitem[{{Barentsen} {et~al}\mbox{.}(2011){Barentsen}, {Vink}, {Drew},
  {Greimel}, {Wright}, {Drake}, {Martin}, {Valdivielso}, \& {Corradi}}]{bare11}
{Barentsen} G. {et~al.}, 2011, \mnras, 415, 103

\bibitem[{{Bastian} {et~al}\mbox{.}(2010){Bastian}, {Covey}, \&
  {Meyer}}]{bast10}
{Bastian} N., {Covey} K.~R., {Meyer} M.~R., 2010, \araa, 48, 339

\bibitem[{{Baume} {et~al}\mbox{.}(2014){Baume}, {Rodr{\'\i}guez}, {Corti},
  {Carraro}, \& {Panei}}]{baum14}
{Baume} G., {Rodr{\'\i}guez} M.~J., {Corti} M.~A., {Carraro} G., {Panei} J.~A.,
  2014, \mnras, 443, 411

\bibitem[{{Baumgardt} \& {Kroupa}(2007)}]{baum07}
{Baumgardt} H., {Kroupa} P., 2007, \mnras, 380, 1589

\bibitem[{{Beccari} {et~al}\mbox{.}(2018){Beccari}, {Boffin}, {Jerabkova},
  {Wright}, {Kalari}, {Carraro}, {De Marchi}, \& {de Wit}}]{becc18}
{Beccari} G., {Boffin} H.~M.~J., {Jerabkova} T., {Wright} N.~J., {Kalari}
  V.~M., {Carraro} G., {De Marchi} G., {de Wit} W.-J., 2018, \mnras, 481, L11

\bibitem[{{Beccari} {et~al}\mbox{.}(2017){Beccari}, {Petr-Gotzens}, {Boffin},
  {Romaniello}, {Fedele}, {Carraro}, {De Marchi}, {de Wit}, {Drew}, {Kalari},
  {Manara}, {Martin}, {Mieske}, {Panagia}, {Testi}, {Vink}, {Walsh}, \&
  {Wright}}]{becc17}
{Beccari} G. {et~al.}, 2017, \aap, 604, A22

\bibitem[{{Belikov} {et~al}\mbox{.}(2002{\natexlab{a}}){Belikov}, {Kharchenko},
  {Piskunov}, {Schilbach}, \& {Scholz}}]{beli02b}
{Belikov} A.~N., {Kharchenko} N.~V., {Piskunov} A.~E., {Schilbach} E., {Scholz}
  R.~D., 2002{\natexlab{a}}, \aap, 387, 117

\bibitem[{{Belikov} {et~al}\mbox{.}(2002{\natexlab{b}}){Belikov}, {Kharchenko},
  {Piskunov}, {Schilbach}, {Scholz}, \& {Yatsenko}}]{beli02a}
{Belikov} A.~N., {Kharchenko} N.~V., {Piskunov} A.~E., {Schilbach} E., {Scholz}
  R.~D., {Yatsenko} A.~I., 2002{\natexlab{b}}, \aap, 384, 145

\bibitem[{{Bell} {et~al}\mbox{.}(2013){Bell}, {Naylor}, {Mayne}, {Jeffries}, \&
  {Littlefair}}]{bell13}
{Bell} C.~P.~M., {Naylor} T., {Mayne} N.~J., {Jeffries} R.~D., {Littlefair}
  S.~P., 2013, \mnras, 434, 806

\bibitem[{{Berlanas} {et~al}\mbox{.}(2018){Berlanas}, {Herrero}, {Comer{\'o}n},
  {Pasquali}, {Bertelli Motta}, \& {Sota}}]{berl18a}
{Berlanas} S.~R., {Herrero} A., {Comer{\'o}n} F., {Pasquali} A., {Bertelli
  Motta} C., {Sota} A., 2018, \aap, 612, A50

\bibitem[{{Berlanas} {et~al}\mbox{.}(2020){Berlanas}, {Herrero}, {Comer{\'o}n},
  {Sim{\'o}n-D{\'\i}az}, {Lennon}, {Pasquali}, {Ma{\'\i}z Apellan{\'\i}z},
  {Sota}, \& {Peller{\'\i}n}}]{berl20}
{Berlanas} S.~R. {et~al.}, 2020, arXiv e-prints, arXiv:2008.09917

\bibitem[{{Berlanas} {et~al}\mbox{.}(2019){Berlanas}, {Wright}, {Herrero},
  {Drew}, \& {Lennon}}]{berl19}
{Berlanas} S.~R., {Wright} N.~J., {Herrero} A., {Drew} J.~E., {Lennon} D.~J.,
  2019, \mnras, 484, 1838

\bibitem[{{Bertoldi}(1989)}]{bert89}
{Bertoldi} F., 1989, \apj, 346, 735

\bibitem[{{Beuther} {et~al}\mbox{.}(2007){Beuther}, {Churchwell}, {McKee}, \&
  {Tan}}]{beut07}
{Beuther} H., {Churchwell} E.~B., {McKee} C.~F., {Tan} J.~C., 2007, in
  Protostars and Planets V, {Reipurth} B., {Jewitt} D., {Keil} K., eds., p. 165

\bibitem[{{Biazzo} {et~al}\mbox{.}(2011){Biazzo}, {Randich}, \&
  {Palla}}]{biaz11}
{Biazzo} K., {Randich} S., {Palla} F., 2011, \aap, 525, A35

\bibitem[{{Bica} {et~al}\mbox{.}(2008){Bica}, {Bonatto}, {Dutra}, \&
  {Santos}}]{bica08}
{Bica} E., {Bonatto} C., {Dutra} C.~M., {Santos} J.~F.~C., 2008, \mnras, 389,
  678

\bibitem[{{Binks} \& {Jeffries}(2014)}]{bink14}
{Binks} A.~S., {Jeffries} R.~D., 2014, \mnras, 438, L11

\bibitem[{{Binks} {et~al}\mbox{.}(2015){Binks}, {Jeffries}, \&
  {Maxted}}]{bink15}
{Binks} A.~S., {Jeffries} R.~D., {Maxted} P.~F.~L., 2015, \mnras, 452, 173

\bibitem[{{Blaauw}(1944)}]{blaa44}
{Blaauw} A., 1944, Bulletin of the Astronomical Institutes of the Netherlands,
  10, 29

\bibitem[{{Blaauw}(1946)}]{blaa46}
{Blaauw} A., 1946, Publications of the Kapteyn Astronomical Laboratory
  Groningen, 52, 1

\bibitem[{{Blaauw}(1952)}]{blaa52}
{Blaauw} A., 1952, {Bulletin of the Astronomical Institutes of the
  Netherlands}, 11, 414

\bibitem[{{Blaauw}(1956)}]{blaa56}
{Blaauw} A., 1956, \apj, 123, 408

\bibitem[{{Blaauw}(1958)}]{blaa58}
{Blaauw} A., 1958, \aj, 63, 186

\bibitem[{{Blaauw}(1961)}]{blaa61}
{Blaauw} A., 1961, {Bulletin of the Astronomical Institutes of the
  Netherlands}, 15, 265

\bibitem[{{Blaauw}(1964{\natexlab{a}})}]{blaa64}
{Blaauw} A., 1964{\natexlab{a}}, \araa, 2, 213

\bibitem[{{Blaauw}(1964{\natexlab{b}})}]{blaa64b}
{Blaauw} A., 1964{\natexlab{b}}, in IAU Symposium, Vol.~20, The Galaxy and the
  Magellanic Clouds, {Kerr} F.~J., ed., p.~50

\bibitem[{{Blaauw}(1978)}]{blaa78}
{Blaauw} A., 1978, {Internal Motions and Age of the Sub-Association Upper
  Scorpio}, Armenian Acad. of Sci., p. 101

\bibitem[{{Blaauw}(1984)}]{blaa84}
{Blaauw} A., 1984, Irish Astronomical Journal, 16, 141

\bibitem[{{Blaauw}(1991)}]{blaa91}
{Blaauw} A., 1991, in NATO ASIC Proc. 342: The Physics of Star Formation and
  Early Stellar Evolution, {Lada} C.~J., {Kylafis} N.~D., eds., pp. 125--+

\bibitem[{{Blaauw} {et~al}\mbox{.}(1959){Blaauw}, {Hiltner}, \&
  {Johnson}}]{blaa59}
{Blaauw} A., {Hiltner} W.~A., {Johnson} H.~L., 1959, \apj, 130, 69

\bibitem[{{Blaauw} \& {Morgan}(1953)}]{blaa53}
{Blaauw} A., {Morgan} W.~W., 1953, \apj, 117, 256

\bibitem[{{Blaha} \& {Humphreys}(1989)}]{blah89}
{Blaha} C., {Humphreys} R.~M., 1989, \aj, 98, 1598

\bibitem[{{Blitz} \& {Thaddeus}(1980)}]{blit80}
{Blitz} L., {Thaddeus} P., 1980, \apj, 241, 676

\bibitem[{{Bobylev} \& {Baykova}(2020)}]{boby20}
{Bobylev} V.~V., {Baykova} A.~T., 2020, Astronomy Reports, 64, 326

\bibitem[{{Bochkarev} \& {Sitnik}(1985)}]{boch85}
{Bochkarev} N.~G., {Sitnik} T.~G., 1985, \apss, 108, 237

\bibitem[{{Boily} \& {Kroupa}(2003)}]{boil03a}
{Boily} C.~M., {Kroupa} P., 2003, \mnras, 338, 665

\bibitem[{{Bok}(1934)}]{bok34}
{Bok} B.~J., 1934, Harvard College Observatory Circular, 384, 1

\bibitem[{{Bonnell} {et~al}\mbox{.}(2007){Bonnell}, {Larson}, \&
  {Zinnecker}}]{bonn07}
{Bonnell} I.~A., {Larson} R.~B., {Zinnecker} H., 2007, in Protostars and
  Planets V, {Reipurth} B., {Jewitt} D., {Keil} K., eds., p. 149

\bibitem[{{Bouy} \& {Alves}(2015)}]{bouy15}
{Bouy} H., {Alves} J., 2015, \aap, 584, A26

\bibitem[{{Bouy} {et~al}\mbox{.}(2014){Bouy}, {Alves}, {Bertin}, {Sarro}, \&
  {Barrado}}]{bouy14}
{Bouy} H., {Alves} J., {Bertin} E., {Sarro} L.~M., {Barrado} D., 2014, \aap,
  564, A29

\bibitem[{{Brand} \& {Zealey}(1975)}]{bran75}
{Brand} P.~W.~J.~L., {Zealey} W.~J., 1975, \aap, 38, 363

\bibitem[{{Brandner} {et~al}\mbox{.}(1996){Brandner}, {Alcala}, {Kunkel},
  {Moneti}, \& {Zinnecker}}]{bran96}
{Brandner} W., {Alcala} J.~M., {Kunkel} M., {Moneti} A., {Zinnecker} H., 1996,
  \aap, 307, 121

\bibitem[{{Bressert} {et~al}\mbox{.}(2010){Bressert}, {Bastian}, {Gutermuth},
  {Megeath}, {Allen}, {Evans}, {Rebull}, {Hatchell}, {Johnstone}, {Bourke},
  {Cieza}, {Harvey}, {Merin}, {Ray}, \& {Tothill}}]{bres10}
{Bressert} E. {et~al.}, 2010, \mnras, 409, L54

\bibitem[{{Brice{\~n}o}(2009)}]{bric09}
{Brice{\~n}o} C., 2009, in Revista Mexicana de Astronomia y Astrofisica
  Conference Series, Vol.~35, pp. 27--32

\bibitem[{{Brice{\~n}o} {et~al}\mbox{.}(2005){Brice{\~n}o}, {Calvet},
  {Hern{\'a}ndez}, {Vivas}, {Hartmann}, {Downes}, \& {Berlind}}]{bric05}
{Brice{\~n}o} C., {Calvet} N., {Hern{\'a}ndez} J., {Vivas} A.~K., {Hartmann}
  L., {Downes} J.~J., {Berlind} P., 2005, \aj, 129, 907

\bibitem[{{Brice{\~n}o} {et~al}\mbox{.}(2019){Brice{\~n}o}, {Calvet},
  {Hern{\'a}ndez}, {Vivas}, {Mateu}, {Jos{\'e} Downes}, {Loerincs},
  {P{\'e}rez-Blanco}, {Berlind}, {Espaillat}, {Allen}, {Hartmann}, {Mateo}, \&
  {Bailey}}]{bric19}
{Brice{\~n}o} C. {et~al.}, 2019, \aj, 157, 85

\bibitem[{{Brice{\~n}o} {et~al}\mbox{.}(2007){Brice{\~n}o}, {Preibisch},
  {Sherry}, {Mamajek}, {Mathieu}, {Walter}, \& {Zinnecker}}]{bric07}
{Brice{\~n}o} C., {Preibisch} T., {Sherry} W.~H., {Mamajek} E.~A., {Mathieu}
  R.~D., {Walter} F.~M., {Zinnecker} H., 2007, in Protostars and Planets V,
  {Reipurth} B., {Jewitt} D., {Keil} K., eds., pp. 345--360

\bibitem[{{Brice{\~n}o} {et~al}\mbox{.}(2001){Brice{\~n}o}, {Vivas}, {Calvet},
  {Hartmann}, {Pacheco}, {Herrera}, {Romero}, {Berlind}, {S{\'a}nchez},
  {Snyder}, \& {Andrews}}]{bric01}
{Brice{\~n}o} C. {et~al.}, 2001, Science, 291, 93

\bibitem[{{Brown}(1998)}]{brow98}
{Brown} A. G.~A., 1998, in Astronomical Society of the Pacific Conference
  Series, Vol. 142, The Stellar Initial Mass Function (38th Herstmonceux
  Conference), {Gilmore} G., {Howell} D., eds., p.~45

\bibitem[{{Brown} {et~al}\mbox{.}(1999){Brown}, {Blaauw}, {Hoogerwerf}, {de
  Bruijne}, \& {de Zeeuw}}]{brow99}
{Brown} A.~G.~A., {Blaauw} A., {Hoogerwerf} R., {de Bruijne} J.~H.~J., {de
  Zeeuw} P.~T., 1999, in NATO Advanced Science Institutes (ASI) Series C,
  {Lada} C.~J., {Kylafis} N.~D., eds., Vol. 540, p. 411

\bibitem[{{Brown} {et~al}\mbox{.}(1994){Brown}, {de Geus}, \& {de
  Zeeuw}}]{brow94}
{Brown} A.~G.~A., {de Geus} E.~J., {de Zeeuw} P.~T., 1994, \aap, 289, 101

\bibitem[{{Brown} {et~al}\mbox{.}(1997){Brown}, {Dekker}, \& {de
  Zeeuw}}]{brow97}
{Brown} A.~G.~A., {Dekker} G., {de Zeeuw} P.~T., 1997, \mnras, 285, 479

\bibitem[{{Brown} {et~al}\mbox{.}(1995){Brown}, {Hartmann}, \&
  {Burton}}]{brow95}
{Brown} A.~G.~A., {Hartmann} D., {Burton} W.~B., 1995, \aap, 300, 903

\bibitem[{{Brown} {et~al}\mbox{.}(1989){Brown}, {Sneden}, {Lambert}, \&
  {Dutchover}}]{brow89}
{Brown} J.~A., {Sneden} C., {Lambert} D.~L., {Dutchover}, Edward J., 1989,
  \apjs, 71, 293

\bibitem[{{Buckner} {et~al}\mbox{.}(2019){Buckner}, {Khorrami}, {Khalaj},
  {Lumsden}, {Joncour}, {Moraux}, {Clark}, {Oudmaijer}, {Blanco}, {de la
  Calle}, {Herrera-Fernand ez}, {Motte}, {Salgado}, \&
  {Valero-Mart{\'\i}n}}]{buck19}
{Buckner} A. S.~M. {et~al.}, 2019, \aap, 622, A184

\bibitem[{{Caballero} \& {Dinis}(2008)}]{caba08b}
{Caballero} J.~A., {Dinis} L., 2008, Astronomische Nachrichten, 329, 801

\bibitem[{{Caballero-Nieves} {et~al}\mbox{.}(2020){Caballero-Nieves}, {Gies},
  {Baines}, {Bouchez}, {Dekany}, {Goodwin}, {Rickman}, {Roberts}, {Taggart},
  {ten Brummelaar}, \& {Turner}}]{caba20}
{Caballero-Nieves} S.~M. {et~al.}, 2020, arXiv e-prints, arXiv:2008.00064

\bibitem[{{Caballero-Nieves} {et~al}\mbox{.}(2014){Caballero-Nieves}, {Nelan},
  {Gies}, {Wallace}, {DeGioia-Eastwood}, {Herrero}, {Jao}, {Mason}, {Massey},
  {Moffat}, \& {Walborn}}]{caba14}
{Caballero-Nieves} S.~M. {et~al.}, 2014, \aj, 147, 40

\bibitem[{{Cai} {et~al}\mbox{.}(2019){Cai}, {Portegies Zwart}, {Kouwenhoven},
  \& {Spurzem}}]{cai19}
{Cai} M.~X., {Portegies Zwart} S., {Kouwenhoven} M.~B.~N., {Spurzem} R., 2019,
  \mnras, 489, 4311

\bibitem[{{Cantat-Gaudin} {et~al}\mbox{.}(2019{\natexlab{a}}){Cantat-Gaudin},
  {Jordi}, {Wright}, {Armstrong}, {Vallenari}, {Balaguer-N{\'u}{\~n}ez},
  {Ramos}, {Bossini}, {Padoan}, \& {Pelkonen}}]{cant19b}
{Cantat-Gaudin} T. {et~al.}, 2019{\natexlab{a}}, \aap, 626, A17

\bibitem[{{Cantat-Gaudin} {et~al}\mbox{.}(2019{\natexlab{b}}){Cantat-Gaudin},
  {Mapelli}, {Balaguer-N{\'u}{\~n}ez}, {Jordi}, {Sacco}, \&
  {Vallenari}}]{cant19a}
{Cantat-Gaudin} T., {Mapelli} M., {Balaguer-N{\'u}{\~n}ez} L., {Jordi} C.,
  {Sacco} G., {Vallenari} A., 2019{\natexlab{b}}, \aap, 621, A115

\bibitem[{{Carpenter} {et~al}\mbox{.}(2001){Carpenter}, {Hillenbrand}, \&
  {Skrutskie}}]{carp01}
{Carpenter} J.~M., {Hillenbrand} L.~A., {Skrutskie} M.~F., 2001, \aj, 121, 3160

\bibitem[{{Carpenter} {et~al}\mbox{.}(2006){Carpenter}, {Mamajek},
  {Hillenbrand}, \& {Meyer}}]{carp06}
{Carpenter} J.~M., {Mamajek} E.~E., {Hillenbrand} L.~A., {Meyer} M.~R., 2006,
  \apjl, 651, L49

\bibitem[{{Cartwright} \& {Whitworth}(2004)}]{cart04}
{Cartwright} A., {Whitworth} A.~P., 2004, \mnras, 348, 589

\bibitem[{{Cash} {et~al}\mbox{.}(1980){Cash}, {Charles}, {Bowyer}, {Walter},
  {Garmire}, \& {Riegler}}]{cash80}
{Cash} W., {Charles} P., {Bowyer} S., {Walter} F., {Garmire} G., {Riegler} G.,
  1980, \apjl, 238, L71

\bibitem[{{Chatterjee} {et~al}\mbox{.}(2004){Chatterjee}, {Cordes},
  {Vlemmings}, {Arzoumanian}, {Goss}, \& {Lazio}}]{chatt04}
{Chatterjee} S., {Cordes} J.~M., {Vlemmings} W.~H.~T., {Arzoumanian} Z., {Goss}
  W.~M., {Lazio} T.~J.~W., 2004, \apj, 604, 339

\bibitem[{{Chen} {et~al}\mbox{.}(2019){Chen}, {D'Onghia}, {Alves}, \&
  {Adamo}}]{chen19}
{Chen} B., {D'Onghia} E., {Alves} J., {Adamo} A., 2019, arXiv e-prints,
  arXiv:1905.11429

\bibitem[{{Chen} \& {Lee}(2008)}]{chen08}
{Chen} W.~P., {Lee} H.~T., 2008, {The Lacerta OB1 Association}, Vol.~4,
  {Handbook of Star Forming Regions}, p. 124

\bibitem[{{Chen} {et~al}\mbox{.}(2011){Chen}, {Pandey}, {Sharma}, {Chen},
  {Chen}, {Sperauskas}, {Ogura}, {Chuang}, \& {Boyle}}]{chen11b}
{Chen} W.~P. {et~al.}, 2011, \aj, 142, 71

\bibitem[{{Chevalier}(1977)}]{chev77}
{Chevalier} R.~A., 1977, \araa, 15, 175

\bibitem[{{Chini} {et~al}\mbox{.}(2012){Chini}, {Hoffmeister}, {Nasseri},
  {Stahl}, \& {Zinnecker}}]{chin12}
{Chini} R., {Hoffmeister} V.~H., {Nasseri} A., {Stahl} O., {Zinnecker} H.,
  2012, \mnras, 424, 1925

\bibitem[{{Churchwell} \& {Felli}(1970)}]{chur70}
{Churchwell} E., {Felli} M., 1970, \aap, 4, 309

\bibitem[{{Clari{\'a}}(1974)}]{clar74b}
{Clari{\'a}} J.~J., 1974, \aap, 37, 229

\bibitem[{{Clark} {et~al}\mbox{.}(2005){Clark}, {Bonnell}, {Zinnecker}, \&
  {Bate}}]{clar05c}
{Clark} P.~C., {Bonnell} I.~A., {Zinnecker} H., {Bate} M.~R., 2005, \mnras,
  359, 809

\bibitem[{{Comer{\'o}n} {et~al}\mbox{.}(2016){Comer{\'o}n}, {Djupvik},
  {Schneider}, \& {Pasquali}}]{come16}
{Comer{\'o}n} F., {Djupvik} A.~A., {Schneider} N., {Pasquali} A., 2016, \aap,
  586, A46

\bibitem[{{Comer{\'o}n} \& {Pasquali}(2007)}]{come07}
{Comer{\'o}n} F., {Pasquali} A., 2007, \aap, 467, L23

\bibitem[{{Comer{\'o}n} \& {Pasquali}(2012)}]{come12}
{Comer{\'o}n} F., {Pasquali} A., 2012, \aap, 543, A101

\bibitem[{{Comer{\'o}n} {et~al}\mbox{.}(2008){Comer{\'o}n}, {Pasquali},
  {Figueras}, \& {Torra}}]{come08}
{Comer{\'o}n} F., {Pasquali} A., {Figueras} F., {Torra} J., 2008, \aap, 486,
  453

\bibitem[{{Comer{\'o}n} {et~al}\mbox{.}(2002){Comer{\'o}n}, {Pasquali},
  {Rodighiero}, {Stanishev}, {De Filippis}, {L{\'o}pez Mart{\'{\i}}},
  {G{\'a}lvez Ortiz}, {Stankov}, \& {Gredel}}]{come02}
{Comer{\'o}n} F. {et~al.}, 2002, \aap, 389, 874

\bibitem[{{Comeron} {et~al}\mbox{.}(1998){Comeron}, {Torra}, \&
  {Gomez}}]{come98}
{Comeron} F., {Torra} J., {Gomez} A.~E., 1998, \aap, 330, 975

\bibitem[{{Costado} {et~al}\mbox{.}(2017){Costado}, {Alfaro}, {Gonz{\'a}lez},
  \& {Sampedro}}]{cost17}
{Costado} M.~T., {Alfaro} E.~J., {Gonz{\'a}lez} M., {Sampedro} L., 2017,
  \mnras, 465, 3879

\bibitem[{{Cottaar} {et~al}\mbox{.}(2012){Cottaar}, {Meyer}, \&
  {Parker}}]{cott12b}
{Cottaar} M., {Meyer} M.~R., {Parker} R.~J., 2012, \aap, 547, A35

\bibitem[{{Cox}(1972)}]{cox72}
{Cox} D.~P., 1972, \apj, 178, 159

\bibitem[{{Crowther} {et~al}\mbox{.}(1995){Crowther}, {Smith}, {Hillier}, \&
  {Schmutz}}]{crow95}
{Crowther} P.~A., {Smith} L.~J., {Hillier} D.~J., {Schmutz} W., 1995, \aap,
  293, 427

\bibitem[{{Crundall} {et~al}\mbox{.}(2019){Crundall}, {Ireland}, {Krumholz},
  {Federrath}, {{\v{Z}}erjal}, \& {Hansen}}]{crun19}
{Crundall} T.~D., {Ireland} M.~J., {Krumholz} M.~R., {Federrath} C.,
  {{\v{Z}}erjal} M., {Hansen} J.~T., 2019, arXiv e-prints, arXiv:1902.07732

\bibitem[{{Cunha} \& {Lambert}(1994)}]{cunh94}
{Cunha} K., {Lambert} D.~L., 1994, \apj, 426, 170

\bibitem[{{Da Rio} {et~al}\mbox{.}(2017){Da Rio}, {Tan}, {Covey}, {Cottaar},
  {Foster}, {Cullen}, {Tobin}, {Kim}, {Meyer}, {Nidever}, {Stassun},
  {Chojnowski}, {Flaherty}, {Majewski}, {Skrutskie}, {Zasowski}, \&
  {Pan}}]{dari17}
{Da Rio} N. {et~al.}, 2017, \apj, 845, 105

\bibitem[{{Da Rio} {et~al}\mbox{.}(2016){Da Rio}, {Tan}, {Covey}, {Cottaar},
  {Foster}, {Cullen}, {Tobin}, {Kim}, {Meyer}, {Nidever}, {Stassun},
  {Chojnowski}, {Flaherty}, {Majewski}, {Skrutskie}, {Zasowski}, \&
  {Pan}}]{dari16}
{Da Rio} N. {et~al.}, 2016, \apj, 818, 59

\bibitem[{{Dale}(2015)}]{dale15b}
{Dale} J.~E., 2015, New Astronomy Reviews, 68, 1

\bibitem[{{Dale} {et~al}\mbox{.}(2015){Dale}, {Ercolano}, \&
  {Bonnell}}]{dale15}
{Dale} J.~E., {Ercolano} B., {Bonnell} I.~A., 2015, \mnras, 451, 987

\bibitem[{{Dale} {et~al}\mbox{.}(2013){Dale}, {Ngoumou}, {Ercolano}, \&
  {Bonnell}}]{dale13}
{Dale} J.~E., {Ngoumou} J., {Ercolano} B., {Bonnell} I.~A., 2013, \mnras, 436,
  3430

\bibitem[{{Dalton} {et~al}\mbox{.}(2018){Dalton}, {Trager}, {Abrams},
  {Bonifacio}, {Aguerri}, {Vallenari}, {Middleton}, {Benn}, {Dee},
  {Say{\`e}de}, {Lewis}, {Pragt}, {Pic{\'o}}, {Walton}, {Rey}, {Allende},
  {Lhom{\'e}}, {Terrett}, {Brock}, {Gilbert}, {Ridings}, {Verheijen}, {Tosh},
  {Steele}, {Stuik}, {Kroes}, {Tromp}, {Kragt}, {Lesman}, {Mottram}, {Bates},
  {Gribbin}, {Burgal}, {Herreros}, {Delgado}, {Martin}, {Cano}, {Navarro},
  {Irwin}, {Lewis}, {Gonzales Solares}, {O'Mahony}, {Bianco}, {Zurita}, {ter
  Horst}, {Molinari}, {Lodi}, {Guerra}, {Baruffolo}, {Carrasco}, {Farkas},
  {Schallig}, {Hill}, {Smith}, {Drew}, {Poggianti}, {Pieri}, {Jin}, {Dominquez
  Palmero}, {Fari{\~n}a}, {Martin}, {Worley}, {Murphy}, {Hidalgo}, {Mignot},
  {Bishop}, {Guest}, {Elswijk}, {de Haan}, {Hanenburg}, {Salasnich}, {Mayya},
  {Izazaga-P{\'e}rez}, \& {Peralta de Arriba}}]{dalt18}
{Dalton} G. {et~al.}, 2018, in Society of Photo-Optical Instrumentation
  Engineers (SPIE) Conference Series, Vol. 10702, Proceedings of SPIE, p.
  107021B

\bibitem[{{Damiani}(2018)}]{dami18}
{Damiani} F., 2018, \aap, 615, A148

\bibitem[{{Damiani} {et~al}\mbox{.}(2017){Damiani}, {Bonito}, {Prisinzano},
  {Zwitter}, {Bayo}, {Kalari}, {Jim{\'e}nez-Esteban}, {Costado}, {Jofr{\'e}},
  {Randich}, {Flaccomio}, {Lanzafame}, {Lardo}, {Morbidelli}, \&
  {Zaggia}}]{dami17}
{Damiani} F. {et~al.}, 2017, \aap, 604, A135

\bibitem[{{Damiani} {et~al}\mbox{.}(2016){Damiani}, {Micela}, \&
  {Sciortino}}]{dami16}
{Damiani} F., {Micela} G., {Sciortino} S., 2016, \aap, 596, A82

\bibitem[{{Damiani} {et~al}\mbox{.}(2014){Damiani}, {Prisinzano}, {Micela},
  {Randich}, {Gilmore}, {Drew}, {Jeffries}, {Fr{\'e}mat}, {Alfaro}, {Bensby},
  {Bragaglia}, {Flaccomio}, {Lanzafame}, {Pancino}, {Recio-Blanco}, {Sacco},
  {Smiljanic}, {Jackson}, {de Laverny}, {Morbidelli}, {Worley}, {Hourihane},
  {Costado}, {Jofr{\'e}}, {Lind}, \& {Maiorca}}]{dami14}
{Damiani} F. {et~al.}, 2014, \aap, 566, A50

\bibitem[{{Damiani} {et~al}\mbox{.}(2019){Damiani}, {Prisinzano}, {Micela}, \&
  {Sciortino}}]{dami19}
{Damiani} F., {Prisinzano} L., {Micela} G., {Sciortino} S., 2019, \aap, 623,
  A25

\bibitem[{{D'Antona} \& {Mazzitelli}(1994)}]{dant94}
{D'Antona} F., {Mazzitelli} I., 1994, \apjs, 90, 467

\bibitem[{{De Becker} {et~al}\mbox{.}(2004){De Becker}, {Rauw}, \&
  {Manfroid}}]{debe04}
{De Becker} M., {Rauw} G., {Manfroid} J., 2004, \aap, 424, L39

\bibitem[{{de Bruijne}(1999{\natexlab{a}})}]{debr99b}
{de Bruijne} J.~H.~J., 1999{\natexlab{a}}, \mnras, 306, 381

\bibitem[{{de Bruijne}(1999{\natexlab{b}})}]{debr99}
{de Bruijne} J.~H.~J., 1999{\natexlab{b}}, \mnras, 310, 585

\bibitem[{{de Geus}(1992)}]{dege92}
{de Geus} E.~J., 1992, \aap, 262, 258

\bibitem[{{de Geus} {et~al}\mbox{.}(1989){de Geus}, {de Zeeuw}, \&
  {Lub}}]{dege89}
{de Geus} E.~J., {de Zeeuw} P.~T., {Lub} J., 1989, \aap, 216, 44

\bibitem[{{de Jong} {et~al}\mbox{.}(2019){de Jong}, {Agertz}, {Berbel}, {Aird},
  {Alexander}, {Amarsi}, {Anders}, {Andrae}, {Ansarinejad}, {Ansorge},
  {Antilogus}, {Anwand -Heerwart}, {Arentsen}, {Arnadottir}, {Asplund},
  {Auger}, {Azais}, {Baade}, {Baker}, {Baker}, {Balbinot}, {Baldry}, {Banerji},
  {Barden}, {Barklem}, {Barth{\'e}l{\'e}my-Mazot}, {Battistini}, {Bauer},
  {Bell}, {Bellido-Tirado}, {Bellstedt}, {Belokurov}, {Bensby}, {Bergemann},
  {Bestenlehner}, {Bielby}, {Bilicki}, {Blake}, {Bland-Hawthorn}, {Boeche},
  {Boland}, {Boller}, {Bongard}, {Bongiorno}, {Bonifacio}, {Boudon}, {Brooks},
  {Brown}, {Brown}, {Br{\"u}ggen}, {Brynnel}, {Brzeski}, {Buchert},
  {Buschkamp}, {Caffau}, {Caillier}, {Carrick}, {Casagrande}, {Case}, {Casey},
  {Cesarini}, {Cescutti}, {Chapuis}, {Chiappini}, {Childress}, {Christlieb},
  {Church}, {Cioni}, {Cluver}, {Colless}, {Collett}, {Comparat}, {Cooper},
  {Couch}, {Courbin}, {Croom}, {Croton}, {Daguis{\'e}}, {Dalton}, {Davies},
  {Davis}, {de Laverny}, {Deason}, {Dionies}, {Disseau}, {Doel}, {D{\"o}scher},
  {Driver}, {Dwelly}, {Eckert}, {Edge}, {Edvardsson}, {Youssoufi}, {Elhaddad},
  {Enke}, {Erfanianfar}, {Farrell}, {Fechner}, {Feiz}, {Feltzing}, {Ferreras},
  {Feuerstein}, {Feuillet}, {Finoguenov}, {Ford}, {Fotopoulou}, {Fouesneau},
  {Frenk}, {Frey}, {Gaessler}, {Geier}, {Fusillo}, {Gerhard}, {Giannantonio},
  {Giannone}, {Gibson}, {Gillingham}, {Gonz{\'a}lez-Fern{\'a}ndez},
  {Gonzalez-Solares}, {Gottloeber}, {Gould}, {Grebel}, {Gueguen}, {Guiglion},
  {Haehnelt}, {Hahn}, {Hansen}, {Hartman}, {Hauptner}, {Hawkins}, {Haynes},
  {Haynes}, {Heiter}, {Helmi}, {Aguayo}, {Hewett}, {Hinton}, {Hobbs}, {Hoenig},
  {Hofman}, {Hook}, {Hopgood}, {Hopkins}, {Hourihane}, {Howes}, {Howlett},
  {Huet}, {Irwin}, {Iwert}, {Jablonka}, {Jahn}, {Jahnke}, {Jarno}, {Jin},
  {Jofre}, {Johl}, {Jones}, {J{\"o}nsson}, {Jordan}, {Karovicova}, {Khalatyan},
  {Kelz}, {Kennicutt}, {King}, {Kitaura}, {Klar}, {Klauser}, {Kneib}, {Koch},
  {Koposov}, {Kordopatis}, {Korn}, {Kosmalski}, {Kotak}, {Kovalev}, {Kreckel},
  {Kripak}, {Krumpe}, {Kuijken}, {Kunder}, {Kushniruk}, {Lam}, {Lamer},
  {Laurent}, {Lawrence}, {Lehmitz}, {Lemasle}, {Lewis}, {Li}, {Lidman}, {Lind},
  {Liske}, {Lizon}, {Loveday}, {Ludwig}, {McDermid}, {Maguire}, {Mainieri},
  {Mali}, {Mandel}, {Mandel}, {Mannering}, {Martell}, {Martinez Delgado},
  {Matijevic}, {McGregor}, {McMahon}, {McMillan}, {Mena}, {Merloni}, {Meyer},
  {Michel}, {Micheva}, {Migniau}, {Minchev}, {Monari}, {Muller}, {Murphy},
  {Muthukrishna}, {Nandra}, {Navarro}, {Ness}, {Nichani}, {Nichol}, {Nicklas},
  {Niederhofer}, {Norberg}, {Obreschkow}, {Oliver}, {Owers}, {Pai},
  {Pankratow}, {Parkinson}, {Paschke}, {Paterson}, {Pecontal}, {Parry},
  {Phillips}, {Pillepich}, {Pinard}, {Pirard}, {Piskunov}, {Plank},
  {Pl{\"u}schke}, {Pons}, {Popesso}, {Power}, {Pragt}, {Pramskiy}, {Pryer},
  {Quattri}, {Queiroz}, {Quirrenbach}, {Rahurkar}, {Raichoor}, {Ramstedt},
  {Rau}, {Recio-Blanco}, {Reiss}, {Renaud}, {Revaz}, {Rhode}, {Richard},
  {Richter}, {Rix}, {Robotham}, {Roelfsema}, {Romaniello}, {Rosario},
  {Rothmaier}, {Roukema}, {Ruchti}, {Rupprecht}, {Rybizki}, {Ryde}, {Saar},
  {Sadler}, {Sahl{\'e}n}, {Salvato}, {Sassolas}, {Saunders}, {Saviauk},
  {Sbordone}, {Schmidt}, {Schnurr}, {Scholz}, {Schwope}, {Seifert}, {Shanks},
  {Sheinis}, {Sivov}, {Sk{\'u}lad{\'o}ttir}, {Smartt}, {Smedley}, {Smith},
  {Smith}, {Sorce}, {Spitler}, {Starkenburg}, {Steinmetz}, {Stilz}, {Storm},
  {Sullivan}, {Sutherland}, {Swann}, {Tamone}, {Taylor}, {Teillon}, {Tempel},
  {ter Horst}, {Thi}, {Tolstoy}, {Trager}, {Traven}, {Tremblay}, {Tresse},
  {Valentini}, {van de Weygaert}, {van den Ancker}, {Veljanoski}, {Venkatesan},
  {Wagner}, {Wagner}, {Walcher}, {Waller}, {Walton}, {Wang}, {Winkler},
  {Wisotzki}, {Worley}, {Worseck}, {Xiang}, {Xu}, {Yong}, {Zhao}, {Zheng},
  {Zscheyge}, \& {Zucker}}]{dejo19}
{de Jong} R.~S. {et~al.}, 2019, The Messenger, 175, 3

\bibitem[{{de Juan Ovelar} {et~al}\mbox{.}(2012){de Juan Ovelar}, {Kruijssen},
  {Bressert}, {Testi}, {Bastian}, \& {C{\'a}novas}}]{deju12}
{de Juan Ovelar} M., {Kruijssen} J.~M.~D., {Bressert} E., {Testi} L., {Bastian}
  N., {C{\'a}novas} H., 2012, \aap, 546, L1

\bibitem[{{de la Reza} {et~al}\mbox{.}(2006){de la Reza}, {Jilinski}, \&
  {Ortega}}]{dela06}
{de la Reza} R., {Jilinski} E., {Ortega} V.~G., 2006, \aj, 131, 2609

\bibitem[{{de Zeeuw} {et~al}\mbox{.}(1999){de Zeeuw}, {Hoogerwerf}, {de
  Bruijne}, {Brown}, \& {Blaauw}}]{deze99}
{de Zeeuw} P.~T., {Hoogerwerf} R., {de Bruijne} J.~H.~J., {Brown} A.~G.~A.,
  {Blaauw} A., 1999, \aj, 117, 354

\bibitem[{{DeGioia-Eastwood} {et~al}\mbox{.}(2001){DeGioia-Eastwood}, {Throop},
  {Walker}, \& {Cudworth}}]{degi01}
{DeGioia-Eastwood} K., {Throop} H., {Walker} G., {Cudworth} K.~M., 2001, \apj,
  549, 578

\bibitem[{{Delore} {et~al}\mbox{.}(1986){Delore}, {Willis}, \&
  {Lascarides}}]{hodg86}
{Delore} C.~W.~H., {Willis} A.~J., {Lascarides} P., eds., 1986, {Luminous Stars
  and Associations in Galaxies}, {Delore} C.~W.~H., {Willis} A.~J.,
  {Lascarides} P., eds. {IAU Symposium}

\bibitem[{{Dias} {et~al}\mbox{.}(2002){Dias}, {Alessi}, {Moitinho}, \&
  {L{\'e}pine}}]{dias02}
{Dias} W.~S., {Alessi} B.~S., {Moitinho} A., {L{\'e}pine} J.~R.~D., 2002, \aap,
  389, 871

\bibitem[{{Diehl} {et~al}\mbox{.}(2004){Diehl}, {Cervi{\~n}o}, {Hartmann}, \&
  {Kretschmer}}]{dieh04}
{Diehl} R., {Cervi{\~n}o} M., {Hartmann} D.~H., {Kretschmer} K., 2004, New
  Astronomy Reviews, 48, 81

\bibitem[{{Downes} \& {Rinehart}(1966)}]{down66}
{Downes} D., {Rinehart} R., 1966, \apj, 144, 937

\bibitem[{{Drew} {et~al}\mbox{.}(2014){Drew}, {Gonzalez-Solares}, {Greimel},
  {Irwin}, {K{\"u}pc{\"u} Yoldas}, {Lewis}, {Barentsen}, {Eisl{\"o}ffel},
  {Farnhill}, {Martin}, {Walsh}, {Walton}, {Mohr-Smith}, {Raddi}, {Sale},
  {Wright}, {Groot}, {Barlow}, {Corradi}, {Drake}, {Fabregat}, {Frew},
  {G{\"a}nsicke}, {Knigge}, {Mampaso}, {Morris}, {Naylor}, {Parker},
  {Phillipps}, {Ruhland}, {Steeghs}, {Unruh}, {Vink}, {Wesson}, \&
  {Zijlstra}}]{drew14}
{Drew} J.~E. {et~al.}, 2014, \mnras, 440, 2036

\bibitem[{{Drew} {et~al}\mbox{.}(2005){Drew}, {Greimel}, {Irwin},
  {Aungwerojwit}, {Barlow}, {Corradi}, {Drake}, {G{\"a}nsicke}, {Groot},
  {Hales}, {Hopewell}, {Irwin}, \& {Knigge}}]{drew05}
{Drew} J.~E. {et~al.}, 2005, \mnras, 362, 753

\bibitem[{{Drew} {et~al}\mbox{.}(2008){Drew}, {Greimel}, {Irwin}, \&
  {Sale}}]{drew08}
{Drew} J.~E., {Greimel} R., {Irwin} M.~J., {Sale} S.~E., 2008, \mnras, 386,
  1761

\bibitem[{{Drew} {et~al}\mbox{.}(2018){Drew}, {Herrero}, {Mohr-Smith},
  {Mongui{\'o}}, {Wright}, {Kupfer}, \& {Napiwotzki}}]{drew18}
{Drew} J.~E., {Herrero} A., {Mohr-Smith} M., {Mongui{\'o}} M., {Wright} N.~J.,
  {Kupfer} T., {Napiwotzki} R., 2018, \mnras, 480, 2109

\bibitem[{{Duch{\^e}ne} {et~al}\mbox{.}(1999){Duch{\^e}ne}, {Bouvier}, \&
  {Simon}}]{duch99}
{Duch{\^e}ne} G., {Bouvier} J., {Simon} T., 1999, \aap, 343, 831

\bibitem[{{Duch{\^e}ne} \& {Kraus}(2013)}]{duch13}
{Duch{\^e}ne} G., {Kraus} A., 2013, \araa, 51, 269

\bibitem[{{Duch{\^e}ne} {et~al}\mbox{.}(2018){Duch{\^e}ne}, {Lacour}, {Moraux},
  {Goodwin}, \& {Bouvier}}]{duch18}
{Duch{\^e}ne} G., {Lacour} S., {Moraux} E., {Goodwin} S., {Bouvier} J., 2018,
  \mnras, 478, 1825

\bibitem[{{Duerr} {et~al}\mbox{.}(1982){Duerr}, {Imhoff}, \& {Lada}}]{duer82}
{Duerr} R., {Imhoff} C.~L., {Lada} C.~J., 1982, \apj, 261, 135

\bibitem[{{Dzib} {et~al}\mbox{.}(2017){Dzib}, {Loinard}, {Rodr{\'\i}guez},
  {G{\'o}mez}, {Forbrich}, {Menten}, {Kounkel}, {Mioduszewski}, {Hartmann},
  {Tobin}, \& {Rivera}}]{dzib17}
{Dzib} S.~A. {et~al.}, 2017, \apj, 834, 139

\bibitem[{{Dzib} {et~al}\mbox{.}(2013){Dzib}, {Rodr{\'{\i}}guez}, {Loinard},
  {Mioduszewski}, {Ortiz-Le{\'o}n}, \& {Araudo}}]{dzib13}
{Dzib} S.~A., {Rodr{\'{\i}}guez} L.~F., {Loinard} L., {Mioduszewski} A.~J.,
  {Ortiz-Le{\'o}n} G.~N., {Araudo} A.~T., 2013, \apj, 763, 139

\bibitem[{{Eddington}(1914)}]{eddi14}
{Eddington} A.~S., 1914, {Stellar movements and the structure of the universe}.
  {}

\bibitem[{{Efremov} {et~al}\mbox{.}(1987){Efremov}, {Ivanov}, \&
  {Nikolov}}]{efre87}
{Efremov} I.~N., {Ivanov} G.~R., {Nikolov} N.~S., 1987, \apss, 135, 119

\bibitem[{{Efremov} \& {Elmegreen}(1998)}]{efre98}
{Efremov} Y.~N., {Elmegreen} B.~G., 1998, \mnras, 299, 588

\bibitem[{{Eggen}(1998)}]{egge98}
{Eggen} O.~J., 1998, \aj, 116, 1314

\bibitem[{{Eisner} {et~al}\mbox{.}(2018){Eisner}, {Arce}, {Ballering}, {Bally},
  {Andrews}, {Boyden}, {Di Francesco}, {Fang}, {Johnstone}, {Kim}, {Mann},
  {Matthews}, {Pascucci}, {Ricci}, {Sheehan}, \& {Williams}}]{eisn18}
{Eisner} J.~A. {et~al.}, 2018, \apj, 860, 77

\bibitem[{{Ekstr{\"o}m} {et~al}\mbox{.}(2012){Ekstr{\"o}m}, {Georgy},
  {Eggenberger}, {Meynet}, {Mowlavi}, {Wyttenbach}, {Granada}, {Decressin},
  {Hirschi}, {Frischknecht}, {Charbonnel}, \& {Maeder}}]{ekst12}
{Ekstr{\"o}m} S. {et~al.}, 2012, \aap, 537, A146

\bibitem[{{Eldridge}(2009)}]{eldr09}
{Eldridge} J.~J., 2009, \mnras, 400, L20

\bibitem[{{Elias} {et~al}\mbox{.}(2006){Elias}, {Cabrera-Ca{\~n}o}, \&
  {Alfaro}}]{elia06a}
{Elias} F., {Cabrera-Ca{\~n}o} J., {Alfaro} E.~J., 2006, \aj, 131, 2700

\bibitem[{{Elmegreen}(2008)}]{elme08}
{Elmegreen} B.~G., 2008, \apj, 672, 1006

\bibitem[{{Elmegreen} \& {Elmegreen}(1978)}]{elme78}
{Elmegreen} B.~G., {Elmegreen} D.~M., 1978, \apj, 220, 1051

\bibitem[{{Elmegreen} \& {Falgarone}(1996)}]{elme96b}
{Elmegreen} B.~G., {Falgarone} E., 1996, \apj, 471, 816

\bibitem[{{Elmegreen} \& {Hunter}(2010)}]{elme10}
{Elmegreen} B.~G., {Hunter} D.~A., 2010, \apj, 712, 604

\bibitem[{{Elmegreen} \& {Lada}(1977)}]{elme77}
{Elmegreen} B.~G., {Lada} C.~J., 1977, \apj, 214, 725

\bibitem[{{Essex} {et~al}\mbox{.}(2020){Essex}, {Basu}, {Prehl}, \&
  {Hoffmann}}]{esse20}
{Essex} C., {Basu} S., {Prehl} J., {Hoffmann} K.~H., 2020, \mnras, 494, 1579

\bibitem[{{Fang} {et~al}\mbox{.}(2017){Fang}, {Kim}, {Pascucci}, {Apai},
  {Zhang}, {Sicilia-Aguilar}, {Alonso-Mart{\'\i}nez}, {Eiroa}, \&
  {Wang}}]{fang17}
{Fang} M. {et~al.}, 2017, \aj, 153, 188

\bibitem[{{Feiden}(2016)}]{feid16}
{Feiden} G.~A., 2016, \aap, 593, A99

\bibitem[{{Feigelson} \& {Babu}(2012)}]{feig12}
{Feigelson} E.~D., {Babu} G.~J., 2012, {Modern Statistical Methods for
  Astronomy}. {Cambridge University Press}

\bibitem[{{Feigelson} {et~al}\mbox{.}(2011){Feigelson}, {Getman}, {Townsley},
  {Broos}, {Povich}, {Garmire}, {King}, {Montmerle}, {Preibisch}, {Smith},
  {Stassun}, {Wang}, {Wolk}, \& {Zinnecker}}]{feig11}
{Feigelson} E.~D. {et~al.}, 2011, \apjs, 194, 9

\bibitem[{{Feigelson} {et~al}\mbox{.}(2004){Feigelson}, {Hornschemeier},
  {Micela}, {Bauer}, {Alexander}, {Brandt}, {Favata}, {Sciortino}, \&
  {Garmire}}]{feig04}
{Feigelson} E.~D. {et~al.}, 2004, \apj, 611, 1107

\bibitem[{{Feinstein} \& {Forte}(1974)}]{fein74}
{Feinstein} A., {Forte} J.~C., 1974, \pasp, 86, 284

\bibitem[{{Fernandes} {et~al}\mbox{.}(2019){Fernandes}, {Montmerle},
  {Santos-Silva}, \& {Gregorio-Hetem}}]{fern19}
{Fernandes} B., {Montmerle} T., {Santos-Silva} T., {Gregorio-Hetem} J., 2019,
  \aap, 628, A44

\bibitem[{{Fern{\'a}ndez} {et~al}\mbox{.}(2008){Fern{\'a}ndez}, {Figueras}, \&
  {Torra}}]{fern08}
{Fern{\'a}ndez} D., {Figueras} F., {Torra} J., 2008, \aap, 480, 735

\bibitem[{{F{\H u}r{\'e}sz} {et~al}\mbox{.}(2008){F{\H u}r{\'e}sz}, {Hartmann},
  {Megeath}, {Szentgyorgyi}, \& {Hamden}}]{fure08}
{F{\H u}r{\'e}sz} G., {Hartmann} L.~W., {Megeath} S.~T., {Szentgyorgyi} A.~H.,
  {Hamden} E.~T., 2008, \apj, 676, 1109

\bibitem[{{Finkbeiner}(2003)}]{fink03}
{Finkbeiner} D.~P., 2003, \apjs, 146, 407

\bibitem[{{Fischer} {et~al}\mbox{.}(2016){Fischer}, {Padgett}, {Stapelfeldt},
  \& {Sewi{\l}o}}]{fisc16}
{Fischer} W.~J., {Padgett} D.~L., {Stapelfeldt} K.~L., {Sewi{\l}o} M., 2016,
  \apj, 827, 96

\bibitem[{{Fitzgerald} {et~al}\mbox{.}(1990){Fitzgerald}, {Harris}, \&
  {Reed}}]{fitz90}
{Fitzgerald} M.~P., {Harris} G.~L., {Reed} B.~C., 1990, \pasp, 102, 865

\bibitem[{{Flaccomio} {et~al}\mbox{.}(1999){Flaccomio}, {Micela}, {Sciortino},
  {Favata}, {Corbally}, \& {Tomaney}}]{flac99}
{Flaccomio} E., {Micela} G., {Sciortino} S., {Favata} F., {Corbally} C.,
  {Tomaney} A., 1999, \aap, 345, 521

\bibitem[{{Forbes}(2000)}]{forb00}
{Forbes} D., 2000, \aj, 120, 2594

\bibitem[{{Franciosini} {et~al}\mbox{.}(2018){Franciosini}, {Sacco},
  {Jeffries}, {Damiani}, {Roccatagliata}, {Fedele}, \& {Rand ich}}]{fran18}
{Franciosini} E., {Sacco} G.~G., {Jeffries} R.~D., {Damiani} F.,
  {Roccatagliata} V., {Fedele} D., {Rand ich} S., 2018, \aap, 616, L12

\bibitem[{{Gaia Collaboration} {et~al}\mbox{.}(2018){Gaia Collaboration},
  {Babusiaux}, {van Leeuwen}, {Barstow}, {Jordi}, {Vallenari}, {Bossini},
  {Bressan}, {Cantat-Gaudin}, {van Leeuwen}, {Brown}, {Prusti}, {de Bruijne},
  {Bailer-Jones}, {Biermann}, {Evans}, {Eyer}, {Jansen}, {Klioner}, {Lammers},
  {Lindegren}, {Luri}, {Mignard}, {Panem}, {Pourbaix}, {Randich}, {Sartoretti},
  {Siddiqui}, {Soubiran}, {Walton}, {Arenou}, {Bastian}, {Cropper}, {Drimmel},
  {Katz}, {Lattanzi}, {Bakker}, {Cacciari}, {Casta{\~n}eda}, {Chaoul}, {Cheek},
  {De Angeli}, {Fabricius}, {Guerra}, {Holl}, {Masana}, {Messineo}, {Mowlavi},
  {Nienartowicz}, {Panuzzo}, {Portell}, {Riello}, {Seabroke}, {Tanga},
  {Th{\'e}venin}, {Gracia-Abril}, {Comoretto}, {Garcia-Reinaldos}, {Teyssier},
  {Altmann}, {Andrae}, {Audard}, {Bellas-Velidis}, {Benson}, {Berthier},
  {Blomme}, {Burgess}, {Busso}, {Carry}, {Cellino}, {Clementini}, {Clotet},
  {Creevey}, {Davidson}, {De Ridder}, {Delchambre}, {Dell'Oro}, {Ducourant},
  {Fern{\'a}ndez-Hern{\'a}ndez}, {Fouesneau}, {Fr{\'e}mat}, {Galluccio},
  {Garc{\'\i}a-Torres}, {Gonz{\'a}lez-N{\'u}{\~n}ez}, {Gonz{\'a}lez-Vidal},
  {Gosset}, {Guy}, {Halbwachs}, {Hambly}, {Harrison}, {Hern{\'a}ndez},
  {Hestroffer}, {Hodgkin}, {Hutton}, {Jasniewicz}, {Jean-Antoine-Piccolo},
  {Jordan}, {Korn}, {Krone-Martins}, {Lanzafame}, {Lebzelter}, {L{\"o}ffler},
  {Manteiga}, {Marrese}, {Mart{\'\i}n-Fleitas}, {Moitinho}, {Mora}, {Muinonen},
  {Osinde}, {Pancino}, {Pauwels}, {Petit}, {Recio-Blanco}, {Richards},
  {Rimoldini}, {Robin}, {Sarro}, {Siopis}, {Smith}, {Sozzetti}, {S{\"u}veges},
  {Torra}, {van Reeven}, {Abbas}, {Abreu Aramburu}, {Accart}, {Aerts},
  {Altavilla}, {{\'A}lvarez}, {Alvarez}, {Alves}, {Anderson}, {Andrei},
  {Anglada Varela}, {Antiche}, {Antoja}, {Arcay}, {Astraatmadja}, {Bach},
  {Baker}, {Balaguer-N{\'u}{\~n}ez}, {Balm}, {Barache}, {Barata}, {Barbato},
  {Barblan}, {Barklem}, {Barrado}, {Barros}, {Bartholom{\'e} Mu{\~n}oz},
  {Bassilana}, {Becciani}, {Bellazzini}, {Berihuete}, {Bertone}, {Bianchi},
  {Bienaym{\'e}}, {Blanco-Cuaresma}, {Boch}, {Boeche}, {Bombrun}, {Borrachero},
  {Bouquillon}, {Bourda}, {Bragaglia}, {Bramante}, {Breddels}, {Brouillet},
  {Br{\"u}semeister}, {Brugaletta}, {Bucciarelli}, {Burlacu}, {Busonero},
  {Butkevich}, {Buzzi}, {Caffau}, {Cancelliere}, {Cannizzaro}, {Carballo},
  {Carlucci}, {Carrasco}, {Casamiquela}, {Castellani}, {Castro-Ginard},
  {Charlot}, {Chemin}, {Chiavassa}, {Cocozza}, {Costigan}, {Cowell}, {Crifo},
  {Crosta}, {Crowley}, {Cuypers}, {Dafonte}, {Damerdji}, {Dapergolas}, {David},
  {David}, {de Laverny}, {De Luise}, {De March}, {de Martino}, {de Souza}, {de
  Torres}, {Debosscher}, {del Pozo}, {Delbo}, {Delgado}, {Delgado}, {Diakite},
  {Diener}, {Distefano}, {Dolding}, {Drazinos}, {Dur{\'a}n}, {Edvardsson},
  {Enke}, {Eriksson}, {Esquej}, {Eynard Bontemps}, {Fabre}, {Fabrizio},
  {Faigler}, {Falc{\~a}o}, {Farr{\`a}s Casas}, {Federici}, {Fedorets},
  {Fernique}, {Figueras}, {Filippi}, {Findeisen}, {Fonti}, {Fraile}, {Fraser},
  {Fr{\'e}zouls}, {Gai}, {Galleti}, {Garabato}, {Garc{\'\i}a-Sedano},
  {Garofalo}, {Garralda}, {Gavel}, {Gavras}, {Gerssen}, {Geyer}, {Giacobbe},
  {Gilmore}, {Girona}, {Giuffrida}, {Glass}, {Gomes}, {Granvik}, {Gueguen},
  {Guerrier}, {Guiraud}, {Guti{\'e}}, {Haigron}, {Hatzidimitriou}, {Hauser},
  {Haywood}, {Heiter}, {Helmi}, {Heu}, {Hilger}, {Hobbs}, {Hofmann}, {Holland},
  {Huckle}, {Hypki}, {Icardi}, {Jan{\ss}en}, {Jevardat de Fombelle}, {Jonker},
  {Juh{\'a}sz}, {Julbe}, {Karampelas}, {Kewley}, {Klar}, {Kochoska}, {Kohley},
  {Kolenberg}, {Kontizas}, {Kontizas}, {Koposov}, {Kordopatis},
  {Kostrzewa-Rutkowska}, {Koubsky}, {Lambert}, {Lanza}, {Lasne}, {Lavigne}, {Le
  Fustec}, {Le Poncin-Lafitte}, {Lebreton}, {Leccia}, {Leclerc},
  {Lecoeur-Taibi}, {Lenhardt}, {Leroux}, {Liao}, {Licata}, {Lindstr{\o}m},
  {Lister}, {Livanou}, {Lobel}, {L{\'o}pez}, {Managau}, {Mann}, {Mantelet},
  {Marchal}, {Marchant}, {Marconi}, {Marinoni}, {Marschalk{\'o}}, {Marshall},
  {Martino}, {Marton}, {Mary}, {Massari}, {Matijevi{\v{c}}}, {Mazeh},
  {McMillan}, {Messina}, {Michalik}, {Millar}, {Molina}, {Molinaro},
  {Moln{\'a}r}, {Montegriffo}, {Mor}, {Morbidelli}, {Morel}, {Morris},
  {Mulone}, {Muraveva}, {Musella}, {Nelemans}, {Nicastro}, {Noval},
  {O'Mullane}, {Ord{\'e}novic}, {Ord{\'o}{\~n}ez-Blanco}, {Osborne}, {Pagani},
  {Pagano}, {Pailler}, {Palacin}, {Palaversa}, {Panahi}, {Pawlak},
  {Piersimoni}, {Pineau}, {Plachy}, {Plum}, {Poggio}, {Poujoulet},
  {Pr{\v{s}}a}, {Pulone}, {Racero}, {Ragaini}, {Rambaux}, {Ramos-Lerate},
  {Regibo}, {Reyl{\'e}}, {Riclet}, {Ripepi}, {Riva}, {Rivard}, {Rixon},
  {Roegiers}, {Roelens}, {Romero-G{\'o}mez}, {Rowell}, {Royer}, {Ruiz-Dern},
  {Sadowski}, {Sagrist{\`a} Sell{\'e}s}, {Sahlmann}, {Salgado}, {Salguero},
  {Sanna}, {Santana-Ros}, {Sarasso}, {Savietto}, {Schultheis}, {Sciacca},
  {Segol}, {Segovia}, {S{\'e}gransan}, {Shih}, {Siltala}, {Silva}, {Smart},
  {Smith}, {Solano}, {Solitro}, {Sordo}, {Soria Nieto}, {Souchay}, {Spagna},
  {Spoto}, {Stampa}, {Steele}, {Steidelm{\"u}ller}, {Stephenson}, {Stoev},
  {Suess}, {Surdej}, {Szabados}, {Szegedi-Elek}, {Tapiador}, {Taris}, {Tauran},
  {Taylor}, {Teixeira}, {Terrett}, {Teyssand ier}, {Thuillot}, {Titarenko},
  {Torra Clotet}, {Turon}, {Ulla}, {Utrilla}, {Uzzi}, {Vaillant}, {Valentini},
  {Valette}, {van Elteren}, {Van Hemelryck}, {Vaschetto}, {Vecchiato},
  {Veljanoski}, {Viala}, {Vicente}, {Vogt}, {von Essen}, {Voss}, {Votruba},
  {Voutsinas}, {Walmsley}, {Weiler}, {Wertz}, {Wevers}, {Wyrzykowski},
  {Yoldas}, {{\v{Z}}erjal}, {Ziaeepour}, {Zorec}, {Zschocke}, {Zucker},
  {Zurbach}, \& {Zwitter}}]{babu18}
{Gaia Collaboration} {et~al.}, 2018, \aap, 616, A10

\bibitem[{{Gaia Collaboration} {et~al}\mbox{.}(2016){Gaia Collaboration},
  {Prusti}, {de Bruijne}, {Brown}, {Vallenari}, {Babusiaux}, {Bailer-Jones},
  {Bastian}, {Biermann}, {Evans}, \& et~al.}]{prus16}
{Gaia Collaboration} {et~al.}, 2016, \aap, 595, A1

\bibitem[{{Gaia Collaboration} {et~al}\mbox{.}(2017){Gaia Collaboration}, {van
  Leeuwen}, {Vallenari}, {Jordi}, {Lindegren}, {Bastian}, {Prusti}, {de
  Bruijne}, {Brown}, {Babusiaux}, {Bailer-Jones}, {Biermann}, {Evans}, {Eyer},
  {Jansen}, {Klioner}, {Lammers}, {Luri}, {Mignard}, {Panem}, {Pourbaix}, {Rand
  ich}, {Sartoretti}, {Siddiqui}, {Soubiran}, {Valette}, {Walton}, {Aerts},
  {Arenou}, {Cropper}, {Drimmel}, {H{\o}g}, {Katz}, {Lattanzi}, {O'Mullane},
  {Grebel}, {Holland }, {Huc}, {Passot}, {Perryman}, {Bramante}, {Cacciari},
  {Casta{\~n}eda}, {Chaoul}, {Cheek}, {De Angeli}, {Fabricius}, {Guerra},
  {Hern{\'a}ndez}, {Jean-Antoine-Piccolo}, {Masana}, {Messineo}, {Mowlavi},
  {Nienartowicz}, {Ord{\'o}{\~n}ez-Blanco}, {Panuzzo}, {Portell}, {Richards},
  {Riello}, {Seabroke}, {Tanga}, {Th{\'e}venin}, {Torra}, {Els},
  {Gracia-Abril}, {Comoretto}, {Garcia-Reinaldos}, {Lock}, {Mercier},
  {Altmann}, {Andrae}, {Astraatmadja}, {Bellas-Velidis}, {Benson}, {Berthier},
  {Blomme}, {Busso}, {Carry}, {Cellino}, {Clementini}, {Cowell}, {Creevey},
  {Cuypers}, {Davidson}, {De Ridder}, {de Torres}, {Delchambre}, {Dell'Oro},
  {Ducourant}, {Fr{\'e}mat}, {Garc{\'\i}a-Torres}, {Gosset}, {Halbwachs},
  {Hambly}, {Harrison}, {Hauser}, {Hestroffer}, {Hodgkin}, {Huckle}, {Hutton},
  {Jasniewicz}, {Jordan}, {Kontizas}, {Korn}, {Lanzafame}, {Manteiga},
  {Moitinho}, {Muinonen}, {Osinde}, {Pancino}, {Pauwels}, {Petit},
  {Recio-Blanco}, {Robin}, {Sarro}, {Siopis}, {Smith}, {Smith}, {Sozzetti},
  {Thuillot}, {van Reeven}, {Viala}, {Abbas}, {Abreu Aramburu}, {Accart},
  {Aguado}, {Allan}, {Allasia}, {Altavilla}, {{\'A}lvarez}, {Alves},
  {Anderson}, {Andrei}, {Anglada Varela}, {Antiche}, {Antoja}, {Ant{\'o}n},
  {Arcay}, {Bach}, {Baker}, {Balaguer-N{\'u}{\~n}ez}, {Barache}, {Barata},
  {Barbier}, {Barblan}, {Barrado y Navascu{\'e}s}, {Barros}, {Barstow},
  {Becciani}, {Bellazzini}, {Bello Garc{\'\i}a}, {Belokurov}, {Bendjoya},
  {Berihuete}, {Bianchi}, {Bienaym{\'e}}, {Billebaud}, {Blagorodnova},
  {Blanco-Cuaresma}, {Boch}, {Bombrun}, {Borrachero}, {Bouquillon}, {Bourda},
  {Bouy}, {Bragaglia}, {Breddels}, {Brouillet}, {Br{\"u}semeister},
  {Bucciarelli}, {Burgess}, {Burgon}, {Burlacu}, {Busonero}, {Buzzi}, {Caffau},
  {Cambras}, {Campbell}, {Cancelliere}, {Cantat-Gaudin}, {Carlucci},
  {Carrasco}, {Castellani}, {Charlot}, {Charnas}, {Chiavassa}, {Clotet},
  {Cocozza}, {Collins}, {Costigan}, {Crifo}, {Cross}, {Crosta}, {Crowley},
  {Dafonte}, {Damerdji}, {Dapergolas}, {David}, {David}, {De Cat}, {de Felice},
  {de Laverny}, {De Luise}, {De March}, {de Martino}, {de Souza}, {Debosscher},
  {del Pozo}, {Delbo}, {Delgado}, {Delgado}, {Di Matteo}, {Diakite},
  {Distefano}, {Dolding}, {Dos Anjos}, {Drazinos}, {Dur{\'a}n}, {Dzigan},
  {Edvardsson}, {Enke}, {Evans}, {Eynard Bontemps}, {Fabre}, {Fabrizio},
  {Faigler}, {Falc{\~a}o}, {Farr{\`a}s Casas}, {Federici}, {Fedorets},
  {Fern{\'a}ndez-Hern{\'a}ndez}, {Fernique}, {Fienga}, {Figueras}, {Filippi},
  {Findeisen}, {Fonti}, {Fouesneau}, {Fraile}, {Fraser}, {Fuchs}, {Gai},
  {Galleti}, {Galluccio}, {Garabato}, {Garc{\'\i}a-Sedano}, {Garofalo},
  {Garralda}, {Gavras}, {Gerssen}, {Geyer}, {Gilmore}, {Girona}, {Giuffrida},
  {Gomes}, {Gonz{\'a}lez-Marcos}, {Gonz{\'a}lez-N{\'u}{\~n}ez},
  {Gonz{\'a}lez-Vidal}, {Granvik}, {Guerrier}, {Guillout}, {Guiraud},
  {G{\'u}rpide}, {Guti{\'e}rrez-S{\'a}nchez}, {Guy}, {Haigron},
  {Hatzidimitriou}, {Haywood}, {Heiter}, {Helmi}, {Hobbs}, {Hofmann}, {Holl},
  {Holland }, {Hunt}, {Hypki}, {Icardi}, {Irwin}, {Jevardat de Fombelle},
  {Jofr{\'e}}, {Jonker}, {Jorissen}, {Julbe}, {Karampelas}, {Kochoska},
  {Kohley}, {Kolenberg}, {Kontizas}, {Koposov}, {Kordopatis}, {Koubsky},
  {Krone-Martins}, {Kudryashova}, {Kull}, {Bachchan}, {Lacoste-Seris}, {Lanza},
  {Lavigne}, {Le Poncin-Lafitte}, {Lebreton}, {Lebzelter}, {Leccia}, {Leclerc},
  {Lecoeur-Taibi}, {Lemaitre}, {Lenhardt}, {Leroux}, {Liao}, {Licata},
  {Lindstr{\o}m}, {Lister}, {Livanou}, {Lobel}, {L{\"o}ffler}, {L{\'o}pez},
  {Lorenz}, {MacDonald}, {Magalh{\~a}es Fernandes}, {Managau}, {Mann},
  {Mantelet}, {Marchal}, {Marchant}, {Marconi}, {Marinoni}, {Marrese},
  {Marschalk{\'o}}, {Marshall}, {Mart{\'\i}n-Fleitas}, {Martino}, {Mary},
  {Matijevi{\v{c}}}, {Mazeh}, {McMillan}, {Messina}, {Michalik}, {Millar},
  {Mirand a}, {Molina}, {Molinaro}, {Molinaro}, {Moln{\'a}r}, {Moniez},
  {Montegriffo}, {Mor}, {Mora}, {Morbidelli}, {Morel}, {Morgenthaler},
  {Morris}, {Mulone}, {Muraveva}, {Musella}, {Narbonne}, {Nelemans},
  {Nicastro}, {Noval}, {Ord{\'e}novic}, {Ordieres-Mer{\'e}}, {Osborne},
  {Pagani}, {Pagano}, {Pailler}, {Palacin}, {Palaversa}, {Parsons}, {Pecoraro},
  {Pedrosa}, {Pentik{\"a}inen}, {Pichon}, {Piersimoni}, {Pineau}, {Plachy},
  {Plum}, {Poujoulet}, {Pr{\v{s}}a}, {Pulone}, {Ragaini}, {Rago}, {Rambaux},
  {Ramos-Lerate}, {Ranalli}, {Rauw}, {Read}, {Regibo}, {Reyl{\'e}}, {Ribeiro},
  {Rimoldini}, {Ripepi}, {Riva}, {Rixon}, {Roelens}, {Romero-G{\'o}mez},
  {Rowell}, {Royer}, {Ruiz-Dern}, {Sadowski}, {Sagrist{\`a} Sell{\'e}s},
  {Sahlmann}, {Salgado}, {Salguero}, {Sarasso}, {Savietto}, {Schultheis},
  {Sciacca}, {Segol}, {Segovia}, {Segransan}, {Shih}, {Smareglia}, {Smart},
  {Solano}, {Solitro}, {Sordo}, {Soria Nieto}, {Souchay}, {Spagna}, {Spoto},
  {Stampa}, {Steele}, {Steidelm{\"u}ller}, {Stephenson}, {Stoev}, {Suess},
  {S{\"u}veges}, {Surdej}, {Szabados}, {Szegedi-Elek}, {Tapiador}, {Taris},
  {Tauran}, {Taylor}, {Teixeira}, {Terrett}, {Tingley}, {Trager}, {Turon},
  {Ulla}, {Utrilla}, {Valentini}, {van Elteren}, {Van Hemelryck}, {vanLeeuwen},
  {Varadi}, {Vecchiato}, {Veljanoski}, {Via}, {Vicente}, {Vogt}, {Voss},
  {Votruba}, {Voutsinas}, {Walmsley}, {Weiler}, {Weingrill}, {Wevers},
  {Wyrzykowski}, {Yoldas}, {{\v{Z}}erjal}, {Zucker}, {Zurbach}, {Zwitter},
  {Alecu}, {Allen}, {Allende Prieto}, {Amorim}, {Anglada-Escud{\'e}},
  {Arsenijevic}, {Azaz}, {Balm}, {Beck}, {Bernstein}, {Bigot}, {Bijaoui},
  {Blasco}, {Bonfigli}, {Bono}, {Boudreault}, {Bressan}, {Brown}, {Brunet},
  {Bunclark}, {Buonanno}, {Butkevich}, {Carret}, {Carrion}, {Chemin},
  {Ch{\'e}reau}, {Corcione}, {Darmigny}, {de Boer}, {de Teodoro}, {de Zeeuw},
  {Delle Luche}, {Domingues}, {Dubath}, {Fodor}, {Fr{\'e}zouls}, {Fries},
  {Fustes}, {Fyfe}, {Gallardo}, {Gallegos}, {Gardiol}, {Gebran}, {Gomboc},
  {G{\'o}mez}, {Grux}, {Gueguen}, {Heyrovsky}, {Hoar}, {Iannicola}, {Isasi
  Parache}, {Janotto}, {Joliet}, {Jonckheere}, {Keil}, {Kim}, {Klagyivik},
  {Klar}, {Knude}, {Kochukhov}, {Kolka}, {Kos}, {Kutka}, {Lainey}, {LeBouquin},
  {Liu}, {Loreggia}, {Makarov}, {Marseille}, {Martayan}, {Martinez-Rubi},
  {Massart}, {Meynadier}, {Mignot}, {Munari}, {Nguyen}, {Nordlander},
  {O'Flaherty}, {Ocvirk}, {Olias Sanz}, {Ortiz}, {Osorio}, {Oszkiewicz},
  {Ouzounis}, {Palmer}, {Park}, {Pasquato}, {Peltzer}, {Peralta},
  {P{\'e}turaud}, {Pieniluoma}, {Pigozzi}, {Poels}, {Prat}, {Prod'homme},
  {Raison}, {Rebordao}, {Risquez}, {Rocca-Volmerange}, {Rosen}, {Ruiz-Fuertes},
  {Russo}, {Sembay}, {Serraller Vizcaino}, {Short}, {Siebert}, {Silva},
  {Sinachopoulos}, {Slezak}, {Soffel}, {Sosnowska}, {Strai{\v{z}}ys}, {ter
  Linden}, {Terrell}, {Theil}, {Tiede}, {Troisi}, {Tsalmantza}, {Tur},
  {Vaccari}, {Vachier}, {Valles}, {Van Hamme}, {Veltz}, {Virtanen}, {Wallut},
  {Wichmann}, {Wilkinson}, {Ziaeepour}, \& {Zschocke}}]{vanl16}
{Gaia Collaboration} {et~al.}, 2017, \aap, 601, A19

\bibitem[{{Galli} {et~al}\mbox{.}(2018){Galli}, {Joncour}, \&
  {Moraux}}]{gall18}
{Galli} P. A.~B., {Joncour} I., {Moraux} E., 2018, \mnras, 477, L50

\bibitem[{{Garcia}(1994)}]{garc94}
{Garcia} B., 1994, \apj, 436, 705

\bibitem[{{Garc{\'\i}a} \& {Mermilliod}(2001)}]{garc01}
{Garc{\'\i}a} B., {Mermilliod} J.~C., 2001, \aap, 368, 122

\bibitem[{{Garmany}(1994)}]{garm94}
{Garmany} C.~D., 1994, \pasp, 106, 25

\bibitem[{{Garmany} {et~al}\mbox{.}(1980){Garmany}, {Conti}, \&
  {Massey}}]{garm80}
{Garmany} C.~D., {Conti} P.~S., {Massey} P., 1980, \apj, 242, 1063

\bibitem[{{Garmany} \& {Stencel}(1992)}]{garm92}
{Garmany} C.~D., {Stencel} R.~E., 1992, \aaps, 94, 211

\bibitem[{{Gerola} \& {Seiden}(1978)}]{gero78}
{Gerola} H., {Seiden} P.~E., 1978, \apj, 223, 129

\bibitem[{{Getman} {et~al}\mbox{.}(2011){Getman}, {Broos}, {Feigelson},
  {Townsley}, {Povich}, {Garmire}, {Montmerle}, {Yonekura}, \&
  {Fukui}}]{getm11}
{Getman} K.~V. {et~al.}, 2011, \apjs, 194, 3

\bibitem[{{Getman} {et~al}\mbox{.}(2006){Getman}, {Feigelson}, {Townsley},
  {Broos}, {Garmire}, \& {Tsujimoto}}]{getm06}
{Getman} K.~V., {Feigelson} E.~D., {Townsley} L., {Broos} P., {Garmire} G.,
  {Tsujimoto} M., 2006, \apjs, 163, 306

\bibitem[{{Gies}(1987)}]{gies87}
{Gies} D.~R., 1987, \apjs, 64, 545

\bibitem[{{Gies} \& {Bolton}(1986)}]{gies86}
{Gies} D.~R., {Bolton} C.~T., 1986, \apjs, 61, 419

\bibitem[{{Girichidis} {et~al}\mbox{.}(2012){Girichidis}, {Federrath},
  {Banerjee}, \& {Klessen}}]{giri12}
{Girichidis} P., {Federrath} C., {Banerjee} R., {Klessen} R.~S., 2012, \mnras,
  420, 613

\bibitem[{{Golden} {et~al}\mbox{.}(2005){Golden}, {Brisken}, {Thorsett},
  {Benjamin}, \& {Goss}}]{gold05}
{Golden} A., {Brisken} W.~F., {Thorsett} S.~E., {Benjamin} R.~A., {Goss} W.~M.,
  2005, Baltic Astronomy, 14, 436

\bibitem[{{Goldman} {et~al}\mbox{.}(2018){Goldman}, {R{\"o}ser}, {Schilbach},
  {Mo{\'o}r}, \& {Henning}}]{gold18}
{Goldman} B., {R{\"o}ser} S., {Schilbach} E., {Mo{\'o}r} A.~C., {Henning} T.,
  2018, \apj, 868, 32

\bibitem[{{Gomez} {et~al}\mbox{.}(1993){Gomez}, {Hartmann}, {Kenyon}, \&
  {Hewett}}]{gome93}
{Gomez} M., {Hartmann} L., {Kenyon} S.~J., {Hewett} R., 1993, \aj, 105, 1927

\bibitem[{{Goodwin}(1997)}]{good97}
{Goodwin} S.~P., 1997, \mnras, 284, 785

\bibitem[{{Goodwin} \& {Bastian}(2006)}]{good06}
{Goodwin} S.~P., {Bastian} N., 2006, \mnras, 373, 752

\bibitem[{{Gouliermis}(2018)}]{goul18}
{Gouliermis} D.~A., 2018, \pasp, 130, 072001

\bibitem[{{Grasdalen} {et~al}\mbox{.}(1973){Grasdalen}, {Strom}, \&
  {Strom}}]{gras73}
{Grasdalen} G.~L., {Strom} K.~M., {Strom} S.~E., 1973, \apjl, 184, L53

\bibitem[{{Gravity Collaboration} {et~al}\mbox{.}(2018){Gravity Collaboration},
  {Karl}, {Pfuhl}, {Eisenhauer}, {Genzel}, {Grellmann}, {Habibi}, {Abuter},
  {Accardo}, {Amorim}, {Anugu}, {{\'A}vila}, {Benisty}, {Berger}, {Blind},
  {Bonnet}, {Bourget}, {Brandner}, {Brast}, {Buron}, {Caratti O Garatti},
  {Chapron}, {Cl{\'e}net}, {Collin}, {Coud{\'e} Du Foresto}, {de Wit}, {de
  Zeeuw}, {Deen}, {Delplancke-Str{\"o}bele}, {Dembet}, {Derie}, {Dexter},
  {Duvert}, {Ebert}, {Eckart}, {Esselborn}, {F{\'e}dou}, {Finger}, {Garcia},
  {Garcia Dabo}, {Garcia Lopez}, {Gao}, {Gendron}, {Gillessen}, {Gont{\'e}},
  {Gordo}, {Gr{\"o}zinger}, {Guajardo}, {Guieu}, {Haguenauer}, {Hans},
  {Haubois}, {Haug}, {Hau{\ss}mann}, {Henning}, {Hippler}, {Horrobin}, {Huber},
  {Hubert}, {Hubin}, {Jakob}, {Jochum}, {Jocou}, {Kaufer}, {Kellner},
  {Kendrew}, {Kern}, {Kervella}, {Kiekebusch}, {Klein}, {K{\"o}hler}, {Kolb},
  {Kulas}, {Lacour}, {Lapeyr{\`e}re}, {Lazareff}, {Le Bouquin}, {L{\'e}na},
  {Lenzen}, {L{\'e}v{\^e}que}, {Lin}, {Lippa}, {Magnard}, {Mehrgan},
  {M{\'e}rand}, {Moulin}, {M{\"u}ller}, {M{\"u}ller}, {Neumann}, {Oberti},
  {Ott}, {Pallanca}, {Pand uro}, {Pasquini}, {Paumard}, {Percheron}, {Perraut},
  {Perrin}, {Pfl{\"u}ger}, {Duc}, {Plewa}, {Popovic}, {Rabien}, {Ram{\'\i}rez},
  {Ramos}, {Rau}, {Riquelme}, {Rodr{\'\i}guez-Coira}, {Rohloff}, {Rosales},
  {Rousset}, {Sanchez-Bermudez}, {Scheithauer}, {Sch{\"o}ller}, {Schuhler},
  {Spyromilio}, {Straub}, {Straubmeier}, {Sturm}, {Suarez}, {Tristram},
  {Ventura}, {Vincent}, {Waisberg}, {Wank}, {Widmann}, {Wieprecht}, {Wiest},
  {Wiezorrek}, {Wittkowski}, {Woillez}, {Wolff}, {Yazici}, {Ziegler}, \&
  {Zins}}]{karl18}
{Gravity Collaboration} {et~al.}, 2018, \aap, 620, A116

\bibitem[{{Gregorio-Hetem}(2008)}]{greg08}
{Gregorio-Hetem} J., 2008, {The Canis Major Star Forming Region}, Vol.~5,
  {Astronomical Society of the Pacific}, p.~1

\bibitem[{{Griffiths} {et~al}\mbox{.}(2018){Griffiths}, {Goodwin}, \&
  {Caballero-Nieves}}]{grif18}
{Griffiths} D.~W., {Goodwin} S.~P., {Caballero-Nieves} S.~M., 2018, \mnras,
  476, 2493

\bibitem[{{Grudi{\'c}} {et~al}\mbox{.}(2020){Grudi{\'c}}, {Kruijssen},
  {Faucher-Gigu{\`e}re}, {Hopkins}, {Ma}, {Quataert}, \&
  {Boylan-Kolchin}}]{grud20}
{Grudi{\'c}} M.~Y., {Kruijssen} J.~M.~D., {Faucher-Gigu{\`e}re} C.-A.,
  {Hopkins} P.~F., {Ma} X., {Quataert} E., {Boylan-Kolchin} M., 2020, arXiv
  e-prints, arXiv:2008.04453

\bibitem[{{Guarcello} {et~al}\mbox{.}(2016){Guarcello}, {Drake}, {Wright},
  {Albacete-Colombo}, {Clarke}, {Ercolano}, {Flaccomio}, {Kashyap}, {Micela},
  {Naylor}, {Schneider}, {Sciortino}, \& {Vink}}]{guar16}
{Guarcello} M.~G. {et~al.}, 2016, ArXiv e-prints

\bibitem[{{Guarcello} {et~al}\mbox{.}(2013){Guarcello}, {Drake}, {Wright},
  {Drew}, {Gutermuth}, {Hora}, {Naylor}, {Aldcroft}, {Fruscione},
  {Garc{\'{\i}}a-Alvarez}, {Kashyap}, \& {King}}]{guar13}
{Guarcello} M.~G. {et~al.}, 2013, \apj, 773, 135

\bibitem[{{Guarcello} {et~al}\mbox{.}(2007){Guarcello}, {Prisinzano}, {Micela},
  {Damiani}, {Peres}, \& {Sciortino}}]{guar07}
{Guarcello} M.~G., {Prisinzano} L., {Micela} G., {Damiani} F., {Peres} G.,
  {Sciortino} S., 2007, \aap, 462, 245

\bibitem[{{Gutermuth} {et~al}\mbox{.}(2008){Gutermuth}, {Myers}, {Megeath},
  {Allen}, {Pipher}, {Muzerolle}, {Porras}, {Winston}, \& {Fazio}}]{gute08}
{Gutermuth} R.~A. {et~al.}, 2008, \apj, 674, 336

\bibitem[{{Hacar} {et~al}\mbox{.}(2016){Hacar}, {Alves}, {Forbrich},
  {Meingast}, {Kubiak}, \& {Gro{\ss}schedl}}]{haca16}
{Hacar} A., {Alves} J., {Forbrich} J., {Meingast} S., {Kubiak} K.,
  {Gro{\ss}schedl} J., 2016, \aap, 589, A80

\bibitem[{{Hanson}(2003)}]{hans03}
{Hanson} M.~M., 2003, \apj, 597, 957

\bibitem[{{Harris}(1956)}]{harr56}
{Harris}, Daniel~L. I., 1956, \apj, 123, 371

\bibitem[{{Hartmann} \& {Burkert}(2007)}]{hart07}
{Hartmann} L., {Burkert} A., 2007, \apj, 654, 988

\bibitem[{{Havlen}(1972)}]{havl72}
{Havlen} R.~J., 1972, \aap, 17, 413

\bibitem[{{Heiles}(1998)}]{heil98}
{Heiles} C., 1998, \apj, 498, 689

\bibitem[{{Herbig}(1998)}]{herb98}
{Herbig} G.~H., 1998, \apj, 497, 736

\bibitem[{{Herbig} \& {Terndrup}(1986)}]{herb86}
{Herbig} G.~H., {Terndrup} D.~M., 1986, \apj, 307, 609

\bibitem[{{Herbst} \& {Assousa}(1977)}]{herb77}
{Herbst} W., {Assousa} G.~E., 1977, \apj, 217, 473

\bibitem[{{Hern{\'a}ndez} {et~al}\mbox{.}(2014){Hern{\'a}ndez}, {Calvet},
  {Perez}, {Brice{\~n}o}, {Olguin}, {Contreras}, {Hartmann}, {Allen},
  {Espaillat}, \& {Hernan}}]{hern14}
{Hern{\'a}ndez} J. {et~al.}, 2014, \apj, 794, 36

\bibitem[{{Heyer} {et~al}\mbox{.}(2001){Heyer}, {Carpenter}, \&
  {Snell}}]{heye01}
{Heyer} M.~H., {Carpenter} J.~M., {Snell} R.~L., 2001, \apj, 551, 852

\bibitem[{{Higdon} \& {Lingenfelter}(2013)}]{higd13}
{Higdon} J.~C., {Lingenfelter} R.~E., 2013, \apj, 775, 110

\bibitem[{{Hillenbrand} {et~al}\mbox{.}(2008){Hillenbrand}, {Bauermeister}, \&
  {White}}]{hill08}
{Hillenbrand} L.~A., {Bauermeister} A., {White} R.~J., 2008, in Astronomical
  Society of the Pacific Conference Series, Vol. 384, 14th Cambridge Workshop
  on Cool Stars, Stellar Systems, and the Sun, {van Belle} G., ed., p. 200

\bibitem[{{Hillenbrand} {et~al}\mbox{.}(1998){Hillenbrand}, {Strom}, {Calvet},
  {Merrill}, {Gatley}, {Makidon}, {Meyer}, \& {Skrutskie}}]{hill98}
{Hillenbrand} L.~A., {Strom} S.~E., {Calvet} N., {Merrill} K.~M., {Gatley} I.,
  {Makidon} R.~B., {Meyer} M.~R., {Skrutskie} M.~F., 1998, \aj, 116, 1816

\bibitem[{{Hills}(1980)}]{hill80}
{Hills} J.~G., 1980, \apj, 235, 986

\bibitem[{{Hoogerwerf} \& {Aguilar}(1999)}]{hoog99}
{Hoogerwerf} R., {Aguilar} L.~A., 1999, \mnras, 306, 394

\bibitem[{{Hoogerwerf} {et~al}\mbox{.}(2000){Hoogerwerf}, {de Bruijne}, \& {de
  Zeeuw}}]{hoog00}
{Hoogerwerf} R., {de Bruijne} J.~H.~J., {de Zeeuw} P.~T., 2000, \apjl, 544,
  L133

\bibitem[{{Hoogerwerf} {et~al}\mbox{.}(2001){Hoogerwerf}, {de Bruijne}, \& {de
  Zeeuw}}]{hoog01}
{Hoogerwerf} R., {de Bruijne} J.~H.~J., {de Zeeuw} P.~T., 2001, \aap, 365, 49

\bibitem[{{Humphreys}(1978)}]{hump78}
{Humphreys} R.~M., 1978, \apjs, 38, 309

\bibitem[{{Humphreys} \& {McElroy}(1984)}]{hump84}
{Humphreys} R.~M., {McElroy} D.~B., 1984, \apj, 284, 565

\bibitem[{{Hur} {et~al}\mbox{.}(2012){Hur}, {Sung}, \& {Bessell}}]{hur12}
{Hur} H., {Sung} H., {Bessell} M.~S., 2012, \aj, 143, 41

\bibitem[{{Hurley} \& {Tout}(1998)}]{hurl98}
{Hurley} J., {Tout} C.~A., 1998, \mnras, 300, 977

\bibitem[{{Ivanov}(1987)}]{ivan87}
{Ivanov} G.~R., 1987, \apss, 136, 113

\bibitem[{{Iwamoto} {et~al}\mbox{.}(1998){Iwamoto}, {Mazzali}, {Nomoto},
  {Umeda}, {Nakamura}, {Patat}, {Danziger}, {Young}, {Suzuki}, {Shigeyama},
  {Augusteijn}, {Doublier}, {Gonzalez}, {Boehnhardt}, {Brewer}, {Hainaut},
  {Lidman}, {Leibundgut}, {Cappellaro}, {Turatto}, {Galama}, {Vreeswijk},
  {Kouveliotou}, {van Paradijs}, {Pian}, {Palazzi}, \& {Frontera}}]{iwam98}
{Iwamoto} K. {et~al.}, 1998, \nat, 395, 672

\bibitem[{{Jackson} \& {Jeffries}(2014)}]{jack14}
{Jackson} R.~J., {Jeffries} R.~D., 2014, \mnras, 441, 2111

\bibitem[{{Janson} {et~al}\mbox{.}(2013){Janson}, {Lafreni{\`e}re},
  {Jayawardhana}, {Bonavita}, {Girard}, {Brandeker}, \& {Gizis}}]{jans13}
{Janson} M., {Lafreni{\`e}re} D., {Jayawardhana} R., {Bonavita} M., {Girard}
  J.~H., {Brandeker} A., {Gizis} J.~E., 2013, \apj, 773, 170

\bibitem[{{Jeffries} {et~al}\mbox{.}(2006){Jeffries}, {Evans}, {Pye}, \&
  {Briggs}}]{jeff06}
{Jeffries} R.~D., {Evans} P.~A., {Pye} J.~P., {Briggs} K.~R., 2006, \mnras,
  367, 781

\bibitem[{{Jeffries} {et~al}\mbox{.}(2014){Jeffries}, {Jackson}, {Cottaar},
  {Koposov}, {Lanzafame}, {Meyer}, {Prisinzano}, {Randich}, {Sacco},
  {Brugaletta}, {Caramazza}, {Damiani}, {Franciosini}, {Frasca}, {Gilmore},
  {Feltzing}, {Micela}, {Alfaro}, {Bensby}, {Pancino}, {Recio-Blanco}, {de
  Laverny}, {Lewis}, {Magrini}, {Morbidelli}, {Costado}, {Jofr{\'e}},
  {Klutsch}, {Lind}, \& {Maiorca}}]{jeff14}
{Jeffries} R.~D. {et~al.}, 2014, \aap, 563, A94

\bibitem[{{Jeffries} {et~al}\mbox{.}(2009){Jeffries}, {Naylor}, {Walter},
  {Pozzo}, \& {Devey}}]{jeff09}
{Jeffries} R.~D., {Naylor} T., {Walter} F.~M., {Pozzo} M.~P., {Devey} C.~R.,
  2009, \mnras, 393, 538

\bibitem[{{Jerabkova} {et~al}\mbox{.}(2019){Jerabkova}, {Beccari}, {Boffin},
  {Petr-Gotzens}, {Manara}, {Prada Moroni}, {Tognelli}, \&
  {Degl'Innocenti}}]{jera19}
{Jerabkova} T., {Beccari} G., {Boffin} H. M.~J., {Petr-Gotzens} M.~G., {Manara}
  C.~F., {Prada Moroni} P.~G., {Tognelli} E., {Degl'Innocenti} S., 2019, \aap,
  627, A57

\bibitem[{{Johnson}(1962)}]{john62}
{Johnson} H.~L., 1962, \apj, 136, 1135

\bibitem[{{Johnson} \& {Morgan}(1953)}]{john53}
{Johnson} H.~L., {Morgan} W.~W., 1953, \apj, 117, 313

\bibitem[{{Johnson} \& {Morgan}(1954)}]{john54}
{Johnson} H.~L., {Morgan} W.~W., 1954, \apj, 119, 344

\bibitem[{{Johnstone} {et~al}\mbox{.}(1998){Johnstone}, {Hollenbach}, \&
  {Bally}}]{john98}
{Johnstone} D., {Hollenbach} D., {Bally} J., 1998, \apj, 499, 758

\bibitem[{{Jones} \& {Walker}(1988)}]{jone88}
{Jones} B.~F., {Walker} M.~F., 1988, \aj, 95, 1755

\bibitem[{{Jones}(1971)}]{jone71}
{Jones} D.~H.~P., 1971, \mnras, 152, 231

\bibitem[{{Jordi} {et~al}\mbox{.}(1996){Jordi}, {Trullols}, \&
  {Galadi-Enriquez}}]{jord96}
{Jordi} C., {Trullols} E., {Galadi-Enriquez} D., 1996, \aap, 312, 499

\bibitem[{{Joubaud} {et~al}\mbox{.}(2019){Joubaud}, {Grenier}, {Ballet}, \&
  {Soler}}]{joub19}
{Joubaud} T., {Grenier} I.~A., {Ballet} J., {Soler} J.~D., 2019, \aap, 631, A52

\bibitem[{{Kalas} {et~al}\mbox{.}(2015){Kalas}, {Rajan}, {Wang},
  {Millar-Blanchaer}, {Duchene}, {Chen}, {Fitzgerald}, {Dong}, {Graham},
  {Patience}, {Macintosh}, {Murray-Clay}, {Matthews}, {Rameau}, {Marois},
  {Chilcote}, {De Rosa}, {Doyon}, {Draper}, {Lawler}, {Ammons}, {Arriaga},
  {Bulger}, {Cotten}, {Follette}, {Goodsell}, {Greenbaum}, {Hibon}, {Hinkley},
  {Hung}, {Ingraham}, {Konapacky}, {Lafreniere}, {Larkin}, {Long}, {Maire},
  {Marchis}, {Metchev}, {Morzinski}, {Nielsen}, {Oppenheimer}, {Perrin},
  {Pueyo}, {Rantakyr{\"o}}, {Ruffio}, {Saddlemyer}, {Savransky}, {Schneider},
  {Sivaramakrishnan}, {Soummer}, {Song}, {Thomas}, {Vasisht}, {Ward-Duong},
  {Wiktorowicz}, \& {Wolff}}]{kala15b}
{Kalas} P.~G. {et~al.}, 2015, \apj, 814, 32

\bibitem[{{Kaltcheva} \& {Georgiev}(1994)}]{kalt94}
{Kaltcheva} N.~T., {Georgiev} L.~N., 1994, \mnras, 269, 289

\bibitem[{{Kaltcheva} \& {Hilditch}(2000)}]{kalt00}
{Kaltcheva} N.~T., {Hilditch} R.~W., 2000, \mnras, 312, 753

\bibitem[{{Kapteyn}(1914)}]{kapt14}
{Kapteyn} J.~C., 1914, \apj, 40, 43

\bibitem[{{Kharchenko} {et~al}\mbox{.}(2005){Kharchenko}, {Piskunov},
  {R{\"o}ser}, {Schilbach}, \& {Scholz}}]{khar05}
{Kharchenko} N.~V., {Piskunov} A.~E., {R{\"o}ser} S., {Schilbach} E., {Scholz}
  R.~D., 2005, \aap, 438, 1163

\bibitem[{{Kharchenko} {et~al}\mbox{.}(2013){Kharchenko}, {Piskunov},
  {Schilbach}, {R{\"o}ser}, \& {Scholz}}]{khar13}
{Kharchenko} N.~V., {Piskunov} A.~E., {Schilbach} E., {R{\"o}ser} S., {Scholz}
  R.~D., 2013, \aap, 558, A53

\bibitem[{{Kim} {et~al}\mbox{.}(2019){Kim}, {Lu}, {Konopacky}, {Chu}, {Toller},
  {Anderson}, {Theissen}, \& {Morris}}]{kim19}
{Kim} D., {Lu} J.~R., {Konopacky} Q., {Chu} L., {Toller} E., {Anderson} J.,
  {Theissen} C.~A., {Morris} M.~R., 2019, \aj, 157, 109

\bibitem[{{Kiminki} \& {Kobulnicky}(2012)}]{kimi12b}
{Kiminki} D.~C., {Kobulnicky} H.~A., 2012, \apj, 751, 4

\bibitem[{{Kiminki} {et~al}\mbox{.}(2012){Kiminki}, {Kobulnicky}, {Ewing},
  {Bagley Kiminki}, {Lundquist}, {Alexander}, {Vargas-Alvarez}, {Choi}, \&
  {Henderson}}]{kimi12}
{Kiminki} D.~C. {et~al.}, 2012, \apj, 747, 41

\bibitem[{{Kiminki} {et~al}\mbox{.}(2009){Kiminki}, {Kobulnicky}, {Gilbert},
  {Bird}, \& {Chunev}}]{kimi09}
{Kiminki} D.~C., {Kobulnicky} H.~A., {Gilbert} I., {Bird} S., {Chunev} G.,
  2009, \aj, 137, 4608

\bibitem[{{Kiminki} {et~al}\mbox{.}(2007){Kiminki}, {Kobulnicky}, {Kinemuchi},
  {Irwin}, {Fryer}, {Berrington}, {Uzpen}, {Monson}, {Pierce}, \&
  {Woosley}}]{kimi07}
{Kiminki} D.~C. {et~al.}, 2007, \apj, 664, 1102

\bibitem[{{Kiminki} {et~al}\mbox{.}(2008){Kiminki}, {Kobulnicky}, {Kinemuchi},
  {Irwin}, {Fryer}, {Berrington}, {Uzpen}, {Monson}, {Pierce}, \&
  {Woosley}}]{kimi08b}
{Kiminki} D.~C. {et~al.}, 2008, \apj, 681, 735

\bibitem[{{Kiminki} \& {Smith}(2018)}]{kimi18}
{Kiminki} M.~M., {Smith} N., 2018, \mnras, 477, 2068

\bibitem[{{Kimura} {et~al}\mbox{.}(2013){Kimura}, {Tsunemi}, {Tomida},
  {Sugizaki}, {Ueno}, {Hanayama}, {Yoshidome}, \& {Sasaki}}]{kimu13}
{Kimura} M., {Tsunemi} H., {Tomida} H., {Sugizaki} M., {Ueno} S., {Hanayama}
  T., {Yoshidome} K., {Sasaki} M., 2013, \pasj, 65, 14

\bibitem[{{King} {et~al}\mbox{.}(2012){King}, {Goodwin}, {Parker}, \&
  {Patience}}]{king12}
{King} R.~R., {Goodwin} S.~P., {Parker} R.~J., {Patience} J., 2012, \mnras,
  427, 2636

\bibitem[{{Kn{\"o}dlseder}(2000)}]{knod00}
{Kn{\"o}dlseder} J., 2000, \aap, 360, 539

\bibitem[{{Kobulnicky} {et~al}\mbox{.}(2014){Kobulnicky}, {Kiminki},
  {Lundquist}, {Burke}, {Chapman}, {Keller}, {Lester}, {Rolen}, {Topel},
  {Bhattacharjee}, {Smullen}, {Vargas {\'A}lvarez}, {Runnoe}, {Dale}, \&
  {Brotherton}}]{kobu14}
{Kobulnicky} H.~A. {et~al.}, 2014, \apjs, 213, 34

\bibitem[{{K{\"o}hler} {et~al}\mbox{.}(2000){K{\"o}hler}, {Kunkel}, {Leinert},
  \& {Zinnecker}}]{kohl00}
{K{\"o}hler} R., {Kunkel} M., {Leinert} C., {Zinnecker} H., 2000, \aap, 356,
  541

\bibitem[{{Kos} {et~al}\mbox{.}(2018){Kos}, {Bland-Hawthorn}, {Asplund},
  {Buder}, {Lewis}, {Lin}, {Martell}, {Ness}, {Sharma}, {De Silva}, {Simpson},
  {Zucker}, {Zwitter}, {{\v{C}}otar}, \& {Spina}}]{kos18}
{Kos} J. {et~al.}, 2018, arXiv e-prints, arXiv:1811.11762

\bibitem[{{Kounkel}(2020)}]{koun20}
{Kounkel} M., 2020, arXiv e-prints, arXiv:2007.09160

\bibitem[{{Kounkel} \& {Covey}(2019)}]{koun19}
{Kounkel} M., {Covey} K., 2019, \aj, 158, 122

\bibitem[{{Kounkel} {et~al}\mbox{.}(2018){Kounkel}, {Covey}, {Su{\'a}rez},
  {Rom{\'a}n-Z{\'u}{\~n}iga}, {Hernandez}, {Stassun}, {Jaehnig}, {Feigelson},
  {Pe{\~n}a Ram{\'\i}rez}, {Roman-Lopes}, {Da Rio}, {Stringfellow}, {Kim},
  {Borissova}, {Fern{\'a}ndez-Trincado}, {Burgasser},
  {Garc{\'\i}a-Hern{\'a}ndez}, {Zamora}, {Pan}, \& {Nitschelm}}]{koun18}
{Kounkel} M. {et~al.}, 2018, \aj, 156, 84

\bibitem[{{Kounkel} {et~al}\mbox{.}(2017){Kounkel}, {Hartmann}, {Mateo}, \&
  {Bailey}}]{koun17}
{Kounkel} M., {Hartmann} L., {Mateo} M., {Bailey}, John~I. I., 2017, \apj, 844,
  138

\bibitem[{{Kouwenhoven} {et~al}\mbox{.}(2007){Kouwenhoven}, {Brown}, {Portegies
  Zwart}, \& {Kaper}}]{kouw07}
{Kouwenhoven} M.~B.~N., {Brown} A.~G.~A., {Portegies Zwart} S.~F., {Kaper} L.,
  2007, \aap, 474, 77

\bibitem[{{Kouwenhoven} {et~al}\mbox{.}(2005){Kouwenhoven}, {Brown},
  {Zinnecker}, {Kaper}, \& {Portegies Zwart}}]{kouw05}
{Kouwenhoven} M.~B.~N., {Brown} A.~G.~A., {Zinnecker} H., {Kaper} L.,
  {Portegies Zwart} S.~F., 2005, \aap, 430, 137

\bibitem[{{Kraus} {et~al}\mbox{.}(2015){Kraus}, {Cody}, {Covey}, {Rizzuto},
  {Mann}, \& {Ireland}}]{krau15}
{Kraus} A.~L., {Cody} A.~M., {Covey} K.~R., {Rizzuto} A.~C., {Mann} A.~W.,
  {Ireland} M.~J., 2015, \apj, 807, 3

\bibitem[{{Krause} {et~al}\mbox{.}(2018){Krause}, {Burkert}, {Diehl},
  {Fierlinger}, {Gaczkowski}, {Kroell}, {Ngoumou}, {Roccatagliata}, {Siegert},
  \& {Preibisch}}]{krau18}
{Krause} M. G.~H. {et~al.}, 2018, \aap, 619, A120

\bibitem[{{Krone-Martins} \& {Moitinho}(2014)}]{kron14}
{Krone-Martins} A., {Moitinho} A., 2014, \aap, 561, A57

\bibitem[{{Kroupa}(2001)}]{krou01a}
{Kroupa} P., 2001, \mnras, 322, 231

\bibitem[{{Kroupa}(2011)}]{krou11}
{Kroupa} P., 2011, in Stellar Clusters and Associations: A RIA Workshop on
  Gaia, pp. 17--27

\bibitem[{{Kroupa} {et~al}\mbox{.}(2001){Kroupa}, {Aarseth}, \&
  {Hurley}}]{krou01}
{Kroupa} P., {Aarseth} S., {Hurley} J., 2001, \mnras, 321, 699

\bibitem[{{Kroupa} \& {Bouvier}(2003)}]{krou03}
{Kroupa} P., {Bouvier} J., 2003, \mnras, 346, 343

\bibitem[{{Kroupa} \& {Burkert}(2001)}]{krou01b}
{Kroupa} P., {Burkert} A., 2001, \apj, 555, 945

\bibitem[{{Kruijssen}(2011)}]{krui11}
{Kruijssen} J.~M.~D., 2011, in Stellar Clusters and Associations: A RIA
  Workshop on Gaia, pp. 137--141

\bibitem[{{Kruijssen}(2012)}]{krui12}
{Kruijssen} J.~M.~D., 2012, \mnras, 426, 3008

\bibitem[{{Kruijssen} {et~al}\mbox{.}(2012){Kruijssen}, {Maschberger},
  {Moeckel}, {Clarke}, {Bastian}, \& {Bonnell}}]{krui12b}
{Kruijssen} J.~M.~D., {Maschberger} T., {Moeckel} N., {Clarke} C.~J., {Bastian}
  N., {Bonnell} I.~A., 2012, \mnras, 419, 841

\bibitem[{{Kubiak} {et~al}\mbox{.}(2017){Kubiak}, {Alves}, {Bouy}, {Sarro},
  {Ascenso}, {Burkert}, {Forbrich}, {Gro{\ss}schedl}, {Hacar}, {Hasenberger},
  {Lombardi}, {Meingast}, {K{\"o}hler}, \& {Teixeira}}]{kubi17}
{Kubiak} K. {et~al.}, 2017, \aap, 598, A124

\bibitem[{{Kuhn} {et~al}\mbox{.}(2019){Kuhn}, {Hillenbrand}, {Sills},
  {Feigelson}, \& {Getman}}]{kuhn19}
{Kuhn} M.~A., {Hillenbrand} L.~A., {Sills} A., {Feigelson} E.~D., {Getman}
  K.~V., 2019, \apj, 870, 32

\bibitem[{{Lada}(1987)}]{lada87}
{Lada} C.~J., 1987, in IAU Symposium, Vol. 115, Star Forming Regions,
  {Peimbert} M., {Jugaku} J., eds., pp. 1--17

\bibitem[{{Lada} \& {Lada}(1991)}]{lada91}
{Lada} C.~J., {Lada} E.~A., 1991, in Astronomical Society of the Pacific
  Conference Series, Vol.~13, The Formation and Evolution of Star Clusters,
  {Janes} K., ed., pp. 3--22

\bibitem[{{Lada} \& {Lada}(2003)}]{lada03}
{Lada} C.~J., {Lada} E.~A., 2003, \araa, 41, 57

\bibitem[{{Lamb} {et~al}\mbox{.}(2010){Lamb}, {Oey}, {Werk}, \&
  {Ingleby}}]{lamb10}
{Lamb} J.~B., {Oey} M.~S., {Werk} J.~K., {Ingleby} L.~D., 2010, \apj, 725, 1886

\bibitem[{{Lamers} \& {Leitherer}(1993)}]{lame93}
{Lamers} H. J.~G.~L.~M., {Leitherer} C., 1993, \apj, 412, 771

\bibitem[{{Larson}(1981)}]{lars81}
{Larson} R.~B., 1981, \mnras, 194, 809

\bibitem[{{Larson}(1985)}]{lars85}
{Larson} R.~B., 1985, \mnras, 214, 379

\bibitem[{{Laughlin} \& {Adams}(1998)}]{laug98}
{Laughlin} G., {Adams} F.~C., 1998, \apjl, 508, L171

\bibitem[{{Lee} {et~al}\mbox{.}(2020){Lee}, {Offner}, {Hennebelle},
  {Andr{\'e}}, {Zinnecker}, {Ballesteros-Paredes}, {Inutsuka}, \&
  {Kruijssen}}]{lee20}
{Lee} Y.-N., {Offner} S. S.~R., {Hennebelle} P., {Andr{\'e}} P., {Zinnecker}
  H., {Ballesteros-Paredes} J., {Inutsuka} S.-i., {Kruijssen} J.~M.~D., 2020,
  Space Science Reviews, 216, 70

\bibitem[{{Lesh}(1968)}]{lesh68}
{Lesh} J.~R., 1968, \apj, 152, 905

\bibitem[{{Lesh}(1969)}]{lesh69}
{Lesh} J.~R., 1969, \aj, 74, 891

\bibitem[{{Levine} {et~al}\mbox{.}(2006){Levine}, {Steinhauer}, {Elston}, \&
  {Lada}}]{levi06}
{Levine} J.~L., {Steinhauer} A., {Elston} R.~J., {Lada} E.~A., 2006, \apj, 646,
  1215

\bibitem[{{Lim} {et~al}\mbox{.}(2019){Lim}, {Naz{\'e}}, {Gosset}, \&
  {Rauw}}]{lim19}
{Lim} B., {Naz{\'e}} Y., {Gosset} E., {Rauw} G., 2019, \mnras, 490, 440

\bibitem[{{Lindegren} {et~al}\mbox{.}(2018){Lindegren}, {Hern{\'a}ndez},
  {Bombrun}, {Klioner}, {Bastian}, {Ramos-Lerate}, {de Torres},
  {Steidelm{\"u}ller}, {Stephenson}, {Hobbs}, {Lammers}, {Biermann}, {Geyer},
  {Hilger}, {Michalik}, {Stampa}, {McMillan}, {Casta{\~n}eda}, {Clotet},
  {Comoretto}, {Davidson}, {Fabricius}, {Gracia}, {Hambly}, {Hutton}, {Mora},
  {Portell}, {van Leeuwen}, {Abbas}, {Abreu}, {Altmann}, {Andrei}, {Anglada},
  {Balaguer-N{\'u}{\~n}ez}, {Barache}, {Becciani}, {Bertone}, {Bianchi},
  {Bouquillon}, {Bourda}, {Br{\"u}semeister}, {Bucciarelli}, {Busonero},
  {Buzzi}, {Cancelliere}, {Carlucci}, {Charlot}, {Cheek}, {Crosta}, {Crowley},
  {de Bruijne}, {de Felice}, {Drimmel}, {Esquej}, {Fienga}, {Fraile}, {Gai},
  {Garralda}, {Gonz{\'a}lez-Vidal}, {Guerra}, {Hauser}, {Hofmann}, {Holl},
  {Jordan}, {Lattanzi}, {Lenhardt}, {Liao}, {Licata}, {Lister}, {L{\"o}ffler},
  {Marchant}, {Martin-Fleitas}, {Messineo}, {Mignard}, {Morbidelli}, {Poggio},
  {Riva}, {Rowell}, {Salguero}, {Sarasso}, {Sciacca}, {Siddiqui}, {Smart},
  {Spagna}, {Steele}, {Taris}, {Torra}, {van Elteren}, {van Reeven}, \&
  {Vecchiato}}]{lind18}
{Lindegren} L. {et~al.}, 2018, \aap, 616, A2

\bibitem[{{Liu} {et~al}\mbox{.}(1981){Liu}, {Zhang}, \& {Kimura}}]{liu81}
{Liu} C.-p., {Zhang} C.-s., {Kimura} H., 1981, Chinese Astronomy and
  Astrophysics, 5, 276

\bibitem[{{Lorenzo-Guti{\'e}rrez} {et~al}\mbox{.}(2019){Lorenzo-Guti{\'e}rrez},
  {Alfaro}, {Ma{\'\i}z Apell{\'a}niz}, {Barb{\'a}}, {Mar{\'\i}n-Franch},
  {Ederoclite}, {Crist{\'o}bal-Hornillos}, {Varela}, {V{\'a}zquez Rami{\'o}},
  {Cenarro}, {Lennon}, \& {Garc{\'\i}a-Lario}}]{lore19}
{Lorenzo-Guti{\'e}rrez} A. {et~al.}, 2019, \mnras, 486, 966

\bibitem[{{Lozinskaia} {et~al}\mbox{.}(1986){Lozinskaia}, {Sitnik}, \&
  {Lomovskii}}]{lozi86}
{Lozinskaia} T.~A., {Sitnik} T.~G., {Lomovskii} A.~I., 1986, \apss, 121, 357

\bibitem[{{Lucke} \& {Hodge}(1970)}]{luck70}
{Lucke} P.~B., {Hodge} P.~W., 1970, \aj, 75, 171

\bibitem[{{MacConnell}(1968)}]{macc68}
{MacConnell} D.~J., 1968, \apjs, 16, 275

\bibitem[{{MacDonald} \& {Mullan}(2013)}]{macd13}
{MacDonald} J., {Mullan} D.~J., 2013, \apj, 765, 126

\bibitem[{{Madsen} {et~al}\mbox{.}(2002){Madsen}, {Dravins}, \&
  {Lindegren}}]{mads02}
{Madsen} S., {Dravins} D., {Lindegren} L., 2002, \aap, 381, 446

\bibitem[{{Maeder}(1987)}]{maed87}
{Maeder} A., 1987, \aap, 173, 247

\bibitem[{{Mahy} {et~al}\mbox{.}(2015){Mahy}, {Rauw}, {De Becker}, {Eenens}, \&
  {Flores}}]{mahy15}
{Mahy} L., {Rauw} G., {De Becker} M., {Eenens} P., {Flores} C.~A., 2015, \aap,
  577, A23

\bibitem[{{Ma{\'\i}z Apell{\'a}niz} {et~al}\mbox{.}(2018){Ma{\'\i}z
  Apell{\'a}niz}, {Pantaleoni Gonz{\'a}lez}, {Barb{\'a}},
  {Sim{\'o}n-D{\'\i}az}, {Negueruela}, {Lennon}, {Sota}, \& {Trigueros
  P{\'a}ez}}]{maiz18}
{Ma{\'\i}z Apell{\'a}niz} J., {Pantaleoni Gonz{\'a}lez} M., {Barb{\'a}} R.~H.,
  {Sim{\'o}n-D{\'\i}az} S., {Negueruela} I., {Lennon} D.~J., {Sota} A.,
  {Trigueros P{\'a}ez} E., 2018, \aap, 616, A149

\bibitem[{{Ma{\'\i}z-Apell{\'a}niz}
  {et~al}\mbox{.}(2004){Ma{\'\i}z-Apell{\'a}niz}, {P{\'e}rez}, \&
  {Mas-Hesse}}]{maiz04}
{Ma{\'\i}z-Apell{\'a}niz} J., {P{\'e}rez} E., {Mas-Hesse} J.~M., 2004, \aj,
  128, 1196

\bibitem[{{Ma{\'\i}z Apell{\'a}niz} {et~al}\mbox{.}(2013){Ma{\'\i}z
  Apell{\'a}niz}, {Sota}, {Morrell}, {Barb{\'a}}, {Walborn}, {Alfaro}, {Gamen},
  {Arias}, \& {Gallego Calvente}}]{maiz13}
{Ma{\'\i}z Apell{\'a}niz} J. {et~al.}, 2013, in Massive Stars: From alpha to
  Omega, p. 198

\bibitem[{{Makarov}(2007)}]{maka07}
{Makarov} V.~V., 2007, \apjs, 169, 105

\bibitem[{{Makarov} \& {Fabricius}(2001)}]{maka01}
{Makarov} V.~V., {Fabricius} C., 2001, \aap, 368, 866

\bibitem[{{Makarov} {et~al}\mbox{.}(2005){Makarov}, {Gaume}, \&
  {Andrievsky}}]{maka05}
{Makarov} V.~V., {Gaume} R.~A., {Andrievsky} S.~M., 2005, \mnras, 362, 1109

\bibitem[{{Makarov} {et~al}\mbox{.}(2004){Makarov}, {Olling}, \&
  {Teuben}}]{maka04}
{Makarov} V.~V., {Olling} R.~P., {Teuben} P.~J., 2004, \mnras, 352, 1199

\bibitem[{{Makarov} \& {Urban}(2000)}]{maka00}
{Makarov} V.~V., {Urban} S., 2000, \mnras, 317, 289

\bibitem[{{Mamajek}(2005)}]{mama05}
{Mamajek} E.~E., 2005, \apj, 634, 1385

\bibitem[{{Mamajek} \& {Bell}(2014)}]{mama14}
{Mamajek} E.~E., {Bell} C.~P.~M., 2014, \mnras, 445, 2169

\bibitem[{{Mamajek} \& {Feigelson}(2001)}]{mama01}
{Mamajek} E.~E., {Feigelson} E.~D., 2001, in Astronomical Society of the
  Pacific Conference Series, Vol. 244, Young Stars Near Earth: Progress and
  Prospects, {Jayawardhana} R., {Greene} T., eds., pp. 104--115

\bibitem[{{Mamajek} {et~al}\mbox{.}(1999){Mamajek}, {Lawson}, \&
  {Feigelson}}]{mama99}
{Mamajek} E.~E., {Lawson} W.~A., {Feigelson} E.~D., 1999, \apjl, 516, L77

\bibitem[{{Mamajek} {et~al}\mbox{.}(2000){Mamajek}, {Lawson}, \&
  {Feigelson}}]{mama00}
{Mamajek} E.~E., {Lawson} W.~A., {Feigelson} E.~D., 2000, \apj, 544, 356

\bibitem[{{Mamajek} {et~al}\mbox{.}(2002){Mamajek}, {Meyer}, \&
  {Liebert}}]{mama02}
{Mamajek} E.~E., {Meyer} M.~R., {Liebert} J., 2002, \aj, 124, 1670

\bibitem[{{Maschberger}(2013)}]{masc13}
{Maschberger} T., 2013, \mnras, 429, 1725

\bibitem[{{Mason} {et~al}\mbox{.}(1998){Mason}, {Gies}, {Hartkopf}, {Bagnuolo},
  {ten Brummelaar}, \& {McAlister}}]{maso98}
{Mason} B.~D., {Gies} D.~R., {Hartkopf} W.~I., {Bagnuolo}, William~G. J., {ten
  Brummelaar} T., {McAlister} H.~A., 1998, \aj, 115, 821

\bibitem[{{Mason} {et~al}\mbox{.}(2009){Mason}, {Hartkopf}, {Gies}, {Henry}, \&
  {Helsel}}]{maso09}
{Mason} B.~D., {Hartkopf} W.~I., {Gies} D.~R., {Henry} T.~J., {Helsel} J.~W.,
  2009, \aj, 137, 3358

\bibitem[{{Massey} \& {Johnson}(1993)}]{mass93}
{Massey} P., {Johnson} J., 1993, \aj, 105, 980

\bibitem[{{Massey} {et~al}\mbox{.}(1995){Massey}, {Johnson}, \&
  {Degioia-Eastwood}}]{mass95}
{Massey} P., {Johnson} K.~E., {Degioia-Eastwood} K., 1995, \apj, 454, 151

\bibitem[{{Massey} \& {Thompson}(1991)}]{mass91}
{Massey} P., {Thompson} A.~B., 1991, \aj, 101, 1408

\bibitem[{{McCaughrean} \& {Stauffer}(1994)}]{mcca94}
{McCaughrean} M.~J., {Stauffer} J.~R., 1994, \aj, 108, 1382

\bibitem[{{McCray} \& {Snow}(1979)}]{mccr79}
{McCray} R., {Snow}, T.~P. J., 1979, \araa, 17, 213

\bibitem[{{McKee} \& {Williams}(1997)}]{mcke97}
{McKee} C.~F., {Williams} J.~P., 1997, \apj, 476, 144

\bibitem[{{Mel'Nik} \& {Dambis}(2009)}]{meln09}
{Mel'Nik} A.~M., {Dambis} A.~K., 2009, \mnras, 400, 518

\bibitem[{{Mel'nik} \& {Dambis}(2017)}]{meln17}
{Mel'nik} A.~M., {Dambis} A.~K., 2017, \mnras, 472, 3887

\bibitem[{{Melnik} \& {Dambis}(2020)}]{meln20}
{Melnik} A.~M., {Dambis} A.~K., 2020, \mnras, 493, 2339

\bibitem[{{Mel'Nik} \& {Efremov}(1995)}]{meln95}
{Mel'Nik} A.~M., {Efremov} Y.~N., 1995, Astronomy Letters, 21, 10

\bibitem[{{Merloni} {et~al}\mbox{.}(2020){Merloni}, {Nandra}, \&
  {Predehl}}]{merl20}
{Merloni} A., {Nandra} K., {Predehl} P., 2020, Nature Astronomy, 4, 634

\bibitem[{{Mikami} \& {Ogura}(2001)}]{mika01}
{Mikami} T., {Ogura} K., 2001, \apss, 275, 441

\bibitem[{{Miller} \& {Scalo}(1978)}]{mill78}
{Miller} G.~E., {Scalo} J.~M., 1978, \pasp, 90, 506

\bibitem[{{Miret-Roig} {et~al}\mbox{.}(2018){Miret-Roig}, {Antoja},
  {Romero-G{\'o}mez}, \& {Figueras}}]{mire18}
{Miret-Roig} N., {Antoja} T., {Romero-G{\'o}mez} M., {Figueras} F., 2018, \aap,
  615, A51

\bibitem[{{Moffat} \& {Vogt}(1973)}]{moff73}
{Moffat} A.~F.~J., {Vogt} N., 1973, \aaps, 11, 3

\bibitem[{{Mohanty} {et~al}\mbox{.}(2004){Mohanty}, {Basri}, {Jayawardhana},
  {Allard}, {Hauschildt}, \& {Ardila}}]{moha04}
{Mohanty} S., {Basri} G., {Jayawardhana} R., {Allard} F., {Hauschildt} P.,
  {Ardila} D., 2004, \apj, 609, 854

\bibitem[{{Mohr-Smith} {et~al}\mbox{.}(2015){Mohr-Smith}, {Drew}, {Barentsen},
  {Wright}, {Napiwotzki}, {Corradi}, {Eisl{\"o}ffel}, {Groot}, {Kalari},
  {Parker}, {Raddi}, {Sale}, {Unruh}, {Vink}, \& {Wesson}}]{mohr15}
{Mohr-Smith} M. {et~al.}, 2015, \mnras, 450, 3855

\bibitem[{{Mohr-Smith} {et~al}\mbox{.}(2017){Mohr-Smith}, {Drew}, {Napiwotzki},
  {Sim{\'o}n-D{\'\i}az}, {Wright}, {Barentsen}, {Eisl{\"o}ffel}, {Farnhill},
  {Greimel}, {Mongui{\'o}}, {Kalari}, {Parker}, \& {Vink}}]{mohr17}
{Mohr-Smith} M. {et~al.}, 2017, \mnras, 465, 1807

\bibitem[{{Morgan} {et~al}\mbox{.}(1952){Morgan}, {Sharpless}, \&
  {Osterbrock}}]{morg52}
{Morgan} W.~W., {Sharpless} S., {Osterbrock} D., 1952, \aj, 57, 3

\bibitem[{{Morgan} {et~al}\mbox{.}(1953){Morgan}, {Whitford}, \&
  {Code}}]{morg53}
{Morgan} W.~W., {Whitford} A.~E., {Code} A.~D., 1953, \apj, 118, 318

\bibitem[{{M{\"u}nch} \& {Morgan}(1953)}]{munc53}
{M{\"u}nch} L., {Morgan} W.~W., 1953, \apj, 118, 161

\bibitem[{{Murray} \& {Rahman}(2010)}]{murr10}
{Murray} N., {Rahman} M., 2010, \apj, 709, 424

\bibitem[{{Myers}(1983)}]{myer83}
{Myers} P.~C., 1983, \apj, 270, 105

\bibitem[{{Naylor} \& {Fabian}(1999)}]{nayl99}
{Naylor} T., {Fabian} A.~C., 1999, \mnras, 302, 714

\bibitem[{{Neuhaeuser} {et~al}\mbox{.}(1995){Neuhaeuser}, {Sterzik}, {Schmitt},
  {Wichmann}, \& {Krautter}}]{neuh95}
{Neuhaeuser} R., {Sterzik} M.~F., {Schmitt} J.~H.~M.~M., {Wichmann} R.,
  {Krautter} J., 1995, \aap, 297, 391

\bibitem[{{Neuh{\"a}user} {et~al}\mbox{.}(2019){Neuh{\"a}user}, {Gie{\ss}ler},
  \& {Hambaryan}}]{neuh19}
{Neuh{\"a}user} R., {Gie{\ss}ler} F., {Hambaryan} V.~V., 2019, \mnras, 2261

\bibitem[{{Nicholson} {et~al}\mbox{.}(2019){Nicholson}, {Parker}, {Church},
  {Davies}, {Fearon}, \& {Walton}}]{nich19}
{Nicholson} R.~B., {Parker} R.~J., {Church} R.~P., {Davies} M.~B., {Fearon}
  N.~M., {Walton} S. R.~J., 2019, \mnras, 485, 4893

\bibitem[{{North} {et~al}\mbox{.}(2007){North}, {Tuthill}, {Tango}, \&
  {Davis}}]{nort07b}
{North} J.~R., {Tuthill} P.~G., {Tango} W.~J., {Davis} J., 2007, \mnras, 377,
  415

\bibitem[{{Ochsendorf} {et~al}\mbox{.}(2015){Ochsendorf}, {Brown}, {Bally}, \&
  {Tielens}}]{ochs15}
{Ochsendorf} B.~B., {Brown} A. G.~A., {Bally} J., {Tielens} A. G.~G.~M., 2015,
  \apj, 808, 111

\bibitem[{{Oliver} {et~al}\mbox{.}(1996){Oliver}, {Masheder}, \&
  {Thaddeus}}]{oliv96b}
{Oliver} R.~J., {Masheder} M.~R.~W., {Thaddeus} P., 1996, \aap, 315, 578

\bibitem[{{Oort}(1954)}]{oort54}
{Oort} J.~H., 1954, {Bulletin of the Astronomical Institutes of the
  Netherlands}, 12, 177

\bibitem[{{Opik}(1953)}]{opik53}
{Opik} E.~J., 1953, Irish Astronomical Journal, 2, 219

\bibitem[{{Ortega} {et~al}\mbox{.}(2002){Ortega}, {de la Reza}, {Jilinski}, \&
  {Bazzanella}}]{orte02}
{Ortega} V.~G., {de la Reza} R., {Jilinski} E., {Bazzanella} B., 2002, \apjl,
  575, L75

\bibitem[{{Ortiz-Le{\'o}n} {et~al}\mbox{.}(2018){Ortiz-Le{\'o}n}, {Loinard},
  {Dzib}, {Galli}, {Kounkel}, {Mioduszewski}, {Rodr{\'\i}guez}, {Torres},
  {Hartmann}, \& {Boden}}]{orti18}
{Ortiz-Le{\'o}n} G.~N. {et~al.}, 2018, \apj, 865, 73

\bibitem[{{Pabst} {et~al}\mbox{.}(2020){Pabst}, {Goicoechea}, {Teyssier},
  {Bern{\'e}}, {Higgins}, {Chambers}, {Kabanovic}, {G{\"u}sten}, {Stutzki}, \&
  {Tielens}}]{pabs20}
{Pabst} C.~H.~M. {et~al.}, 2020, \aap, 639, A2

\bibitem[{{Pannekoek}(1929)}]{pann29}
{Pannekoek} A., 1929, Publications of the Astronomical Institute of the
  University of Amsterdam, 2, 1

\bibitem[{{Parker}(2020)}]{park20}
{Parker} R.~J., 2020, arXiv e-prints, arXiv:2007.07890

\bibitem[{{Parker} \& {Wright}(2016)}]{park16}
{Parker} R.~J., {Wright} N.~J., 2016, \mnras, 457, 3430

\bibitem[{{Parker} {et~al}\mbox{.}(2014){Parker}, {Wright}, {Goodwin}, \&
  {Meyer}}]{park14}
{Parker} R.~J., {Wright} N.~J., {Goodwin} S.~P., {Meyer} M.~R., 2014, \mnras,
  438, 620

\bibitem[{{Patten} \& {Simon}(1996)}]{patt96}
{Patten} B.~M., {Simon} T., 1996, \apjs, 106, 489

\bibitem[{{Pecaut} \& {Mamajek}(2016)}]{peca16}
{Pecaut} M.~J., {Mamajek} E.~E., 2016, \mnras, 461, 794

\bibitem[{{Pecaut} {et~al}\mbox{.}(2012){Pecaut}, {Mamajek}, \&
  {Bubar}}]{peca12}
{Pecaut} M.~J., {Mamajek} E.~E., {Bubar} E.~J., 2012, \apj, 746, 154

\bibitem[{{Pellizza} {et~al}\mbox{.}(2005){Pellizza}, {Mignani}, {Grenier}, \&
  {Mirabel}}]{pell05}
{Pellizza} L.~J., {Mignani} R.~P., {Grenier} I.~A., {Mirabel} I.~F., 2005,
  \aap, 435, 625

\bibitem[{{Perryman} {et~al}\mbox{.}(1997){Perryman}, {Lindegren},
  {Kovalevsky}, {Hog}, {Bastian}, {Bernacca}, {Creze}, {Donati}, {Grenon},
  {Grewing}, {van Leeuwen}, {van der Marel}, {Mignard}, {Murray}, {Le Poole},
  {Schrijver}, {Turon}, {Arenou}, {Froeschle}, \& {Petersen}}]{perr97}
{Perryman} M.~A.~C. {et~al.}, 1997, \aap, 500, 501

\bibitem[{{Pettersson} \& {Reipurth}(2019)}]{pett19}
{Pettersson} B., {Reipurth} B., 2019, \aap, 630, A90

\bibitem[{{Pfalzner}(2009)}]{pfal09}
{Pfalzner} S., 2009, \aap, 498, L37

\bibitem[{{Portegies Zwart} {et~al}\mbox{.}(2010){Portegies Zwart}, {McMillan},
  \& {Gieles}}]{port10}
{Portegies Zwart} S.~F., {McMillan} S.~L.~W., {Gieles} M., 2010, \araa, 48, 431

\bibitem[{{Povich} {et~al}\mbox{.}(2013){Povich}, {Kuhn}, {Getman}, {Busk},
  {Feigelson}, {Broos}, {Townsley}, {King}, \& {Naylor}}]{povi13}
{Povich} M.~S. {et~al.}, 2013, \apjs, 209, 31

\bibitem[{{Povich} {et~al}\mbox{.}(2011){Povich}, {Townsley}, {Broos},
  {Gagn{\'e}}, {Babler}, {Indebetouw}, {Majewski}, {Meade}, {Getman},
  {Robitaille}, \& {Townsend}}]{povi11}
{Povich} M.~S. {et~al.}, 2011, \apjs, 194, 6

\bibitem[{{Pozzo} {et~al}\mbox{.}(2000){Pozzo}, {Jeffries}, {Naylor}, {Totten},
  {Harmer}, \& {Kenyon}}]{pozz00}
{Pozzo} M., {Jeffries} R.~D., {Naylor} T., {Totten} E.~J., {Harmer} S.,
  {Kenyon} M., 2000, \mnras, 313, L23

\bibitem[{{Preibisch} {et~al}\mbox{.}(1999){Preibisch}, {Balega}, {Hofmann},
  {Weigelt}, \& {Zinnecker}}]{prei99b}
{Preibisch} T., {Balega} Y., {Hofmann} K.-H., {Weigelt} G., {Zinnecker} H.,
  1999, New Astronomy, 4, 531

\bibitem[{{Preibisch} {et~al}\mbox{.}(2002){Preibisch}, {Brown}, {Bridges},
  {Guenther}, \& {Zinnecker}}]{prei02}
{Preibisch} T., {Brown} A.~G.~A., {Bridges} T., {Guenther} E., {Zinnecker} H.,
  2002, \aj, 124, 404

\bibitem[{{Preibisch} \& {Feigelson}(2005)}]{prei05}
{Preibisch} T., {Feigelson} E.~D., 2005, \apjs, 160, 390

\bibitem[{{Preibisch} {et~al}\mbox{.}(2017){Preibisch}, {Flaischlen},
  {Gaczkowski}, {Townsley}, \& {Broos}}]{prei17}
{Preibisch} T., {Flaischlen} S., {Gaczkowski} B., {Townsley} L., {Broos} P.,
  2017, \aap, 605, A85

\bibitem[{{Preibisch} {et~al}\mbox{.}(1998){Preibisch}, {Guenther},
  {Zinnecker}, {Sterzik}, {Frink}, \& {Roeser}}]{prei98}
{Preibisch} T., {Guenther} E., {Zinnecker} H., {Sterzik} M., {Frink} S.,
  {Roeser} S., 1998, \aap, 333, 619

\bibitem[{{Preibisch} \& {Mamajek}(2008)}]{prei08}
{Preibisch} T., {Mamajek} E., 2008, {The Nearest OB Association:
  Scorpius-Centaurus (Sco OB2)}, {Handbook of Star Forming Regions}, p. 235

\bibitem[{{Preibisch} {et~al}\mbox{.}(2014){Preibisch}, {Mehlhorn}, {Townsley},
  {Broos}, \& {Ratzka}}]{prei14}
{Preibisch} T., {Mehlhorn} M., {Townsley} L., {Broos} P., {Ratzka} T., 2014,
  \aap, 564, A120

\bibitem[{{Preibisch} {et~al}\mbox{.}(2011){Preibisch}, {Ratzka}, {Kuderna},
  {Ohlendorf}, {King}, {Hodgkin}, {Irwin}, {Lewis}, {McCaughrean}, \&
  {Zinnecker}}]{prei11}
{Preibisch} T. {et~al.}, 2011, \aap, 530, A34

\bibitem[{{Preibisch} \& {Zinnecker}(1999)}]{prei99}
{Preibisch} T., {Zinnecker} H., 1999, \aj, 117, 2381

\bibitem[{{Preibisch} \& {Zinnecker}(2007)}]{prei07}
{Preibisch} T., {Zinnecker} H., 2007, in IAU Symposium, Vol. 237, Triggered
  Star Formation in a Turbulent ISM, {Elmegreen} B.~G., {Palous} J., eds., pp.
  270--277

\bibitem[{{Prisinzano} {et~al}\mbox{.}(2016){Prisinzano}, {Damiani}, {Micela},
  {Jeffries}, {Franciosini}, {Sacco}, {Frasca}, {Klutsch}, {Lanzafame},
  {Alfaro}, {Biazzo}, {Bonito}, {Bragaglia}, {Caramazza}, {Vallenari},
  {Carraro}, {Costado}, {Flaccomio}, {Jofr{\'e}}, {Lardo}, {Monaco},
  {Morbidelli}, {Mowlavi}, {Pancino}, {Randich}, \& {Zaggia}}]{pris16}
{Prisinzano} L. {et~al.}, 2016, \aap, 589, A70

\bibitem[{{Racine}(1968)}]{raci68}
{Racine} R., 1968, \aj, 73, 233

\bibitem[{{Rahman} {et~al}\mbox{.}(2011){Rahman}, {Matzner}, \&
  {Moon}}]{rahm11}
{Rahman} M., {Matzner} C., {Moon} D.-S., 2011, \apjl, 728, L37

\bibitem[{{Randich} {et~al}\mbox{.}(1997){Randich}, {Aharpour}, {Pallavicini},
  {Prosser}, \& {Stauffer}}]{rand97}
{Randich} S., {Aharpour} N., {Pallavicini} R., {Prosser} C.~F., {Stauffer}
  J.~R., 1997, \aap, 323, 86

\bibitem[{{Rate} {et~al}\mbox{.}(2020){Rate}, {Crowther}, \& {Parker}}]{rate20}
{Rate} G., {Crowther} P.~A., {Parker} R.~J., 2020, \mnras, 495, 1209

\bibitem[{{Reddish} {et~al}\mbox{.}(1967){Reddish}, {Lawrence}, \&
  {Pratt}}]{redd67}
{Reddish} V.~C., {Lawrence} L.~C., {Pratt} N.~N., 1967, \mnras, 136, 428

\bibitem[{{Reipurth} \& {Schneider}(2008)}]{reip08}
{Reipurth} B., {Schneider} N., 2008, {Star Formation and Young Clusters in
  Cygnus}, {Handbook of Star Forming Regions}, p.~36

\bibitem[{{Reiter} \& {Parker}(2019)}]{reit19}
{Reiter} M., {Parker} R.~J., 2019, \mnras, 486, 4354

\bibitem[{{Reynolds}(1976)}]{reyn76}
{Reynolds} R.~J., 1976, \apj, 206, 679

\bibitem[{{Reynolds} \& {Ogden}(1978)}]{reyn78}
{Reynolds} R.~J., {Ogden} P.~M., 1978, \apj, 224, 94

\bibitem[{{Reynolds} \& {Ogden}(1979)}]{reyn79}
{Reynolds} R.~J., {Ogden} P.~M., 1979, \apj, 229, 942

\bibitem[{{Ribas} {et~al}\mbox{.}(2005){Ribas}, {Guinan}, {G{\"u}del}, \&
  {Audard}}]{riba05}
{Ribas} I., {Guinan} E.~F., {G{\"u}del} M., {Audard} M., 2005, \apj, 622, 680

\bibitem[{{Richert} {et~al}\mbox{.}(2015){Richert}, {Feigelson}, {Getman}, \&
  {Kuhn}}]{rich15}
{Richert} A. J.~W., {Feigelson} E.~D., {Getman} K.~V., {Kuhn} M.~A., 2015,
  \apj, 811, 10

\bibitem[{{Ricker} {et~al}\mbox{.}(2015){Ricker}, {Winn}, {Vanderspek},
  {Latham}, {Bakos}, {Bean}, {Berta-Thompson}, {Brown}, {Buchhave}, {Butler},
  {Butler}, {Chaplin}, {Charbonneau}, {Christensen-Dalsgaard}, {Clampin},
  {Deming}, {Doty}, {De Lee}, {Dressing}, {Dunham}, {Endl}, {Fressin}, {Ge},
  {Henning}, {Holman}, {Howard}, {Ida}, {Jenkins}, {Jernigan}, {Johnson},
  {Kaltenegger}, {Kawai}, {Kjeldsen}, {Laughlin}, {Levine}, {Lin}, {Lissauer},
  {MacQueen}, {Marcy}, {McCullough}, {Morton}, {Narita}, {Paegert}, {Palle},
  {Pepe}, {Pepper}, {Quirrenbach}, {Rinehart}, {Sasselov}, {Sato}, {Seager},
  {Sozzetti}, {Stassun}, {Sullivan}, {Szentgyorgyi}, {Torres}, {Udry}, \&
  {Villasenor}}]{rick15}
{Ricker} G.~R. {et~al.}, 2015, Journal of Astronomical Telescopes, Instruments,
  and Systems, 1, 014003

\bibitem[{{Rizzuto} {et~al}\mbox{.}(2011){Rizzuto}, {Ireland}, \&
  {Robertson}}]{rizz11}
{Rizzuto} A.~C., {Ireland} M.~J., {Robertson} J.~G., 2011, \mnras, 416, 3108

\bibitem[{{Rizzuto} {et~al}\mbox{.}(2013){Rizzuto}, {Ireland}, {Robertson},
  {Kok}, {Tuthill}, {Warrington}, {Haubois}, {Tango}, {Norris}, {ten
  Brummelaar}, {Kraus}, {Jacob}, \& {Laliberte-Houdeville}}]{rizz13}
{Rizzuto} A.~C. {et~al.}, 2013, \mnras, 436, 1694

\bibitem[{{Roberts}(1957)}]{robe57}
{Roberts} M.~S., 1957, \pasp, 69, 59

\bibitem[{{Robitaille} {et~al}\mbox{.}(2018){Robitaille}, {Scaife}, {Carretti},
  {Haverkorn}, {Crocker}, {Kesteven}, {Poppi}, \& {Staveley-Smith}}]{robi18}
{Robitaille} J.~F., {Scaife} A.~M.~M., {Carretti} E., {Haverkorn} M., {Crocker}
  R.~M., {Kesteven} M.~J., {Poppi} S., {Staveley-Smith} L., 2018, \aap, 617,
  A101

\bibitem[{{R{\"o}ser} {et~al}\mbox{.}(2018){R{\"o}ser}, {Schilbach}, {Goldman},
  {Henning}, {Moor}, \& {Derekas}}]{rose18}
{R{\"o}ser} S., {Schilbach} E., {Goldman} B., {Henning} T., {Moor} A.,
  {Derekas} A., 2018, \aap, 614, A81

\bibitem[{{Rosotti} {et~al}\mbox{.}(2014){Rosotti}, {Dale}, {de Juan Ovelar},
  {Hubber}, {Kruijssen}, {Ercolano}, \& {Walch}}]{roso14}
{Rosotti} G.~P., {Dale} J.~E., {de Juan Ovelar} M., {Hubber} D.~A., {Kruijssen}
  J.~M.~D., {Ercolano} B., {Walch} S., 2014, \mnras, 441, 2094

\bibitem[{{Ruprecht}(1966)}]{rupr66}
{Ruprecht} J., 1966, IAU Trans., 12, 348

\bibitem[{{Rygl} {et~al}\mbox{.}(2012){Rygl}, {Brunthaler}, {Sanna}, {Menten},
  {Reid}, {van Langevelde}, {Honma}, {Torstensson}, \& {Fujisawa}}]{rygl12}
{Rygl} K.~L.~J. {et~al.}, 2012, \aap, 539, A79

\bibitem[{{Sacco} {et~al}\mbox{.}(2015){Sacco}, {Jeffries}, {Randich},
  {Franciosini}, {Jackson}, {Cottaar}, {Spina}, {Palla}, {Mapelli}, {Alfaro},
  {Bonito}, {Damiani}, {Frasca}, {Klutsch}, {Lanzafame}, {Bayo}, {Barrado},
  {Jim{\'e}nez-Esteban}, {Gilmore}, {Micela}, {Vallenari}, {Allende Prieto},
  {Flaccomio}, {Carraro}, {Costado}, {Jofr{\'e}}, {Lardo}, {Magrini},
  {Morbidelli}, {Prisinzano}, \& {Sbordone}}]{sacc15}
{Sacco} G.~G. {et~al.}, 2015, \aap, 574, L7

\bibitem[{{Sahu}(1992)}]{sahu92}
{Sahu} M.~S., 1992, PhD thesis, Kapteyn Institute, Postbus 800 9700 AV
  Groningen, The Netherlands

\bibitem[{{Sana} {et~al}\mbox{.}(2012){Sana}, {de Mink}, {de Koter}, {Langer},
  {Evans}, {Gieles}, {Gosset}, {Izzard}, {Le Bouquin}, \& {Schneider}}]{sana12}
{Sana} H. {et~al.}, 2012, Science, 337, 444

\bibitem[{{Sana} \& {Evans}(2011)}]{sana11}
{Sana} H., {Evans} C.~J., 2011, in IAU Symposium, Vol. 272, IAU Symposium,
  {Neiner} C., {Wade} G., {Meynet} G., {Peters} G., eds., pp. 474--485

\bibitem[{{Sana} {et~al}\mbox{.}(2017){Sana}, {Ram{\'\i}rez-Tannus}, {de
  Koter}, {Kaper}, {Tramper}, \& {Bik}}]{sana17}
{Sana} H., {Ram{\'\i}rez-Tannus} M.~C., {de Koter} A., {Kaper} L., {Tramper}
  F., {Bik} A., 2017, \aap, 599, L9

\bibitem[{{Sancisi} {et~al}\mbox{.}(1974){Sancisi}, {Goss}, {Anderson},
  {Johansson}, \& {Winnberg}}]{sanc74}
{Sancisi} R., {Goss} W.~M., {Anderson} C., {Johansson} L.~E.~B., {Winnberg} A.,
  1974, \aap, 35, 445

\bibitem[{{Sandford} {et~al}\mbox{.}(1982){Sandford}, {Whitaker}, \&
  {Klein}}]{sand82}
{Sandford}, M.~T. I., {Whitaker} R.~W., {Klein} R.~I., 1982, \apj, 260, 183

\bibitem[{{Santos-Silva} {et~al}\mbox{.}(2018){Santos-Silva}, {Gregorio-Hetem},
  {Montmerle}, {Fernandes}, \& {Stelzer}}]{sant18}
{Santos-Silva} T., {Gregorio-Hetem} J., {Montmerle} T., {Fernandes} B.,
  {Stelzer} B., 2018, \aap, 609, A127

\bibitem[{{Sartori} {et~al}\mbox{.}(2003){Sartori}, {L{\'e}pine}, \&
  {Dias}}]{sart03}
{Sartori} M.~J., {L{\'e}pine} J.~R.~D., {Dias} W.~S., 2003, \aap, 404, 913

\bibitem[{{Scally} \& {Clarke}(2001)}]{scal01}
{Scally} A., {Clarke} C., 2001, \mnras, 325, 449

\bibitem[{{Scalo}(1985)}]{scal85}
{Scalo} J.~M., 1985, in Protostars and Planets II, {Black} D.~C., {Matthews}
  M.~S., eds., pp. 201--296

\bibitem[{{Schaerer} \& {de Koter}(1997)}]{scha97}
{Schaerer} D., {de Koter} A., 1997, \aap, 322, 598

\bibitem[{{Schild} \& {Maeder}(1985)}]{schi85}
{Schild} H., {Maeder} A., 1985, \aap, 143, L7

\bibitem[{{Schmeja} {et~al}\mbox{.}(2008){Schmeja}, {Kumar}, \&
  {Ferreira}}]{schm08}
{Schmeja} S., {Kumar} M.~S.~N., {Ferreira} B., 2008, \mnras, 389, 1209

\bibitem[{{Schmidt}(1958)}]{schm58}
{Schmidt} K.~H., 1958, Astronomische Nachrichten, 284, 76

\bibitem[{{Schulte}(1956)}]{schu56}
{Schulte} D.~H., 1956, \apj, 124, 530

\bibitem[{{Schulz} {et~al}\mbox{.}(1997){Schulz}, {Berghoefer}, \&
  {Zinnecker}}]{schu97}
{Schulz} N.~S., {Berghoefer} T.~W., {Zinnecker} H., 1997, \aap, 325, 1001

\bibitem[{{Sewi{\l}o} {et~al}\mbox{.}(2019){Sewi{\l}o}, {Whitney}, {Yung},
  {Robitaille}, {Elia}, {Indebetouw}, {Schisano}, {Szczerba}, {Karska}, \&
  {Wiseman}}]{sewi19}
{Sewi{\l}o} M. {et~al.}, 2019, \apjs, 240, 26

\bibitem[{{Shatsky} \& {Tokovinin}(2002)}]{shat02}
{Shatsky} N., {Tokovinin} A., 2002, \aap, 382, 92

\bibitem[{{Shobbrook}(1983)}]{shob83}
{Shobbrook} R.~R., 1983, \mnras, 205, 1229

\bibitem[{{Sim{\'o}n-D{\'\i}az}(2010)}]{simo10}
{Sim{\'o}n-D{\'\i}az} S., 2010, \aap, 510, A22

\bibitem[{{Simoudis} {et~al}\mbox{.}(1996){Simoudis}, {Han}, \&
  {Fayyad}}]{este96}
{Simoudis} E., {Han} J., {Fayyad} U., eds., 1996, {DBSCAN}, {Simoudis} E.,
  {Han} J., {Fayyad} U., eds. {AAAI Press}

\bibitem[{{Sitnik} {et~al}\mbox{.}(2019){Sitnik}, {Egorov}, {Lozinskaya},
  {Moiseev}, {Tatarnikov}, {Vozyakova}, \& {Wiebe}}]{sitn19}
{Sitnik} T.~G., {Egorov} O.~V., {Lozinskaya} T.~A., {Moiseev} A.~V.,
  {Tatarnikov} A.~M., {Vozyakova} O.~V., {Wiebe} D.~S., 2019, \mnras, 486, 2449

\bibitem[{{Sivan}(1974)}]{siva74}
{Sivan} J.~P., 1974, \aaps, 16, 163

\bibitem[{{Slesnick} {et~al}\mbox{.}(2006){Slesnick}, {Carpenter}, \&
  {Hillenbrand }}]{sles06}
{Slesnick} C.~L., {Carpenter} J.~M., {Hillenbrand } L.~A., 2006, \aj, 131, 3016

\bibitem[{{Smartt}(2009)}]{smar09}
{Smartt} S.~J., 2009, \araa, 47, 63

\bibitem[{{Smith}(2006)}]{smit06}
{Smith} N., 2006, \apj, 644, 1151

\bibitem[{{Smith} \& {Brooks}(2008)}]{smit08b}
{Smith} N., {Brooks} K.~J., 2008, {The Carina Nebula: A Laboratory for Feedback
  and Triggered Star Formation}, {Reipurth} B., ed., Vol.~5, p. 138

\bibitem[{{Smith} {et~al}\mbox{.}(2011){Smith}, {Fellhauer}, {Goodwin}, \&
  {Assmann}}]{smit11}
{Smith} R., {Fellhauer} M., {Goodwin} S., {Assmann} P., 2011, \mnras, 414, 3036

\bibitem[{{Soderblom}(2010)}]{sode10}
{Soderblom} D.~R., 2010, \araa, 48, 581

\bibitem[{{Soderblom} {et~al}\mbox{.}(2014{\natexlab{a}}){Soderblom},
  {Hillenbrand}, {Jeffries}, {Mamajek}, \& {Naylor}}]{burn05}
{Soderblom} D.~R., {Hillenbrand} L.~A., {Jeffries} R.~D., {Mamajek} E.~E.,
  {Naylor} T., 2014{\natexlab{a}}, in Protostars and Planets VI, {Beuther} H.,
  {Klessen} R.~S., {Dullemond} C.~P., {Henning} T., eds., p. 219

\bibitem[{{Soderblom} {et~al}\mbox{.}(2014{\natexlab{b}}){Soderblom},
  {Hillenbrand}, {Jeffries}, {Mamajek}, \& {Naylor}}]{sode14}
{Soderblom} D.~R., {Hillenbrand} L.~A., {Jeffries} R.~D., {Mamajek} E.~E.,
  {Naylor} T., 2014{\natexlab{b}}, in Protostars and Planets VI, {Beuther} H.,
  {Klessen} R.~S., {Dullemond} C.~P., {Henning} T., eds., p. 219

\bibitem[{{Soderblom} {et~al}\mbox{.}(1993){Soderblom}, {Stauffer},
  {MacGregor}, \& {Jones}}]{sode93}
{Soderblom} D.~R., {Stauffer} J.~R., {MacGregor} K.~B., {Jones} B.~F., 1993,
  \apj, 409, 624

\bibitem[{{Sota} {et~al}\mbox{.}(2014){Sota}, {Ma{\'{\i}}z Apell{\'a}niz},
  {Morrell}, {Barb{\'a}}, {Walborn}, {Gamen}, {Arias}, \& {Alfaro}}]{sota14}
{Sota} A., {Ma{\'{\i}}z Apell{\'a}niz} J., {Morrell} N.~I., {Barb{\'a}} R.~H.,
  {Walborn} N.~R., {Gamen} R.~C., {Arias} J.~I., {Alfaro} E.~J., 2014, \apjs,
  211, 10

\bibitem[{{Stauffer} \& {Hartmann}(1986)}]{stau86}
{Stauffer} J.~R., {Hartmann} L.~W., 1986, \apjs, 61, 531

\bibitem[{{Steenbrugge} {et~al}\mbox{.}(2003){Steenbrugge}, {de Bruijne},
  {Hoogerwerf}, \& {de Zeeuw}}]{stee03}
{Steenbrugge} K.~C., {de Bruijne} J.~H.~J., {Hoogerwerf} R., {de Zeeuw} P.~T.,
  2003, \aap, 402, 587

\bibitem[{{Strai{\v{z}}ys} \& {Laugalys}(2007)}]{stra07}
{Strai{\v{z}}ys} V., {Laugalys} V., 2007, Baltic Astronomy, 16, 167

\bibitem[{{Stutzki} {et~al}\mbox{.}(1998){Stutzki}, {Bensch}, {Heithausen},
  {Ossenkopf}, \& {Zielinsky}}]{stut98}
{Stutzki} J., {Bensch} F., {Heithausen} A., {Ossenkopf} V., {Zielinsky} M.,
  1998, \aap, 336, 697

\bibitem[{{Sung} {et~al}\mbox{.}(1997){Sung}, {Bessell}, \& {Lee}}]{sung97}
{Sung} H., {Bessell} M.~S., {Lee} S.-W., 1997, \aj, 114, 2644

\bibitem[{{Sung} {et~al}\mbox{.}(1998){Sung}, {Bessell}, \& {Lee}}]{sung98}
{Sung} H., {Bessell} M.~S., {Lee} S.-W., 1998, \aj, 115, 734

\bibitem[{{Testori} {et~al}\mbox{.}(2006){Testori}, {Arnal}, {Morras},
  {Bajaja}, {P{\"o}ppel}, \& {Reich}}]{test06}
{Testori} J.~C., {Arnal} E.~M., {Morras} R., {Bajaja} E., {P{\"o}ppel}
  W.~G.~L., {Reich} P., 2006, \aap, 458, 163

\bibitem[{{Tetzlaff} {et~al}\mbox{.}(2010){Tetzlaff}, {Neuh{\"a}user}, {Hohle},
  \& {Maciejewski}}]{tetz10}
{Tetzlaff} N., {Neuh{\"a}user} R., {Hohle} M.~M., {Maciejewski} G., 2010,
  \mnras, 402, 2369

\bibitem[{{Throop} \& {Bally}(2005)}]{thro05}
{Throop} H.~B., {Bally} J., 2005, \apjl, 623, L149

\bibitem[{{Tobin} {et~al}\mbox{.}(2009){Tobin}, {Hartmann}, {Furesz}, {Mateo},
  \& {Megeath}}]{tobi09}
{Tobin} J.~J., {Hartmann} L., {Furesz} G., {Mateo} M., {Megeath} S.~T., 2009,
  \apj, 697, 1103

\bibitem[{{Tokovinin} \& {Brice{\~n}o}(2020)}]{toko20}
{Tokovinin} A., {Brice{\~n}o} C., 2020, \aj, 159, 15

\bibitem[{{Tokovinin} {et~al}\mbox{.}(2020){Tokovinin}, {Petr-Gotzens}, \&
  {Brice{\~n}o}}]{toko20b}
{Tokovinin} A., {Petr-Gotzens} M.~G., {Brice{\~n}o} C., 2020, \aj, submitted

\bibitem[{{Torres} {et~al}\mbox{.}(2008){Torres}, {Quast}, {Melo}, \&
  {Sterzik}}]{torr08}
{Torres} C.~A.~O., {Quast} G.~R., {Melo} C.~H.~F., {Sterzik} M.~F., 2008,
  {Young Nearby Loose Associations}, {Astronomical Society of the Pacific}, p.
  757

\bibitem[{{Torres-Dodgen} {et~al}\mbox{.}(1991){Torres-Dodgen}, {Carroll}, \&
  {Tapia}}]{torr91}
{Torres-Dodgen} A.~V., {Carroll} M., {Tapia} M., 1991, \mnras, 249, 1

\bibitem[{{Turner}(1976)}]{turn76}
{Turner} D.~G., 1976, \apj, 210, 65

\bibitem[{{Turner}(1979)}]{turn79}
{Turner} D.~G., 1979, \aap, 76, 350

\bibitem[{{Turner} {et~al}\mbox{.}(1980){Turner}, {Grieve}, {Herbst}, \&
  {Harris}}]{turn80}
{Turner} D.~G., {Grieve} G.~R., {Herbst} W., {Harris} W.~E., 1980, \aj, 85,
  1193

\bibitem[{{Tutukov}(1978)}]{tutu78}
{Tutukov} A.~V., 1978, \aap, 70, 57

\bibitem[{{Uyan{\i}ker} {et~al}\mbox{.}(2001){Uyan{\i}ker}, {F{\"u}rst},
  {Reich}, {Aschenbach}, \& {Wielebinski}}]{uyan01}
{Uyan{\i}ker} B., {F{\"u}rst} E., {Reich} W., {Aschenbach} B., {Wielebinski}
  R., 2001, \aap, 371, 675

\bibitem[{{van Altena} {et~al}\mbox{.}(1988){van Altena}, {Lee}, {Lee}, {Lu},
  \& {Upgren}}]{vana88}
{van Altena} W.~F., {Lee} J.~T., {Lee} J.~F., {Lu} P.~K., {Upgren} A.~R., 1988,
  \aj, 95, 1744

\bibitem[{{van den Bergh}(1964)}]{vand64}
{van den Bergh} S., 1964, \apjs, 9, 65

\bibitem[{{van Elteren} {et~al}\mbox{.}(2019){van Elteren}, {Portegies Zwart},
  {Pelupessy}, {Cai}, \& {McMillan}}]{vane19}
{van Elteren} A., {Portegies Zwart} S., {Pelupessy} I., {Cai} M.~X., {McMillan}
  S.~L.~W., 2019, \aap, 624, A120

\bibitem[{{van Genderen} {et~al}\mbox{.}(1984){van Genderen}, {Bijleveld}, \&
  {van Groningen}}]{vang84}
{van Genderen} A.~M., {Bijleveld} W., {van Groningen} E., 1984, \aaps, 58, 537

\bibitem[{{V{\'a}zquez} {et~al}\mbox{.}(2008){V{\'a}zquez}, {May}, {Carraro},
  {Bronfman}, {Moitinho}, \& {Baume}}]{vazq08}
{V{\'a}zquez} R.~A., {May} J., {Carraro} G., {Bronfman} L., {Moitinho} A.,
  {Baume} G., 2008, \apj, 672, 930

\bibitem[{{V{\'a}zquez-Semadeni} {et~al}\mbox{.}(2017){V{\'a}zquez-Semadeni},
  {Gonz{\'a}lez-Samaniego}, \& {Col{\'{\i}}n}}]{vazq17}
{V{\'a}zquez-Semadeni} E., {Gonz{\'a}lez-Samaniego} A., {Col{\'{\i}}n} P.,
  2017, \mnras, 467, 1313

\bibitem[{{Villa V{\'e}lez} {et~al}\mbox{.}(2018){Villa V{\'e}lez}, {Brown}, \&
  {Kenworthy}}]{vill18}
{Villa V{\'e}lez} J.~A., {Brown} A. G.~A., {Kenworthy} M.~A., 2018, Research
  Notes of the American Astronomical Society, 2, 58

\bibitem[{{Vincke} \& {Pfalzner}(2016)}]{vinc16}
{Vincke} K., {Pfalzner} S., 2016, \apj, 828, 48

\bibitem[{{Vink} {et~al}\mbox{.}(2008){Vink}, {Drew}, {Steeghs}, {Wright},
  {Martin}, {G{\"a}nsicke}, {Greimel}, \& {Drake}}]{vink08}
{Vink} J.~S., {Drew} J.~E., {Steeghs} D., {Wright} N.~J., {Martin} E.~L.,
  {G{\"a}nsicke} B.~T., {Greimel} R., {Drake} J., 2008, \mnras, 387, 308

\bibitem[{{Walborn}(1973)}]{walb73}
{Walborn} N.~R., 1973, \apjl, 180, L35

\bibitem[{{Walborn}(1995)}]{walb95}
{Walborn} N.~R., 1995, in Revista Mexicana de Astronomia y Astrofisica
  Conference Series, Vol.~2, Revista Mexicana de Astronomia y Astrofisica
  Conference Series, {Niemela} V., {Morrell} N., {Feinstein} A., eds., p.~51

\bibitem[{{Walborn}(2002)}]{walb02}
{Walborn} N.~R., 2002, \aj, 124, 507

\bibitem[{{Walborn} {et~al}\mbox{.}(2002){Walborn}, {Howarth}, {Lennon},
  {Massey}, {Oey}, {Moffat}, {Skalkowski}, {Morrell}, {Drissen}, \&
  {Parker}}]{walb02b}
{Walborn} N.~R. {et~al.}, 2002, \aj, 123, 2754

\bibitem[{{Walter} {et~al}\mbox{.}(1988){Walter}, {Brown}, {Mathieu}, {Myers},
  \& {Vrba}}]{walt88}
{Walter} F.~M., {Brown} A., {Mathieu} R.~D., {Myers} P.~C., {Vrba} F.~J., 1988,
  \aj, 96, 297

\bibitem[{{Walter} {et~al}\mbox{.}(2008){Walter}, {Sherry}, {Wolk}, \&
  {Adams}}]{walt08}
{Walter} F.~M., {Sherry} W.~H., {Wolk} S.~J., {Adams} N.~R., 2008, {The
  {\ensuremath{\sigma}} Orionis Cluster}, Vol.~4, {Handbook of Star Forming
  Regions}, p. 732

\bibitem[{{Ward} \& {Kruijssen}(2018)}]{ward18}
{Ward} J.~L., {Kruijssen} J.~M.~D., 2018, \mnras, 475, 5659

\bibitem[{{Ward} {et~al}\mbox{.}(2020){Ward}, {Kruijssen}, \& {Rix}}]{ward20}
{Ward} J.~L., {Kruijssen} J.~M.~D., {Rix} H.-W., 2020, \mnras, 495, 663

\bibitem[{{Wareing} {et~al}\mbox{.}(2018){Wareing}, {Pittard}, {Wright}, \&
  {Falle}}]{ware18}
{Wareing} C.~J., {Pittard} J.~M., {Wright} N.~J., {Falle} S.~A.~E.~G., 2018,
  \mnras, 475, 3598

\bibitem[{{Warren} \& {Hesser}(1978)}]{warr78}
{Warren}, W.~H. J., {Hesser} J.~E., 1978, \apjs, 36, 497

\bibitem[{{Weaver} {et~al}\mbox{.}(1977){Weaver}, {McCray}, {Castor},
  {Shapiro}, \& {Moore}}]{weav77}
{Weaver} R., {McCray} R., {Castor} J., {Shapiro} P., {Moore} R., 1977, \apj,
  218, 377

\bibitem[{{Webb} {et~al}\mbox{.}(1999){Webb}, {Zuckerman}, {Platais},
  {Patience}, {White}, {Schwartz}, \& {McCarthy}}]{webb99}
{Webb} R.~A., {Zuckerman} B., {Platais} I., {Patience} J., {White} R.~J.,
  {Schwartz} M.~J., {McCarthy} C., 1999, \apjl, 512, L63

\bibitem[{{Weigelt} {et~al}\mbox{.}(1999){Weigelt}, {Balega}, {Preibisch},
  {Schertl}, {Sch{\"o}ller}, \& {Zinnecker}}]{weig99}
{Weigelt} G., {Balega} Y., {Preibisch} T., {Schertl} D., {Sch{\"o}ller} M.,
  {Zinnecker} H., 1999, \aap, 347, L15

\bibitem[{{Weinberger} {et~al}\mbox{.}(2013){Weinberger}, {Anglada-Escud{\'e}},
  \& {Boss}}]{wein13}
{Weinberger} A.~J., {Anglada-Escud{\'e}} G., {Boss} A.~P., 2013, \apj, 762, 118

\bibitem[{{Westerlund}(1963)}]{west63}
{Westerlund} B.~E., 1963, \mnras, 127, 71

\bibitem[{{Whitworth}(1979)}]{whit79}
{Whitworth} A., 1979, \mnras, 186, 59

\bibitem[{{Wilking} \& {Lada}(1983)}]{wilk83}
{Wilking} B.~A., {Lada} C.~J., 1983, \apj, 274, 698

\bibitem[{{Wilkinson} {et~al}\mbox{.}(2018){Wilkinson}, {Mer{\'\i}n}, \&
  {Riviere-Marichalar}}]{wilk18}
{Wilkinson} S., {Mer{\'\i}n} B., {Riviere-Marichalar} P., 2018, \aap, 618, A12

\bibitem[{{Williams} \& {Cieza}(2011)}]{will11}
{Williams} J.~P., {Cieza} L.~A., 2011, \araa, 49, 67

\bibitem[{{Winter} {et~al}\mbox{.}(2018){Winter}, {Clarke}, {Rosotti}, {Ih},
  {Facchini}, \& {Haworth}}]{wint18b}
{Winter} A.~J., {Clarke} C.~J., {Rosotti} G., {Ih} J., {Facchini} S., {Haworth}
  T.~J., 2018, \mnras, 478, 2700

\bibitem[{{Winter} {et~al}\mbox{.}(2019){Winter}, {Clarke}, \&
  {Rosotti}}]{wint19}
{Winter} A.~J., {Clarke} C.~J., {Rosotti} G.~P., 2019, \mnras, 485, 1489

\bibitem[{{Wiramihardja} {et~al}\mbox{.}(1989){Wiramihardja}, {Kogure},
  {Yoshida}, {Ogura}, \& {Nakano}}]{wira89}
{Wiramihardja} S.~D., {Kogure} T., {Yoshida} S., {Ogura} K., {Nakano} M., 1989,
  \pasj, 41, 155

\bibitem[{{Wolk} {et~al}\mbox{.}(2013){Wolk}, {Rice}, \& {Aspin}}]{wolk13}
{Wolk} S.~J., {Rice} T.~S., {Aspin} C., 2013, \apj, 773, 145

\bibitem[{{Woolley} \& {Eggen}(1958)}]{wool58}
{Woolley} R., {Eggen} O.~J., 1958, The Observatory, 78, 149

\bibitem[{{Wright} {et~al}\mbox{.}(2010{\natexlab{a}}){Wright}, {Eisenhardt},
  {Mainzer}, {Ressler}, {Cutri}, {Jarrett}, {Kirkpatrick}, {Padgett},
  {McMillan}, \& {Skrutskie}}]{wrig10c}
{Wright} E.~L. {et~al.}, 2010{\natexlab{a}}, \aj, 140, 1868

\bibitem[{{Wright} {et~al}\mbox{.}(2016){Wright}, {Bouy}, {Drew}, {Sarro},
  {Bertin}, {Cuillandre}, \& {Barrado}}]{wrig16}
{Wright} N.~J., {Bouy} H., {Drew} J.~E., {Sarro} L.~M., {Bertin} E.,
  {Cuillandre} J.-C., {Barrado} D., 2016, \mnras, 460, 2593

\bibitem[{{Wright} \& {Drake}(2009)}]{wrig09a}
{Wright} N.~J., {Drake} J.~J., 2009, \apjs, 184, 84

\bibitem[{{Wright} {et~al}\mbox{.}(2010{\natexlab{b}}){Wright}, {Drake}, \&
  {Civano}}]{wrig10b}
{Wright} N.~J., {Drake} J.~J., {Civano} F., 2010{\natexlab{b}}, \apj, 725, 480

\bibitem[{{Wright} {et~al}\mbox{.}(2010{\natexlab{c}}){Wright}, {Drake},
  {Drew}, \& {Vink}}]{wrig10a}
{Wright} N.~J., {Drake} J.~J., {Drew} J.~E., {Vink} J.~S., 2010{\natexlab{c}},
  \apj, 713, 871

\bibitem[{{Wright} {et~al}\mbox{.}(2014{\natexlab{a}}){Wright}, {Drake},
  {Guarcello}, {Aldcroft}, {Kashyap}, {Damiani}, {DePasquale}, \&
  {Fruscione}}]{wrig14c}
{Wright} N.~J., {Drake} J.~J., {Guarcello} M.~G., {Aldcroft} T.~L., {Kashyap}
  V.~L., {Damiani} F., {DePasquale} J., {Fruscione} A., 2014{\natexlab{a}},
  ArXiv e-prints 1408.6579

\bibitem[{{Wright} {et~al}\mbox{.}(2011){Wright}, {Drake}, {Mamajek}, \&
  {Henry}}]{wrig11b}
{Wright} N.~J., {Drake} J.~J., {Mamajek} E.~E., {Henry} G.~W., 2011, \apj, 743,
  48

\bibitem[{{Wright} {et~al}\mbox{.}(2015){Wright}, {Drew}, \&
  {Mohr-Smith}}]{wrig15a}
{Wright} N.~J., {Drew} J.~E., {Mohr-Smith} M., 2015, \mnras, 449, 741

\bibitem[{{Wright} {et~al}\mbox{.}(2019){Wright}, {Jeffries}, {Jackson},
  {Bayo}, {Bonito}, {Damiani}, {Kalari}, {Lanzafame}, {Pancino}, {Parker},
  {Prisinzano}, {Randich}, {Vink}, {Alfaro}, {Bergemann}, {Franciosini},
  {Gilmore}, {Gonneau}, {Hourihane}, {Jofr{\'e}}, {Koposov}, {Lewis},
  {Magrini}, {Micela}, {Morbidelli}, {Sacco}, {Worley}, \& {Zaggia}}]{wrig19}
{Wright} N.~J. {et~al.}, 2019, \mnras, 486, 2477

\bibitem[{{Wright} \& {Mamajek}(2018)}]{wrig18}
{Wright} N.~J., {Mamajek} E.~E., 2018, \mnras, 476, 381

\bibitem[{{Wright} \& {Parker}(2019)}]{wrig19b}
{Wright} N.~J., {Parker} R.~J., 2019, \mnras, 2232

\bibitem[{{Wright} {et~al}\mbox{.}(2014{\natexlab{b}}){Wright}, {Parker},
  {Goodwin}, \& {Drake}}]{wrig14b}
{Wright} N.~J., {Parker} R.~J., {Goodwin} S.~P., {Drake} J.~J.,
  2014{\natexlab{b}}, \mnras, 438, 639

\bibitem[{{Yalyalieva} {et~al}\mbox{.}(2020){Yalyalieva}, {Carraro}, {Vazquez},
  {Rizzo}, {Glushkova}, \& {Costa}}]{yaly20}
{Yalyalieva} L., {Carraro} G., {Vazquez} R., {Rizzo} L., {Glushkova} E.,
  {Costa} E., 2020, \mnras, 495, 1349

\bibitem[{{Zamora-Avil{\'e}s} {et~al}\mbox{.}(2019){Zamora-Avil{\'e}s},
  {Ballesteros-Paredes}, {Hern{\'a}ndez}, {Rom{\'a}n-Z{\'u}{\~n}iga}, {Lora},
  \& {Kounkel}}]{zamo19}
{Zamora-Avil{\'e}s} M., {Ballesteros-Paredes} J., {Hern{\'a}ndez} J.,
  {Rom{\'a}n-Z{\'u}{\~n}iga} C., {Lora} V., {Kounkel} M., 2019, \mnras, 488,
  3406

\bibitem[{{Zari} {et~al}\mbox{.}(2017){Zari}, {Brown}, {de Bruijne}, {Manara},
  \& {de Zeeuw}}]{zari17}
{Zari} E., {Brown} A.~G.~A., {de Bruijne} J., {Manara} C.~F., {de Zeeuw} P.~T.,
  2017, \aap, 608, A148

\bibitem[{{Zari} {et~al}\mbox{.}(2019){Zari}, {Brown}, \& {de Zeeuw}}]{zari19}
{Zari} E., {Brown} A.~G.~A., {de Zeeuw} P.~T., 2019, arXiv e-prints,
  arXiv:1906.07002

\bibitem[{{Zari} {et~al}\mbox{.}(2018){Zari}, {Hashemi}, {Brown}, {Jardine}, \&
  {de Zeeuw}}]{zari18}
{Zari} E., {Hashemi} H., {Brown} A.~G.~A., {Jardine} K., {de Zeeuw} P.~T.,
  2018, \aap, 620, A172

\bibitem[{{Zinnecker} \& {Yorke}(2007)}]{zinn07}
{Zinnecker} H., {Yorke} H.~W., 2007, \araa, 45, 481

\bibitem[{{Zucker} {et~al}\mbox{.}(2020){Zucker}, {Speagle}, {Schlafly},
  {Green}, {Finkbeiner}, {Goodman}, \& {Alves}}]{zuck20}
{Zucker} C., {Speagle} J.~S., {Schlafly} E.~F., {Green} G.~M., {Finkbeiner}
  D.~P., {Goodman} A., {Alves} J., 2020, \aap, 633, A51

\end{thebibliography}

\bsp

\end{document}